\documentclass[1p]{elsarticle}

\usepackage{varioref}
\usepackage{wrapfig}
\usepackage{nomencl}
\usepackage{subfigure}
\usepackage[colorlinks]{hyperref}
\usepackage{amsmath}
\usepackage{amsthm}
\usepackage{amssymb}
\usepackage{booktabs}
\usepackage{cleveref}
\usepackage{bm}
\usepackage[ruled,vlined]{algorithm2e}
\usepackage[dvipsnames]{xcolor}
\usepackage{tikz}
\usetikzlibrary{positioning,arrows,arrows.meta}
\usepackage{stackengine}

\setstackgap{L}{12pt}

\definecolor{PaleGreen}{HTML}{87CBAC}
\definecolor{DeepPurple}{HTML}{3B1C32}
\definecolor{PaleYellow}{HTML}{FFFBBD}
\definecolor{SalmonPink}{HTML}{E16F7C}
\definecolor{DenimBlue}{HTML}{336699}
\definecolor{MacFolderBlue}{RGB}{94,174,213}

\tikzstyle{node}=[circle,draw=blue!50,fill=blue!20,thick,
                   inner sep=0pt,minimum size=1.0mm]
\tikzstyle{face}=[rectangle,draw=black!50,fill=black!20,thick,
                        inner sep=0pt,minimum size=1.0mm]

\DeclareMathOperator{\sign}{\mathrm{sign}}

\newcommand{\half}{\frac{1}{2}}
\newcommand{\dt}{\Delta t}
\newcommand{\dx}{\Delta x}
\newcommand{\dy}{\Delta y}

\newcommand{\grad}{\nabla}
\newcommand{\divergence}{\nabla \cdot}

\newcommand{\Fv}{\mathbf{F}}
\newcommand{\Gv}{\mathbf{G}}
\newcommand{\Wv}{\mathbf{W}}
\newcommand{\Vv}{\mathbf{V}}
\newcommand{\Uv}{\mathbf{U}}
\newcommand{\Rv}{\mathbf{R}}
\newcommand{\Av}{\mathbf{A}}
\newcommand{\Bv}{\mathbf{B}}

\newcommand{\Fxv}{\mathbf{F}^x}
\newcommand{\Fyv}{\mathbf{F}^y}

\newcommand{\Sigmav}{\mathbf{\Sigma}}

\newcommand{\Gxv}{\mathbf{G}^x}
\newcommand{\Gyv}{\mathbf{G}^y}
\newcommand{\Ghv}{\hat{\mathbf{G}}}
\newcommand{\Gxhv}{\hat{\mathbf{G}}^x}
\newcommand{\Gyhv}{\hat{\mathbf{G}}^y}
\newcommand{\Gtv}{\tilde{\mathbf{G}}}
\newcommand{\Gxtv}{\tilde{\mathbf{G}}^x}

\newcommand{\Sv}{\mathbf{S}}
\newcommand{\Shv}{\hat{\mathbf{S}}}

\newcommand{\vel}{\bm{u}}

\newtheorem{mylemma}{Lemma}
\newtheorem{mythm}{Theorem}

\title{A positivity-preserving high-order weighted compact nonlinear scheme for compressible gas-liquid flows}

 \author[stc]{Man Long Wong\corref{cor1}\fnref{fn1}}
 \ead{manlong.wong@nasa.gov}
 \author[nasa]{Jordan B. Angel\fnref{fn1}}
 \ead{jordan.b.angel@nasa.gov}
 \author[nasa]{Michael F. Barad}
 \ead{michael.f.barad@nasa.gov}
 \author[nasa]{Cetin C. Kiris}
 \ead{cetin.c.kiris@nasa.gov}

 \fntext[fn1]{These authors contributed equally to the development of positivity- and boundedness-preserving limiters.}
 \cortext[cor1]{Corresponding author.}

\address[stc]{Science and Technology Corporation, Moffett Field, CA 94035, United States}
\address[nasa]{NASA Ames Research Center, Moffett Field, CA 94035, United States}

\begin{document}

\begin{abstract}
We present a robust, highly accurate, and efficient positivity- and
boundedness-preserving diffuse interface method for the simulations of
compressible gas-liquid two-phase flows with the five-equation model by \citet{allaire2002five} using high-order finite difference weighted compact
nonlinear scheme (WCNS) in the explicit form. The equation of states of gas and
liquid are given by the ideal gas and stiffened gas laws respectively. Under a
mild assumption on the relative magnitude between the ratios of specific heats 
of the gas and liquid, we can
construct limiting procedures for the fifth order incremental-stencil WCNS
(WCNS-IS)  with the first order Harten--Lax--van Leer contact (HLLC) flux such that positive partial densities and squared speed of sound can
be ensured in the solutions, together with bounded volume fractions and mass
fractions. The limiting procedures are discretely conservative for all
conservative equations in the five-equation model and can also be easily
extended for any other conservative finite difference or finite volume scheme.
Numerical tests with liquid water and air are reported to demonstrate the
robustness and high accuracy of the WCNS-IS with the positivity- and
boundedness-preserving limiters even under extreme conditions.

\end{abstract}


\journal{Journal of Computational Physics}

\maketitle

\section{Introduction}

In the numerical computations of compressible flows, simulations cannot proceed when 
negative density or squared speed of sound appears because the system of equations becomes 
ill-posed. This problem is more pronounced for extreme applications such as those in astrophysics where 
strong shocks, rarefactions, or blast waves may exist in the simulations. While one can have
successful simulations with some robust (or even positivity-preserving) first or second 
order shock-capturing schemes~\cite{einfeldt1991godunov,linde1997robust,batten1997choice,liou1996sequel, gressier1999positivity},
these schemes are numerically very dissipative and are inefficient for scale-resolving 
simulations, such as large eddy simulations (LESs) or direct numerical simulations (DNSs). 
Over the decades, many high-order accurate shock-capturing schemes with more localized 
numerical dissipation and higher resolution were developed for scale-resolving simulations
~\cite{shu1988efficient,jiang1996efficient,deng2000developing,borges2008improved,kawai2010assessment,hu2010adaptive,johnsen2010assessment,fu2016family,wong2017high,subramaniam2019high}. 
While these high-order shock-capturing schemes have a certain degree of robustness for 
problems involving shocks and other kinds of discontinuities, there is still no guarantee of
having successful simulations for severe problems using those schemes. Replacing the 
negative density or squared sound speed with positive ones is not conservative and may 
trigger other numerical issues such as spurious oscillations.

In recent years, many positivity-preserving
limiters~\cite{zhang2010positivity,zhang2011positivity,zhang2012positivity,hu2013positivity}
have been developed for high-order shock-capturing schemes. These limiters can
preserve positivity of density and squared speed of sound for compressible
flows. Motivated by the positivity-preserving technique
in~\citet{perthame1996positivity}, \citet{zhang2010positivity} developed
positivity-preserving high-order discontinuous Galerkin (DG) schemes for Euler
equations. The method was later extended to Euler equations with source
terms~\cite{zhang2011positivity}. Positivity-preserving limiters specifically
designed for high-order finite difference schemes didn't appear until the works
by~\citet{zhang2012positivity} and \citet{hu2013positivity} where extreme
problems could be successfully simulated with the finite difference weighted
essentially non-oscillatory (WENO) schemes using positivity-preserving
limiters. These methods are conservative as the limiters are applied to the
fluxes directly in the conservation form.
All of the positivity-preserving limiters aforementioned were designed for 
compressible single-phase flows. The appearance of non-physical states also 
happens in compressible multi-phase simulations. In general it is more likely for 
non-physical states to appear due to higher density gradients across material 
interfaces in the related applications such as supersonic combustion, cavitation erosion, break-up of high-speed liquid jets, water-based acoustic suppression systems, etc. In addition to 
having negative density and squared speed of sound, the solutions are also 
considered non-physical if mass fractions or volume fractions are not bounded 
between zero and one. In order to address the numerical issues, some 
boundedness-preserving diffuse interface 
methods~\cite{shen2017maximum,jain2020conservative} in the Eulerian framework have
been proposed. \Citet{shen2017maximum} adopted the maximum-principle-satisfying 
limiter by~\citet{zhang2010maximum} for a space-time conservation element and 
solution element (CE/SE) scheme. The scheme can ensure the boundedness of volume 
fractions in the five-equation model by~\citet{allaire2002five} for multi-phase 
flows. Another thermodynamics-consistent boundedness-preserving scheme 
by~\citet{jain2020conservative} for the same flow model was developed with the use
of interface-regularization terms. Although both methods can preserve boundedness 
of volume fractions, non-physical states can still appear in the multi-phase 
simulations since partial densities or squared sound speed can still become 
negative. A positivity-preserving high-order method by~\citet{cheng2014positivity}
was proposed for multi-phase simulations in the Lagrangian framework. Compared to 
Eulerian diffuse interface methods, Lagrangian methods can be more accurate at 
material interfaces since the computational mesh moves with the fluids. On the 
other hand, diffuse interface methods in the Eulerian framework is more attractive
for flows involving large deformations as the degree of deformations is limited by
mesh distortions in Lagrangian methods~\cite{saurel2018diffuse}.
Motivated by the need for the simulations of
water sound suppression systems in rocket launch environments that involve interactions between strong
shocks and complex air-water interfaces, we 
propose a high-order positivity-preserving diffuse interface method in the 
Eulerian framework targeting gas-liquid two-phase flows with large deformations, 
where the gas and liquid are described by the ideal gas and stiffened gas equation
of states respectively. 
Unlike the previous works~\cite{housman2009time1,housman2009time2} that are based on the homogeneous relaxation model, the flows in this work are described by the five-equation model by~\citet{allaire2002five}.
Under a necessary but generally valid assumption that the 
ratio of specific heat of the gas is smaller than that of the liquid, the numerical method 
can ensure physically admissible states with positive partial density of each 
phase, positive squared sound speed, and bounded volume fractions and mass 
fractions.

The high-order shock-capturing scheme used in this work is based on the
explicit finite difference formulation of weighted compact nonlinear schemes
(WCNSs)~\cite{deng2000developing,nonomura2007increasing,zhang2008development,nonomura2009effects,deng2011new,nonomura2013robust,wong2017high}
and the nonlinear weighting technique of the incremental-stencil WENO (WENO-IS)
scheme~\cite{wang2018incremental}. The use of WCNS as a diffuse interface method for the five-equation
model~\cite{allaire2002five} has already been demonstrated
by~\citet{wong2017high} but it is only applied for single-phase flows with mixture of ideal gases. The WENO-IS scheme was originally designed as a finite volume scheme by~\citet{wang2018incremental} for compressible multi-phase flows with shocks and material interfaces using the same five-equation model. Although the robustness of the scheme was demonstrated
in that paper, the finite volume approach is computationally more expensive than
the finite difference WCNS for multi-dimensional
problems, while the orders of accuracy are
similar~\cite{sebastian2003multidomain}. The WCNS with the WENO-IS nonlinear weighting technique, WCNS-IS, presented in this work is more efficient and has similar robustness in minimizing spurious oscillations in simulations.

In this work, we first show the convexity of the physically admissible set of 
solution states, under the mild assumption on the relative magnitude between the 
ratios of specific heats of the ideal gas and the liquid. We then prove the 
positivity-preserving and boundedness-preserving properties of the first order 
Harten–Lax–van Leer contact (HLLC) flux for gas-liquid flows with our choice of 
the advection velocity of the material interface using the convexity of the 
admissible set. Based on the positivity- and boundedness-preserving properties of 
the first order HLLC flux, we propose a limiter to blend a flux from any Cartesian 
conservative high-order shock-capturing schemes with the HLLC flux. The flux 
limiter together with a limiter for WENO interpolation can ensure the 
positivity-preserving and boundedness-preserving properties of the overall scheme.
The fifth order accurate WCNS-IS formulation presented in this work is also proved 
mathematically and shown numerically to be high-order accurate in smooth advection
problems, while robust because of the use of lower order interpolation 
near discontinuities such as shocks or material interfaces. We have demonstrated 
that the WCNS-IS scheme with the positivity- and boundedness-preserving limiters, 
PP-WCNS-IS, can successfully simulate very intense one-dimensional (1D) and 
two-dimensional (2D) air-water problems such as Mach 10 shock-water column 
interaction and Mach 100 water jet problems. The results also show that smaller errors are produced at shocks and material interfaces, and fine-scale flow 
features such as vortices are better captured with the high-order scheme 
compared to the first order HLLC scheme due to more localized numerical 
dissipation and higher resolution of the former method. All of the numerical tests
highlight the robustness of the overall positivity- and 
boundedness-preserving finite difference scheme, and demonstrate the method as a highly accurate diffuse interface method for 
scale-resolving compressible gas-liquid simulations.

\section{Governing equations}

The five-equation single-velocity, single-pressure model proposed by~\citet{allaire2002five} for compressible two-phase flows is considered in this work. The flow model has the following form:
\begin{align}
    \partial_t \left( \alpha_1 \rho_1 \right) + \divergence \left( \alpha_1 \rho_1 \vel \right) &= 0 , \\
    \partial_t \left( \alpha_2 \rho_2 \right) + \divergence \left( \alpha_2 \rho_2 \vel \right) &= 0 , \\
    \partial_t \left( \rho \vel \right) + \divergence \left( \rho \vel \otimes \vel \right) + \grad p &= 0 , \\
    \partial_t E + \divergence \left[ \left( E + p \right) \vel \right] &= 0 , \\
   \partial_t {\alpha_1} + \vel \cdot \grad \alpha_1 &= 0 ,
\end{align}
where $\vel$ is the velocity vector and $p$ is the mixture pressure. $\vel = u$
and $\vel = ( u \ v )^{T}$ for 1D and 2D cases respectively (similar extension for the three-dimensional case). $\alpha_k$ and
$\rho_k$ are respectively the volume fraction and phasic density of phase $k$,
where $k=1,2$. $\alpha_k \rho_k = \rho Y_k$ is called partial density of phase
$k$, where $\rho =  \alpha_1 \rho_1 + \alpha_2 \rho_2$ is the mixture density
and $Y_k$ is mass fraction of phase $k$. $E = \rho ( e + \lvert \vel \rvert^2 / 2 )$ 
is the mixture total energy per unit volume, where $e$ is the mixture
specific internal energy. Also, $\alpha_1 + \alpha_2 = 1$ and 
$\rho e = \alpha_1 \rho_1 e_1 + \alpha_2 \rho_2 e_2$, where $e_k$ is the phasic
specific internal energy of phase $k$.
The system is closed by the mechanical equilibrium and equation of
state of each phase\footnote{In this work, each phase consists of one species.}. The stiffened gas equation of state is chosen in this
work because of its popularity for gases and liquids. The equation of state was
first proposed by~\citet{harlow1971fluid}. The stiffened gas equation of state
of each phase is given by:
\begin{equation}
  \frac{p_k}{\gamma_k - 1} + \frac{\gamma_k p_k^\infty}{\gamma_k - 1} = \rho_k e_k, \label{eq:phase_EOS}
\end{equation}
where $\gamma_k$ and $p_k^\infty$ are fitting parameters for each of the fluids. $\gamma_k$ is the ratio of specific heats that is greater than one and $p_k^\infty$ is non-negative. The stiffened gas equation of state is reduced to the ideal gas equation of state if $p_k^\infty = 0$. With the mechanical equilibrium assumption, i.e.\ $p_1 = p_2 = p$, we can obtain the mixture equation of state by multiplying equation~\eqref{eq:phase_EOS} by $\alpha_k$ for each phase and then summing over all the phases:
\begin{equation}
  \frac{p}{\overline{\gamma} - 1} + \frac{\overline{\gamma} ~ \overline{p^\infty}}{\overline{\gamma} - 1} = \rho e,
\end{equation}
where $\overline{\gamma}$ and $\overline{p^\infty}$ are properties of the mixture. They can be defined by the following relations:
\begin{align}
  \frac{1}{\overline{\gamma} - 1} &= \frac{\alpha_1}{\gamma_1 - 1} +  \frac{\alpha_2}{\gamma_2 - 1} , \\
  \frac{\overline{\gamma} ~ \overline{p^\infty}}{\overline{\gamma} - 1} &= \frac{\alpha_1\gamma_1 p_1^\infty}{\gamma_1 - 1} +  \frac{\alpha_2\gamma_2 p_2^\infty}{\gamma_2 - 1} .
\end{align}
 
We define the conservative variable vector as $\Wv = \left( \rho_1 \alpha_1 \ \rho_2 \alpha_2 \ \rho \vel \ E \ \alpha_1 \right)^T$\footnote{Strictly speaking, $\alpha_1$ is not a conservative variable.}. In this work, we also define the set of \emph{admissible states} as,
\begin{equation}
  G = \left\{\left.\Wv = 
  \begin{pmatrix}
    \alpha_1 \rho_1 \\
    \alpha_2 \rho_2 \\
    \rho \vel \\
    E      \\
    \alpha_1
  \end{pmatrix}
 \ \right| \  0 \leq \alpha_1 \leq 1, \ \alpha_1 \rho_1 \geq 0, \ \alpha_2 \rho_2 \geq 0, \ \rho c^2 > 0 \right\}.
\end{equation}
This requires boundedness of the volume fractions, positivity of the partial densities, and positivity of the squared speed of sound $c^2$. The positive squared speed of sound implies that the system of equations remains hyperbolic with real wave speeds.
Note that the positive partial densities also mean that all mass fractions $Y_k$ are bounded between zero and one.
The flow model has a mixture speed of sound $c$ given by~\cite{allaire2002five}:
\begin{equation}
\begin{split} 
  \rho c^2 &= \overline{\gamma} \left(p + \overline{ p^\infty} \right) \\
           &= \overline{\gamma} (\overline{\gamma} - 1)\left(\rho e - \overline{p^\infty} \right)
           = \overline{\gamma} (\overline{\gamma} - 1) \left( E - \frac{1}{2} \frac{\lvert \rho \vel \rvert ^2}{\rho} - \overline{p^\infty} \right) .
\end{split}
\end{equation}
Both $\overline{\gamma}$ and $\overline{p^\infty}$ only depend on $\alpha_1$ in $\Wv$.

It is obvious that $\alpha_k$ and $\alpha_k \rho_k$ are both concave functions of the conserved variables $\Wv$. Since all $\gamma_k$ are greater than one, $\overline{\gamma} > 1$ if $0 \leq \alpha_1 \leq 1$. Therefore, an equivalent physically admissible set is,
\begin{equation}
  G = \left\{\left.\Wv = 
  \begin{pmatrix}
    \alpha_1 \rho_1 \\
    \alpha_2 \rho_2 \\
    \rho \vel \\
    E      \\
    \alpha_1
  \end{pmatrix}
 \ \right| \ 0 \leq \alpha_1 \leq 1, \ \alpha_1 \rho_1 \geq 0, \ \alpha_2 \rho_2 \geq 0, \ \rho e - \overline p^\infty > 0 \right\}.
\end{equation}

\begin{mylemma}
  If $p_2^\infty=0$ ($p_1^\infty=0$), $\gamma_1 \geq \gamma_2$ ($\gamma_2 \geq \gamma_1$) and $0 \leq \alpha_1 \leq 1$, the function $\rho e - \overline p^\infty$ is a concave function of the conserved variables $\Wv$.
  \label{lem:concave}
\end{mylemma}
\begin{proof}
  The non-zero eigenvalues of the Hessian matrix of the function are:
  \begin{equation}
    \left\{-\frac{1}{\alpha_1\rho_1+\alpha_2\rho_2}, \ -\frac{2\lvert \rho \vel \rvert ^2 + (\alpha_1\rho_1+\alpha_2\rho_2)^2}
          {(\alpha_1\rho_1 + \alpha_2 \rho_2)^3}, \ -\frac{\partial^2 \overline p^\infty}{\partial \alpha_1^2} \right\} .
  \end{equation}
  The first non-zero eigenvalue does not exist for the 1D case. If the last non-zero eigenvalue is non-positive, then the lemma is proved since all eigenvalues are non-positive. Here, we consider the $p_2^\infty=0$ case and the other case is symmetric. For $p_2^\infty=0$ we have,
  \begin{equation}
    \frac{\partial^2 \overline p^\infty}{\partial \alpha_1^2} = 
    2\frac{\gamma_1 p_1^\infty}{\gamma_1-1}
    \frac{\frac{1}{\gamma_1-1}-\frac{1}{\gamma_2-1}}{\left(\frac{\alpha_1}{\gamma_1-1}+\frac{\alpha_2}{\gamma_2-1}+1\right)^2}
    \left[ \frac{\alpha_1\left(\frac{1}{\gamma_1-1}-\frac{1}{\gamma_2-1}\right)}{\left(\frac{\alpha_1}{\gamma_1-1}+\frac{\alpha_2}{\gamma_2-1}+1\right)}-1 \right],
  \end{equation}
  and it is clear that if $\gamma_1 \geq \gamma_2$, the last eigenvalue is less than or equal to zero. Note that if $\gamma_2 > \gamma_1$, the function cannot be concave. In this work, we assume that one of the phases is an ideal gas with $p_k^\infty=0$.
\end{proof}
\begin{mythm}
  The set $G$ is a convex set.
\end{mythm}
\begin{proof}
  This is an immediate consequence of Lemma~\ref{lem:concave} and Jensen's inequality.
\end{proof}

The proof of convexity of $G$ relies on the assumption that the ratio of specific heats of the liquid ($p_k^\infty > 0$) is larger than that of the ideal gas ($p_k^\infty = 0$). However, this is a very mild assumption based on most gas-liquid test problems found in previous literature. The ideal gases considered in the previous works are usually either monatomic gases or air, where the ratios of specific heats are around 1.67 and 1.4 respectively. When the stiffened gas equation of state is chosen for the liquid in tests, the most popular liquid used is water, where the ratio of specific heats mostly ranges between 4--7~\cite{saurel1999simple,shyue1999fluid,coralic2014finite,wang2018incremental,perigaud2005compressible}. A rare but justified choice of $\gamma = 1.932$ is utilized in~\cite{chang2007robust} but that ratio of specific heats is still larger than those of monatomic gases or air. Other liquids commonly found in the literature include ethanol~\cite{perigaud2005compressible} with $\gamma = 2.1$ and mercury~\cite{takahira2008numerical} with $\gamma = 8.2$. Both liquids also have larger specific heat ratios than the ideal gases considered.

In the following sections, we make the assumption that ratio of specific heats of liquid is larger than that of ideal gas and formulate our numerical method by taking advantage of the convexity of $G$ such that the solution update is a convex combination of states already in $G$, thus is also in $G$.

\section{First order positivity- and boundedness-preserving scheme with HLLC Riemann solver}

The flux given by the HLLC Riemann solver is illustrated in this section. For simplicity, a 2D case with domain $[x_a, x_b]\times[y_a, y_b]$ is considered with the following equation in compact form:
\begin{equation}
    \partial_t \Wv + \partial_x \Fxv \left( \Wv \right) + \partial_y \Fyv \left( \Wv \right) + \Sigmav \left( \Wv, \grad \Wv \right) = \mathbf{0}, \label{eq:W_F_eqn}
\end{equation}
where
\begin{equation}
\begin{split}
    \Wv &= 
        \begin{pmatrix}
            \alpha_1 \rho_1 \\
            \alpha_2 \rho_2 \\
            \rho u          \\
            \rho v          \\
            E               \\
            \alpha_1
        \end{pmatrix}, \quad
    \Fxv \left( \Wv \right) =
        \begin{pmatrix}
            \alpha_1 \rho_1 u       \\
            \alpha_2 \rho_2 u       \\
            \rho u^2 + p              \\
            \rho v u                  \\
            \left( E + p \right) u  \\
            0
        \end{pmatrix}, \quad
    \Fyv \left( \Wv \right) =
        \begin{pmatrix}
            \alpha_1 \rho_1 v       \\
            \alpha_2 \rho_2 v       \\
            \rho u v                \\
            \rho v^2 + p            \\
            \left( E + p \right) v  \\
            0
        \end{pmatrix}, \\
    \Sigmav \left( \Wv, \grad \Wv \right) &=
        \begin{pmatrix}
            0 \\
            0 \\
            0 \\
            0 \\
            0 \\
            \vel \cdot \grad \alpha_1
        \end{pmatrix} .
\end{split}
\end{equation}
If the domain is discretized uniformly into a Cartesian grid with $N_x \times N_y$ grid points, we have the domain covered by cells $I_{i,j}=\left[x_{i-1/2},\ x_{i+1/2}\right]\times\left[y_{j-1/2},\ y_{j+1/2}\right]$ for $1 \leq i \leq N_x$, $1 \leq j \leq N_y$, where the grid midpoints are given by:
\begin{equation}
  x_{i+\half} = x_a + i \Delta x, \quad
  y_{j+\half} = y_a + j \Delta y
\end{equation}
\noindent and
\begin{equation}
    \Delta x = \frac{x_b - x_a}{N_x}, \quad \Delta y = \frac{y_b - y_a}{N_y} .
\end{equation}

To obtain the numerical scheme with an exact or approximate Riemann solver, the numerical discretizations in different directions are treated independently. The flux in $y$ direction has similar formulation. Therefore, only the numerical discretization in $x$ direction is discussed in details in this section.

\subsection{Approximate solutions}

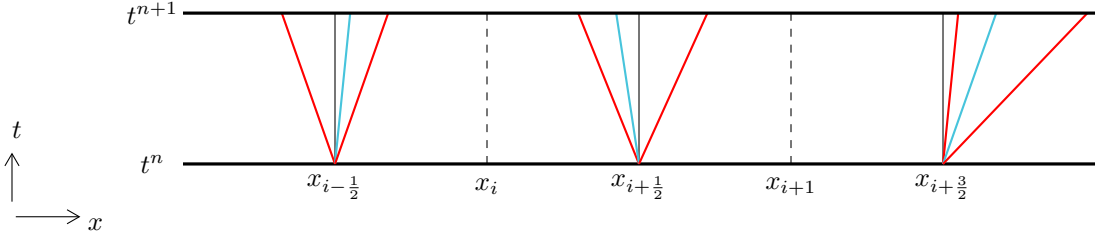
\begin{figure}[hbt]
  \centering
  \begin{tikzpicture}
    \useasboundingbox (0cm,-1cm) rectangle (12cm,3cm);
    \draw[very thick] (0cm,0cm) -- (12cm,0cm);
    \begin{scope}[xshift=-2cm]
      \foreach \x in {1,...,3}
        \draw[] (4*\x,0cm) -- (4*\x,2cm);
      \foreach \x in {1.5,...,2.5}
        \draw[dashed] (4*\x,0cm) -- (4*\x,2cm);
      \draw[red,thick] (4cm,0cm) -- (3.3cm,2.0cm);
      \draw[color=SkyBlue,thick] (4cm,0cm) -- (4.2cm,2.0cm);
      \draw[red,thick] (4cm,0cm) -- (4.7cm,2.0cm);
      \draw[] (4cm,-0.3cm) node{$x_{i-\half}$};
      \draw[red,thick] (8cm,0cm) -- (7.2cm,2.0cm);
      \draw[color=SkyBlue,thick] (8cm,0cm) -- (7.7cm,2.0cm);
      \draw[red,thick] (8cm,0cm) -- (8.9cm,2.0cm);
      \draw[] (8cm,-0.3cm) node{$x_{i+\half}$};
      \draw[red,thick] (12cm,0cm) -- (12.2cm,2.0cm);
      \draw[color=SkyBlue,thick] (12cm,0cm) -- (12.7cm,2.0cm);
      \draw[red,thick] (12cm,0cm) -- (13.9cm,2.0cm);
      \draw[] (12cm,-0.3cm) node{$x_{i+\frac{3}{2}}$};
      \draw[] (6cm,-0.3cm) node{$x_i$};
      \draw[] (10cm,-0.3cm) node{$x_{i+1}$};
      %
    \end{scope}
    \draw[very thick] (0cm,2cm) -- (12cm,2cm);
    \draw[] (-0.4cm,0cm)  node[]{$t^n$};
    \draw[] (-0.4cm,2cm)  node[]{$t^{n+1}$};

    \draw[-{Straight Barb[angle'=60,scale=2]}] (-2.2cm,-0.7cm) -- (-1.35cm,-0.7cm);
    \node[text width=3cm] at (0.25cm,-0.8cm) {$x$};
    \draw[-{Straight Barb[angle'=60,scale=2]}] (-2.25cm,-0.5cm) -- (-2.25cm,0.15cm);
    \node[text width=1cm] at (-1.75cm,0.45cm) {$t$};

  \end{tikzpicture}
  \caption{Illustration of a scheme with an approximate Riemann solver. Waves are generated from discontinuities initially located at grid midpoints.}
  \label{fig:Riemann_problems}
\end{figure}

The discretization in $x$ direction is conducted by considering a generalized Riemann problem with a planar discontinuity initially at each grid midpoint in the $x$ direction, as shown in figure~\ref{fig:Riemann_problems}. The generalized Riemann problem is reduced to a quasi-1D problem due to assumed homogeneity in other directions. The reduced quasi-1D generalized Riemann problem at midpoint $x = x_{i+1/2}$ between grid cells located at $(x_i, y_j)$ and $(x_{i+1}, y_j)$ is hence formulated as:
\begin{equation}
    \left\{ \begin{array}{ll} 
        \partial_t \Wv + \partial_x \Fv \left( \Wv \right) + \Sigmav \left( \Wv, \partial_x \Wv \right) = 0, \\ 
        \Wv \left(x, \ t=0 \right) = 
            \begin{cases}
                \Wv_{L}, &\mbox{$x_{i+\half} < 0$}, \\
                \Wv_{R}, &\mbox{$x_{i+\half} \geq 0$}. \\
            \end{cases}
    \end{array}\right . \label{eq:Riemann_problem}
\end{equation}
where superscript ``$x$" in $\Fxv$ and index ``$j$" are dropped for convenience. For first order accurate spatial approximation, $\Wv_{L} = \Wv_{i}$ and $\Wv_{R} = \Wv_{i+1}$. Higher order spatial approximation can be obtained from high-order interpolation to construct a high-order scheme which will be discussed in another section. The first order solution of Riemann problem given by equation~\eqref{eq:Riemann_problem} is self-similar: $\Wv \left(x, \ t \right) = R \left( x/t, \ \Wv_{i}, \ \Wv_{i+1} \right)$.

\begin{figure}[hbt]
  \centering
  \begin{tikzpicture}
    \useasboundingbox (0cm,-1cm) rectangle (12cm,3cm);
    \draw[very thick] (3cm,0cm) -- (3cm,3cm);
    \draw[very thick] (9cm,0cm) -- (9cm,3cm);
    \draw[] (3cm,-0.3cm) node[]{$x_{i-\half}$};
    \draw[] (6cm,-0.3cm) node[]{$x_{i}$};
    \draw[] (9cm,-0.3cm) node[]{$x_{i+\half}$};
    \draw[draw=none,fill=MacFolderBlue!60]  (3cm, 0.0cm) rectangle (6cm,3cm);
    \draw[draw=none,fill=PaleYellow!80] (6cm, 0.0cm) rectangle (9cm,3cm);
    \draw[very thick] (0cm,0cm) -- (12cm,0cm);
    \draw[very thick] (0cm,3cm) -- (12cm,3cm);
    \draw[thick,dashed] (6cm,0cm) -- (6cm,3cm);
    \draw[thick,red] (3cm,0cm) -- (4.2cm,3cm);
    \draw[thick,dashed,color=ForestGreen] (3cm,0cm) -- (3.8cm,3cm);
    \draw[thick,red] (3cm,0cm) -- (2.2cm,3cm);
    \draw[thick,red] (9cm,0cm) -- (8.2cm,3cm);
    \draw[thick,dashed,color=ForestGreen] (9cm,0cm) -- (8.8cm,3cm);
    \draw[thick,red] (9cm,0cm) -- (10.2cm,3cm);
    \draw[] (-0.4cm,0cm)  node[]{$t^n$};
    \draw[] (-0.4cm,3cm)  node[]{$t^{n+1}$};

    \draw[-{Straight Barb[angle'=60,scale=2]}] (-2.2cm,-0.7cm) -- (-1.35cm,-0.7cm);
    \node[text width=3cm] at (0.25cm,-0.8cm) {$x$};
    \draw[-{Straight Barb[angle'=60,scale=2]}] (-2.25cm,-0.5cm) -- (-2.25cm,0.15cm);
    \node[text width=1cm] at (-1.75cm,0.45cm) {$t$};

  \end{tikzpicture}
  \caption{Approximate waves generated at the left and right edges of a cell $I_i = \left[ x_{i-1/2}, \ x_{i+1/2} \right]$.  The red solid lines and green dotted lines represent the locations of the approximate acoustic waves and contact waves respectively.} \label{fig:approximated_soln_cell}
\end{figure}
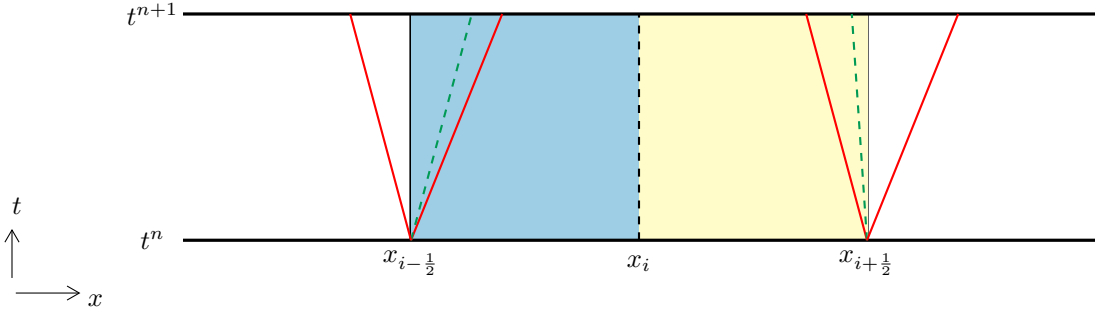

The exact solution to the generalized Riemann problem is computationally expensive and
challenging to obtain. Approximate Riemann solvers can be used to provide approximate solutions in a less expensive way in Godunov-type schemes. Assuming the approximate waves generated at the two midpoints $x_{i \pm 1/2}$ 
do not interact under suitable Courant–Friedrichs–Lewy (CFL) condition as
shown in figure~\ref{fig:approximated_soln_cell}, the approximate numerical
solution at grid cell $i$ is the cell-averaged value:

\begin{equation}
\begin{split}
    \Wv_i^{n+1} &= \frac{1}{\dx} \int_{x_{i-\half}}^{x_{i}} R\left( \frac{x - x_{i-\half}}{\dt}, \Wv_{i-1}^{n}, \Wv_i^{n} \right) dx \\
    &\quad + \frac{1}{\dx} \int_{x_{i}}^{x_{i+\half}}R\left(\frac{x - x_{i+\half}}{\dt}, \Wv_{i}^{n}, \Wv_{i+1}^{n} \right) dx .
\end{split} \label{eq:soln_convex_averaging}
\end{equation}

Specifically since the last equation is an advection equation, the approximate solution of the volume fraction is given by:
\begin{align}
\begin{split}
    {\alpha_1}_i^{n+1} &= \left[ \half {\alpha_1}_i^n - \frac{u_{+,i-\half} \dt}{\dx}\left({\alpha_1}^n_i - {\alpha_1}^n_{i-1}\right) \right] \\
    &\quad + \left[ \half {\alpha_1}_i^n - \frac{u_{-,i+\half} \dt}{\dx}\left({\alpha_1}^n_{i+1} - {\alpha_1}^n_i\right) \right]
\end{split} \\
    {\alpha_1}_i^{n+1} &= {\alpha_1}_i^n - \frac{u_{+,i-\half} \dt}{\dx}\left({\alpha_1}^n_i - {\alpha_1}^n_{i-1}\right) - \frac{u_{-,i+\half} \dt}{\dx}\left({\alpha_1}^n_{i+1} - {\alpha_1}^n_i\right) ,
  \label{eq:1st_order_HLLC_advection}
\end{align}
where $u_{+,i-1/2} = \max\{0,u_{*,i-1/2}\}$ and $u_{-,i+1/2} = \min\{0,u_{*,i+1/2}\}$. $u_{*, i \pm 1/2}$ are the approximate material wave speeds at edges $x_{i \pm 1/2}$ for the advection equation.

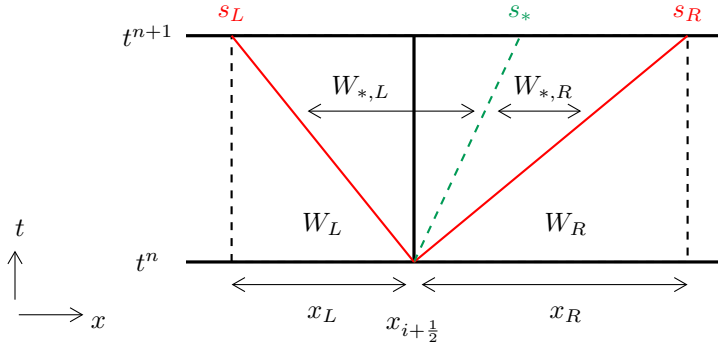
\begin{figure}[hbt]
  \centering
  \begin{tikzpicture}
    \useasboundingbox (1.0cm,-1cm) rectangle (12cm,4cm);
    \draw[thick, dashed]  (3.6cm, 0.0cm) rectangle (9.6cm,3cm);
    \draw[very thick] (6cm,0cm) -- (6cm,3cm);
    \draw[very thick] (3cm,0cm) -- (10cm,0cm);
    \draw[very thick] (3cm,3cm) -- (10cm,3cm);
    \draw[thick,red] (6cm,0cm) -- (9.6cm,3cm);
    \draw[thick,dashed,color=ForestGreen] (6cm,0cm) -- (7.4cm,3cm);
    \draw[thick,red] (6cm,0cm) -- (3.6cm,3cm);

    \draw[{Straight Barb[angle'=60,scale=2]}-{Straight Barb[angle'=60,scale=2]}] (3.6cm,-0.3cm) -- (5.9cm,-0.3cm);
    \draw[{Straight Barb[angle'=60,scale=2]}-{Straight Barb[angle'=60,scale=2]}] (6.1cm,-0.3cm) -- (9.6cm,-0.3cm);
    \draw[] (6.0cm,-0.9cm) node[]{$x_{i+\half}$};
    \draw[] (4.8cm,-0.7cm) node[]{$x_L$};
    \draw[] (8.0cm,-0.7cm) node[]{$x_R$};
    \draw[] (4.8cm,0.5cm) node[]{$W_L$};
    \draw[] (8.0cm,0.5cm) node[]{$W_R$};

    \draw[{Straight Barb[angle'=60,scale=2]}-{Straight Barb[angle'=60,scale=2]}] (4.6cm,2.0cm) -- (6.8cm,2.0cm);
    \draw[{Straight Barb[angle'=60,scale=2]}-{Straight Barb[angle'=60,scale=2]}] (7.1cm,2.0cm) -- (8.2cm,2.0cm);
    \draw[] (5.3cm,2.3cm) node[]{$W_{*,L}$};
    \draw[] (7.7cm,2.3cm) node[]{$W_{*,R}$};
    \draw[red] (3.6cm,3.3cm) node[]{$s_{L}$};
    \draw[red] (9.6cm,3.3cm) node[]{$s_{R}$};
    \draw[color=ForestGreen] (7.4cm,3.3cm) node[]{$s_{*}$};

    \draw[] (2.5cm,0cm)  node[]{$t^n$};
    \draw[] (2.5cm,3cm)  node[]{$t^{n+1}$};

    \draw[-{Straight Barb[angle'=60,scale=2]}] (0.8cm,-0.7cm) -- (1.65cm,-0.7cm);
    \node[text width=3cm] at (3.25cm,-0.8cm) {$x$};
    \draw[-{Straight Barb[angle'=60,scale=2]}] (0.75cm,-0.5cm) -- (0.75cm,0.15cm);
    \node[text width=1cm] at (1.25cm,0.45cm) {$t$};

  \end{tikzpicture}
  \caption{HLLC wave diagram at interface $x_{i+1/2}$. The red solid lines represent the locations of the approximate acoustic waves and the green dotted line represent the location of the contact wave.}
  \label{fig:HLLC_wave_diagram}
\end{figure}

The HLLC discretization in a particular direction for a multi-dimensional problem can be approximated by the solutions of a quasi-1D generalized Riemann problem in that direction. The approximate solutions, $\Wv^{\mathrm{HLLC}}$, of a quasi-1D generalized Riemann problem in $x$ direction with an initial planar discontinuity, is illustrated in figure~\ref{fig:HLLC_wave_diagram}. The approximate solution to the generalized Riemann problem contains three discontinuities: one contact wave and two acoustic waves. The speed of the contact wave is denoted by $s_{*}$ while the smallest and largest acoustic wave speeds are represented by $s_{L}$ and $s_{R}$ respectively. The material wave speed for the advection equation is chosen as the contact wave speed $u_{*} = s_{*}$.

The HLLC approximate solutions in the four different regions separated by the three discontinuities are given by:
\begin{equation}
  \Wv^{\mathrm{HLLC}} 
  = \left\{ \begin{array}{ll} 
      \Wv_L,     &  \text{if }  s_L > 0,\\ 
      \Wv_{*,L}, &  \text{if }  s_L \leq 0 < s_*, \\
      \Wv_{*,R}, &  \text{if }  s_* \leq 0 \leq s_R,\\
      \Wv_R,     &  \text{if }  s_R < 0 , 
             \end{array}\right  .
\end{equation}
where $L$ and $R$ are the left and right states respectively at a midpoint. With $K = L$ or $R$, the star state for the five-equation model for a 2D problem is given by:
\begin{equation}
  \Wv_{*,K}  = 
  \begin{pmatrix}
    \chi_{*,K} \left(\alpha_1 \rho_1\right)_K \\
    \chi_{*,K} \left(\alpha_2 \rho_2\right)_K \\
    \chi_{*,K} \rho_K s_*                     \\
    \chi_{*,K} \rho_K v_K                     \\
    \chi_{*,K}\left[ E_K + (s_*-u_K)\left(\rho_K s_* + \frac{p_K}{s_K-u_K}\right) \right] \\
    \alpha_{1,K}
  \end{pmatrix}
\end{equation}

\noindent $\chi_{*K}$ is defined as:
\begin{equation}
    \chi_{*K} = \frac{s_K - u_K}{s_K - s_*}.
\end{equation}

\noindent We use the wave speeds suggested by~\citet{einfeldt1991godunov}:
\begin{equation}
    s_{L} = \min{\left( \bar{u} - \bar{c}, u_L - c_L \right)}, \quad s_{R} = \max{\left( \bar{u} + \bar{c}, u_R + c_R \right)},
\end{equation}
where $\bar{u}$ and $\bar{c}$ are the arithmetic averages from the left and right states. For instance, $\bar{c}$ is the average of $c_L$ and $c_R$. Following~\citet{batten1997choice}, the wave speed in the star region is given by:
\begin{equation}
    s_{*} = \frac{p_R - p_L + \rho_L u_L \left( s_L - u_L \right) - \rho_R u_R \left( s_R - u_R \right)}{\rho_L \left( s_L - u_L \right) - \rho_R \left( s_R - u_R \right) } . \label{eq:s_star_definition}
\end{equation}

\subsection{Approximate fluxes}

We now introduce a flux-source form that is convenient for  the derivation of the flux-based numerical discretization for the non-conservative system of equations and also the extension for high-order methods. The equation given by \eqref{eq:W_F_eqn} can be rewritten as:
\begin{equation}
    \partial_t \Wv + \partial_x \Gxv \left( \Wv \right) + \partial_y \Gyv \left( \Wv \right) = \Sv \left( \Wv, \grad \Wv \right), \label{eq:W_G_eqn}
\end{equation}
where
\begin{equation}
\begin{split}
    \Gxv \left( \Wv \right) =
        \begin{pmatrix}
            \alpha_1 \rho_1 u       \\
            \alpha_2 \rho_2 u       \\
            \rho u^2 + p              \\
            \rho v u                  \\
            \left( E + p \right) u  \\
            \alpha_1 u
        \end{pmatrix}, \quad
    \Gyv \left( \Wv \right) =
        \begin{pmatrix}
            \alpha_1 \rho_1 v       \\
            \alpha_2 \rho_2 v       \\
            \rho u v                \\
            \rho v^2 + p            \\
            \left( E + p \right) v  \\
            \alpha_1 v
        \end{pmatrix}, \quad
    \Sv \left( \Wv, \grad \Wv \right) =
        \begin{pmatrix}
            0 \\
            0 \\
            0 \\
            0 \\
            0 \\
            \alpha_1 \divergence \vel
        \end{pmatrix} .
\end{split}
\end{equation}
The relation between fluxes $\Fxv$ and $\Gxv$ and that between fluxes $\Fyv$ and $\Gyv$ are given by:
\begin{align}
    \Gxv &= \Fxv + \left( 0\ 0\ 0\ 0\ 0\ f^x_{\alpha} \right)^T, \\
    \Gyv &= \Fyv + \left( 0\ 0\ 0\ 0\ 0\ f^y_{\alpha} \right)^T,
\end{align}
where $f^x_{\alpha} = \alpha_1 u$ and $f^y_{\alpha} = \alpha_1 v$.

The fully discretized form of equation~\eqref{eq:W_G_eqn} with first order accurate forward Euler time integration is given by:
\begin{equation}
    \frac{\Wv^{n+1}_{i,j} - \Wv^{n}_{i,j}}{\dt}
    + \frac{\Gxhv_{i+\half,j}
            - \Gxhv_{i-\half,j}}{\dx} 
      + \frac{\Gyhv_{i,j+\half}
            - \Gyhv_{i,j-\half}}{\dy} =
      \Shv_{i,j} , \label{eq:fully_discretized_W_G_eqn}
\end{equation}
where
\begin{equation}
    \Shv_{i,j} =
      \begin{pmatrix}
              0 \\
              0 \\
              0 \\
              0 \\
              0 \\
              \alpha_{1,i,j}^n \left(
        \frac{\hat{u}_{i+\half,j}
            - \hat{u}_{i-\half,j}}{\dx} 
      + \frac{\hat{v}_{i,j+\half}
            - \hat{v}_{i,j-\half}}{\dy}
            \right)
          \end{pmatrix} . \label{eq:discretized_S}
\end{equation}
Only numerical approximation in $x$ direction is discussed in the following part as discretization in $y$ direction is similar. The first order accurate approximation of $\hat{\mathbf{G}}^{x}_{i\pm\half,j}$ with the HLLC solutions is given by:
\begin{align}
\begin{split}
  \Gxhv_{i-\half,j} &=
  \Gv^{x,\mathrm{HLLC}}_{i-\half,j}
  \left( \Wv_{i-1,j}, \Wv_{i,j} \right) \\
      &=
      \Fv^{x,\mathrm{HLLC}}_{i-\half,j} \left(\Wv_{i-1,j}, \Wv_{i,j}\right)
      + ( 0 \ 0 \ 0 \ 0 \ 0 \ f^{x, \mathrm{HLLC}}_{\alpha,i-\half,j} \left( \Wv_{i-1,j}, \Wv_{i,j} \right) )^T ,
\end{split}
  \\
\begin{split}
  \Gxhv_{i+\half,j} &=
  \Gv^{x,\mathrm{HLLC}}_{i+\half,j}
  \left( \Wv_{i,j}, \Wv_{i+1,j} \right) \\
      &=
      \Fv^{x,\mathrm{HLLC}}_{i+\half,j} \left(\Wv_{i,j}, \Wv_{i+1,j}\right)
      + ( 0 \ 0 \ 0 \ 0 \ 0 \ f^{x, \mathrm{HLLC}}_{\alpha,i+\half,j} \left( \Wv_{i,j}, \Wv_{i+1,j} \right) )^T ,
\end{split}
\end{align}
where
\begin{align}
    \Fv^{x,\mathrm{HLLC}}_{i-\half,j} \left(\Wv_{i-1,j}, \Wv_{i,j}\right)  &= \Fv^{x,\mathrm{HLLC}} \left(R^{\mathrm{HLLC}}\left(0, \Wv_{i-1,j}, \Wv_{i,j}\right)\right) , \\
    \Fv^{x,\mathrm{HLLC}}_{i+\half,j} \left(\Wv_{i,j}, \Wv_{i+1,j}\right) &= \Fv^{x,\mathrm{HLLC}} \left(R^{\mathrm{HLLC}}\left(0, \Wv_{i,j}, \Wv_{i+1,j}\right)\right) .
\end{align}
The conservative HLLC fluxes $\Fv^{x,\mathrm{HLLC}}_{i\pm\half,j}$ can be obtained with the divergence theorem~\cite{batten1997choice}:
\begin{equation}
  \begin{split}
    & \Fv^{x,\mathrm{HLLC}}\left(R^{\mathrm{HLLC}}\left(0, \Wv_L, \Wv_R\right)\right) = \\
    &\quad \frac{1+\sign(s_*)}{2}\left(\Fv_L + s_-\left(\Wv_{*,L}-\Wv_L\right)\right) 
     + \frac{1-\sign(s_*)}{2}\left(\Fv_R + s_+\left(\Wv_{*,R}-\Wv_R\right)\right) , 
  \end{split}
\end{equation}
where
\begin{equation}
    s_{-} = \min{\left( 0, s_L \right)}, \quad s_{+} = \max{\left( 0, s_R \right)} .
\end{equation}
Note that the last component of $\Fv^{x,\mathrm{HLLC}}$ for the advection equation is zero. The discretization of the advection equation is contributed by $\hat{u}_{i\pm\half,j}$ and $f^{x, \mathrm{HLLC}}_{\alpha,i\pm\half,j}$ which are given by the first order accurate approximations as:
\begin{align}
    \hat{u}_{i-\half,j} &= u_{*,i-\half,j} =
    u^{\mathrm{HLLC}}_{*} \left( \Wv_{i-1,j}, \Wv_{i,j} \right) = s_{*} \left( \Wv_{i-1,j}, \Wv_{i,j} \right) , \\
    \hat{u}_{i+\half,j} &= u_{*,i+\half,j} =
    u^{\mathrm{HLLC}}_{*} \left( \Wv_{i,j}, \Wv_{i+1,j} \right) = s_{*} \left( \Wv_{i,j}, \Wv_{i+1,j} \right) ,
\end{align}
and
\begin{align}
\begin{split}
    f^{x,\mathrm{HLLC}}_{\alpha,i-\half,j} \left( \Wv_{i-1,j}, \Wv_{i,j} \right) &=
        \frac{1+\sign(s_{*,i-\half,j})}{2}(\alpha_{1,i-1,j} s_{*,i-\half,j}) \\
        &\quad + \frac{1-\sign(s_{*,i-\half,j})}{2}(\alpha_{1,i,j} s_{*,i-\half,j}) ,
\end{split}
    \\
\begin{split}
    f^{x,\mathrm{HLLC}}_{\alpha,i+\half,j} \left( \Wv_{i,j}, \Wv_{i+1,j} \right) &=
        \frac{1+\sign(s_{*,i+\half,j})}{2}(\alpha_{1,i,j} s_{*,i+\half,j}) \\
        &\quad + \frac{1-\sign(s_{*,i+\half,j})}{2}(\alpha_{1,i+1,j} s_{*,i+\half,j}) .
\end{split}
\end{align}
The expressions given above form the first order accurate solution of volume fraction given by equation~\eqref{eq:1st_order_HLLC_advection}.

Finally, the non-conservative flux $\hat{\mathbf{G}}^{x,\pm}_{i\mp\half,j}$ (similarly for $\hat{\mathbf{G}}^{y,\pm}_{i,j\mp\half}$) is introduced:
\begin{equation}
  \hat{\mathbf{G}}^{x,\pm}_{i\mp\half,j} =
    \Gxhv_{i\mp\half,j}
    - \alpha_{1,i,j} \ (0\ 0\ 0\ 0\ 0\ \hat{u}_{i\mp\half,j} )^T , \label{eq:G_pm}
\end{equation}
where equation~\eqref{eq:fully_discretized_W_G_eqn} can be simplified to:
\begin{equation}
    \frac{\Wv^{n+1}_{i,j} - \Wv^{n}_{i,j}}{\dt}
    + \frac{\hat{\mathbf{G}}^{x,-}_{i+\half,j}
          - \hat{\mathbf{G}}^{x,+}_{i-\half,j}}{\dx} 
    + \frac{\hat{\mathbf{G}}^{y,-}_{i,j+\half}
          - \hat{\mathbf{G}}^{y,+}_{i,j-\half}}{\dy} =
    \mathbf{0} . \label{eq:fully_discretized_W_G_pm_eqn}
\end{equation}
Note that the components of $\hat{\mathbf{G}}^{x,\pm}_{i\mp\half,j}$ (or $\hat{\mathbf{G}}^{x,\mathrm{HLLC},\pm}_{i\mp\half,j}$ more precisely) for all conservative equations, except the last advection equation, are conservative numerical fluxes and are equivalent to the corresponding components of $\Fv^{x,\mathrm{HLLC}}_{i\mp\half,j}$.

\subsection{Proof of positivity- and bounded-preserving preservation of first order HLLC solver}

For a quasi-1D problem, it is shown in equation~\eqref{eq:soln_convex_averaging} that the solution update is the convex averaging of the exact or approximate solutions to the generalized Riemann problem. Therefore, the HLLC Riemann solver gives physically admissible solution if all states generated are physically admissible using Jensen's inequality for integral equations. Here, the left star state is considered and the right star state can be proved to be physically admissible by symmetry. 

With the definition of $s_{*}$ given by equation~\eqref{eq:s_star_definition}, it can be shown that $s_L < s_*$~\cite{batten1997choice}. Also, since $s_L = \min{\left( \bar{u} - \bar{c}, u_L - c_L \right)}$, $s_L < u_L$. As a result, the partial densities in the star state are positive:
\begin{align}
  \left( \alpha_1 \rho_1 \right)_{*,L} &= \frac{s_L-u_L}{s_L-s_*} \left( \alpha_1 \rho_1 \right)_L \geq 0, \\
  \left( \alpha_2 \rho_2 \right)_{*,L} &= \frac{s_L-u_L}{s_L-s_*} \left( \alpha_2 \rho_2 \right)_L \geq 0.
\end{align}
Since $\rho_{*,L} = \left( \alpha_1 \rho_1 \right)_{*,L} + \left( \alpha_2 \rho_2 \right)_{*,L}$,
\begin{equation}
  \rho_{*,L} = \frac{s_L-u_L}{s_L-s_*}\rho_L \geq 0 .
\end{equation}
The positivity of mixture and partial densities implies all mass fractions are bounded between zero and one. As for the volume fraction, since $\alpha_{1,*,L} = \alpha_{1,L}$,
\begin{equation}
    0 \leq \alpha_{1,*,L} \leq 1 .
\end{equation}
The only remaining requirement is $(\rho e)_{*,L} > \overline p^\infty_L$ for positive squared speed of sound. From the definition of $(\rho e)_{*,L}$:
\begin{align}
  \left( \rho e \right)_{*,L} &=
    E_{*,L} - \half \frac{\lvert \rho \vel_{*,L} \rvert^2}{\rho_{*,L}} =
    E_{*,L} - \half \frac{ \left(\rho u \right)_{*,L}^2 + \left( \rho v \right)_{*,L}^2}{\rho_{*,L}}  \\
\begin{split}
  \left( \rho e \right)_{*,L} &= \frac{s_L-u_L}{s_L-s_*}E_L  + \frac{s_L-u_L}{s_L-s_*}(s_*-u_L)\left(\rho_L s_* + \frac{p_L}{s_L-u_L}\right) \\
  &\quad - \half \frac{s_L-u_L}{s_L-s_*}\rho_L \left( s_*^2 + v_L^2 \right) .
\end{split}
\end{align}
Therefore, we require the following inequality:
\begin{align}
\begin{split}
&\frac{s_L - u_L}{s_L - s_*}E_L + \frac{s_L - u_L}{s_L - s_*}(s_* - u_L)
  \left( \rho_L s_* + \frac{p_L}{s_L - u_L} \right)
  - \half \frac{s_L - u_L}{s_L - s_*}\rho_L \left( s_*^2 + v_L^2 \right) \\
  &\qquad > \overline p^\infty_L
\end{split} \\
\begin{split}
&E_L + (s_* - u_L) \left(\rho_L s_* + \frac{p_L}{s_L - u_L}\right)
  - \half \rho_L \left( s_*^2 + v_L^2 \right) \\
  &\qquad > \overline p^\infty_L \frac{(s_L - u_L - s_* + u_L)}{(s_L - u_L)}
\end{split} \\
\begin{split}
&(\rho e)_L + \half \rho_L \left( u_L^2 + v_L^2 \right) +
  (s_* - u_L) \left( \rho_L s_* + \frac{p_L}{s_L - u_L} \right)
  - \half \rho_L \left( s_*^2 + v_L^2 \right) \\
  &\qquad > \overline p^\infty_L \frac{(s_L - u_L - s_* + u_L)}{(s_L - u_L)}
\end{split} \\
&\half \rho_L(s_* - u_L)^2 -
  \frac{p_L + \overline p^\infty_L}{u_L - s_L}(s_* - u_L) +
  (\rho_L e_L - \overline p^\infty_L) > 0 .
\end{align}
Let $\beta = s_* - u_L$, then the inequality above is a quadratic function of $\beta$. We can show that this quadratic has no real roots by ensuring the discriminant is negative. That is,
\begin{equation}
  \left(\frac{p_L + \overline p^\infty_L}{u_L-s_L}\right)^2 - 2\rho_L(\rho_L e_L - \overline p^\infty_L) < 0.
\end{equation}
This implies that we require:
\begin{equation}
  s_L < u_L - \frac{p_L + \overline p^\infty_L}{\sqrt{2 \rho_L\left(\rho_L e_L - \overline p^\infty_L\right)}} . \label{eq:s_L_ineqaulity_constraint}
\end{equation}
It should be noted that:
\begin{align*}
  \frac{p_L + \overline p^\infty_L}{\sqrt{2 \rho_L\left(\rho_L e_L - \overline p^\infty_L\right)}} &= 
  \sqrt{\frac{(\overline \gamma_L - 1)(p_L + \overline p^\infty_L)}{2 \rho_L}}  \\
  &< \sqrt{\frac{\overline \gamma_L (p_L + \overline p^\infty_L)}{\rho_L}}  \\
  &= c_L.
\end{align*}
Since $s_L = \min{(\bar{u} - \bar{c}, u_L - c_L )}$, the constraint on $s_L$ given by equation~\eqref{eq:s_L_ineqaulity_constraint} is already satisfied and we have proved the solutions given by the left star state have positive partial densities and squared speed of sound. Besides, the volume fractions are bounded. Thus, the HLLC Riemann solver is positivity- and boundedness-preserving.

Equation~\eqref{eq:fully_discretized_W_G_pm_eqn} can be re-written as:
\begin{equation}
\begin{split}
  \Wv_{i,j}^{n+1} &=
    \sigma_x \left[ \Wv_{i,j}^{n} + \lambda_x \left(
      \hat{\mathbf{G}}^{x,+}_{i-\half,j}
      - \hat{\mathbf{G}}^{x,-}_{i+\half,j}
    \right) \right] \\
    &\quad + \sigma_y \left[ \Wv_{i,j}^{n} + \lambda_y \left( 
      \hat{\mathbf{G}}^{y,+}_{i,j-\half}
      - \hat{\mathbf{G}}^{y,-}_{i,j+\half}
    \right) \right] ,
  \end{split} \label{eq:convex_split_2D}
\end{equation}
where $\lambda_x = \dt/ (\dx \sigma_x)$ and $\lambda_y = \dt/ (\dy \sigma_y)$. $\sigma_x$ and $\sigma_y$ are partitions of the contribution in the $x$ and $y$ directions respectively where $\sigma_x + \sigma_y = 1$. They can be defined as~\cite{hu2013positivity}:
\begin{equation}
  \sigma_x = \frac{\tau_x}{\tau_x + \tau_y}, \quad
  \sigma_y = \frac{\tau_y}{\tau_x + \tau_y}, \quad
  \tau_x = \frac{\left( \left| u \right| + c \right)_{\mathrm{max}}}{\dx}, \quad
  \tau_y = \frac{\left( \left| v \right| + c \right)_{\mathrm{max}}}{\dy} ,
\end{equation}
such that $0 < \sigma_x < 1$ and $0 < \sigma_y < 1$. If the time-step size $\dt$ is given by a chosen CFL number, $CFL$, with the following equation:
\begin{equation}
  \dt = \frac{CFL}{\tau_x + \tau_y},
\end{equation}
one has the relations for the equivalent 1D time step sizes in different directions, $\dt^x$ and $\dt^y$:
\begin{align}
  \lambda_x &= \frac{CFL}{\left( \left| u \right| + c \right)_{\mathrm{max}}} = \frac{\dt^x}{\dx} , \\
  \lambda_y &= \frac{CFL}{\left( \left| v \right| + c \right)_{\mathrm{max}}} = \frac{\dt^y}{\dy} .
\end{align}
We can define:
\begin{align}
  \Wv_{i,j}^{x} &= \Wv_{i,j}^{n} + \lambda_x \left(
      \hat{\mathbf{G}}^{x,+}_{i-\half,j}
      - \hat{\mathbf{G}}^{x,-}_{i+\half,j}
    \right) , \\
  \Wv_{i,j}^{y} &= \Wv_{i,j}^{n} + \lambda_y \left( 
      \hat{\mathbf{G}}^{y,+}_{i,j-\half}
      - \hat{\mathbf{G}}^{y,-}_{i,j+\half}
    \right) .
\end{align}
Since $\hat{\mathbf{G}}^{x,\pm}_{i\mp\half,j}$ are obtained from the quasi-1D HLLC solutions in the $x$ direction, the approximate waves from the edges at $x_{i\pm\half}$ do not interact if the CFL condition, $CFL \leq 0.5$, is satisfied (same for $\hat{\mathbf{G}}^{y,\pm}_{i,j\mp\half}$ in the $y$ direction). Thus, equation~\eqref{eq:soln_convex_averaging} is satisfied for $\Wv_{i,j}^{x}$ (similar for $\Wv_{i,j}^{y}$ in $y$ direction). The first part of the RHS of equation~\eqref{eq:soln_convex_averaging} is the solution in the half cell from $x_{i-\half}$ to $x_i$ at $t+\dt^x$ and the second part is the solution in another half cell. Using the finite volume approach on the left and right half cells, we will get:
\begin{align}
    \frac{1}{\dx} \int_{x_{i-\half}}^{x_{i}}
      R \left( \frac{x - x_{i-\half}}{\dt^x}, \Wv_{i-1,j}^{n}, \Wv_{i,j}^{n} \right) dx &= \frac{1}{2} \left[
      \Wv_{i,j}^{n} + 2 \lambda_x \left(
        \hat{\mathbf{G}}^{x,+}_{i-\half,j}
        - \Fxv_{i,j}
      \right) \right] , \\
    \frac{1}{\dx} \int_{x_{i}}^{x_{i+\half}}
      R\left(\frac{x - x_{i+\half}}{\dt^x}, \Wv_{i,j}^{n}, \Wv_{i+1,j}^{n} \right) dx &= \frac{1}{2} \left[
      \Wv_{i,j}^{n} - 2 \lambda_x \left(
        \hat{\mathbf{G}}^{x,-}_{i+\half,j}
        - \Fxv_{i,j}
      \right) \right] .
\end{align}
If we define:
\begin{align}
    \Wv_{i,j}^{x,-} &= \frac{2}{\dx} \int_{x_{i-\half}}^{x_{i}}
      R \left( \frac{x - x_{i-\half}}{\dt}, \Wv_{i-1,j}^{n}, \Wv_{i,j}^{n} \right) dx , \\
    \Wv_{i,j}^{x,+} &= \frac{2}{\dx} \int_{x_{i}}^{x_{i+\half}}
      R\left(\frac{x - x_{i+\half}}{\dt}, \Wv_{i,j}^{n}, \Wv_{i+1,j}^{n} \right) dx ,
\end{align}
we will get:
\begin{equation}
  \Wv_{i,j}^{x} = \frac{1}{2} \Wv_{i,j}^{x,-} + \frac{1}{2} \Wv_{i,j}^{x,+} .
\end{equation}
$\Wv_{i,j}^{x,\pm}$ are at physically admissible states since they are convex averaging of the approximate HLLC solutions. This also means $\Wv_{i,j}^{x}$ is also physically admissible. Equation~\eqref{eq:convex_split_2D} becomes:
\begin{equation}
  \Wv_{i,j}^{n+1} =
    \sigma_x \Wv_{i,j}^{x} + \sigma_y \Wv_{i,j}^{y} .
\end{equation}
Therefore, $\Wv_{i,j}^{n+1}$ is also physically admissible since it is a convex combination of $\Wv_{i,j}^{x}$ and $\Wv_{i,j}^{y}$.

\section{Incremental-stencil WCNS}

In this section, a high-order finite difference scheme for discretizing equation~\eqref{eq:W_G_eqn} is introduced. The high-order scheme belongs to the family of weighted compact nonlinear schemes (WCNSs) which is a variant of the WENO schemes for discontinuity-capturing. It was first proposed by~\citet{deng2000developing} in which compact (spatially implicit) finite difference schemes are combined with WENO interpolation. Since then, WCNSs are extended to higher order of accuracy~\cite{nonomura2007increasing,zhang2008development}. In principle, WCNSs can be used with both explicit or compact finite difference schemes. \citet{nonomura2009effects} suggested that explicit finite difference schemes are more efficient and later also proposed a family of robust explicit midpoint-and-node-to-node finite difference schemes~\cite{nonomura2013robust}. In this section, an explicit WCNS with the explicit hybrid cell-midpoint and cell-node finite difference scheme~\cite{deng2011new,yan2016new} and nonlinear interpolation adapted from incremental-stencil reconstruction~\cite{wang2018incremental} for finite volume WENO scheme is presented.

The semi-discretized finite difference form of equation~\eqref{eq:W_G_eqn} is given by:
\begin{equation}
    \frac{\partial \Wv}{\partial t} \bigg|_{i,j}
      + \widehat{ \frac{\partial \Gxv}{\partial x} } \bigg|_{i,j}
      + \widehat{ \frac{\partial \Gyv}{\partial y} } \bigg|_{i,j} = \Shv_{i,j} , \label{eq:semi_W_G_eqn}
\end{equation}
where $\hat{\Sv}_{i,j}$ is given by:
\begin{equation}
    \Shv_{i,j} = \begin{pmatrix}
      0 \\
      0 \\
      0 \\
      0 \\
      0 \\
      \alpha_{1,i,j}
       \left( \left. \widehat{ \frac{\partial u}{\partial x} } \right|_{i,j} +
       \left. \widehat{ \frac{\partial v}{\partial y} } \right|_{i,j}
      \right)
      \end{pmatrix} .
\end{equation}
A high-order discretization is considered consistent (and conservative for any conservative equation) if:
\begin{alignat}{2}
    \widehat{ \frac{\partial \Gv^{x}}{\partial x} } \bigg|_{i,j} &=
      \frac{\Gxhv_{i+\half,j}
            - \Gxhv_{i-\half,j}}{\dx} , \quad
    &\widehat{ \frac{\partial \Gv^{y}}{\partial y} } \bigg|_{i,j} &=
      \frac{\Gyhv_{i,j+\half}
            - \Gyhv_{i,j-\half}}{\dy},
            \label{eq:flux_diff_form} \\
    \widehat{ \frac{\partial u}{\partial x} } \bigg|_{i,j} &=
      \frac{\hat{u}_{i+\half,j} - \hat{u}_{i-\half,j}}{\dx}, \quad
    &\widehat{ \frac{\partial v}{\partial y} } \bigg|_{i,j} &=
      \frac{\hat{v}_{i,j+\half} - \hat{v}_{i,j-\half}}{\dy} .
\end{alignat}
Therefore, equation~\eqref{eq:semi_W_G_eqn} becomes:
\begin{equation}
    \frac{\partial \Wv}{\partial t} \bigg|_{i,j}
      + \frac{\Gxhv_{i+\half,j}
            - \Gxhv_{i-\half,j}}{\dx} 
      + \frac{\Gyhv_{i,j+\half}
            - \Gyhv_{i,j-\half}}{\dy} =
      \Shv_{i,j} , \label{eq:semi_W_G_eqn_cons}
\end{equation}
Equation~\eqref{eq:semi_W_G_eqn_cons} looks as same as equation~\eqref{eq:fully_discretized_W_G_eqn} if it is further discretized in time with forward Euler method but $\Gxhv_{i\pm\half,j}$, $\Gyhv_{i,j\pm\half}$ and $\Shv_{i,j}$ (composed of $\hat{u}_{i\pm\half,j}$ and $\hat{v}_{i,j\pm\half}$) are in high-order accurate approximations and are given by the WCNS introduced in this section. It should be noted that equation~\eqref{eq:semi_W_G_eqn_cons} is a conservative discretization for all conservative equations, except the last advection equation that is given by:
\begin{equation}
\begin{split}
    \frac{\partial \alpha_1}{\partial t} \bigg|_{i,j}
    + \frac{\hat{f}^{x}_{\alpha,i+\half,j}
          - \hat{f}^{x}_{\alpha,i-\half,j}}{\dx} 
    + \frac{\hat{f}^{y}_{\alpha,i,j+\half}
          - \hat{f}^{y}_{\alpha,i,j-\half}}{\dy} = \\
    \alpha_{1,i,j} \left(
      \frac{\hat{u}_{i+\half,j}
          - \hat{u}_{i-\half,j}}{\dx} 
    + \frac{\hat{v}_{i,j+\half}
          - \hat{v}_{i,j-\half}}{\dy}
    \right) .
\end{split}
\end{equation}

\subsection{Explicit hybrid cell-midpoint and cell-node scheme}

The sixth order accurate explicit scheme from the hybrid cell-midpoint and cell-node compact scheme (HCS)~\cite{deng2011new} family is used for the approximation of the first order derivatives $\partial \Gxv / \partial x |_{i,j}$ and $\partial \Gyv / \partial y |_{i,j}$. The sixth order explicit HCS formulation is given by:
\begin{equation}
\begin{split}
    \frac{\partial \Gxv}{\partial x} \bigg|_{i,j} &\approx
    \widehat{ \frac{\partial \Gxv}{\partial x} } \bigg|_{i,j} \\
    &= \frac{1}{\dx}  \left[
      \psi \left(\Gxtv_{i+\half,j} - \Gxtv_{i-\half,j} \right)
      - \frac{175 \psi - 192}{256} \left(\Gxv_{i+1,j} - \Gxv_{i-1,j} \right) \right. \\
      &\quad \left. + \frac{35 \psi - 48}{320} \left(\Gxv_{i+2,j} - \Gxv_{i-2,j} \right) - \frac{45 \psi - 64}{3840} \left(\Gxv_{i+3,j} - \Gxv_{i-3,j} \right) \right]. \label{eq:HCS6}
\end{split}
\end{equation}
If we replace $\Gxtv_{i \pm \half,j}$ with the exact fluxes,
\begin{equation}
\begin{split}
    \widehat{ \frac{\partial \Gxv}{\partial x} } \bigg|_{i,j} &= \frac{\partial \Gxv}{\partial x} \bigg|_{i,j} - \left( \frac{5 \psi}{1024} - \frac{1}{140} \right) \frac{\partial^7 \Gxv}{\partial x^7} \bigg|_{i,j} \dx^6 \\
    &\quad - \left( \frac{95 \psi}{98304} - \frac{1}{720} \right) \frac{\partial^9 \Gxv}{\partial x^9} \bigg|_{i,j} \dx^8
    + \mathcal{O} \left( \dx^{10} \right) . \label{eq:HCS6_Taylor_expand}
\end{split}
\end{equation}
If $\psi = 256/175$, the scheme becomes eighth order accurate:
\begin{equation}
    \widehat{ \frac{\partial \Gxv}{\partial x} } \bigg|_{i,j} = \frac{\partial \Gxv}{\partial x} \bigg|_{i,j} - \frac{1}{40320} \frac{\partial^9 \Gxv}{\partial x^9} \bigg|_{i,j} \dx^8 + \mathcal{O} \left( \dx^{10} \right) . \label{eq:HCS6_Taylor_expand_8th_order}
\end{equation}
$\psi = 256/175$ is adopted in this work.

Any central explicit or compact finite difference scheme can be rewritten into the flux-difference forms given by equation~\eqref{eq:flux_diff_form} and it is derived in~\cite{subramaniam2019high}. Following that work, the HCS given by equation~\eqref{eq:HCS6} has implied reconstructed flux $\Gxhv_{i+1/2,j}$ given by:
\begin{equation}
\begin{split}
    \Gxhv_{i+\frac{1}{2},j} &= \psi \Gxtv_{i+\frac{1}{2},j} - \left( \frac{75 \psi}{128} - \frac{37}{60} \right) \left( \Gxv_{i,j} + \Gxv_{i+1,j} \right) \\
    &\quad + \left( \frac{25 \psi}{256} - \frac{2}{15} \right) \left( \Gxv_{i-1,j} + \Gxv_{i+2,j} \right)
    - \left( \frac{3 \psi}{256} - \frac{1}{60} \right) \left( \Gxv_{i-2,j} + \Gxv_{i+3,j} \right) \label{eq:reconstructed_G}.
\end{split}
\end{equation}
Note that the equation above is also used for reconstructing the flux $\hat{f}^{x}_{\alpha,i+\half,j}$ in the advection equation.

High-order finite difference approximations of the velocity components are also required for $\Shv_{i,j}$. Following the idea of \cite{wong2017high}, the numerical derivatives of the velocity components are also given by the same finite difference scheme as the flux derivatives:
\begin{equation}
\begin{split}
    \frac{\partial u}{\partial x} \bigg|_{i,j} \approx
    \widehat{ \frac{\partial u}{\partial x} } \bigg|_{i,j} &=
    \frac{1}{\dx}  \left[
      \psi \left(\tilde{u}_{i+\half,j} - \tilde{u}_{i-\half,j} \right)
      - \frac{175 \psi - 192}{256} \left(u_{i+1,j} - u_{i-1,j} \right) \right. \\
      &\quad \left. + \frac{35 \psi - 48}{320} \left(u_{i+2,j} - u_{i-2,j} \right) - \frac{45 \psi - 64}{3840} \left(u_{i+3,j} - u_{i-3,j} \right) \right].
\end{split}
\end{equation}
The implied reconstructed velocity component $\hat{u}_{i+1/2,j}$ is given by:
\begin{equation}
\begin{split}
    \hat{u}_{i+\frac{1}{2},j} &= \psi \tilde{u}_{i+\frac{1}{2},j} - \left( \frac{75 \psi}{128} - \frac{37}{60} \right) \left( u_{i,j} + u_{i+1,j} \right) + \left( \frac{25 \psi}{256} - \frac{2}{15} \right) \left( u_{i-1,j} + u_{i+2,j} \right) \\
    &\quad - \left( \frac{3 \psi}{256} - \frac{1}{60} \right) \left( u_{i-2,j} + u_{i+3,j} \right) \label{eq:reconstructed_vel} .
\end{split}
\end{equation}

The discretizations for flux and velocity component derivatives in $y$ direction are similar. High-order accurate approximations are required for $\Gxtv_{i+1/2,j}$ and $\tilde{u}_{i+1/2,j}$ to form high-order discretization for equation~\eqref{eq:semi_W_G_eqn_cons}, which are discussed in the following sections. Finally, by using equation~\eqref{eq:G_pm}, equation~\eqref{eq:semi_W_G_eqn} can be rewritten as:
\begin{equation}
    \frac{\partial \Wv}{\partial t} \bigg|_{i}
      + \frac{\hat{\mathbf{G}}^{x,-}_{i+\half,j}
          - \hat{\mathbf{G}}^{x,+}_{i-\half,j}}{\dx} 
      + \frac{\hat{\mathbf{G}}^{y,-}_{i,j+\half}
          - \hat{\mathbf{G}}^{y,+}_{i,j-\half}}{\dy} =
    \mathbf{0} . \label{eq:semi_W_G_pm_eqn}
\end{equation}

\subsection{Methodology of WCNS}

In WCNSs, the fluxes at the midpoints are obtained with aid of explicit nonlinear interpolations, which can also be interpreted as nonlinear filtering processes to avoid spurious oscillations near shocks and other discontinuities. For simplicity, the implementation details of a WCNS is explained with a 1D version of equation~\eqref{eq:semi_W_G_pm_eqn} in this and the following sub-section. Thus, superscript ``$x$" in $\Gxv$ is dropped for convenience.

For the 1D five-equation model, the algorithm to obtain the high-order fluxes $\hat{\mathbf{G}}^{\pm}_{i+1/2}$ with the WCNS approach is given below:

\begin{enumerate}
\item Convert all $\Wv_{i}$ in the stencils of left-biased and right-biased nonlinear WENO interpolations to primitive variable vectors $\Vv_{i}$.
\item Perform characteristic decomposition by transforming all $\Vv_{i}$ in the stencils of interpolations to characteristic variable vectors $\Uv_{i}$ with the projection matrix $\Rv^{-1}_{i+1/2}$: $\Uv_{i} = \Rv^{-1}_{i+1/2} \Vv_{i}$.
\item Compute $\tilde{\Uv}_L$ and $\tilde{\Uv}_R$ at each midpoint with $\Uv_{i}$ using left-biased and right-biased nonlinear WENO interpolations respectively.
\item Transform $\tilde{\Uv}_L$ and $\tilde{\Uv}_R$ to $\tilde{\Vv}_L$ and $\tilde{\Vv}_R$ with the projection matrix $\Rv_{i+1/2}$: $\tilde{\Vv}_L = \Rv_{i+1/2} \tilde{\Uv}_L$, and $\tilde{\Vv}_R = \Rv_{i+1/2} \tilde{\Uv}_R$.
\item Convert $\tilde{\Vv}_L$ and $\tilde{\Vv}_R$ to $\tilde{\Wv}_L$ and $\tilde{\Wv}_R$.
\item Compute the high-order flux and velocity at each midpoint with the Riemann solver. If the HLLC Riemann solver is used: $\Gtv_{i+\frac{1}{2}} = \Gv^{\mathrm{HLLC}} \left (\tilde{\Wv}_L,\tilde{\Wv}_R \right)$ and $\tilde{u}_{i+\frac{1}{2}} = u^{\mathrm{HLLC}}_{*} \left( \tilde{\Wv}_L, \tilde{\Wv}_R \right)$.
\item Compute the flux and velocity at the nodes: $\Gv_{i} = \Gv(\Wv_i)$ and $u_{i} = u(\Wv_i)$.
\item Reconstruct the flux $\Ghv_{i+\frac{1}{2}}$ and velocity $\hat{u}_{i+\frac{1}{2}}$ at the midpoints using the flux differencing approach (equations~\eqref{eq:reconstructed_G} and \eqref{eq:reconstructed_vel} in this work).
\item Compute $\hat{\mathbf{G}}^{\pm}_{i+1/2}$ using equation~\eqref{eq:G_pm}:
  \begin{alignat*}{2}
  \hat{\mathbf{G}}^{-}_{i+\half} &=
      \Ghv_{i+\half} -
      \alpha_{1,i} \ &(0\ 0\ 0\ 0\ \hat{u}_{i+\half} )^T , \\
  \hat{\mathbf{G}}^{+}_{i+\half} &=
      \Ghv_{i+\half} -
      \alpha_{1,i+1} \ &(0\ 0\ 0\ 0\ \hat{u}_{i+\half} )^T .
  \end{alignat*}
\end{enumerate}

In this work, only the incremental-stencil WENO interpolation of left-biased midpoint values is presented. The interpolation of right-biased midpoint values is similar due to symmetry and can be obtained by flipping the stencils and corresponding coefficients. The projection matrices for transformation between primitive variables and characteristic variables can be found in \ref{appendix:char_decomp}. The projection matrices $\Rv_{i+1/2}$ are computed at midpoints $x_{i+1/2}$ with the arithmetic averages of partial densities, mixture density and speed of sound at $x_{i}$ and $x_{i+1}$.

\subsection{Incremental-stencil WENO interpolation}

The finite volume WENO scheme with the incremental-stencil reconstruction (WENO-IS) was proposed by~\citet{wang2018incremental}. The finite volume WENO-IS is robust for compressible multi-phase problems with shocks and is also accurate for those problems due to high-order WENO reconstruction with the use of HLLC Riemann solver, which is well-known for its accuracy in capturing material interfaces. However, in general a finite volume WENO scheme is more expensive compared with finite difference WCNS and WENO schemes with similar orders of accuracy for multi-dimensional simulations. This is due to the fact that a finite volume scheme requires multi-dimensional reconstructions to obtain point values at many Gaussian points on the cell boundaries from cell averages when one desires third or high order of accuracy~\cite{shu2003high,titarev2004finite,coralic2014finite}. Generally, finite difference WENO schemes or WCNSs in explicit forms are four times cheaper in 2D and nine times cheaper in 3D compared to finite volume WENO schemes since multi-dimensional reconstructions are not required for the former schemes~\cite{sebastian2003multidomain}. While the costs of WENO reconstruction in finite difference WENO schemes and WENO interpolation in finite difference WCNS methods are similar, a finite difference WENO scheme can only be used with flux-vector splitting methods, such as Lax--Friedrichs flux splitting, when high-order of accuracy is desired. The use of flux-difference splitting methods such as Riemann solvers in a finite difference WENO scheme for multi-dimensional simulations degenerates to only second order but a WCNS can still maintain high order of accuracy when used with a Riemann solver. Therefore, in this work we propose a WCNS with a WENO interpolation adapted from the robust incremental-stencil WENO reconstruction such that the HLLC Riemann solver can be applied for upwinding while the overall scheme is still high-order accurate and efficient.

\begin{figure}[hbt]
  \centering
  \begin{tikzpicture}
    \useasboundingbox ( 0.0cm,-1cm) rectangle (12cm,4.5cm);
    \draw[thin]       ( 0.0cm,3.0cm) -- (12.0cm,3.0cm);
    \draw[very thick] ( 6.0cm,2.7cm) -- ( 6.0cm,3.3cm) node[above=0.1cm]{\large $\bm{i+\frac{1}{2}}$};
    \draw[thin]       ( 2.0cm,2.7cm) -- ( 2.0cm,3.3cm) node[above=0.1cm]{\large $i-\frac{3}{2}$};
    \draw[thin]       ( 4.0cm,2.7cm) -- ( 4.0cm,3.3cm) node[above=0.1cm]{\large $i-\frac{1}{2}$};
    \draw[thin]       ( 8.0cm,2.7cm) -- ( 8.0cm,3.3cm) node[above=0.1cm]{\large $i+\frac{3}{2}$};
    \draw[thin]       (10.0cm,2.7cm) -- (10.0cm,3.3cm) node[above=0.1cm]{\large $i+\frac{5}{2}$};
    
    \fill ( 1.0cm,3.0cm) circle(2.0pt) node[below=0.2cm]{\large $i-2$};
    \fill ( 3.0cm,3.0cm) circle(2.0pt) node[below=0.2cm]{\large $i-1$};
    \fill ( 5.0cm,3.0cm) circle(2.0pt) node[below=0.2cm]{\large $i$};
    \fill ( 7.0cm,3.0cm) circle(2.0pt) node[below=0.2cm]{\large $i+1$};
    \fill ( 9.0cm,3.0cm) circle(2.0pt) node[below=0.2cm]{\large $i+2$};
    \fill (11.0cm,3.0cm) circle(2.0pt) node[below=0.2cm]{\large $i+3$};
 
    \draw[thin] (5.0cm,2.0cm) node[left]{\large $S_{0}$} -- (7.0cm,2.0cm);
    \draw[black,fill=white] (6.0cm,2.0cm) circle(2.0pt);
    \fill (5.0cm,2.0cm) circle(2.0pt);
    \fill (7.0cm,2.0cm) circle(2.0pt);
    
    \draw[thin] (3.0cm,1.5cm) node[left]{\large $S_{1}$} -- (6.0cm,1.5cm);
    \draw[black,fill=white] (6.0cm,1.5cm) circle(2.0pt);
    \fill (3.0cm,1.5cm) circle(2.0pt);
    \fill (5.0cm,1.5cm) circle(2.0pt);
    
    \draw[thin] (5.0cm,1.0cm) node[left]{\large $S_{2}$} -- (9.0cm,1.0cm);
    \draw[black,fill=white] (6.0cm,1.0cm) circle(2.0pt);
    \fill (5.0cm,1.0cm) circle(2.0pt);
    \fill (7.0cm,1.0cm) circle(2.0pt);
    \fill (9.0cm,1.0cm) circle(2.0pt);
    
    \draw[thin] (1.0cm,0.5cm) node[left]{\large $S_{3}$} -- (6.0cm,0.5cm);
    \draw[black,fill=white] (6.0cm,0.5cm) circle(2.0pt);
    \fill (1.0cm,0.5cm) circle(2.0pt);
    \fill (3.0cm,0.5cm) circle(2.0pt);
    \fill (5.0cm,0.5cm) circle(2.0pt);
    
    \draw[thin] (1.0cm,0.0cm) node[left]{\large $S_{5}$} -- (9.0cm,0.0cm);
    \draw[black,fill=white] (6.0cm,0.0cm) circle(2.0pt);
    \fill (1.0cm,0.0cm) circle(2.0pt);
    \fill (3.0cm,0.0cm) circle(2.0pt);
    \fill (5.0cm,0.0cm) circle(2.0pt);
    \fill (7.0cm,0.0cm) circle(2.0pt);
    \fill (9.0cm,0.0cm) circle(2.0pt);
    
    \draw[gray,thin] (3.0cm,-0.5cm) node[left]{\large $S_{01}$} -- (7.0cm,-0.5cm);
    \draw[gray,fill=white] (6.0cm,-0.5cm) circle(2.0pt);
    \draw[gray,fill] (3.0cm,-0.5cm) circle(2.0pt);
    \draw[gray,fill] (5.0cm,-0.5cm) circle(2.0pt);
    \draw[gray,fill] (7.0cm,-0.5cm) circle(2.0pt);

  \end{tikzpicture}
  \caption{Sub-stencils of incremental-stencil WENO interpolation. The solid circles represent nodes used in the interpolation stencils, while empty circles represent the midpoint for the interpolated values. The stencil $S_{01}$ is in grey since it is not considered in the actual interpolation.}
  \label{fig:stencil_WCNS}
\end{figure}
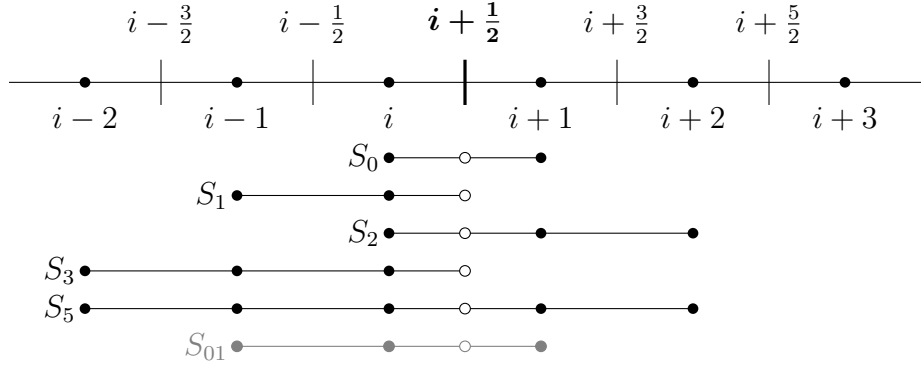

The incremental-stencil (IS) interpolation approximates the midpoint values by nonlinear combination of linearly interpolated values from four different sub-stencils, $S_0$--$S_3$ (shown in figure~\ref{fig:stencil_WCNS}). The interpolated values at the midpoints $\tilde{u}_{j+\half}$ from the four different sub-stencils are given by:
\begin{align}
    S_0: \quad \tilde{u}_{i+\half}^{0} =&
      \frac{1}{2}\left(u_i + u_{i+1} \right),
      \label{eq:interpolation_stencil_0} \\
    S_1: \quad \tilde{u}_{i+\half}^{1} =&
      \frac{1}{2}\left(-u_{i-1} + 3u_{i} \right),
      \label{eq:interpolation_stencil_1} \\
    S_2: \quad \tilde{u}_{i+\half}^{2} =&
      \frac{1}{8}\left(3u_{i} + 6u_{i+1} - u_{i+2} \right),
      \label{eq:interpolation_stencil_2} \\
    S_3: \quad \tilde{u}_{i+\half}^{3} =&
      \frac{1}{8}\left(3u_{i-2} - 10u_{i-1} + 15u_{i} \right).
      \label{eq:interpolation_stencil_3}
\end{align}

\noindent The interpolated values from stencils $S_0$ and $S_1$ are second order accurate and those from stencils $S_2$ and $S_3$ are third order accurate. The variable $u$ can either be fluxes, conservative variables, primitive variables or variables that are projected to the characteristic fields. In this work, the primitive variables projected to the characteristic fields are employed in the interpolation process.

A fifth order 5-point linear interpolation from $S_5$ can be obtained from linear combination of the lower order interpolations:
\begin{equation}
    S_5: \quad \tilde{u}_{i+\half}^{5} =
      \sum_{k=0}^{3} d_k \tilde{u}_{i+\half}^{k},
      \label{eq:WCNS_stencil_5}
\end{equation}
where the linear weights are given by:
\begin{equation}
    d_0 = \frac{15}{32}, \quad
    d_1 = \frac{5}{32}, \quad
    d_2 = \frac{5}{16}, \quad
    d_3 = \frac{1}{16}.
\end{equation}

\noindent The expanded form of the linear interpolation is given by:
\begin{equation}
    S_5: \quad \tilde{u}_{i+\half}^{5} =
      \frac{1}{128} \left(3u_{i-2} - 20u_{i-1} +
      90u_i + 60u_{i+1} - 5u_{i+2} \right) .
      \label{eq:WCNS_stencil_5_epxanded}
\end{equation}

The 5-point linear interpolation scheme may generate spurious oscillations due to Gibbs phenomenon or is even unstable near shocks or discontinuities. To improve robustness, a nonlinear WENO interpolation is suggested in this work by replacing the linear weights with the incremental-stencil (IS) nonlinear weights~\cite{wang2018incremental}. The nonlinear weights have the following form:
\begin{equation}
    \tilde{u}_{i+\half} =
    \sum_{k=0}^{3} \omega_k \tilde{u}_{i+\half}^{k} .
    \label{eq:WENO_nonlinear_interpolation}
\end{equation}

\noindent The IS nonlinear weights are given by~\cite{wang2018incremental}:
\begin{align}
    \omega_k &= \frac{\eta_k}{\sum_{s=0}^{3} \eta_s}, \\
    \eta_k &=
      \left\{ \begin{array}{ll} 
        d_k \left( 1 + \frac{\tau_5}{\beta_k + \epsilon}
          \cdot \frac{\tau_5}{\beta_{01} + \epsilon} \right),
          & \text{if }  k<2,  \\ 
        d_k \left( 1 + \frac{\tau_5}{\beta_k + \epsilon} \right),
          & \text{otherwise} .
      \end{array}\right.
\end{align}

\noindent The smoothness indicators are defined by:
\begin{equation}
    \beta_k =
    \left\{
      \begin{array}{ll} 
        \int^{x_{i+\half}}_{x_{i-\half}} \dx
        \left( \frac{\partial}{\partial x} \tilde{u}^{k}(x) \right)^2 dx,
        &  k = 0, 1, \\ 
        \sum^{2}_{l=1} \int^{x_{i+\half}}_{x_{i-\half}} \dx^{2l-1}
        \left( \frac{\partial^{l}}{\partial x^l} \tilde{u}^{k}(x) \right)^2 dx,
        &  \text{otherwise (including $k=01$)},
      \end{array}
    \right.
\end{equation}
where $\tilde{u}^{k}(x)$ are the Lagrange interpolating polynomials from stencils $S_k$. The nonlinear weights require smoothness indicator computed with the Lagrange interpolation polynomial from stencil $S_{01}$ besides those from $S_0$--$S_3$.
The integrated forms of smoothness indicators are given by:
\begin{align}
    \beta_0 &= \left( u_{i} - u_{i+1} \right)^2, \label{eq:beta_0_IS} \\
    \beta_1 &= \left( u_{i-1} - u_{i} \right)^2, \label{eq:beta_1_IS} \\
    \beta_{01} &= \frac{13}{12} \left( u_{i-1} - 2u_{i} + u_{i+1} \right)^2 +
        \frac{1}{4} \left( u_{i-1} - u_{i+1} \right)^2, \label{eq:beta_01_IS} \\
    \beta_2 &= \frac{13}{12} \left( u_{i} - 2u_{i+1} + u_{i+2} \right)^2 +
        \frac{1}{4} \left( 3u_{i} - 4u_{i+1} + u_{i+2} \right)^2, \label{eq:beta_2_IS} \\
    \beta_3 &= \frac{13}{12} \left( u_{i-2} - 2u_{i-1} + u_{i} \right)^2 +
        \frac{1}{4} \left( u_{i-2} - 4u_{i-1} + 3u_{i} \right)^2 . \label{eq:beta_3_IS} 
\end{align}

\noindent The reference smoothness indicator, $\tau_5$, is defined by~\cite{fu2016family}:
\begin{equation}
    \tau_5 = \sum^{4}_{l=3} \int^{x_{i+\half}}_{x_{i-\half}} \dx^{2l-1}
      \left( \frac{\partial^{l}}{\partial x^l} \tilde{u}^{5}(x) \right)^2 dx
\end{equation}
where $\tilde{u}^{5}(x)$ is the Lagrange interpolating polynomial from stencil $S_5$. The integrated form of the reference smoothness indicator is given by:
\begin{equation}
\begin{split}
    \tau_5 &= \frac{13}{12} \left( u_{i+2} - 4u_{i+1} + 6u_{i} - 4u_{i-1} + u_{i-2} \right)^2 \\
      &\quad + \frac{1}{4} \left( u_{i+2} - 2u_{i+1} + 2u_{i-1} - u_{i-2} \right)^2 . \label{eq:tau_5_IS}
\end{split}
\end{equation}

The robustness of incremental-stencil WENO interpolation presented above is as high as the WENO incremental-stencil reconstruction, as both of them can choose one of the 2-point stencils when there are closely located discontinuities, as explained in~\cite{wang2018incremental}. In \ref{appendix:convergence}, it is proved that the IS nonlinear interpolation with the HCS finite differencing is fifth order accurate for a 1D scalar hyperbolic conservation law with perfect upwinding, provided not at critical points. This WCNS is termed WCNS-IS in this work.

\section{Positivity- and boundedness-preserving limiting procedures}

Similar to the finite volume WENO-IS, the finite difference WCNS-IS is generally robust in capturing shocks and material interfaces in multi-phase flow simulations. However, it cannot be guaranteed that these high-order schemes are free from numerical failures due to negative squared speed of sound, partial densities (hence also out-of-bounds mass fractions), and out-of-bounds volume fractions. In this section, positivity- and boundedness-preserving limiting procedures are introduced to improve the robustness of the WCNS-IS. The procedures are conservative for the corresponding equations in the system that are conservative, i.e. except the volume fraction advection equation.

In the algorithm of WCNS, there are two stages where positivity and boundedness can be violated. The first stage is the WENO interpolation step for the left-biased and right-biased interpolated conservative variable vectors ($\tilde{\Wv}_L$ and $\tilde{\Wv}_R$), where the interpolated conservative variables may not be physically admissible. Another stage is the flux reconstruction step of $\hat{\mathbf{G}}^{\pm}_{i+1/2}$ using the Riemann solver and the high-order HCS finite difference scheme. The positivity- and boundedness-preserving interpolation limiter and flux limiter are introduced in this section to respectively deal with the two issues mentioned.

\subsection{Positivity- and boundedness-preserving interpolation limiter}

The incremental-stencil WENO interpolation is robust but it is still not positivity-preserving for partial densities and squared speed of sound. It is also not boundedness-preserving for volume fractions. However, the first order interpolation with the left and right node values are in the admissible state set and are positivity-preserving and boundedness-preserving. Therefore, the high-order WENO interpolation can be limited with a convex combination of itself and the first order interpolation. For simplicity, this sub-section only discusses the limiting procedures for the left-biased WENO interpolation in the $x$ direction for the governing equations. Hence, the subscripts ``$L$" and ``$j$" are omitted.

The first stage of the interpolation limiting procedures is to obtain a limited conservative variable vector with positive partial densities at each midpoint, $\tilde{\Wv}_{i+\half}^{*}$. As described in algorithm~\ref{alg:interp_limit_densities}, $\tilde{\Wv}_{i+\half}^{*}$ is first initialized as the WENO interpolated conservative variable vector. It is then limited for positive partial densities through repeated convex combination of $\tilde{\Wv}_{i+\half}^{*}$ with the first order interpolated conservative variable vector for each phase using a user-defined small tolerance $\epsilon_{\alpha_k \rho_k}$. In the next stage, limiting procedure on the volume fractions can be applied similarly through another set of successive convex combinations given by algorithm~\ref{alg:interp_limit_vol_frac} with another user-defined small threshold $\epsilon_{\alpha_k}$. After this stage, the limited vector $\tilde{\Wv}_{i+\half}^{**}$ should have all partial densities (including mixture density) positive and all volume fractions bounded between $\epsilon_{\alpha_k}$ and $1 - \epsilon_{\alpha_k}$. This also means that all mass fractions are bounded between 0 and 1. The remaining quantity to limit is the squared speed of sound.

We can define a helper variable $\tilde{c}$ where $\tilde{c}^2 = ( \rho e - \overline p^\infty) / \rho$ such that squared sound speed and $\tilde{c}$ are related through $c^2 = \overline{\gamma} \left( \overline{\gamma} - 1 \right) \tilde{c}^2$.
As the final stage of the limiting procedures, $\tilde{\Wv}_{i+\half}^{***}$ with positive $\rho \tilde{c}^2$ can be obtained using algorithm~\ref{alg:interp_limit_sound}. This is carried out through convex combination of $\tilde{\Wv}_{i+\half}^{**}$ and the first order interpolated conservative variable vector with tolerance $\epsilon_{\rho \tilde{c}^2}$ by utilizing the convexity of the admissible set proved with the Jensen's inequality.
The squared speed of sound of the final limited conservative variable vector, $\tilde{\Wv}_{i+\half}^{***}$, is also positive as $c^2 = \overline \gamma(\overline \gamma -1) (\rho \tilde{c}^2) / \rho$. Note that $\overline \gamma$ is larger than 1 since volume fractions of $\tilde{\Wv}_{i+\half}^{**}$ are already bounded between 0 and 1. The positivity and boundedness limiting procedures for the right-biased WENO interpolation in the $x$ direction can be performed with $\Wv_{i+1}$ instead of $\Wv_i$ in a similar way. We have chosen the tolerances for the limiting procedures as $\epsilon_{\alpha_k \rho_k} = 1.0\mathrm{e}{-10}$, $\epsilon_{\alpha_k} = 1.0\mathrm{e}{-10}$, and $\epsilon_{\rho \tilde{c}^2} = 1.0\mathrm{e}{-8}$. Note that in practice, user should make sure the tolerances are chosen to be smaller than the minimum values of the corresponding initial fields.

Due to numerical round-off, the $\tilde{\Wv}_{i+\half}^{***}$ may still have negative partial densities, negative squared speed of sound, or out-of-bounds volume fractions. Therefore, a hard switch is suggested by setting $\tilde{\Wv}_{i+\half}^{***} = \Wv_i$ when $\alpha_k \rho_k \left( \tilde{\Wv}_{i+\half}^{***} \right) < \epsilon^{\mathrm{HS}}_{\alpha_k \rho_k}$, $\alpha_k \left( \tilde{\Wv}_{i+\half}^{***} \right) < \epsilon^{\mathrm{HS}}_{\alpha_k}$, or $\rho \tilde{c}^2 \left( \tilde{\Wv}_{i+\half}^{***} \right) < \epsilon^{\mathrm{HS}}_{\rho \tilde{c}^2 }$. The tolerances of the hard switch are chosen as $\epsilon^{\mathrm{HS}}_{\alpha_k \rho_k} = 1.0\mathrm{e}{-11}$,  $\epsilon^{\mathrm{HS}}_{\alpha_k} = 1.0\mathrm{e}{-11}$, and $\epsilon^{\mathrm{HS}}_{\rho \tilde{c}^2 } = 1.0\mathrm{e}{-9}$.

\begin{algorithm}[!ht]
\SetAlgoLined
  Set $\tilde{\Wv}_{i+\half}^{*} = \tilde{\Wv}_{i+\half}$\;
  \For{$k = 1,2$}{
    \For{all midpoints }{
      \uIf{$\alpha_k \rho_k \left( \Wv_i \right) < \epsilon_{\alpha_k \rho_k}$}{
        $\theta_{i+\half}=0$\; 
      }
      \uElseIf{$\alpha_k \rho_k \left( \tilde{\Wv}_{i+\half}^{*} \right) < \epsilon_{\alpha_k \rho_k}$}{
        Solve $\theta_{i+\half}$ from the formula:
        \begin{equation*}
          \left( 1 - \theta_{i+\half} \right) \alpha_k \rho_k \left( \Wv_i \right) + \theta_{i+\half} \alpha_k \rho_k \left( \tilde{\Wv}_{i+\half}^{*} \right) = \epsilon_{\alpha_k \rho_k}; 
        \end{equation*}
      }
      \Else{
        $\theta_{i+\half} = 1$\;
      }
      Perform convex combination:
      \begin{equation*}
        \tilde{\Wv}_{i+\half}^{*} = \left( 1 - \theta_{i+\half} \right) \Wv_i + \theta_{i+\half} \tilde{\Wv}_{i+\half}^{*} ;
      \end{equation*}
    }
  }
 \caption{Left-biased interpolation limiting procedure for positive partial densities.}
 \label{alg:interp_limit_densities}
\end{algorithm}

\begin{algorithm}[!hbt]
\SetAlgoLined
  Set $\tilde{\Wv}_{i+\half}^{**} = \tilde{\Wv}_{i+\half}^{*}$\;
  \For{$k = 1,2$}{
    \For{all midpoints }{
      \uIf{$\alpha_k \left( \Wv_i \right) < \epsilon_{\alpha_k}$}{
        $\theta_{i+\half}=0$\; 
      }
      \uElseIf{$\alpha_k \left( \tilde{\Wv}_{i+\half}^{**} \right) < \epsilon_{\alpha_k}$}{
        Solve $\theta_{i+\half}$ from the formula:
        \begin{equation*}
          \left( 1 - \theta_{i+\half} \right) \alpha_k \left( \Wv_i \right) + \theta_{i+\half} \alpha_k \left( \tilde{\Wv}_{i+\half}^{**} \right) = \epsilon_{\alpha_k};
        \end{equation*}
      }
      \Else{
        $\theta_{i+\half} = 1$\;
      }
      Perform convex combination:
      \begin{equation*}
        \tilde{\Wv}_{i+\half}^{**} = \left( 1 - \theta_{i+\half} \right) \Wv_i + \theta_{i+\half} \tilde{\Wv}_{i+\half}^{**} ;
      \end{equation*}
    }
  }
 \caption{Left-biased interpolation limiting procedure for bounded volume fractions.}
 \label{alg:interp_limit_vol_frac}
\end{algorithm}

\begin{algorithm}[!hbt]
\SetAlgoLined
    \For{all midpoints }{
      \uIf{$\rho \tilde{c}^2 \left( \Wv_i \right) < \epsilon_{\rho \tilde{c}^2 }$
      }{
        $\theta_{i+\half}=0$\; 
      }
      \uElseIf{$\rho \tilde{c}^2 \left( \tilde{\Wv}_{i+\half}^{**} \right) < \epsilon_{\rho \tilde{c}^2 }$}{
        Solve $\theta_{i+\half}$ from the formula:
        \begin{equation*}
          \left( 1 - \theta_{i+\half} \right) \rho \tilde{c}^2 \left( \Wv_i \right) + \theta_{i+\half} \rho \tilde{c}^2 \left( \tilde{\Wv}_{i+\half}^{**} \right) = \epsilon_{\rho \tilde{c}^2};
        \end{equation*}
      }
      \Else{
        $\theta_{i+\half} = 1$\;
      }
      Perform convex combination (applying Jensen's inequality):
      \begin{equation*}
        \tilde{\Wv}_{i+\half}^{***} = \left( 1 - \theta_{i+\half} \right) \Wv_i + \theta_{i+\half} \tilde{\Wv}_{i+\half}^{**} ;
      \end{equation*}
    }
 \caption{Left-biased interpolation limiting procedure for positive $\rho \tilde{c}^2$.}
 \label{alg:interp_limit_sound}
\end{algorithm}


\subsection{Positivity- and boundedness-preserving flux limiter}

The high-order flux reconstruction step of $\hat{\mathbf{G}}^{\pm}_{i+1/2}$ using a Riemann solver and the HCS finite difference scheme is not positivity- and boundedness-preserving in general and may cause numerical failures. Therefore, a flux limiter is critical to make sure that the flux used for time stepping gives physically admissible state. Following the same splitting idea introduced in the section of HLLC Riemann solver, if first order forward Euler time stepping is used for the WCNS, equation~\eqref{eq:semi_W_G_pm_eqn} can be rewritten as:
\begin{equation}
  \Wv_{i,j}^{n+1} =
    \sigma_x \underbrace{ \left( \frac{1}{2} \Wv_{i,j}^{x,-} + \frac{1}{2} \Wv_{i,j}^{x,+} \right) }_{\Wv_{i,j}^{x}} + \sigma_y \underbrace{ \left( \frac{1}{2} \Wv_{i,j}^{y,-} + \frac{1}{2} \Wv_{i,j}^{y,+} \right) }_{\Wv_{i,j}^{y}} ,
\end{equation}
where
\begin{align}
  \Wv_{i,j}^{x,\mp} &= 
    \Wv_{i,j}^{n} \pm 2 \lambda_x \left(
      \hat{\mathbf{G}}^{x,\pm}_{i\mp\half,j}
      - \Fxv_{i,j}
    \right) , \\
  \Wv_{i,j}^{y,\mp} &= 
    \Wv_{i,j}^{n} \pm 2 \lambda_y \left(
      \hat{\mathbf{G}}^{y,\pm}_{i,j\mp\half}
      - \Fyv_{i,j}
    \right) .
\end{align}
Note that $\hat{\mathbf{G}}^{x,\pm}_{i\mp\half,j}$ and $\hat{\mathbf{G}}^{y,\pm}_{i,j\mp\half}$ here are the high-order reconstructed fluxes in contrast to the first order HLLC fluxes, $\hat{\mathbf{G}}^{x,\mathrm{HLLC},\pm}_{i\mp\half,j}$ and $\hat{\mathbf{G}}^{y,\mathrm{HLLC},\pm}_{i,j\mp\half}$.
Since both $\sigma_x$ and $\sigma_y$ are positive and $\sigma_x + \sigma_y = 1$, $\Wv_{i,j}^{n+1}$ is a convex combination of $\Wv_{i,j}^{x,\pm}$ and $\Wv_{i,j}^{y,\pm}$. If all four conservative variable vectors are in the physically admissible set, $\Wv_{i,j}^{n+1}$ is also in the physically admissible set. For simplicity, only positivity- and boundedness-preserving flux limiting in $x$ direction for $\Wv_{i,j}^{x,\pm}$ is discussed here and hence ``$x$" superscript and ``$j$" subscript are dropped in the following part.

If the first order flux from the approximate Riemann solver is positively preserving such that all intermediate states given by the approximate Riemann solutions are in the admissible state set, such as the HLLC Riemann solver presented in this work, we can first construct the positivity flux limiting procedure for partial densities through the convex combination of the high-order solution $\Wv_i^{\pm}$ and the first order solution $\Wv_i^{\mathrm{HLLC}, \pm}$.
This first stage of the flux limiting procedures to obtain the limited flux $\hat{\Gv}_{i+\half}^{*,\pm}$ is detailed in algorithm~\ref{alg:flux_limit_densities} with tolerance $\epsilon_{\alpha_k \rho_k}$. After this stage, both solutions $\Wv_i^{*, \pm}$ time-advanced with the limited fluxes have all partial densities (including mixture density) positive, where
\begin{equation}
  \Wv_{i}^{*,\mp} = 
    \Wv_{i}^{n} \pm 2 \lambda \left(
      \hat{\mathbf{G}}^{*,\pm}_{i\mp\half}
      - \Fv_{i}
    \right) .
\end{equation}
Note that in the last step of the algorithm, the intention is to hybridize $( \hat{\Gv}_{i + \half}^{*,\pm} - \Fv_i )$ with $( \hat{\Gv}_{i + \half}^{\mathrm{HLLC},\pm} - \Fv_i )$ but the $\Fv_i$ on both sides of the equation cancel each other. Also, the flux limiting process is conservative for all equations except the last advection equation of volume fraction.

In the next stage, we can apply the boundedness flux limiting for volume fractions similarly to obtain the limited flux $\hat{\Gv}_{i+\half}^{**,\pm}$, which is given in algorithm~\ref{alg:flux_limit_vol_frac} with threshold $\epsilon_{\alpha_k}$. After this step, the limited solutions $\Wv_i^{**, \pm}$ should have all volume fractions bounded in addition to positive partial densities (including mixture density), where
\begin{equation}
  \Wv_{i}^{**,\mp} = 
    \Wv_{i}^{n} \pm 2 \lambda \left(
      \hat{\mathbf{G}}^{**,\pm}_{i\mp\half}
      - \Fv_{i}
    \right) .
\end{equation}

Finally, the positivity flux limiting for squared sound speed is conducted through the helper variable $\rho \tilde{c}^2$ similarly with the previous sub-section. This step is described in algorithm~\ref{alg:flux_limit_sound} with tolerance $\epsilon_{\rho \tilde{c}^2}$ and makes use of the fact that the admissible set of the conservative variable vector is convex to obtain the final limited flux $\hat{\Gv}_{i+\half}^{***,\pm}$. The squared speeds of sound $c^2$ of $\Wv_{i}^{***,\pm}$, where
\begin{equation}
  \Wv_{i}^{***,\mp} = 
    \Wv_{i}^{n} \pm 2 \lambda \left(
      \hat{\mathbf{G}}^{***,\pm}_{i\mp\half}
      - \Fv_{i}
    \right) ,
\end{equation}
are limited to be positive since $c^2 = \overline \gamma(\overline \gamma -1) (\rho \tilde{c}^2) / \rho$.

Note that the tolerance values ($\epsilon_{\alpha_k \rho_k}$,  $\epsilon_{\alpha_k}$, and $\epsilon_{\rho \tilde{c}^2 }$) are as same as those used in the positivity- and boundedness-preserving interpolation limiter.
Similar to the interpolation limiter, a hard switch is used by setting $\hat{\Gv}_{i+\half}^{***,\pm} = \hat{\Gv}_{i+\half}^{\mathrm{HLLC,\pm}}$ if either $\Wv_{i}^{***,+}$ or $\Wv_{i+1}^{***,-}$ are at states that are not bounded by smaller tolerances ($\epsilon^{\mathrm{HS}}_{\alpha_k \rho_k} = 1.0\mathrm{e}{-11}$,  $\epsilon^{\mathrm{HS}}_{\alpha_k} = 1.0\mathrm{e}{-11}$, and $\epsilon^{\mathrm{HS}}_{\rho \tilde{c}^2 } = 1.0\mathrm{e}{-9}$).

\begin{algorithm}[!ht]
\SetAlgoLined
  Set $\hat{\Gv}_{i+\half}^{*,\pm} = \hat{\Gv}_{i+\half}^{\pm}$ ($\Wv_i^{*, +} = \Wv_i^{+}$, $\Wv_{i+1}^{*, -} = \Wv_{i+1}^{-}$)\;
  \For{$k = 1,2$}{
    \For{all midpoints }{
      \uIf{$\alpha_k \rho_k \left( \Wv_i^{\mathrm{HLLC}, +} \right) < \epsilon_{\alpha_k \rho_k}$}{
        $\theta_{i+\half}^{+}=0$\; 
      }
      \uElseIf{$\alpha_k \rho_k \left( \Wv_i^{*, +} \right) < \epsilon_{\alpha_k \rho_k}$}{
        Solve $\theta_{i+\half}^{+}$ from the formula:
        \begin{equation*}
          \left( 1 - \theta_{i+\half}^{+} \right) \alpha_k \rho_k \left( \Wv_i^{\mathrm{HLLC}, +} \right) + \theta_{i+\half}^{+} \alpha_k \rho_k \left( \Wv_i^{*, +} \right) = \epsilon_{\alpha_k \rho_k}; 
        \end{equation*}
      }
      \Else{
        $\theta_{i+\half}^{+}=1$\;
      }
      
      \uIf{$\alpha_k \rho_k \left( \Wv_{i+1}^{\mathrm{HLLC}, -} \right) < \epsilon_{\alpha_k \rho_k}$}{
        $\theta_{i+\half}^{-}=0$\; 
      }
      \uElseIf{$\alpha_k \rho_k \left( \Wv_{i+1}^{*, -} \right) < \epsilon_{\alpha_k \rho_k}$}{
        Solve $\theta_{i+\half}^{-}$ from the formula:
        \begin{equation*}
          \left( 1 - \theta_{i+\half}^{-} \right) \alpha_k \rho_k \left( \Wv_{i+1}^{\mathrm{HLLC}, -} \right) + \theta_{i+\half}^{-} \alpha_k \rho_k \left( \Wv_{i+1}^{*, -} \right) = \epsilon_{\alpha_k \rho_k}; 
        \end{equation*}
      }
      \Else{
        $\theta_{i+\half}^{-}=1$\;
      }
      Set $\theta_{i+\half}=\min\left( \theta_{i+\half}^{+}, \theta_{i+\half}^{-} \right)$\;
      Perform convex combination:
      \begin{align*}
        \hat{\Gv}_{i+\half}^{*,+} = \left( 1 - \theta_{i+\half} \right) \hat{\Gv}_{i+\half}^{\mathrm{HLLC,+}} + \theta_{i+\half} \hat{\Gv}_{i+\half}^{*,+}; \\
        \hat{\Gv}_{i+\half}^{*,-} = \left( 1 - \theta_{i+\half} \right) \hat{\Gv}_{i+\half}^{\mathrm{HLLC,-}} + \theta_{i+\half} \hat{\Gv}_{i+\half}^{*,-};
      \end{align*}
    }
  }
 \caption{Flux limiting procedure for positive partial densities.}
 \label{alg:flux_limit_densities}
\end{algorithm}

\begin{algorithm}[!ht]
\SetAlgoLined
  Set $\hat{\Gv}_{i+\half}^{**,\pm} = \hat{\Gv}_{i+\half}^{*,\pm}$ ($\Wv_i^{**, +} = \Wv_i^{*, +}$, $\Wv_{i+1}^{**, -} = \Wv_{i+1}^{*, -}$)\;
  \For{$k = 1,2$}{
    \For{all midpoints }{
      \uIf{$\alpha_k \left( \Wv_i^{\mathrm{HLLC}, +} \right) < \epsilon_{\alpha_k}$}{
        $\theta_{i+\half}^{+}=0$\; 
      }
      \uElseIf{$\alpha_k \left( \Wv_i^{**, +} \right) < \epsilon_{\alpha_k}$}{
        Solve $\theta_{i+\half}^{+}$ from the formula:
        \begin{equation*}
          \left( 1 - \theta_{i+\half}^{+} \right) \alpha_k \left( \Wv_i^{\mathrm{HLLC}, +} \right) + \theta_{i+\half}^{+} \alpha_k \left( \Wv_i^{**, +} \right) = \epsilon_{\alpha_k}; 
        \end{equation*}
      }
      \Else{
        $\theta_{i+\half}^{+}=1$\;
      }
      
      \uIf{$\alpha_k \left( \Wv_{i+1}^{\mathrm{HLLC}, -} \right) < \epsilon_{\alpha_k}$}{
        $\theta_{i+\half}^{-}=0$\; 
      }
      \uElseIf{$\alpha_k \left( \Wv_{i+1}^{**, -} \right) < \epsilon_{\alpha_k}$}{
        Solve $\theta_{i+\half}^{-}$ from the formula:
        \begin{equation*}
          \left( 1 - \theta_{i+\half}^{-} \right) \alpha_k \left( \Wv_{i+1}^{\mathrm{HLLC}, -} \right) + \theta_{i+\half}^{-} \alpha_k \left( \Wv_{i+1}^{**, -} \right) = \epsilon_{\alpha_k}; 
        \end{equation*}
      }
      \Else{
        $\theta_{i+\half}^{-}=1$\;
      }
      Set $\theta_{i+\half}=\min\left( \theta_{i+\half}^{+}, \theta_{i+\half}^{-} \right)$\;
      Perform convex combination:
    \begin{align*}
      \hat{\Gv}_{i+\half}^{**,+} = \left( 1 - \theta_{i+\half} \right) \hat{\Gv}_{i+\half}^{\mathrm{HLLC,+}} + \theta_{i+\half} \hat{\Gv}_{i+\half}^{**,+}; \\
      \hat{\Gv}_{i+\half}^{**,-} = \left( 1 - \theta_{i+\half} \right) \hat{\Gv}_{i+\half}^{\mathrm{HLLC,-}} + \theta_{i+\half} \hat{\Gv}_{i+\half}^{**,-};
    \end{align*}
    }
  }
 \caption{Flux limiting procedure for bounded volume fractions.}
 \label{alg:flux_limit_vol_frac}
\end{algorithm}

\begin{algorithm}[!ht]
\SetAlgoLined
    \For{all midpoints }{
      \uIf{$\rho \tilde{c}^2 \left( \Wv_i^{\mathrm{HLLC}, +} \right) < \epsilon_{\rho \tilde{c}^2 }$}{
        $\theta_{i+\half}^{+}=0$\; 
      }
      \uElseIf{$\rho \tilde{c}^2 \left( \Wv_i^{**, +} \right) < \epsilon_{\rho \tilde{c}^2 }$}{
        Solve $\theta_{i+\half}^{+}$ from the formula:
        \begin{equation*}
          \left( 1 - \theta_{i+\half}^{+} \right) \rho \tilde{c}^2 \left( \Wv_i^{\mathrm{HLLC}, +} \right) + \theta_{i+\half}^{+} \rho \tilde{c}^2 \left( \Wv_i^{**, +} \right) = \epsilon_{\rho \tilde{c}^2 }; 
        \end{equation*}
      }
      \Else{
        $\theta_{i+\half}^{+}=1$\;
      }
      
      \uIf{$\rho \tilde{c}^2 \left( \Wv_{i+1}^{\mathrm{HLLC}, -} \right) < \epsilon_{\rho \tilde{c}^2 }$}{
        $\theta_{i+\half}^{-}=0$\; 
      }
      \uElseIf{$\rho \tilde{c}^2 \left( \Wv_{i+1}^{**, -} \right) < \epsilon_{\rho \tilde{c}^2 }$}{
        Solve $\theta_{i+\half}^{-}$ from the formula:
        \begin{equation*}
          \left( 1 - \theta_{i+\half}^{-} \right) \rho \tilde{c}^2 \left( \Wv_{i+1}^{\mathrm{HLLC}, -} \right) + \theta_{i+\half}^{-} \rho \tilde{c}^2 \left( \Wv_{i+1}^{**, -} \right) = \epsilon_{\rho \tilde{c}^2 }; 
        \end{equation*}
      }
      \Else{
        $\theta_{i+\half}^{-}=1$\;
      }
      Set $\theta_{i+\half}=\min\left( \theta_{i+\half}^{+}, \theta_{i+\half}^{-} \right)$\;
      Perform convex combination (applying Jensen's inequality):
      \begin{align*}
        \hat{\Gv}_{i+\half}^{***,+} &= \left( 1 - \theta_{i+\half} \right) \hat{\Gv}_{i+\half}^{\mathrm{HLLC,+}} + \theta_{i+\half} \hat{\Gv}_{i+\half}^{**,+}; \\
        \hat{\Gv}_{i+\half}^{***,-} &= \left( 1 - \theta_{i+\half} \right) \hat{\Gv}_{i+\half}^{\mathrm{HLLC,-}} + \theta_{i+\half} \hat{\Gv}_{i+\half}^{**,-};
      \end{align*}
    }
 \caption{Flux limiting procedure for positive $\rho \tilde{c}^2$.}
 \label{alg:flux_limit_sound}
\end{algorithm}

\subsection{Extension to strong stability preserving Runge--Kutta time stepping scheme}

The extension of positivity- and boundedness-preserving limiters for high-order flux with the high-order strong stability preserving Runge--Kutta (SSPRK) time stepping methods~\cite{shu1988total,gottlieb2009high,gottlieb2001strong} is trivial since SSPRK time stepping schemes are convex combinations of Euler forward steps. However, the upper limit of CFL number is still constrained by 0.5.

\section{Test problems}

1D and 2D test problems are conducted with the first order HLLC scheme and the high-order WCNS-IS with the positivity- and boundedness-preserving limiters. The two schemes are termed HLLC and PP-WCNS-IS respectively in this work. All tests involve liquid water and air as an ideal gas. The properties of the fluids are given in table~\ref{table:fluid_properties}. Note that the fluids satisfy the requirement from the positivity- and boundedness-preserving limiters as the ratio of specific heats of water is larger than that of air. The three-stage third order SSPRK scheme (TVDRK3) \cite{shu1988total} is used for time stepping for both schemes.
The CFL number is chosen to be 0.5 unless constant time step size is used. When constant time step size is used, the corresponding CFL number is always less than 0.5 until the end of simulations.

\begin{table}[!ht]
  \begin{center}
  \begin{tabular}{@{}c | ccc@{}}\toprule
    Fluid & \textnormal{Phase number} & $\gamma$ & \addstackgap{$p_{\infty}\ (\mathrm{Pa})$} \\ \midrule
    Liquid water & 1 & $6.12$ & $3.43\mathrm{e}{8}$ \\
    Air   & 2 & $1.40$ & $0$ \\ \bottomrule
  \end{tabular}
  \caption{Properties of the fluids.}
  \label{table:fluid_properties}
  \end{center}
\end{table}

\subsection{Convergence study}

To verify the formal order of accuracy of each scheme, advection of volume fraction disturbance in a 2D periodic domain $[-1, 1) \ \mathrm{m} \times[-1, 1) \ \mathrm{m}$ is used as the test problem similar to that in~\cite{wong2017high}. The initial conditions are given by table~\ref{table:IC_2D_convergence} and the exact solutions are given by table~\ref{table:exact_2D_convergence}.
\begin{table}[!ht]
  \begin{center}
    \begin{tabular}{@{}cccccc@{}}\toprule
    \addstackgap{\stackanchor{$\rho_1$}{$(\mathrm{kg\ m^{-3}})$}} &
    \stackanchor{$\rho_2$}{$(\mathrm{kg\ m^{-3}})$} &
    \stackanchor{$u$}{$(\mathrm{m\ s^{-1}})$} &
    \stackanchor{$v$}{$(\mathrm{m\ s^{-1}})$} &
    \stackanchor{$p$}{$(\mathrm{Pa})$} &
    $\alpha_1$ \\ \midrule
    \addstackgap{1000} & 1 & 10 & 10 & 101325 & $0.5 + 0.25  \sin \left[ \pi (x + y) \right]$ \\\bottomrule
    \end{tabular}
  \end{center}
  \caption{Initial conditions of 2D convergence problem.}
  \label{table:IC_2D_convergence}
\end{table}

\begin{table}[!ht]
  \begin{center}
    \begin{tabular}{@{}cccccc@{}}\toprule
    \addstackgap{\stackanchor{$\rho_1$}{$(\mathrm{kg\ m^{-3}})$}} &
    \stackanchor{$\rho_2$}{$(\mathrm{kg\ m^{-3}})$} &
    \stackanchor{$u$}{$(\mathrm{m\ s^{-1}})$} &
    \stackanchor{$v$}{$(\mathrm{m\ s^{-1}})$} &
    \stackanchor{$p$}{$(\mathrm{Pa})$} &
    $\alpha_1$ \\ \midrule
    \addstackgap{1000} & 1 & 10 & 10 & 101325 & $0.5 + 0.25 \sin \left[ \pi (x + y - 20t) \right]$ \\\bottomrule
    \end{tabular}
  \end{center}
  \caption{Exact solutions of 2D convergence problem.}
  \label{table:exact_2D_convergence}
\end{table}

\noindent Simulations using different schemes are conducted up to $t = 0.1\ \mathrm{ms}$ with mesh refinements from $N_x = N_y = 8$ to $N_x = N_y = 256$. All simulations are run with very small constant time steps in order to observe the spatial orders of accuracy of different numerical schemes. $\Delta t / \Delta x = 4.0e-5 \ \mathrm{s\ m^{-1}}$ is used.

Table~\ref{table:2D_L2_error_and_rate_of_convergence_advection_volume_fraction} shows the $L_2$ errors and the computed rates of convergence of volume fraction $\alpha_1$ respectively by the two schemes at $t = 0.1\ \mathrm{ms}$. The $L_2$ error of volume fraction is computed as:
\begin{align}
    L_2\ \mathrm{error}&= \sqrt{ \sum_{i=0}^{N-1} \sum_{j=0}^{N-1} \Delta x \Delta y \left( {\alpha_1}_{i, j} - {\alpha_1}_{i, j}^{\mathrm{exact}} \right)^2 / \sum_{i=0}^{N-1} \sum_{j=0}^{N-1} \Delta x \Delta y },
\end{align}
where ${\alpha_1}_{i, j}^{\mathrm{exact}}$ is the exact solution of volume fraction at the corresponding grid point. It can be seen from the table that all schemes achieve the expected rates of convergence.

\begin{table}[!ht]
\centering
\small
\begin{tabular}{@{}c | cccc@{}}\toprule
    Number of & \multicolumn{2}{c}{HLLC} & \multicolumn{2}{c}{PP-WCNS-IS} \\
    \cmidrule(r){2-5}
    grid points & error & order & error & order \\ \midrule
      $8^2$ & $4.283\mathrm{e}{-04}$ &      & $3.136\mathrm{e}{-05}$ &      \\
     $16^2$ & $2.170\mathrm{e}{-04}$ & 0.98 & $1.517\mathrm{e}{-07}$ & 7.69 \\
     $32^2$ & $1.089\mathrm{e}{-04}$ & 0.99 & $4.855\mathrm{e}{-09}$ & 4.97 \\
     $64^2$ & $5.450\mathrm{e}{-05}$ & 1.00 & $1.679\mathrm{e}{-10}$ & 4.85 \\
    $128^2$ & $2.726\mathrm{e}{-05}$ & 1.00 & $5.382\mathrm{e}{-12}$ & 4.96 \\
    $256^2$ & $1.363\mathrm{e}{-05}$ & 1.00 & $1.714\mathrm{e}{-13}$ & 4.97 \\ \bottomrule
\end{tabular}
\caption{$L_2$ errors and orders of convergence of volume fraction for the 2D advection problem from different schemes at $t = 0.1\ \mathrm{ms}$.}
\label{table:2D_L2_error_and_rate_of_convergence_advection_volume_fraction}
\end{table}

\subsection{One-dimensional material interface advection}

The next multi-phase problem is a 1D problem with the advection of two material interfaces. The settings of this problem are similar to those in \cite{coralic2014finite,wong2017high,aslani2018localized}. The initial conditions are given by table~\ref{table:IC_1D_material_interface_advection}. Periodic conditions are applied at both boundaries. The spatial domain is $x \in \left[0, 1 \right) \ \mathrm{m}$ and the final time is at $t = 0.01 \ \mathrm{s}$. Simulations are evolved with constant time steps $\dt = 1.25\mathrm{e}{-6} \ \mathrm{s}$ on a uniform grid with 200 grid points where $\dx = 0.005 \ \mathrm{m}$. The two material interfaces have exactly advected one period at the end of the simulations.

\begin{table}[!ht]
  \begin{center}
    \begin{tabular}{@{}c | ccccc@{}}\toprule
     &
    \addstackgap{\stackanchor{$\alpha_1 \rho_1$}{$(\mathrm{kg\ m^{-3}})$}} &
    \stackanchor{$\alpha_2 \rho_2$}{$(\mathrm{kg\ m^{-3}})$} &
    \stackanchor{$u$}{$(\mathrm{m\ s^{-1}})$} &
    \stackanchor{$p$}{$(\mathrm{Pa})$} &
    $\alpha_1$ \\ \midrule
    \addstackgap{$0.25 \leq x < 0.75$} & 1000 & $1.0\mathrm{e}{-8}$ & 100 & 101325 & $1 - 1.0\mathrm{e}{-8}$ \\
    \addstackgap{otherwise} & $1.0\mathrm{e}{-8}$ & 1.204 & 100 & 101325 & $1.0\mathrm{e}{-8}$ \\ \bottomrule
    \end{tabular}
  \end{center}
  \caption{Initial conditions of 1D material interface advection problem.}
  \label{table:IC_1D_material_interface_advection}
\end{table}

\begin{figure}[!ht]
\centering
\subfigure[Global density profile]{%
\includegraphics[width=0.48\textwidth]{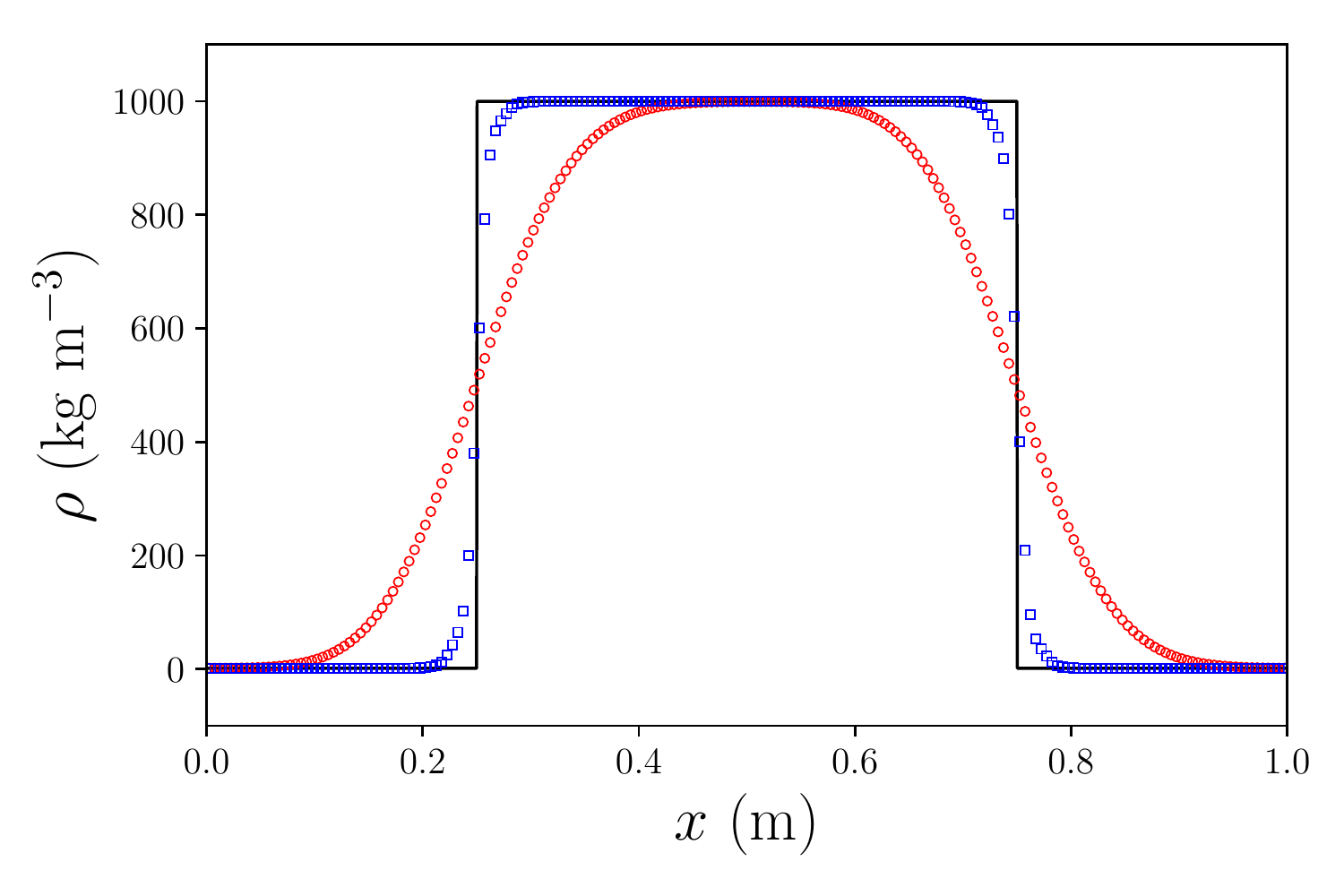}
\label{fig:compare_material_interface_advection_rho_global}}
\subfigure[Local density profile]{%
\includegraphics[width=0.48\textwidth]{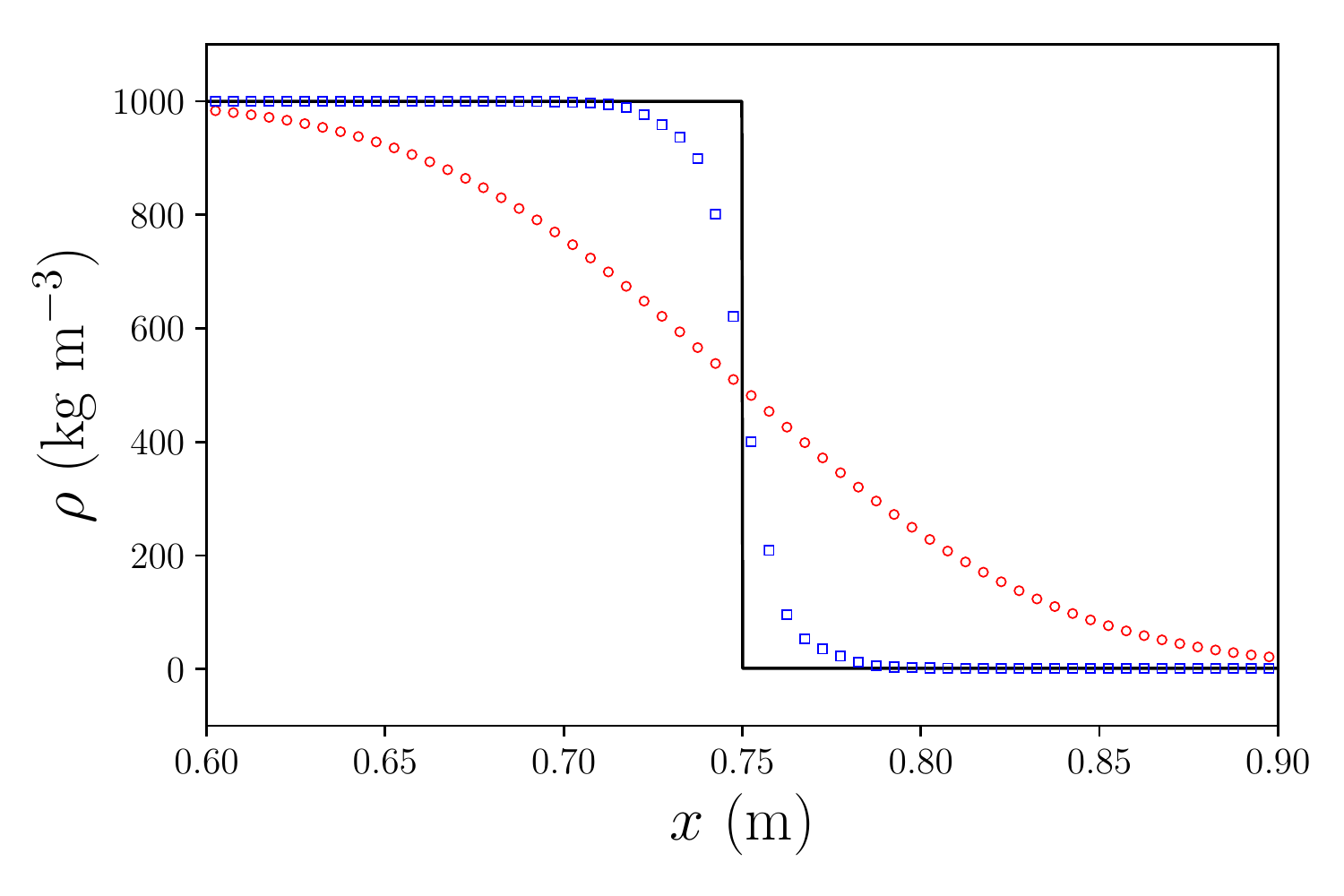}
\label{fig:compare_material_interface_advection_rho_local}}
\caption{Material interface advection problem at $t = 0.01 \ \mathrm{s}$ using different schemes. Black solid line: exact; red circles: HLLC; blue squares: PP-WCNS-IS.}
\label{fig:compare_material_interface_advection_rho}
\end{figure}

\begin{figure}[!ht]
\centering
\subfigure[Velocity]{%
\includegraphics[width=0.48\textwidth]{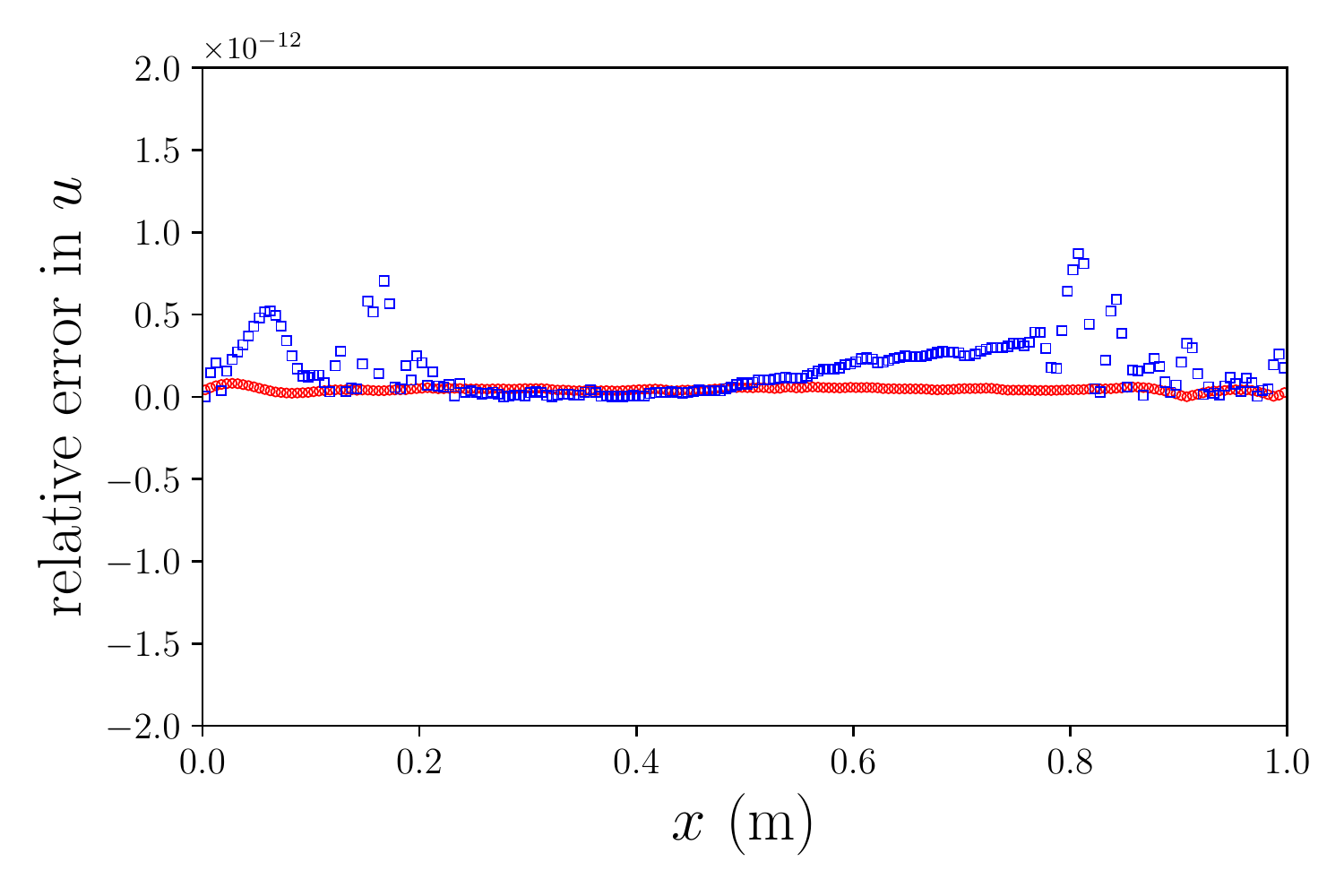}
\label{fig:compare_material_interface_advection_u_global}}
\subfigure[Pressure]{%
\includegraphics[width=0.48\textwidth]{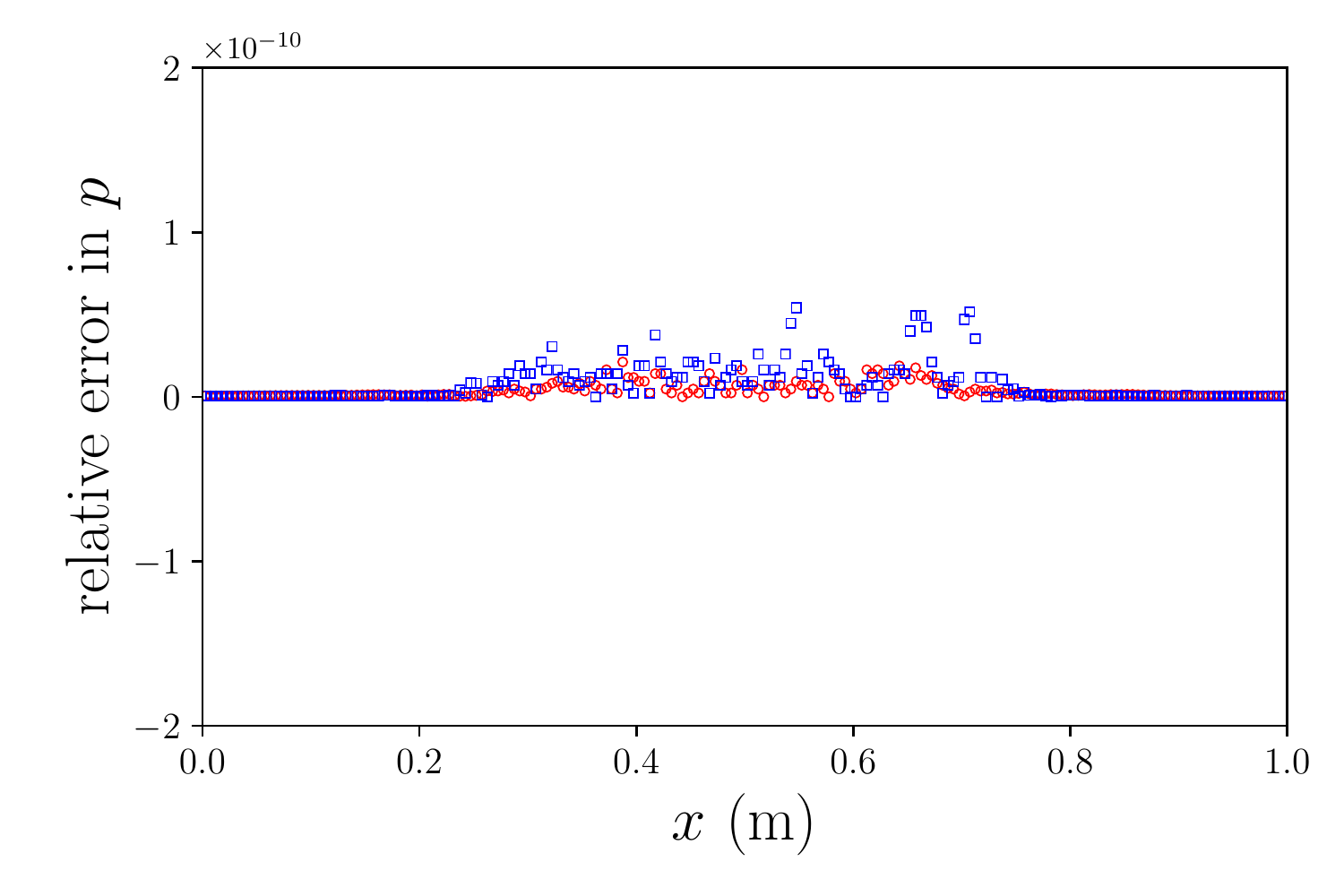}
\label{fig:compare_material_interface_advection_p_global}}
\caption{Relative errors for the material interface advection problem at $t = 0.01 \ \mathrm{s}$ using different schemes. Red circles: HLLC; blue squares: PP-WCNS-IS.}
\label{fig:compare_material_interface_advection_error}
\end{figure}

The density fields obtained with the two schemes at the final simulation time are compared with the exact solution in figure~\ref{fig:compare_material_interface_advection_rho}. It can be seen that the high-order PP-WCNS-IS can capture the material interfaces with much smaller numerical widths compared to the first order HLLC scheme and no spurious oscillations are observed at the two material interfaces for PP-WCNS-IS. Both the velocity and pressure fields are uniform and constant in this advection problem. In figure~\ref{fig:compare_material_interface_advection_error}, it can be seen that the relative errors in velocity and pressure fields for both schemes are insignificantly small and the uniform and constant fields are maintained well over time.

\subsection{One-dimensional gas/liquid Sod shock tube problem}

This gas/liquid shock tube problem is taken from~\citet{chen2008flow} and \citet{wang2018incremental}. The initial conditions are given by table~\ref{table:IC_1D_gas_liquid_Sod_shock_tube_problem}. Extrapolations are applied at both boundaries. The spatial domain is $x \in \left[0, 1.5 \right] \ \mathrm{m}$ and the final time is at $t = 3\mathrm{e}{-4} \ \mathrm{s}$. Simulations are evolved with constant time steps $\Delta t = 1.25\mathrm{e}{-6} \ \mathrm{s}$ on a uniform grid with 200 grid points. The reference solutions are obtained using PP-WCNS-IS with 4000 grid points.

\begin{table}[!ht]
  \begin{center}
    \begin{tabular}{@{}c | ccccc@{}}\toprule
     &
    \addstackgap{\stackanchor{$\alpha_1 \rho_1$}{$(\mathrm{kg\ m^{-3}})$}} &
    \stackanchor{$\alpha_2 \rho_2$}{$(\mathrm{kg\ m^{-3}})$} &
    \stackanchor{$u$}{$(\mathrm{m\ s^{-1}})$} &
    \stackanchor{$p$}{$(\mathrm{Pa})$} &
    $\alpha_1$ \\ \midrule
    \addstackgap{$x < 0.8$} & 1000 & $1.0\mathrm{e}{-8}$ & 0 & $1.0\mathrm{e}{9}$ & $1 - 1.0\mathrm{e}{-8}$ \\
    \addstackgap{$x \geq 0.8$} & $1.0\mathrm{e}{-8}$ & 20 & 0 & $1.0\mathrm{e}{5}$ & $1.0\mathrm{e}{-8}$ \\ \bottomrule
    \end{tabular}
  \end{center}
  \caption{Initial conditions of 1D gas/liquid Sod shock tube problem.}
  \label{table:IC_1D_gas_liquid_Sod_shock_tube_problem}
\end{table}

\begin{figure}[!ht]
\centering
\subfigure[Global density profile]{%
\includegraphics[width=0.45\textwidth]{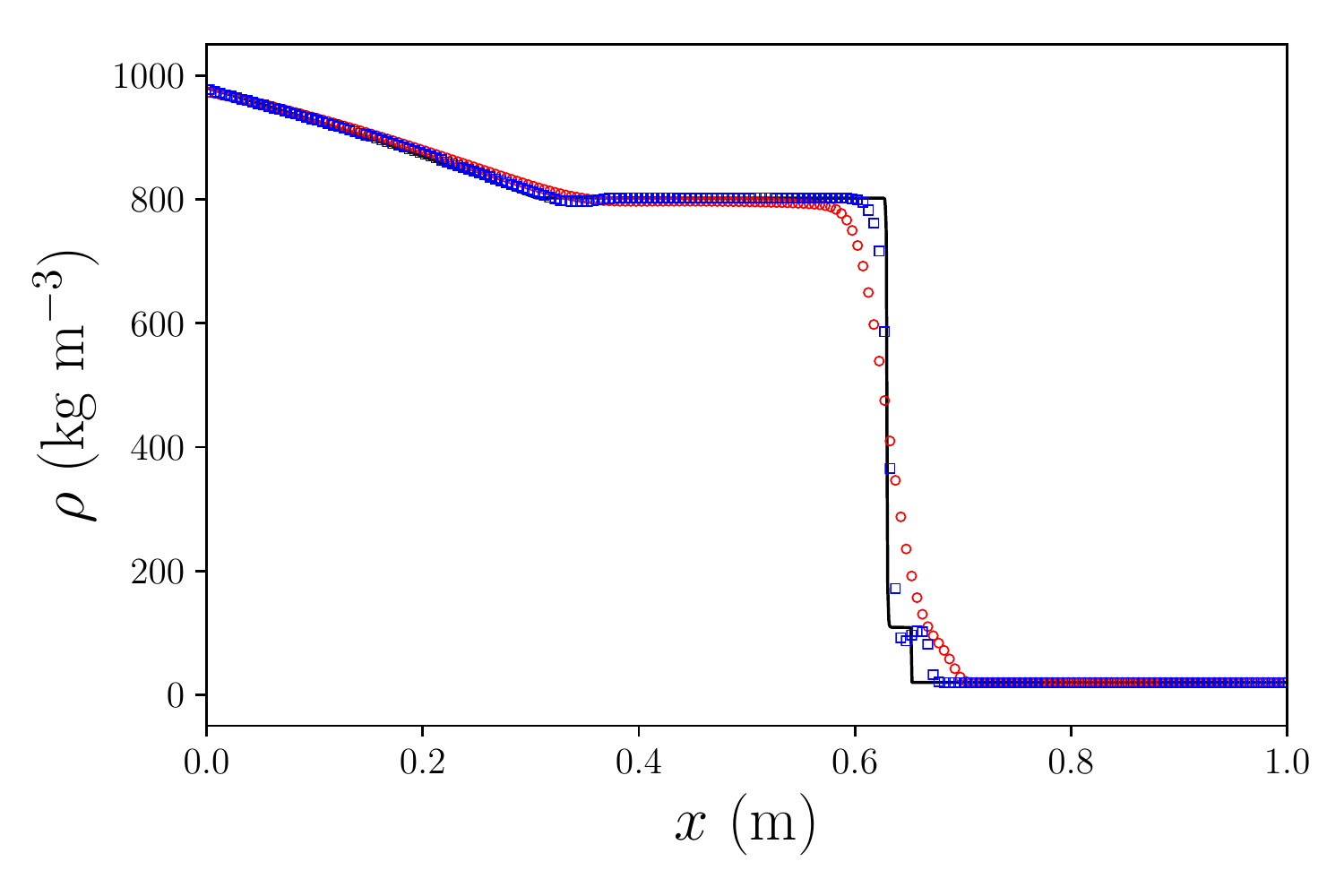}
\label{fig:compare_gas_liquid_Sod_rho_global}}
\subfigure[Global velocity profile]{%
\includegraphics[width=0.45\textwidth]{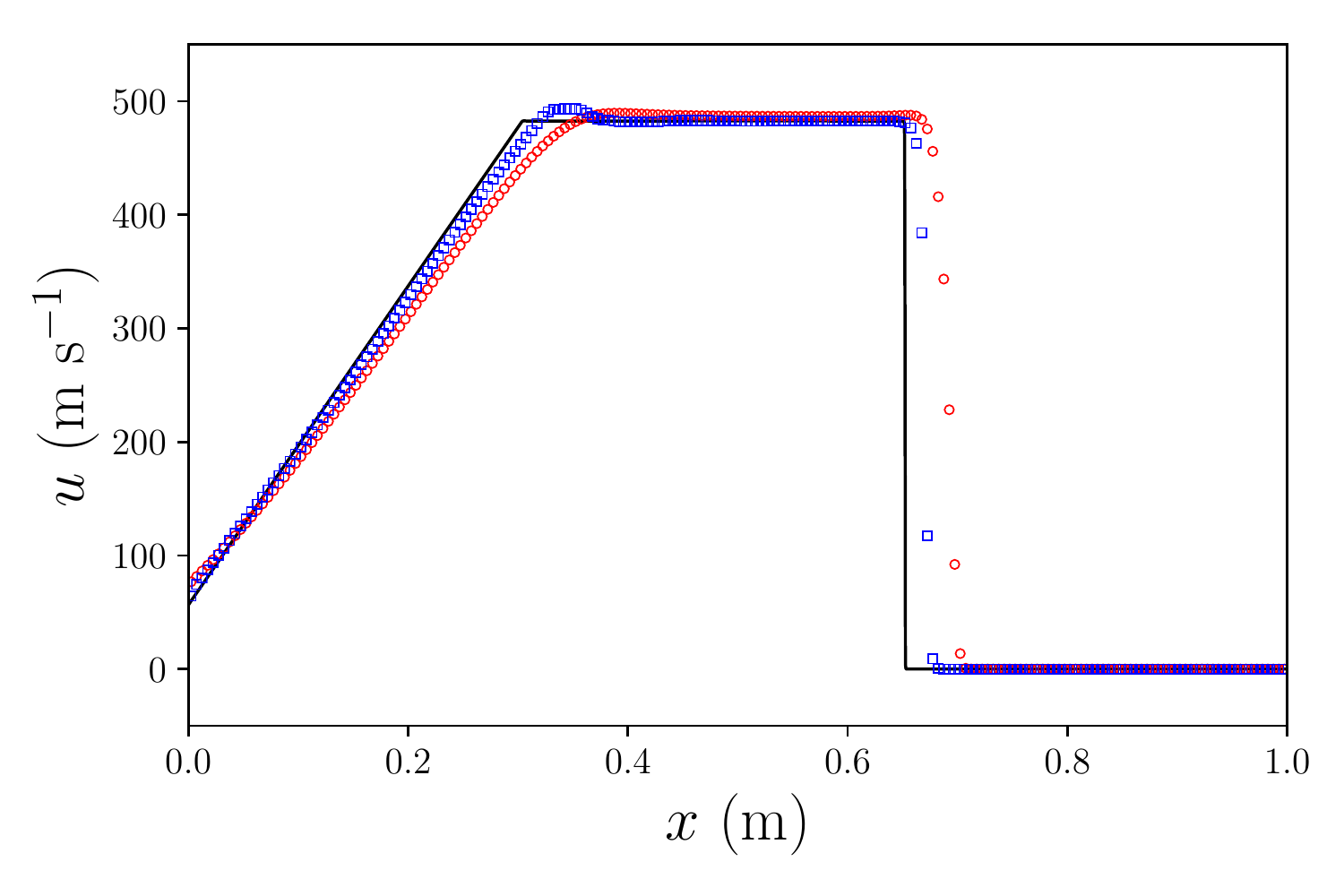}
\label{fig:compare_gas_liquid_Sod_u_global}}
\caption{Gas/liquid Sod shock tube problem at $t = 3\mathrm{e}{-4} \ \mathrm{s}$ using different schemes. Black solid line: reference; red circles: HLLC; blue squares: PP-WCNS-IS.}
\label{fig:compare_gas_liquid_Sod}
\end{figure}

\begin{figure}[!ht]
\centering
\subfigure[Global pressure profile]{%
\includegraphics[width=0.45\textwidth]{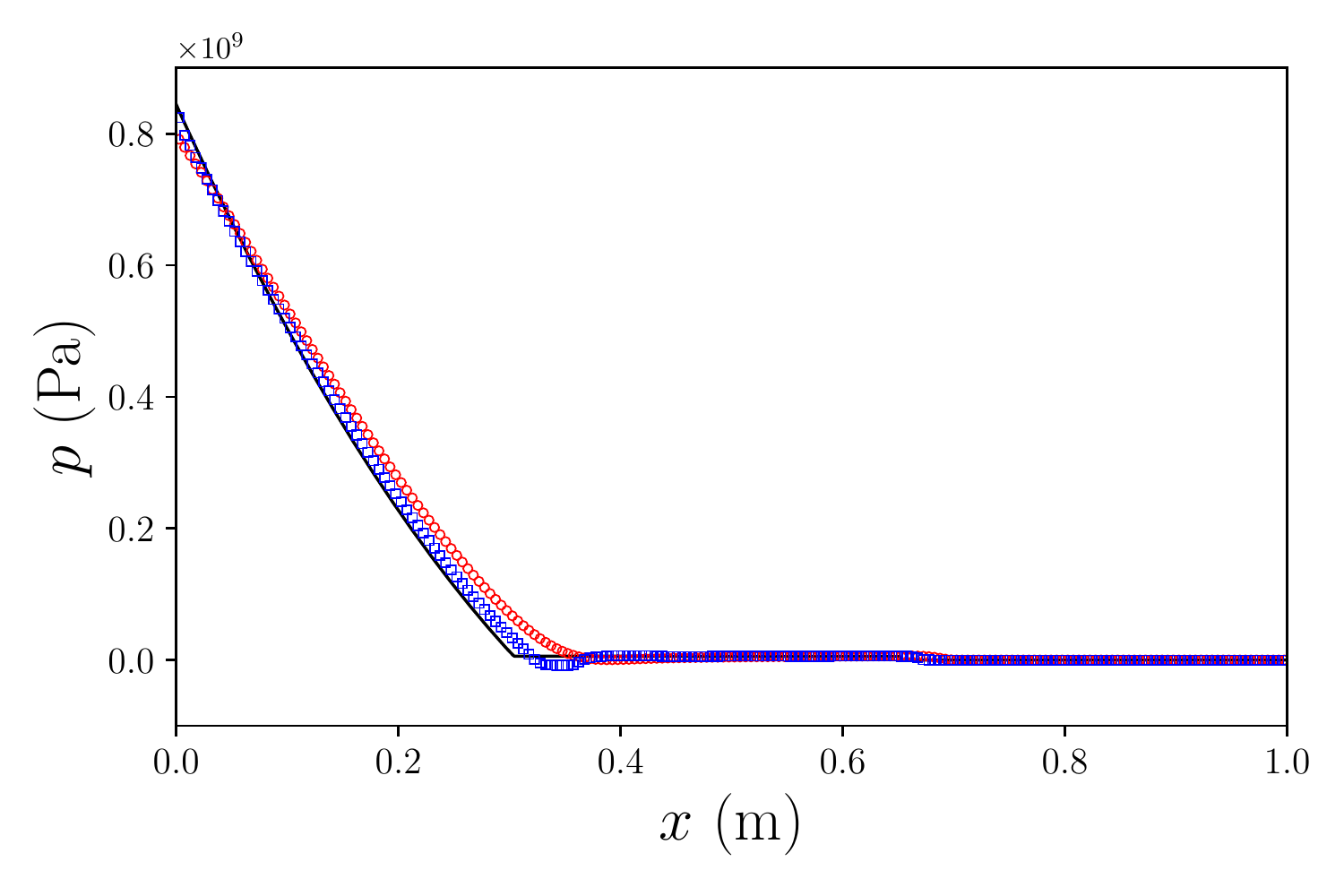}
\label{fig:compare_gas_liquid_Sod_p_global}}
\subfigure[Local pressure profile]{%
\includegraphics[width=0.45\textwidth]{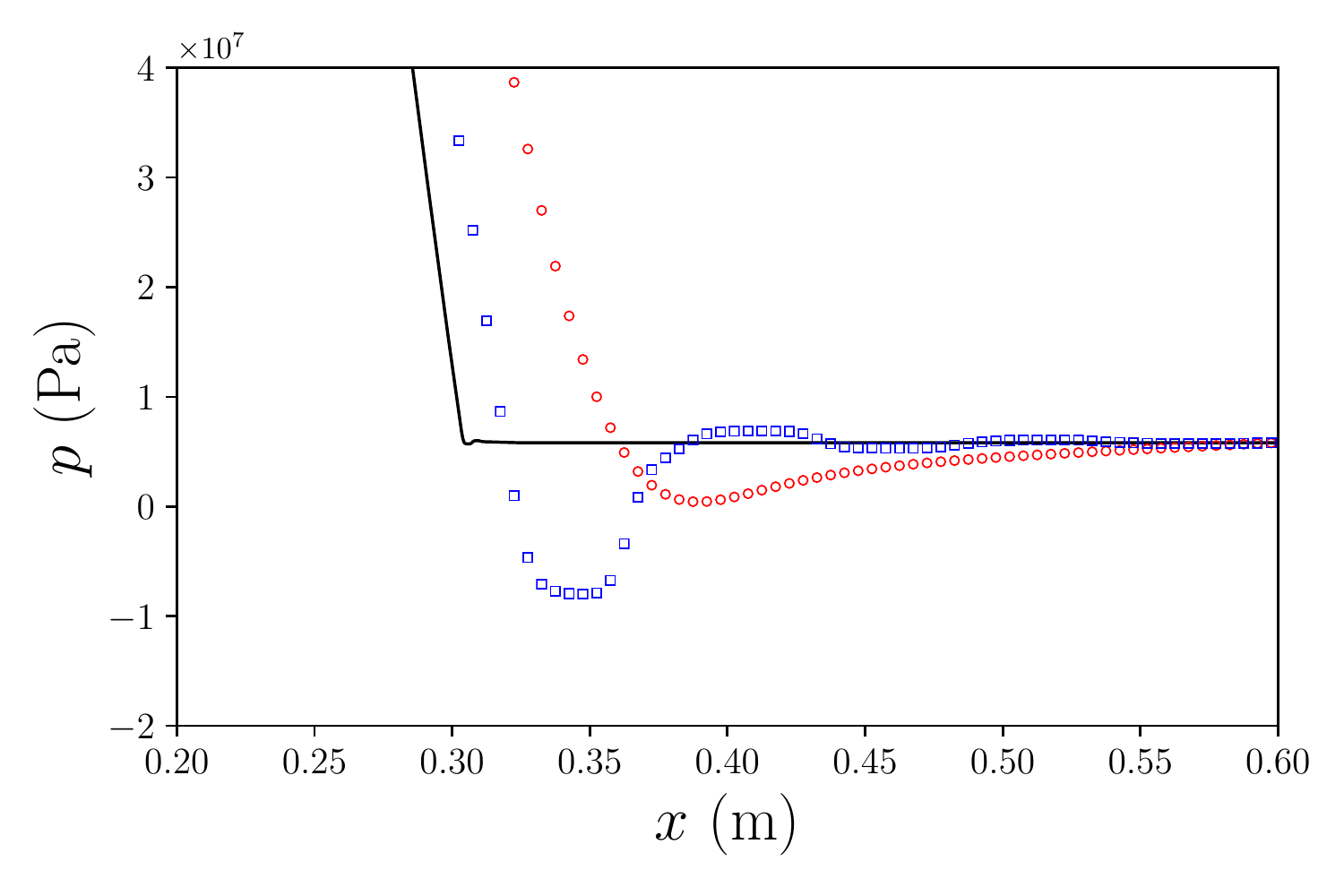}
\label{fig:compare_gas_liquid_Sod_p_local}}
\caption{Gas/liquid Sod shock tube problem at $t = 3\mathrm{e}{-4} \ \mathrm{s}$ using different schemes. Black solid line: reference; red circles: HLLC; blue squares: PP-WCNS-IS.}
\label{fig:compare_gas_liquid_Sod_p}
\end{figure}

Figures~\ref{fig:compare_gas_liquid_Sod} and \ref{fig:compare_gas_liquid_Sod_p} compare the numerical solutions from the two schemes with the reference solutions. In figure~\ref{fig:compare_gas_liquid_Sod_rho_global}, it can be seen that PP-WCNS-IS can capture both the material interface (the left larger density jump) and the shock (the right smaller density jump) with only a few grid points. On the other hand, the first order HLLC scheme is too dissipative that both the material interface and shock are smeared out severely, and hence the two discontinuities cannot be distinguished at this grid resolution. In figures~\ref{fig:compare_gas_liquid_Sod_u_global} and \ref{fig:compare_gas_liquid_Sod_p_global}, it can be seen that PP-WCNS-IS can give more accurate solutions in velocity and pressure fields compared to first order HLLC scheme. However, as seen in figure~\ref{fig:compare_gas_liquid_Sod_p_local}, PP-WCNS-IS produces a slightly larger undershoot and overshoot at the expansion fan. 

\subsection{One-dimensional planar multi-material Sedov blast wave problem}

This is a multi-material version modified from the well-known single-phase 1D planar Sedov blast wave problem~\cite{sedov1993similarity,zhang2012positivity,hu2013positivity}. Initially there is a singularity of highly pressurized air at the center of the domain filled with very low pressure water. Blast waves are created at the original position of the singularity and propagate towards the domain boundaries. The initial conditions are given by table~\ref{table:IC_1D_multmaterial_Sedov}. Extrapolations are applied at both boundaries. The spatial domain is $x \in \left[0, 4.0 \right] \ \mathrm{m}$ and the final time is at $t = 1\mathrm{e}{-3} \ \mathrm{s}$. Simulations are evolved with constant time steps $\Delta t = 2.5\mathrm{e}{-7} \ \mathrm{s}$ on a uniform grid with 401 grid points. The reference solutions are generated using PP-WCNS-IS with 4001 grid points.

\begin{table}[!ht]
  \begin{center}
    \begin{tabular}{@{}c | ccccc@{}}\toprule
     &
    \addstackgap{\stackanchor{$\alpha_1 \rho_1$}{$(\mathrm{kg\ m^{-3}})$}} &
    \stackanchor{$\alpha_2 \rho_2$}{$(\mathrm{kg\ m^{-3}})$} &
    \stackanchor{$u$}{$(\mathrm{m\ s^{-1}})$} &
    \stackanchor{$p$}{$(\mathrm{Pa})$} &
    $\alpha_1$ \\ \midrule
    \addstackgap{\stackanchor{$x < 2 - 0.5 \Delta x$}{or $x > 2 + 0.5 \Delta x$}} & 1000 & $1.0\mathrm{e}{-8}$ & 0 & $-3.0\mathrm{e}{7}$ & $1 - 1.0\mathrm{e}{-8}$ \\
    \addstackgap{otherwise} & $1.0\mathrm{e}{-8}$ & 1 & 0 & $1.28\mathrm{e}{6}/\Delta x$ & $1.0\mathrm{e}{-8}$ \\ \bottomrule
    \end{tabular}
  \end{center}
  \caption{Initial conditions of 1D planar multi-material Sedov blast wave problem.}
  \label{table:IC_1D_multmaterial_Sedov}
\end{table}

\begin{figure}[!ht]
\centering
\subfigure[Global density profile]{%
\includegraphics[width=0.45\textwidth]{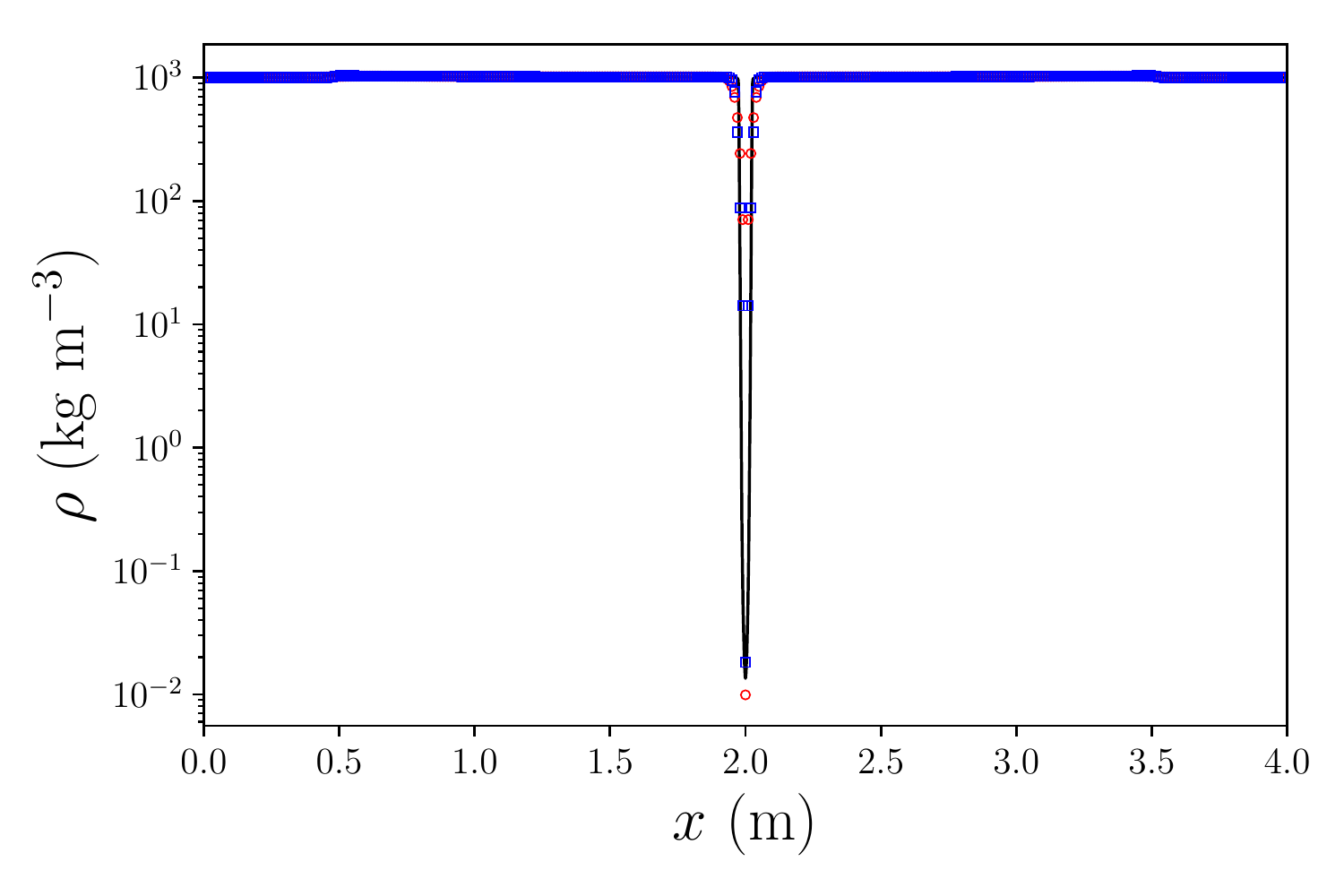}
\label{fig:compare_multi_material_Sedov_1D_rho_global}}
\subfigure[Global velocity profile]{%
\includegraphics[width=0.45\textwidth]{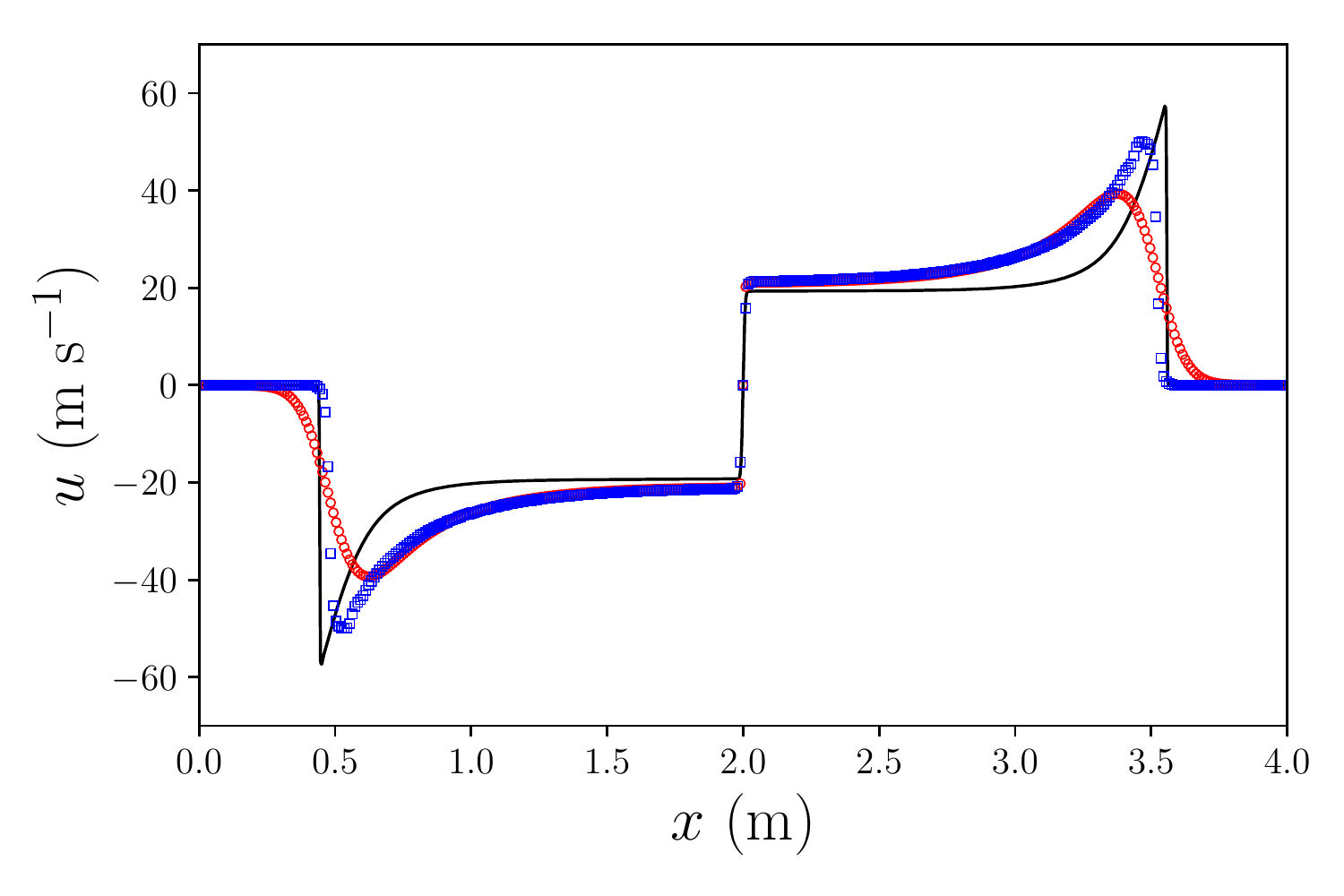}
\label{fig:compare_multi_material_Sedov_1D_u_global}}
\caption{1D multi-material Sedov problem at $t = 1\mathrm{e}{-3} \ \mathrm{s}$ using different schemes. Black solid line: reference; red circles: HLLC; blue squares: PP-WCNS-IS.}
\label{fig:compare_multi_material_Sedov_1D_1}
\end{figure}

\begin{figure}[!ht]
\centering
\subfigure[Global pressure profile]{%
\includegraphics[width=0.45\textwidth]{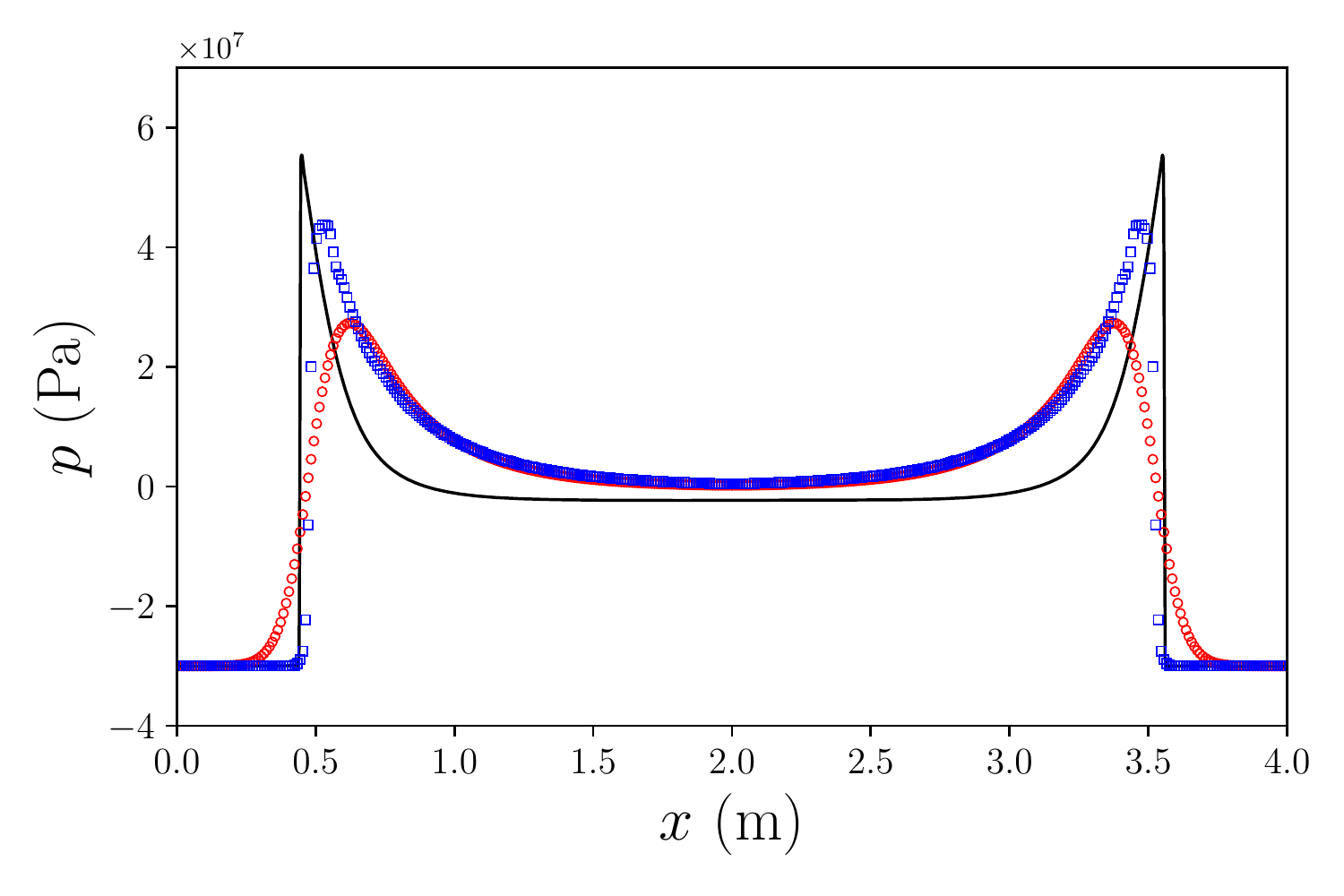}
\label{fig:compare_multi_material_Sedov_1D_p_global}}
\subfigure[Global water volume fraction profile]{%
\includegraphics[width=0.45\textwidth]{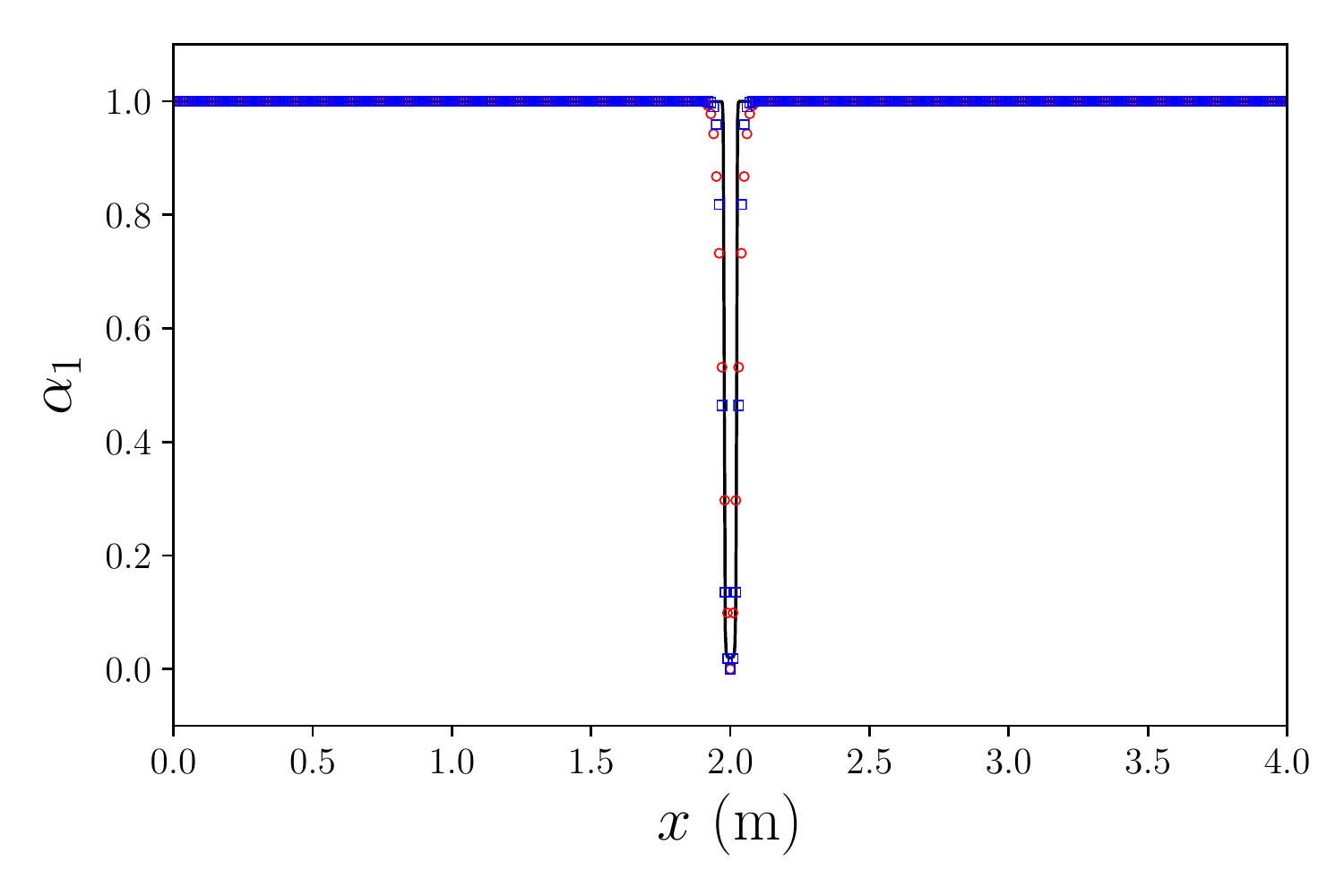}
\label{fig:compare_multi_material_Sedov_1D_Z0_global}}
\caption{1D multi-material Sedov problem at $t = 1\mathrm{e}{-3} \ \mathrm{s}$ using different schemes. Black solid line: reference; red circles: HLLC; blue squares: PP-WCNS-IS.}
\label{fig:compare_multi_material_Sedov_1D_2}
\end{figure}

Figures~\ref{fig:compare_multi_material_Sedov_1D_u_global} and \ref{fig:compare_multi_material_Sedov_1D_2} show the density, velocity, pressure and volume fraction profiles respectively obtained using the two schemes at the final simulation time. It can be seen that both schemes can capture the blast waves without spurious oscillations. However, the velocity and pressure profiles computed with PP-WCNS-IS are much sharper at the shock fronts while the shocks captured with first order HLLC are severely smeared out due to excess numerical dissipation.

\subsection{Two-dimensional Mach 2.4 shock water cylinder interaction problem by~\citet{sembian2016plane}}

\begin{figure}[hbt]
  \centering
  \begin{tikzpicture}[thick,scale=0.8, every node/.style={transform shape}]
    \useasboundingbox (0cm,-1cm)  rectangle (8cm,7cm);
    \draw[black]        (-0.5cm,0.0cm) rectangle ++(9cm,6cm);
    \draw[black]        ( 3.4cm,2.9cm) circle (1.2cm);
    \draw[black, thick] ( 2.2cm,0.0cm) -- (2.2cm,6cm);

    \node[text width=3cm] at (1.3cm,4.75cm) {Post-shock air};
    \node[text width=3cm] at (4.5cm,4.75cm) {Pre-shock air};
    \node[text width=3cm] at (4.5cm,3.3cm) {Water};

    \draw[{Straight Barb[angle'=60,scale=3]}-{Straight Barb[angle'=60,scale=3]}] ( 7.0cm,0.0cm) -- (7.0cm,6cm);
    \draw[{Straight Barb[angle'=60,scale=3]}-{Straight Barb[angle'=60,scale=3]}] (-0.5cm,6.5cm) -- (8.5cm,6.5cm);
    \draw[{Straight Barb[angle'=60,scale=3]}-{Straight Barb[angle'=60,scale=3]}] ( 2.2cm,3.0cm) -- (4.6cm,3.0cm);

    \node[text width=3cm] at (5.2cm,6.8cm) {$11.1\ \mathrm{cm}$};
    \node[text width=3cm] at (8.7cm,3.0cm) {$7.4\  \mathrm{cm}$};
    \node[text width=3cm] at (4.5cm,2.7cm) {$2.2\  \mathrm{cm}$};

    \draw[-{Straight Barb[angle'=60,scale=3]}] (-0.5cm,-0.5cm) -- (1.0cm,-0.5cm);
    \node[text width=3cm] at (2.6cm,-0.6cm) {$x$};
    \draw[black] (-0.5cm,-0.75cm) -- (-0.5cm,-0.25cm);
    \draw[-{Straight Barb[angle'=60,scale=3]}] (-1.0cm,3.0cm) -- (-1.0cm,4.5cm);
    \node[text width=1cm] at (-0.5cm,4.75cm) {$y$};
    \draw[black] (-1.25cm,3.0cm) -- (-0.75cm,3.0cm);
  \end{tikzpicture}
  \caption{Schematic diagram of 2D Mach 2.4 shock water cylinder interaction problem by~\citet{sembian2016plane}.} \label{fig:schematic_2D_Mach_2_4_shock_water_cylinder_Sembian}
\end{figure}
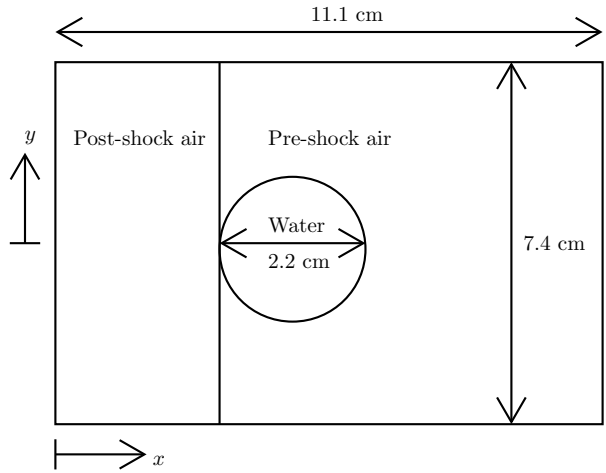

The case of a Mach 2.4 planar shock interacting with a water cylinder in the paper by~\citet{sembian2016plane} is simulated. The purpose of this test case is to investigate the reliability of the flow model with the high-order diffuse interface method for simulating two-phase flows with shocks. Figure~\ref{fig:schematic_2D_Mach_2_4_shock_water_cylinder_Sembian} shows the schematic of the initial flow field and domain. The water cylinder is initially placed at location $\left[ 4\ \mathrm{cm}, 0\ \mathrm{cm} \right]$. The initial conditions are given by table~\ref{table:IC_2D_Sembian_shock_water_cylinder}. Constant extrapolation is used at all domain boundaries. The computations is performed with PP-WCNS-IS on a $3072 \times 2048$ mesh.

\begin{table}[!ht]
  \begin{center}
    \begin{tabular}{@{}c | cccccc@{}}\toprule
     &
    \addstackgap{\stackanchor{$\alpha_1 \rho_1$}{$(\mathrm{kg\ m^{-3}})$}} &
    \stackanchor{$\alpha_2 \rho_2$}{$(\mathrm{kg\ m^{-3}})$} &
    \stackanchor{$u$}{$(\mathrm{m\ s^{-1}})$} &
    \stackanchor{$v$}{$(\mathrm{m\ s^{-1}})$} &
    \stackanchor{$p$}{$(\mathrm{Pa})$} &
    $\alpha_1$ \\ \midrule
    \addstackgap{pre-shock air}  & $1.0\mathrm{e}{-8}$ & 1.17 & 0 & 0 & $1.01\mathrm{e}{5}$ & $1.0\mathrm{e}{-8}$ \\
    \addstackgap{post-shock air} & $1.0\mathrm{e}{-8}$ & 3.7579 & 574.57 & 0 & $6.6189\mathrm{e}{5}$ & $1.0\mathrm{e}{-8}$ \\
    \addstackgap{water cylinder} & 1000 & $1.0\mathrm{e}{-8}$ & 0 & 0 & $1.01\mathrm{e}{5}$ & $1 - 1.0\mathrm{e}{-8}$ \\
    \bottomrule
    \end{tabular}
  \end{center}
  \caption{Initial conditions of 2D Mach 2.4 shock water cylinder interaction problem~\cite{sembian2016plane}.}
  \label{table:IC_2D_Sembian_shock_water_cylinder}
\end{table}

\begin{figure}[!ht]
\centering
\vspace{-1cm}
\subfigure[Experiment]{%
\includegraphics[trim=0 -1.5cm 0 0,width=0.27\textwidth]{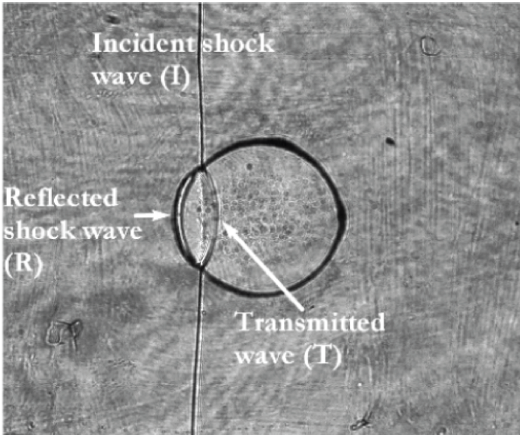}
\label{fig:compare_2D_Mach_2_4_shock_water_cylinder_Sembian_t1_experiment}}
\subfigure[$t = 4\ \mu\mathrm{s}$, PP-WCNS-IS]{%
\includegraphics[width=0.46\textwidth]{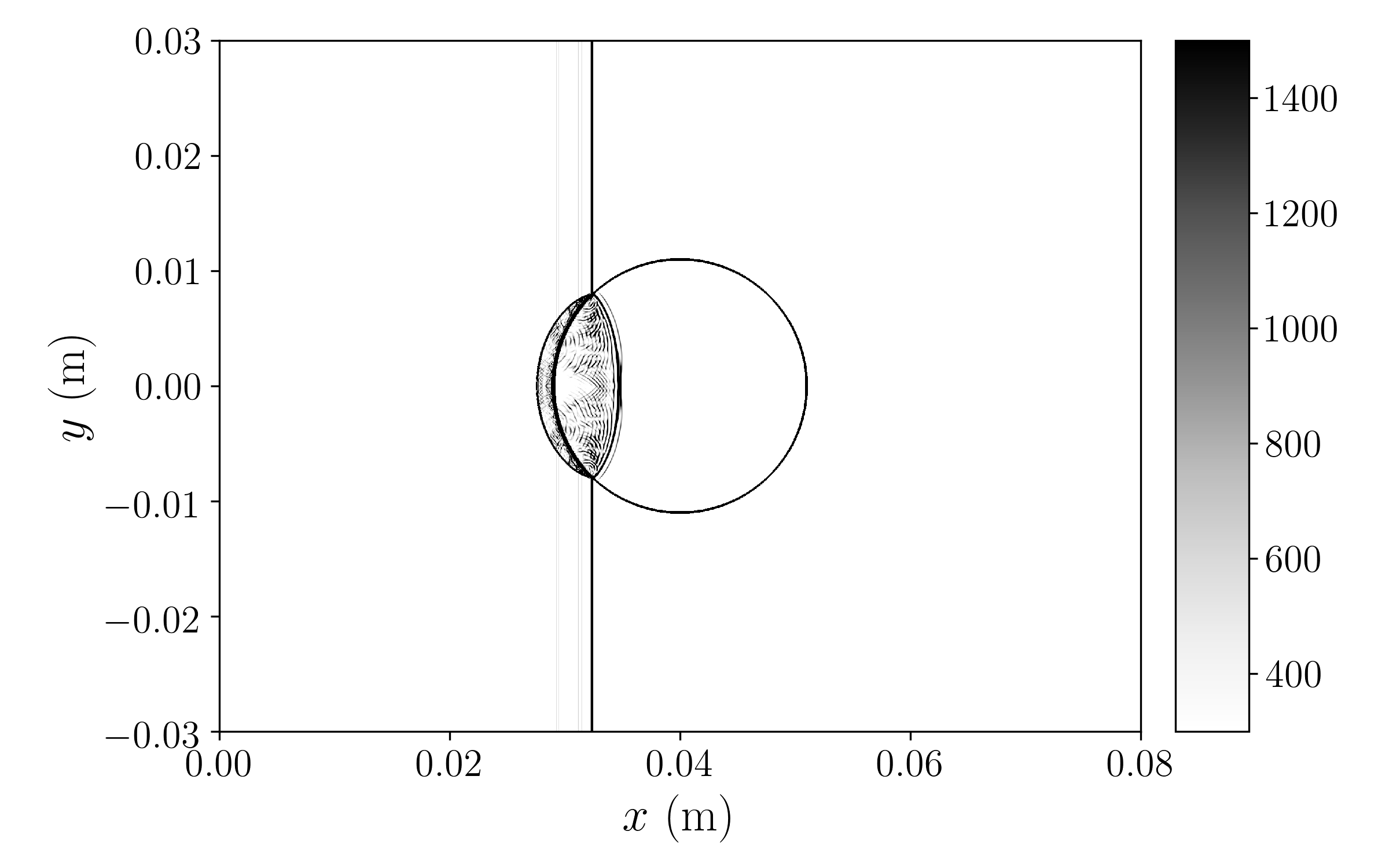}
\label{fig:compare_2D_Mach_2_4_shock_water_cylinder_Sembian_t1_WCNS5_IS_PP}}
\newline
\subfigure[Experiment]{%
\includegraphics[trim=0 -1.5cm 0 0,width=0.27\textwidth]{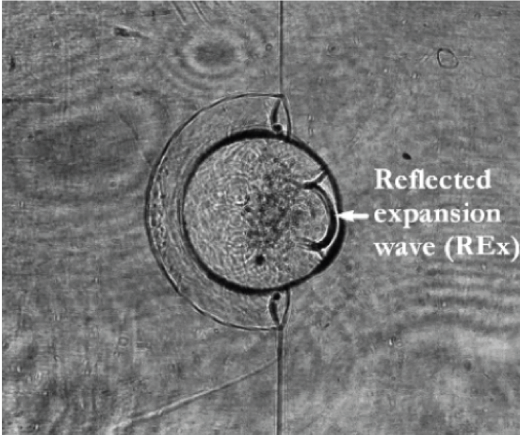}
\label{fig:compare_2D_Mach_2_4_shock_water_cylinder_Sembian_t2_experiment}}
\subfigure[$t = 17\ \mu\mathrm{s}$, PP-WCNS-IS]{%
\includegraphics[width=0.46\textwidth]{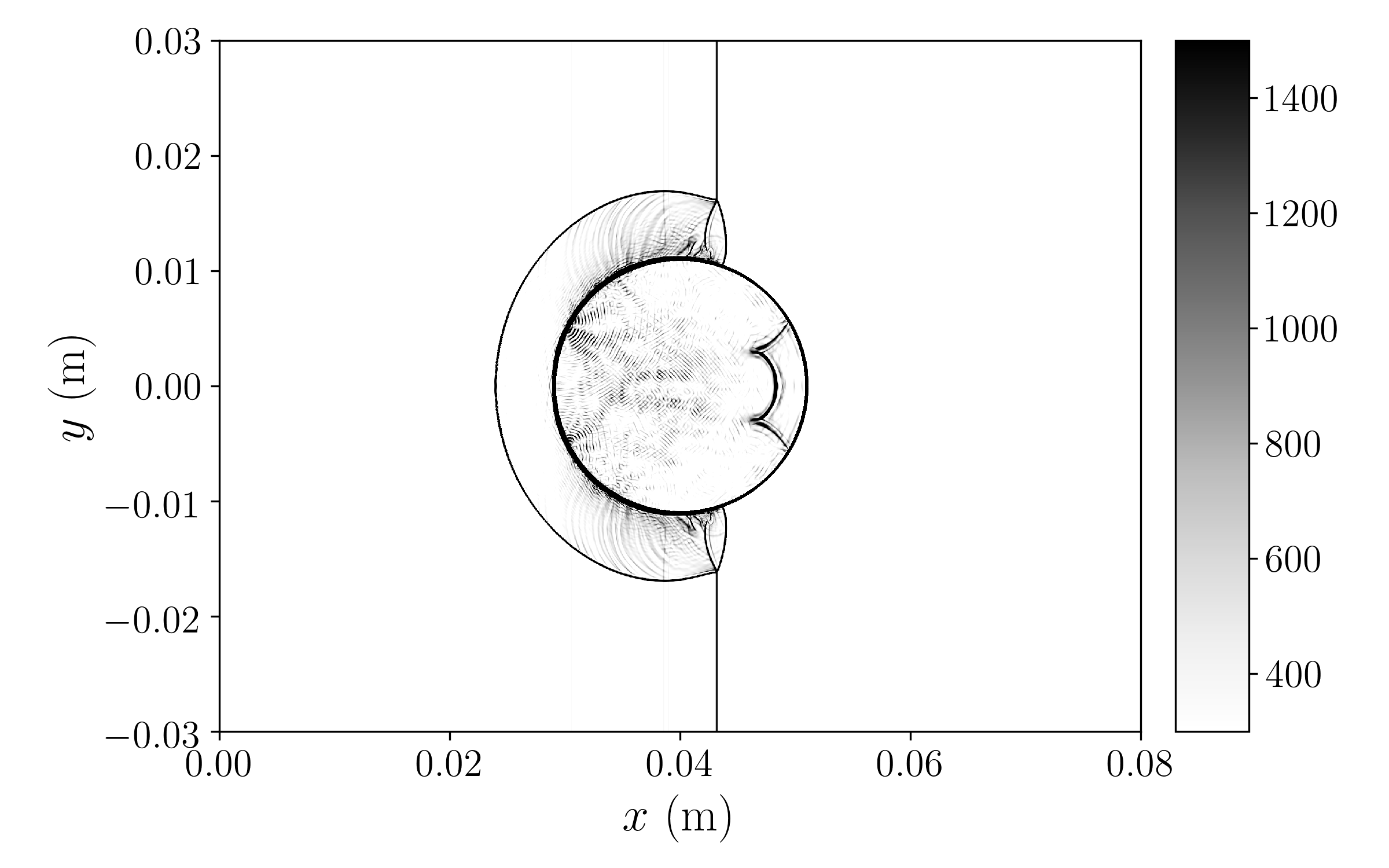}
\label{fig:compare_2D_Mach_2_4_shock_water_cylinder_Sembian_t2_WCNS5_IS_PP}}
\newline
\subfigure[Experiment]{%
\includegraphics[trim=0 -1.5cm 0 0,width=0.27\textwidth]{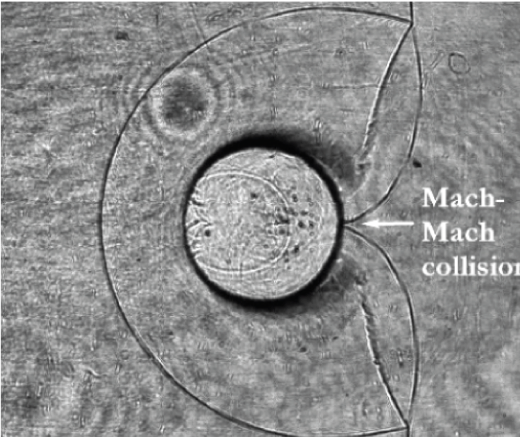}
\label{fig:compare_2D_Mach_2_4_shock_water_cylinder_Sembian_t3_experiment}}
\subfigure[$t = 40\ \mu\mathrm{s}$, PP-WCNS-IS]{%
\includegraphics[width=0.46\textwidth]{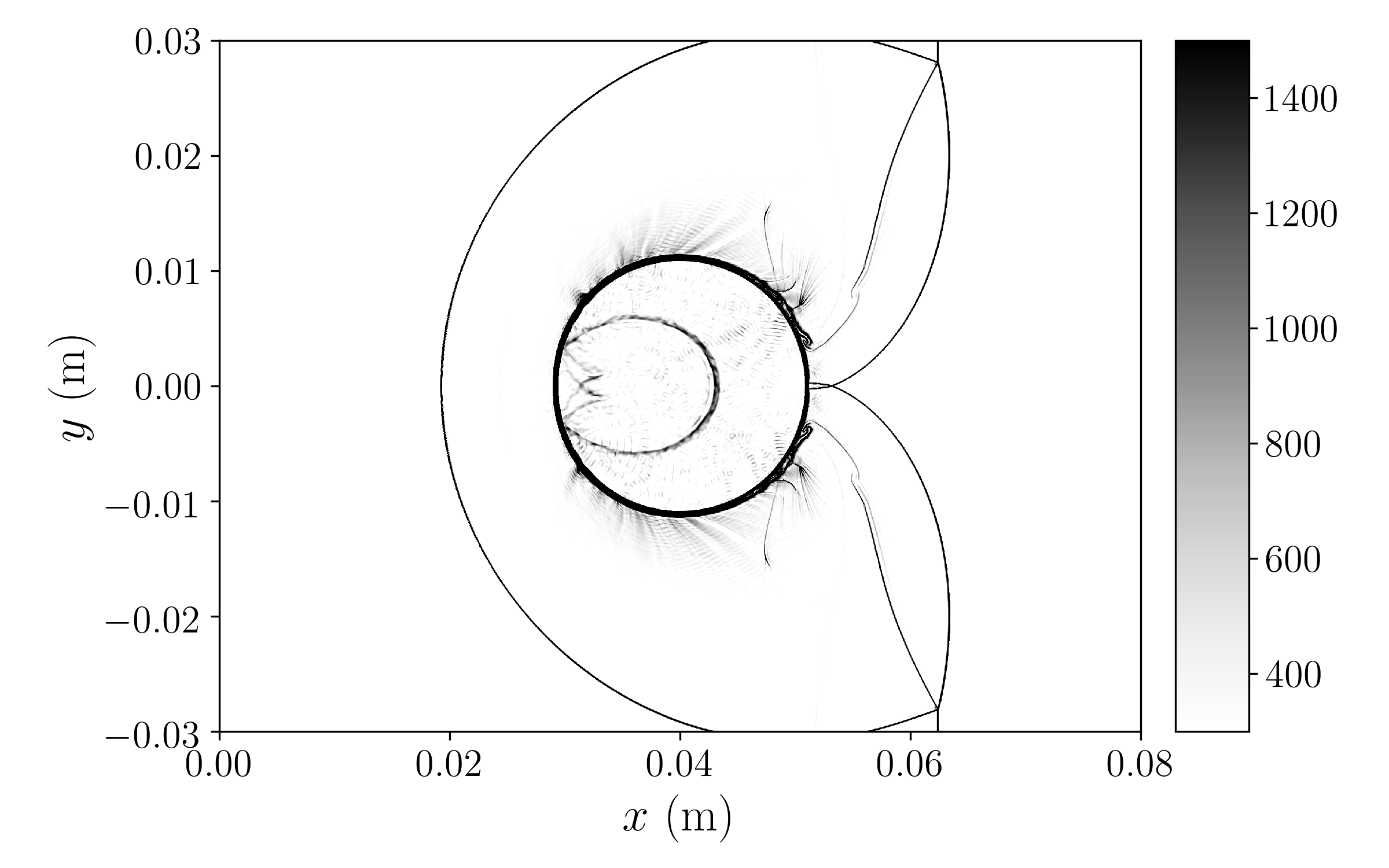}
\label{fig:compare_2D_Mach_2_4_shock_water_cylinder_Sembian_t3_WCNS5_IS_PP}}
\newline
\subfigure[Experiment]{%
\includegraphics[trim=0 -1.5cm 0 0,width=0.27\textwidth]{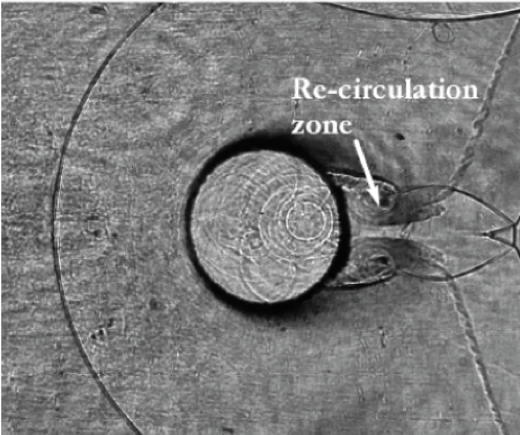}
\label{fig:compare_2D_Mach_2_4_shock_water_cylinder_Sembian_t4_experiment}}
\subfigure[$t = 67\ \mu\mathrm{s}$, PP-WCNS-IS]{%
\includegraphics[width=0.46\textwidth]{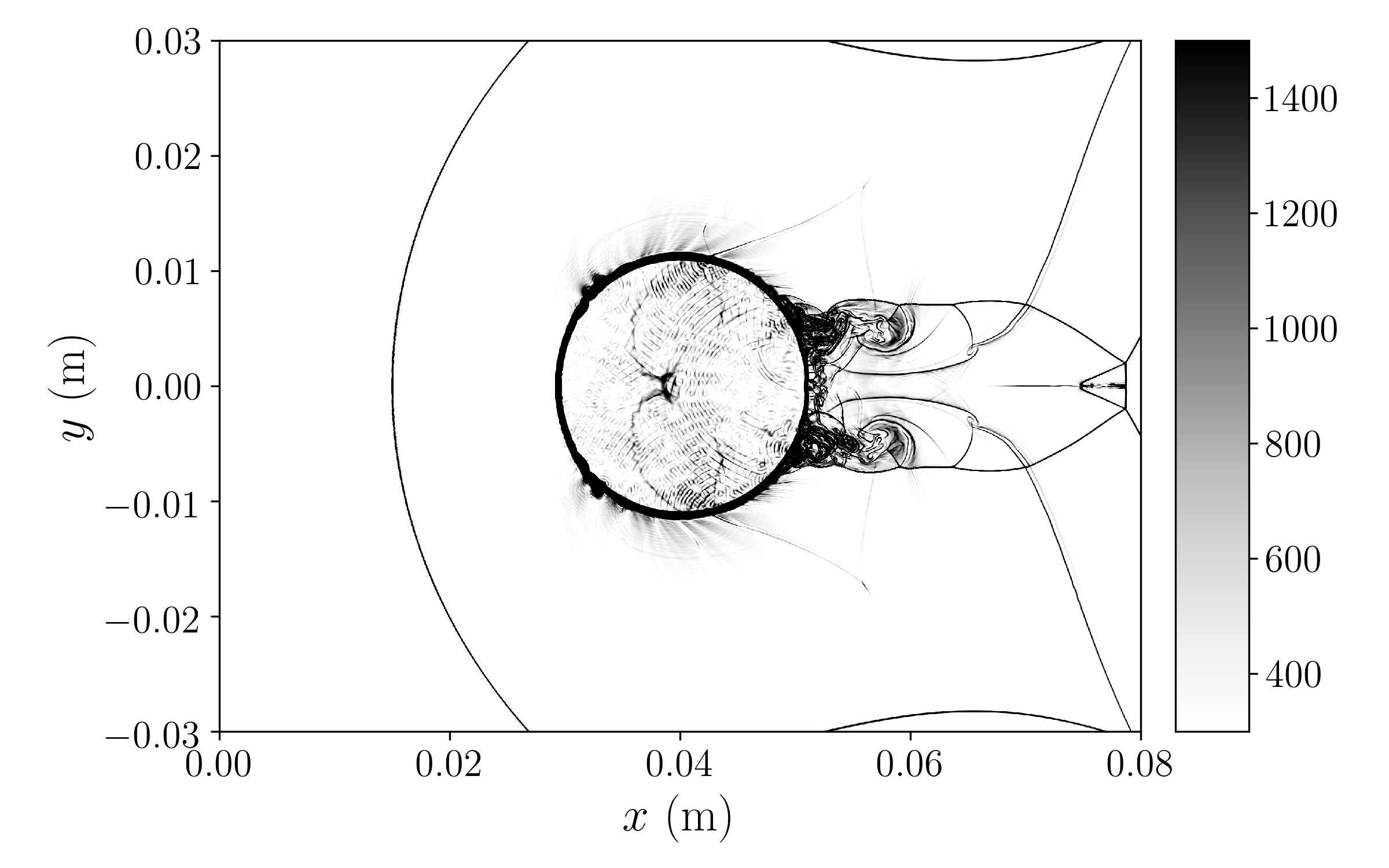}
\label{fig:compare_2D_Mach_2_4_shock_water_cylinder_Sembian_t4_WCNS5_IS_PP}}
\newline
\caption{Comparision of 2D Mach 2.4 shock water cylinder interaction problem by~\citet{sembian2016plane}. Left: shadowgraph images from~\cite{sembian2016plane}; right: density gradient from simulation with PP-WCNS-IS.}
\label{fig:compare_2D_Mach_2_4_shock_water_cylinder_Sembian}
\end{figure}

The grey schlieren images from the experiment~\cite{sembian2016plane} are shown in the left column of figure~\ref{fig:compare_2D_Mach_2_4_shock_water_cylinder_Sembian}. At the instance when the incident shock interacts with the water column, a shock is reflected upstream since the acoustic impedance of water is higher than that of air. The reflected shock interacts with the incident shock to generate a triple point where the reflected shock, incident shock and a Mach stem along with its slip line conincide. Meanwhile, there is also a shock transmitted into the water column. The transmitted shock travels faster than the shocks outside the water column and it gets reflected as an expansion wave when the transmitted shock reaches the downstream water-air interface. The reflected expansion wave focuses at a point due to the column's downstream concave geometry, where negative presure is produced due to tensile stresses. The reflected expansion wave forms a ``horse-shoe" structure after focusing and is reflected again at the upstream water-air interface. The expansion wave continues to get reflected inside the water column repeatedly. As the water column is a buff body, the surrounding air separates in the adverse pressure gradient region on the water column surface. Therefore, recirculation regions are created and two counter-rotating vortices are formed downstream of the flow. In the right column of figure~\ref{fig:compare_2D_Mach_2_4_shock_water_cylinder_Sembian}, the density gradients computed with the simulation results using the high-order PP-WCNS-IS scheme at different times are displayed. Compared with the schlieren images from the experiment, it can be seen that most of the wave features, such as the incident, reflected and transmitted shocks, and expansion waves are captured accurately. Also, the two counter-rotating vortices are reproduced in the simulation. 

\subsection{Two-dimensional Mach 10 shock water cylinder interaction problem}

\begin{figure}[hbt]
  \centering
  \begin{tikzpicture}[thick,scale=0.8, every node/.style={transform shape}]
    \useasboundingbox (0cm,-1cm)  rectangle (8cm,7cm);
    \draw[black]        (-0.5cm,0.0cm) rectangle ++(9cm,6cm);
    \draw[black]        ( 3.4cm,2.9cm) circle (1.2cm);
    \draw[black, thick] ( 2.2cm,0.0cm) -- (2.2cm,6cm);

    \node[text width=3cm] at (1.3cm,4.75cm) {Post-shock air};
    \node[text width=3cm] at (4.5cm,4.75cm) {Pre-shock air};
    \node[text width=3cm] at (4.5cm,3.3cm) {Water};

    \draw[{Straight Barb[angle'=60,scale=3]}-{Straight Barb[angle'=60,scale=3]}] ( 7.5cm,0.0cm) -- (7.5cm,6cm);
    \draw[{Straight Barb[angle'=60,scale=3]}-{Straight Barb[angle'=60,scale=3]}] (-0.5cm,6.5cm) -- (8.5cm,6.5cm);
    \draw[{Straight Barb[angle'=60,scale=3]}-{Straight Barb[angle'=60,scale=3]}] ( 2.2cm,3.0cm) -- (4.6cm,3.0cm);

    \node[text width=3cm] at (5.2cm,6.8cm) {$30L$};
    \node[text width=3cm] at (9.2cm,3.0cm) {$20L$};
    \node[text width=3cm] at (4.7cm,2.7cm) {$8L$};

    \draw[-{Straight Barb[angle'=60,scale=3]}] (-0.5cm,-0.5cm) -- (1.0cm,-0.5cm);
    \node[text width=3cm] at (2.6cm,-0.6cm) {$x$};
    \draw[black] (-0.5cm,-0.75cm) -- (-0.5cm,-0.25cm);
    \draw[-{Straight Barb[angle'=60,scale=3]}] (-1.0cm,3.0cm) -- (-1.0cm,4.5cm);
    \node[text width=1cm] at (-0.5cm,4.75cm) {$y$};
    \draw[black] (-1.25cm,3.0cm) -- (-0.75cm,3.0cm);
  \end{tikzpicture}
  \caption{Schematic diagram of 2D Mach 10 shock water cylinder interaction problem.} \label{fig:schematic_2D_Mach_10_shock_water_cylinder}
\end{figure}
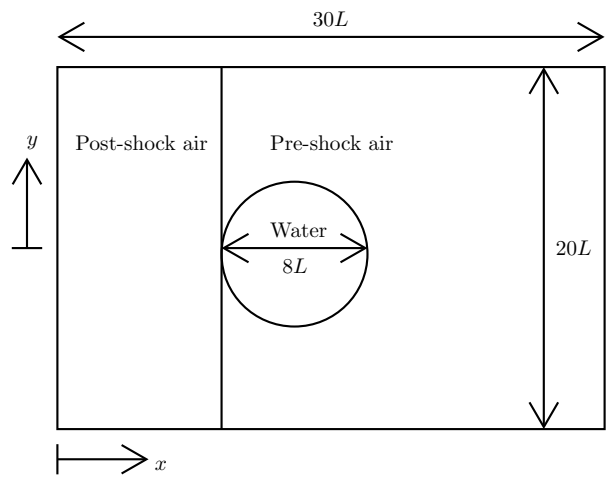

This is the more extreme case of the previous test case. In this problem, a Mach 10 shock wave in air interacts with a water cylinder with diameter of $8L$ in a domain of $\left[ 0, 30L \right] \times \left[ -10L, 10L \right]$, where $L = 1\ \mathrm{mm}$ is chosen. Figure~\ref{fig:schematic_2D_Mach_10_shock_water_cylinder} shows the schematic of the initial flow field and domain. The water cylinder is initially placed at location $\left[ 13L, 0 \right]$. The initial conditions are given by table~\ref{table:IC_2D_Mach_10_shock_water_cylinder}. Constant extraploation is used at all domain boundaries. All computations are performed on a $1152 \times 768$ mesh.

\begin{table}[!ht]
  \begin{center}
    \begin{tabular}{@{}c | cccccc@{}}\toprule
     &
    \addstackgap{\stackanchor{$\alpha_1 \rho_1$}{$(\mathrm{kg\ m^{-3}})$}} &
    \stackanchor{$\alpha_2 \rho_2$}{$(\mathrm{kg\ m^{-3}})$} &
    \stackanchor{$u$}{$(\mathrm{m\ s^{-1}})$} &
    \stackanchor{$v$}{$(\mathrm{m\ s^{-1}})$} &
    \stackanchor{$p$}{$(\mathrm{Pa})$} &
    $\alpha_1$ \\ \midrule
    \addstackgap{pre-shock air}  & $1.0\mathrm{e}{-8}$ & 1.2 & 0 & 0 & $1.0\mathrm{e}{5}$ & $1.0\mathrm{e}{-8}$ \\
    \addstackgap{post-shock air} & $1.0\mathrm{e}{-8}$ & 6.8571 & $2.8179\mathrm{e}{3}$ & 0 & $1.165\mathrm{e}{7}$ & $1.0\mathrm{e}{-8}$ \\
    \addstackgap{water cylinder} & 1000 & $1.0\mathrm{e}{-8}$ & 0 & 0 & $1.0\mathrm{e}{5}$ & $1 - 1.0\mathrm{e}{-8}$ \\ \bottomrule
    \end{tabular}
  \end{center}
  \caption{Initial conditions of 2D Mach 10 shock water cylinder interaction problem.}
  \label{table:IC_2D_Mach_10_shock_water_cylinder}
\end{table}

\begin{figure}[!ht]
\centering
\subfigure[$t = 1\ \mu\mathrm{s}$, HLLC]{%
\includegraphics[width=0.45\textwidth]{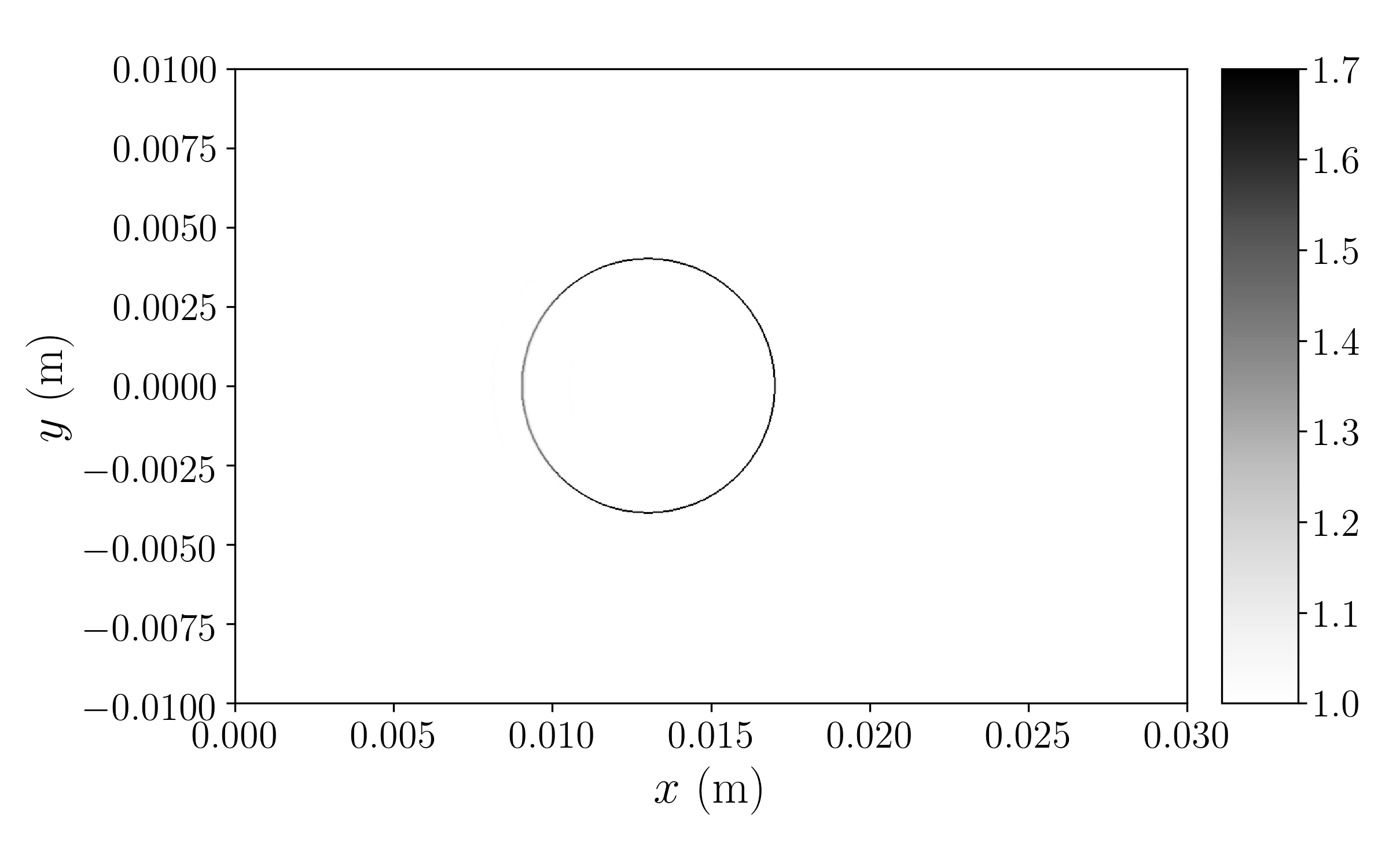}
\label{fig:compare_2D_Mach_10_shock_water_cylinder_schl_t1_HLLC}}
\subfigure[$t = 1\ \mu\mathrm{s}$, PP-WCNS-IS]{%
\includegraphics[width=0.45\textwidth]{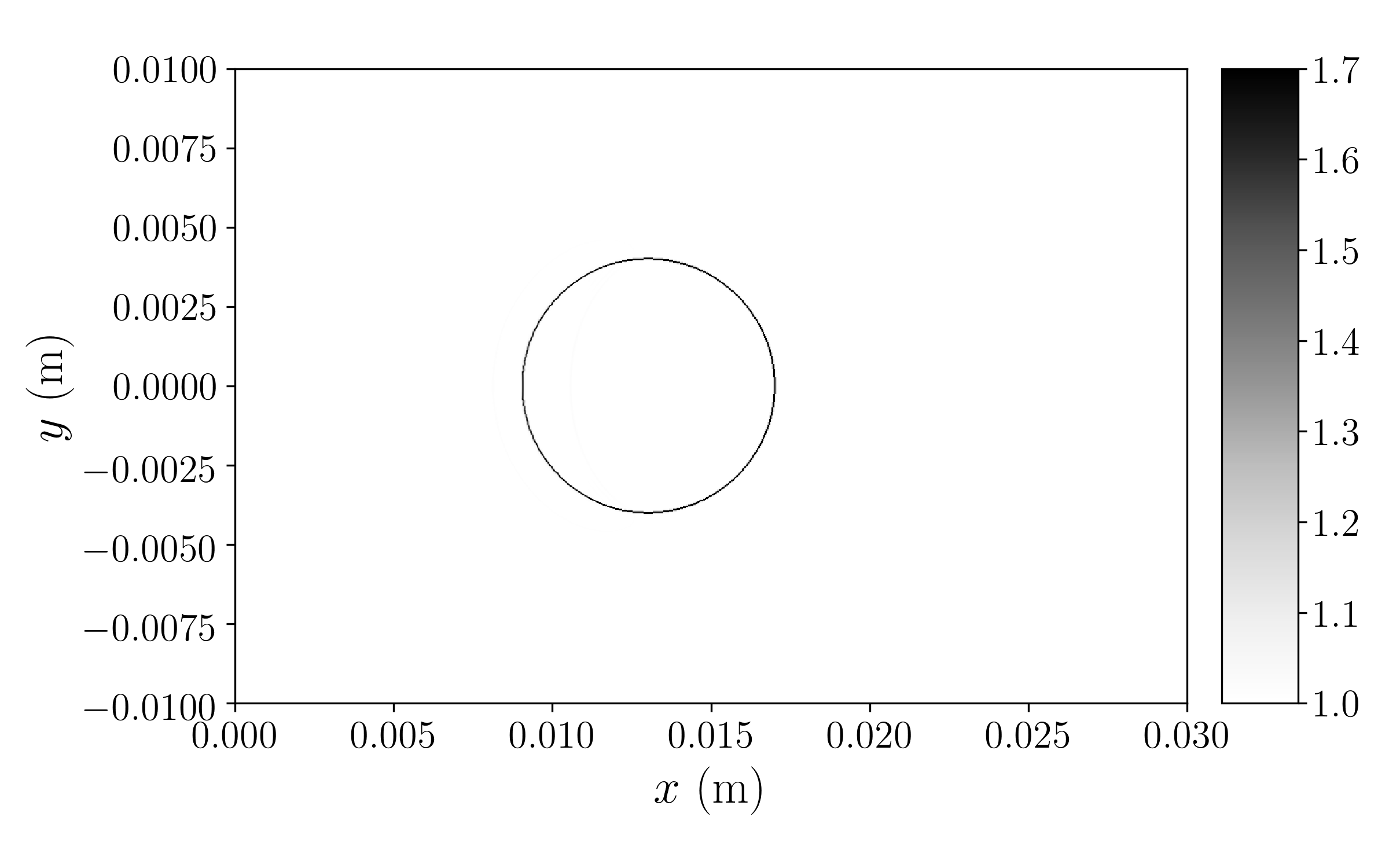}
\label{fig:compare_2D_Mach_10_shock_water_cylinder_schl_t1_WCNS5_IS_PP}}
\subfigure[$t = 4\ \mu\mathrm{s}$, HLLC]{%
\includegraphics[width=0.45\textwidth]{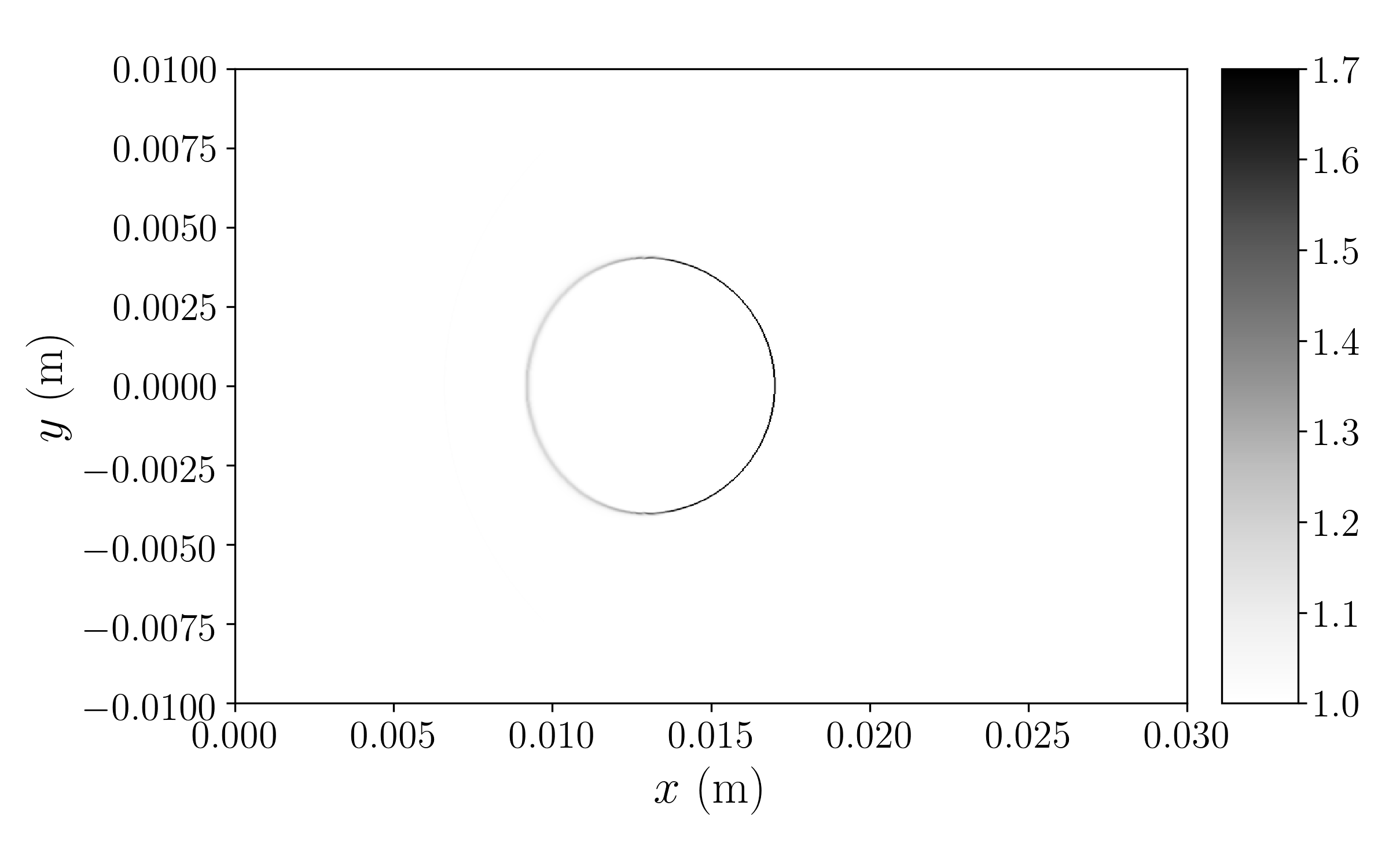}
\label{fig:compare_2D_Mach_10_shock_water_cylinder_schl_t2_HLLC}}
\subfigure[$t = 4\ \mu\mathrm{s}$, PP-WCNS-IS]{%
\includegraphics[width=0.45\textwidth]{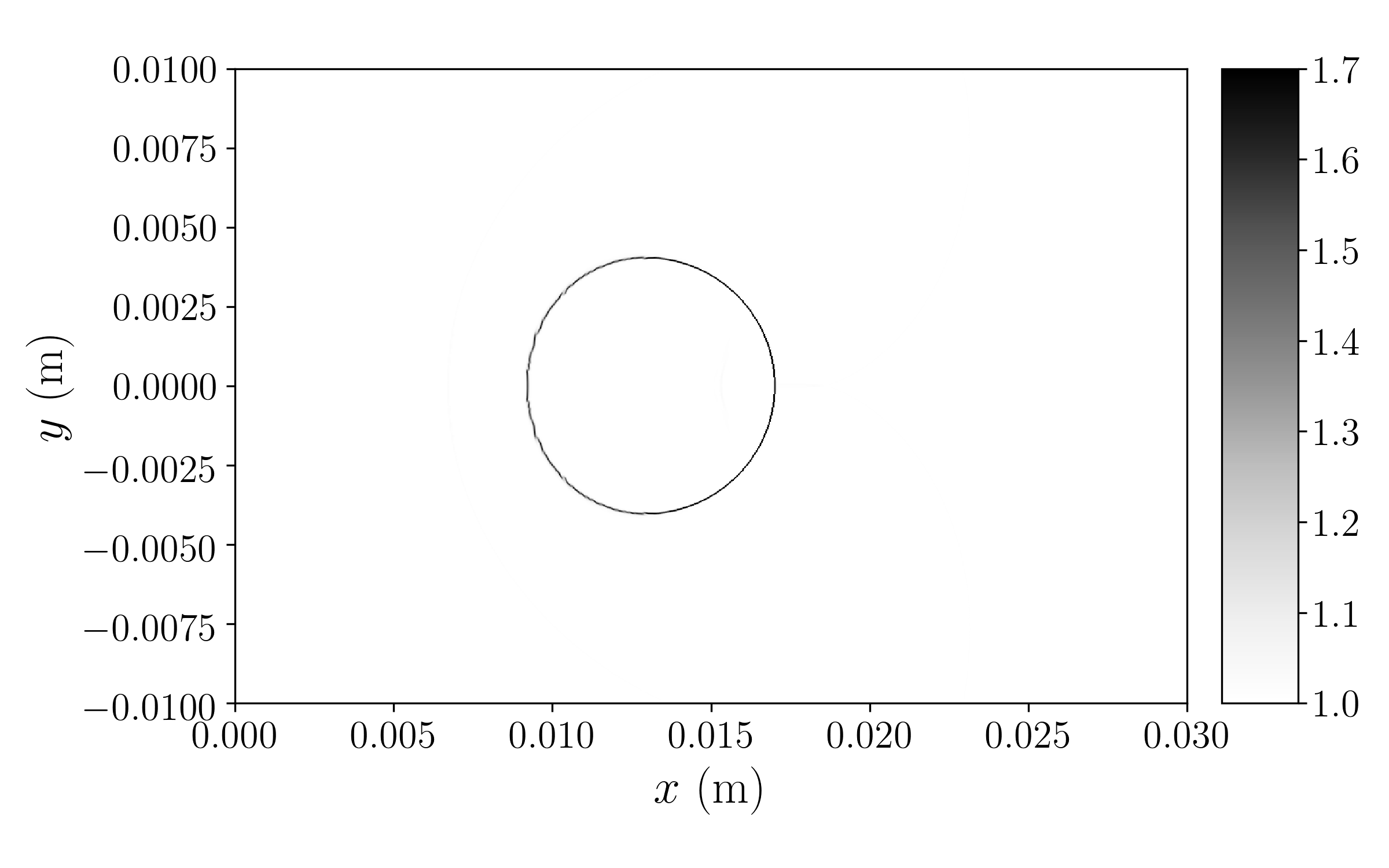}
\label{fig:compare_2D_Mach_10_shock_water_cylinder_schl_t2_WCNS5_IS_PP}}
\subfigure[$t = 8\ \mu\mathrm{s}$, HLLC]{%
\includegraphics[width=0.45\textwidth]{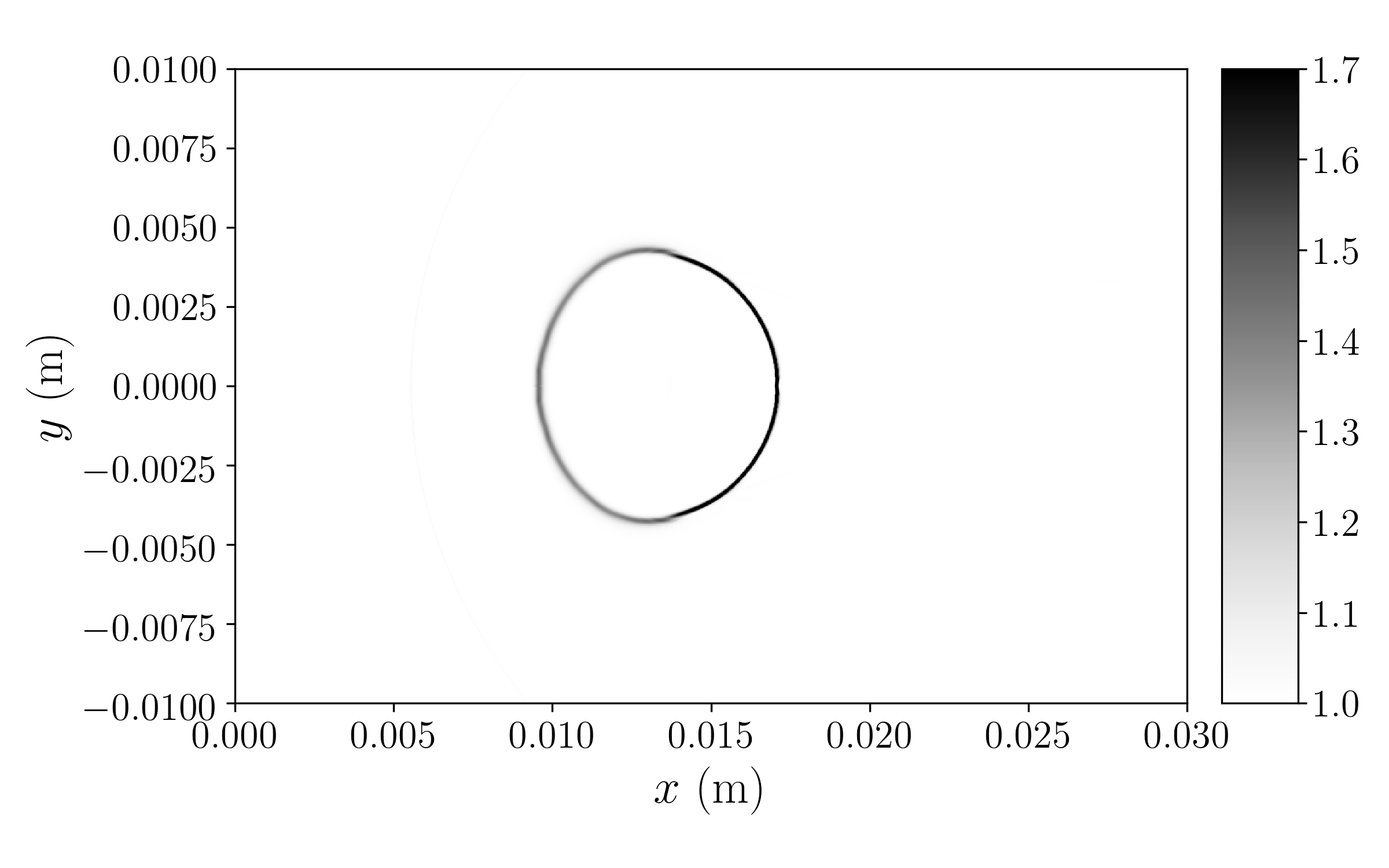}
\label{fig:compare_2D_Mach_10_shock_water_cylinder_schl_t3_HLLC}}
\subfigure[$t = 8\ \mu\mathrm{s}$, PP-WCNS-IS]{%
\includegraphics[width=0.45\textwidth]{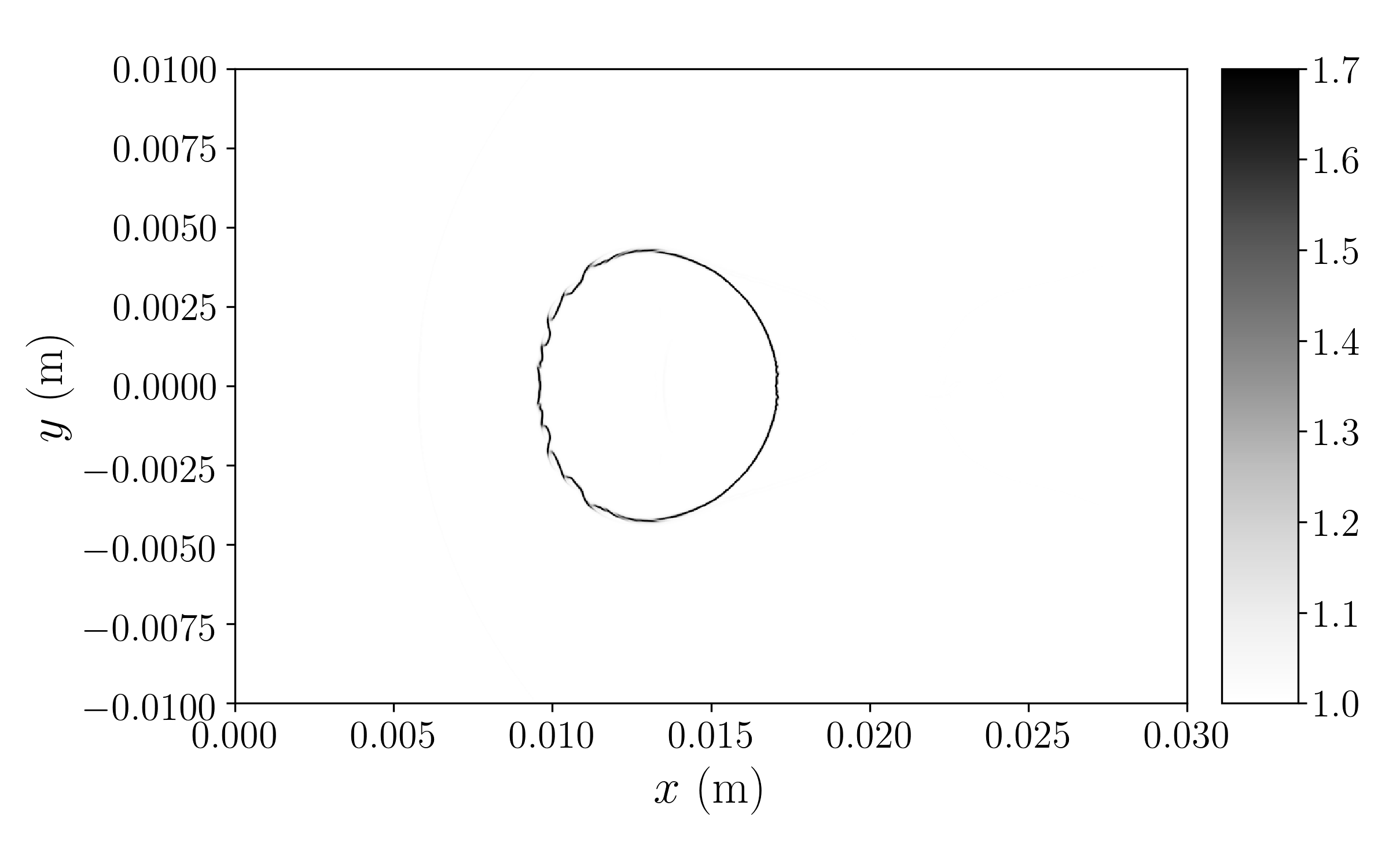}
\label{fig:compare_2D_Mach_10_shock_water_cylinder_schl_t3_WCNS5_IS_PP}}
\subfigure[$t = 16\ \mu\mathrm{s}$, HLLC]{%
\includegraphics[width=0.45\textwidth]{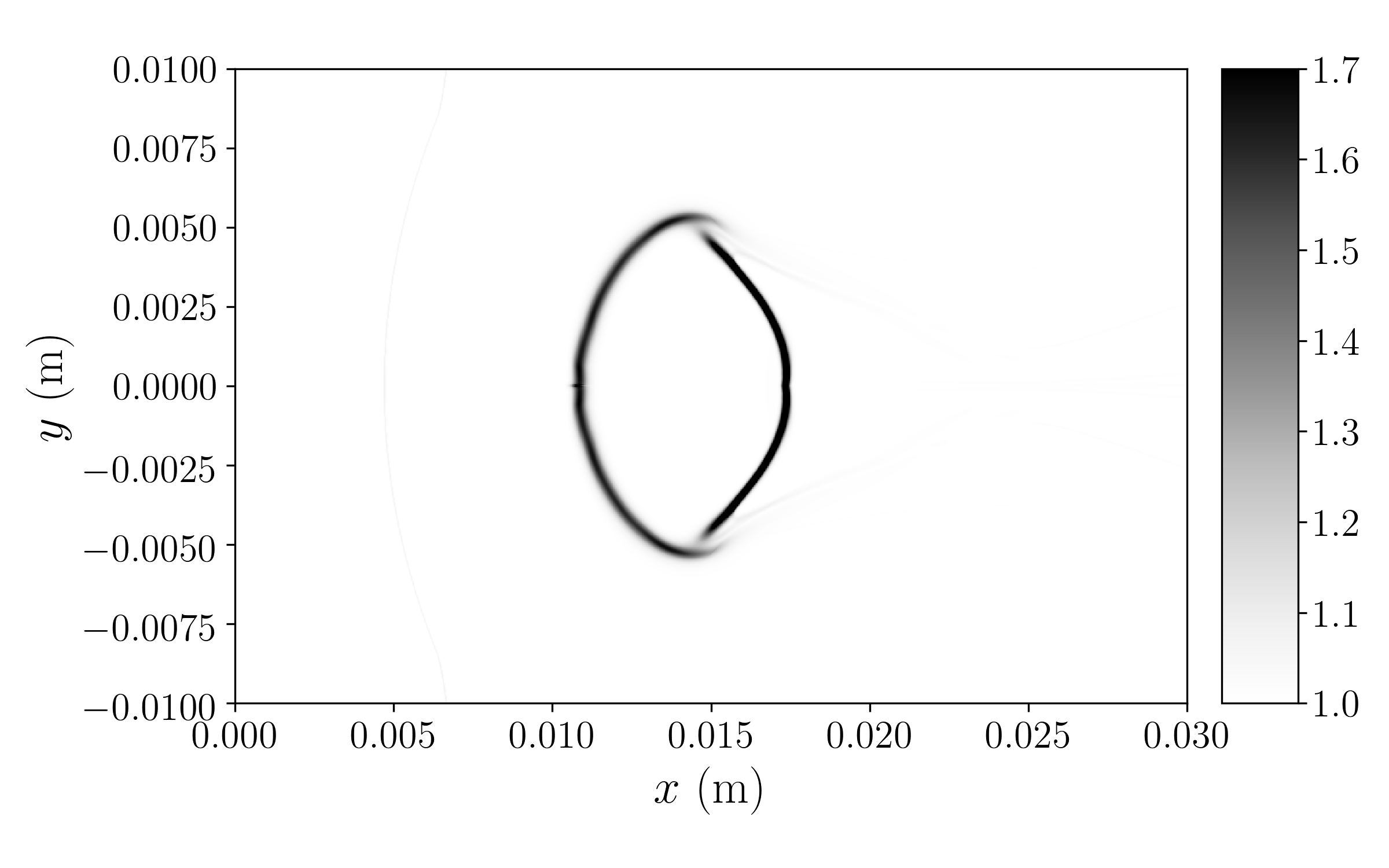}
\label{fig:compare_2D_Mach_10_shock_water_cylinder_schl_t4}}
\subfigure[$t = 16\ \mu\mathrm{s}$, PP-WCNS-IS]{%
\includegraphics[width=0.45\textwidth]{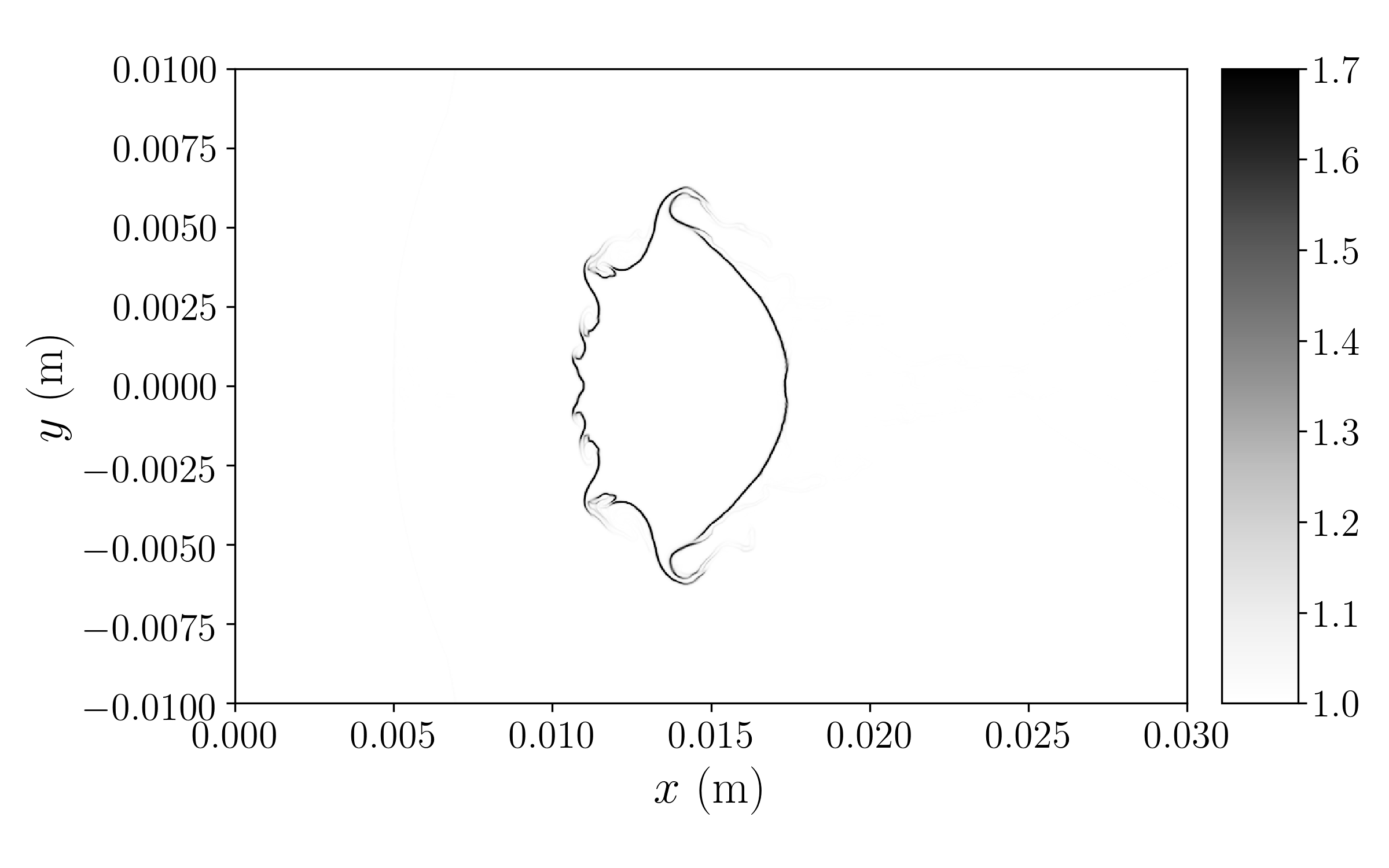}
\label{fig:compare_2D_Mach_10_shock_water_cylinder_schl_t4_WCNS5_IS_PP}}
\caption{Numerical schlieren ($\exp{\left( \left| \nabla \rho \right| / \left| \nabla \rho \right|_{\mathrm{max}} \right)}$) of 2D Mach 10 shock water cylinder interaction problem.}
\label{fig:compare_2D_Mach_10_shock_water_cylinder_schl}
\end{figure}

\begin{figure}[!ht]
\centering
\subfigure[$t = 1\ \mu\mathrm{s}$, HLLC]{%
\includegraphics[width=0.45\textwidth]{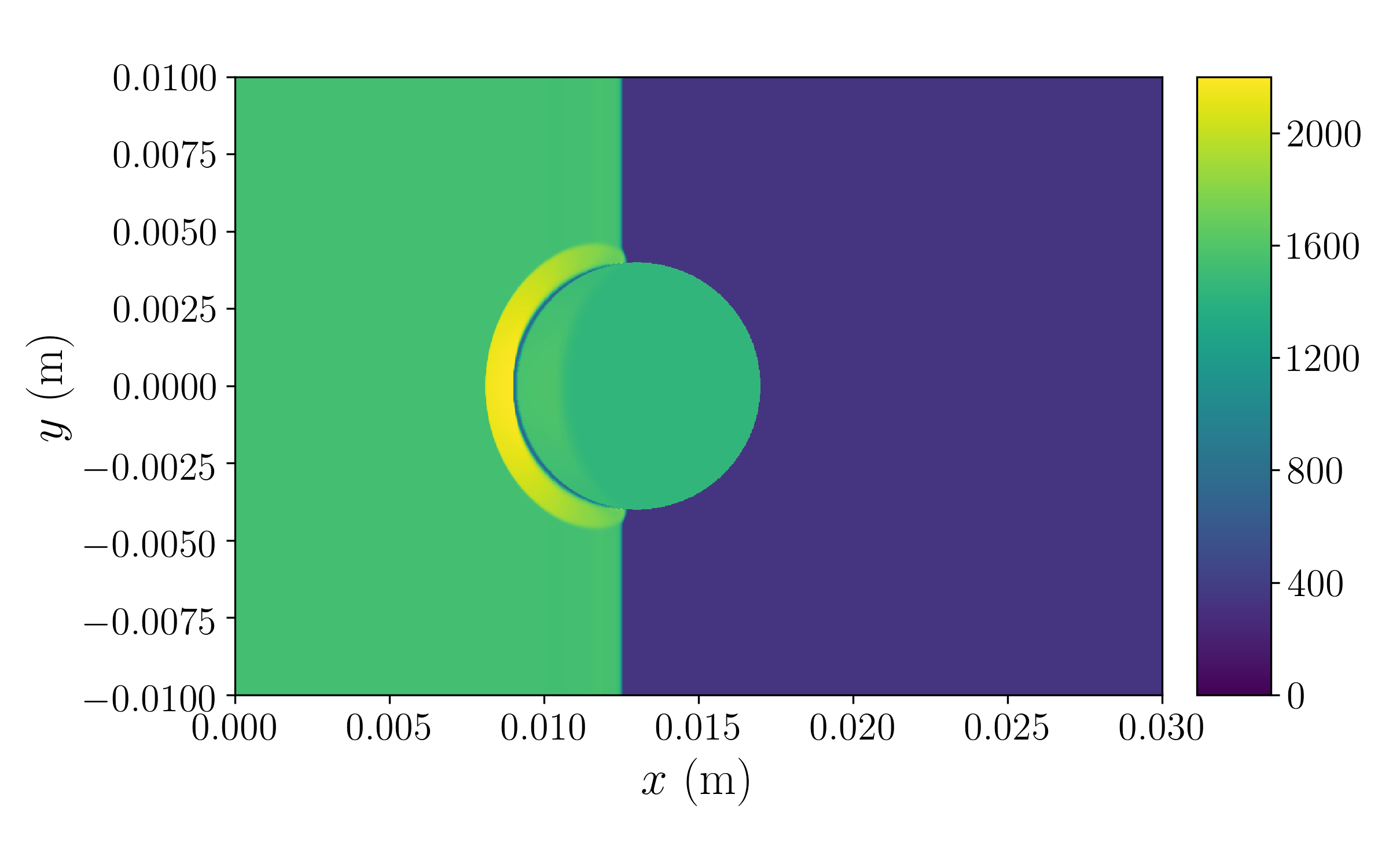}
\label{fig:compare_2D_Mach_10_shock_water_cylinder_sos_t1_HLLC}}
\subfigure[$t = 1\ \mu\mathrm{s}$, PP-WCNS-IS]{%
\includegraphics[width=0.45\textwidth]{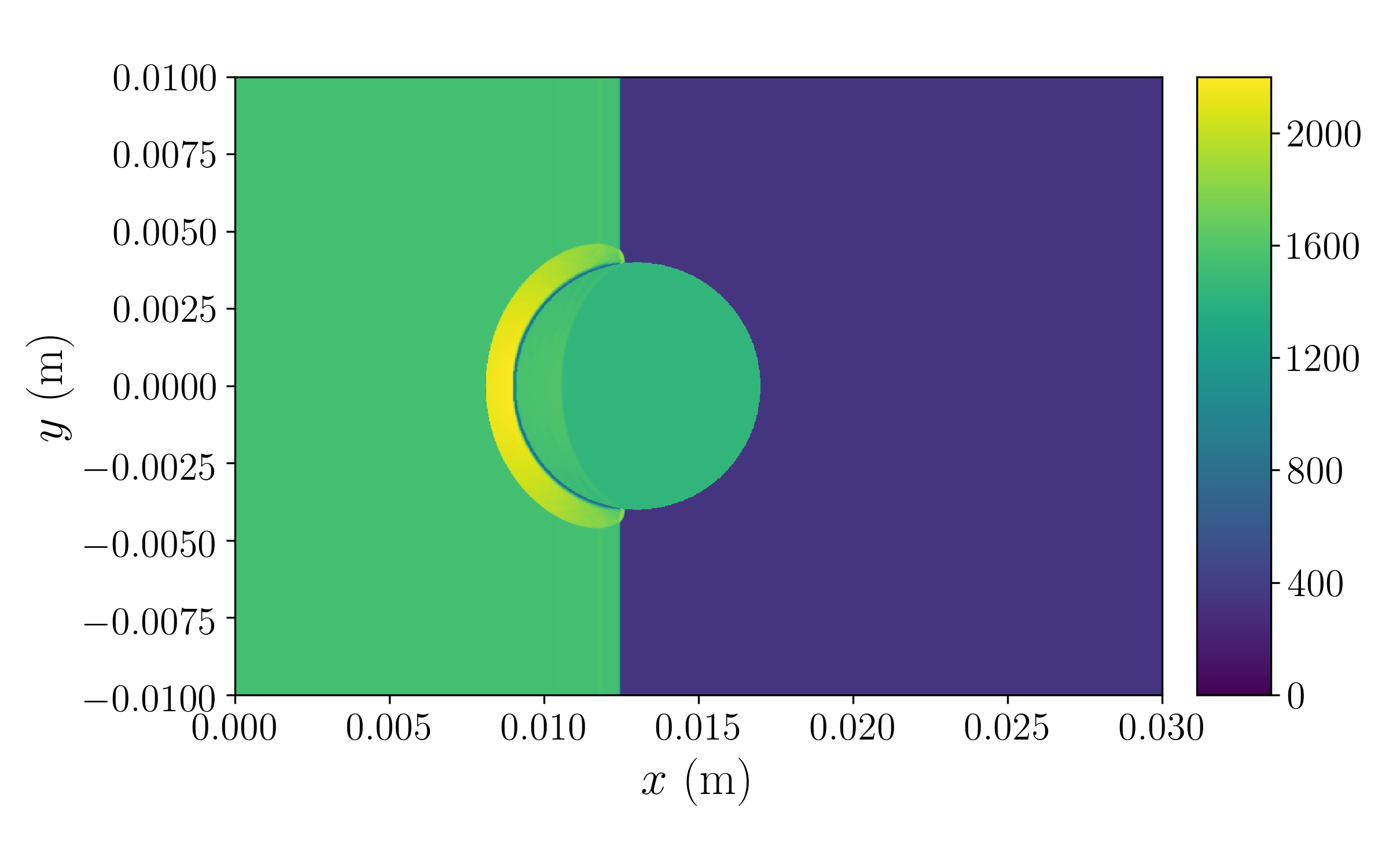}
\label{fig:compare_2D_Mach_10_shock_water_cylinder_sos_t1_WCNS5_IS_PP}}
\subfigure[$t = 4\ \mu\mathrm{s}$, HLLC]{%
\includegraphics[width=0.45\textwidth]{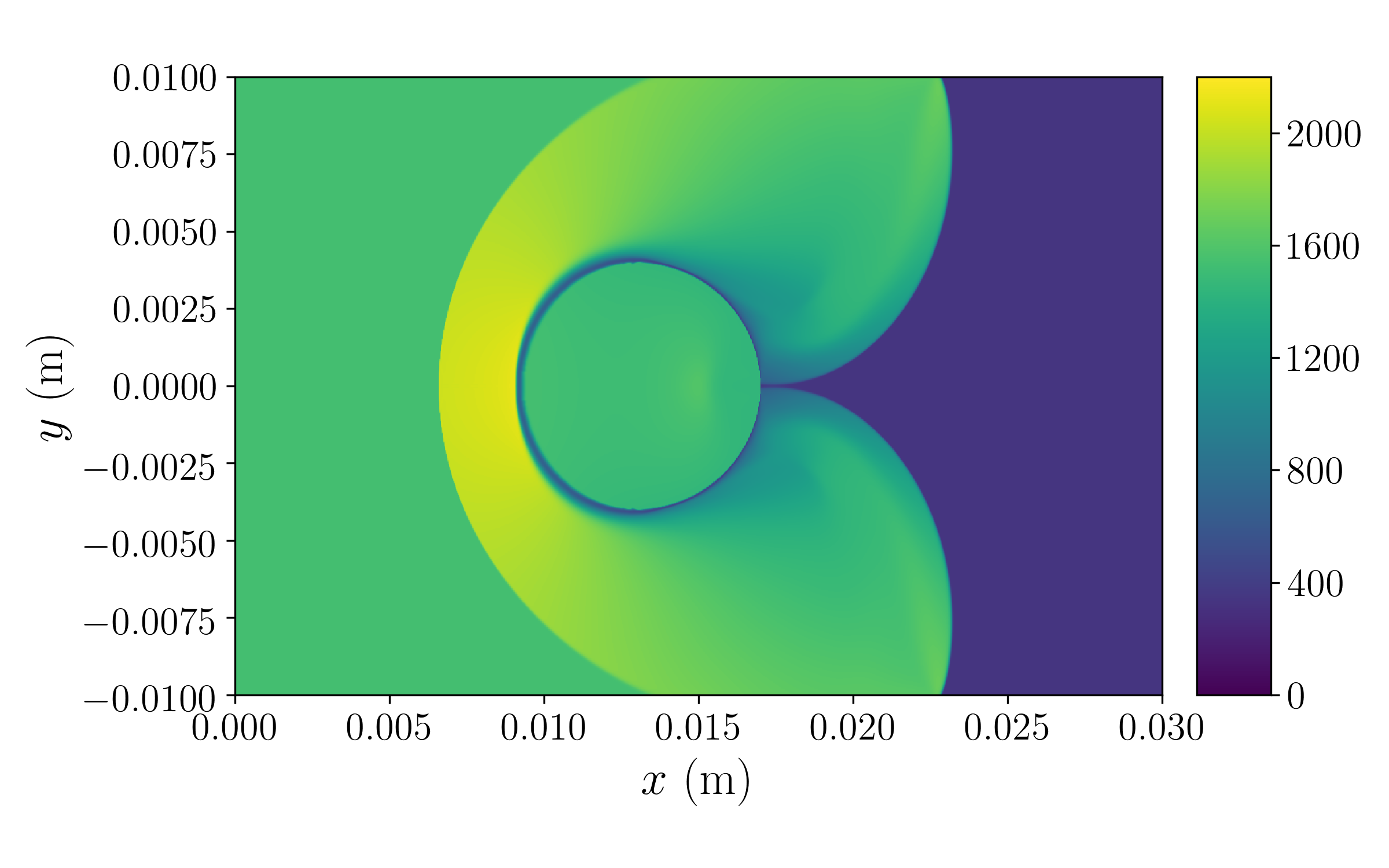}
\label{fig:compare_2D_Mach_10_shock_water_cylinder_sos_t2_HLLC}}
\subfigure[$t = 4\ \mu\mathrm{s}$, PP-WCNS-IS]{%
\includegraphics[width=0.45\textwidth]{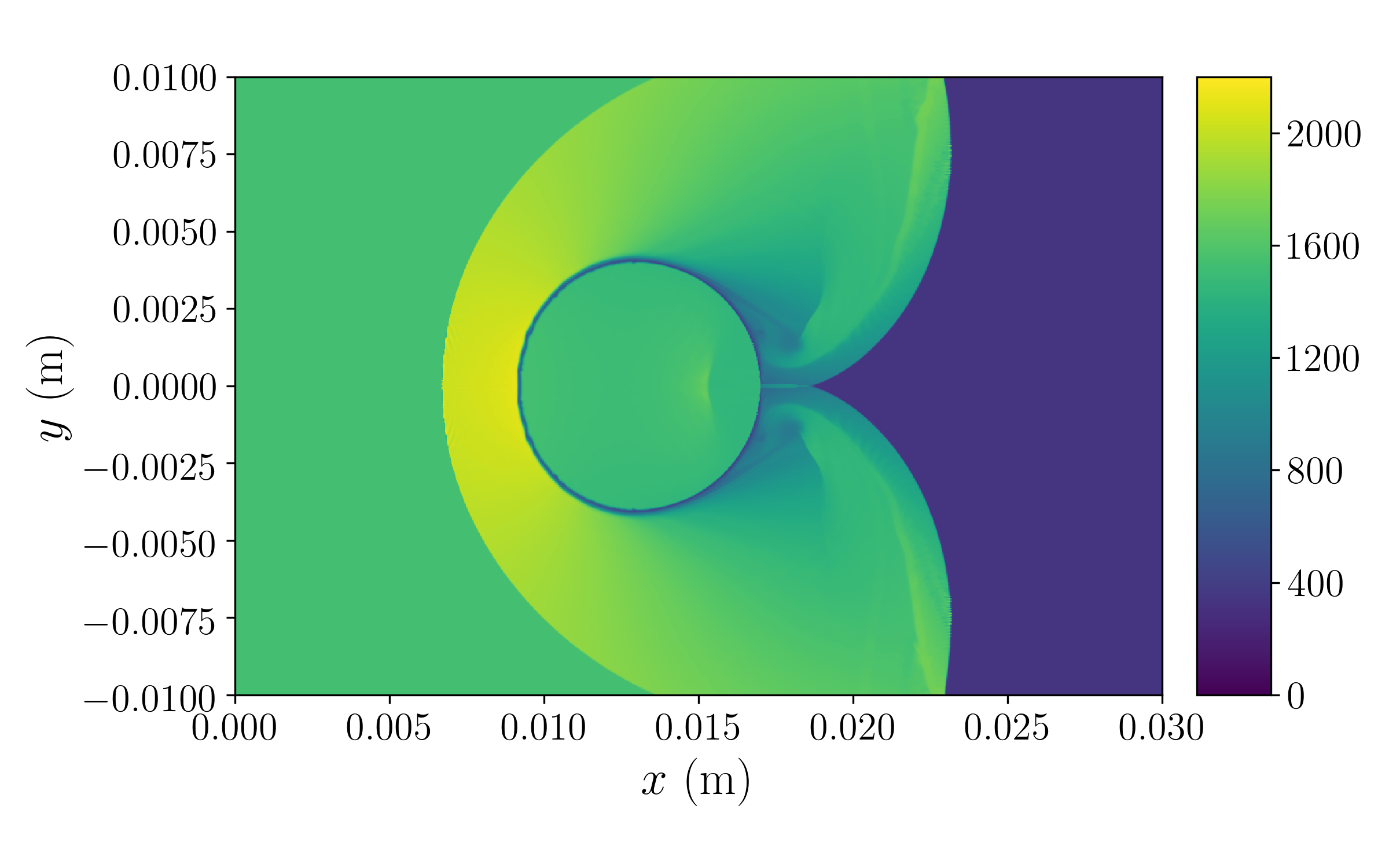}
\label{fig:compare_2D_Mach_10_shock_water_cylinder_sos_t2_WCNS5_IS_PP}}
\subfigure[$t = 8\ \mu\mathrm{s}$, HLLC]{%
\includegraphics[width=0.45\textwidth]{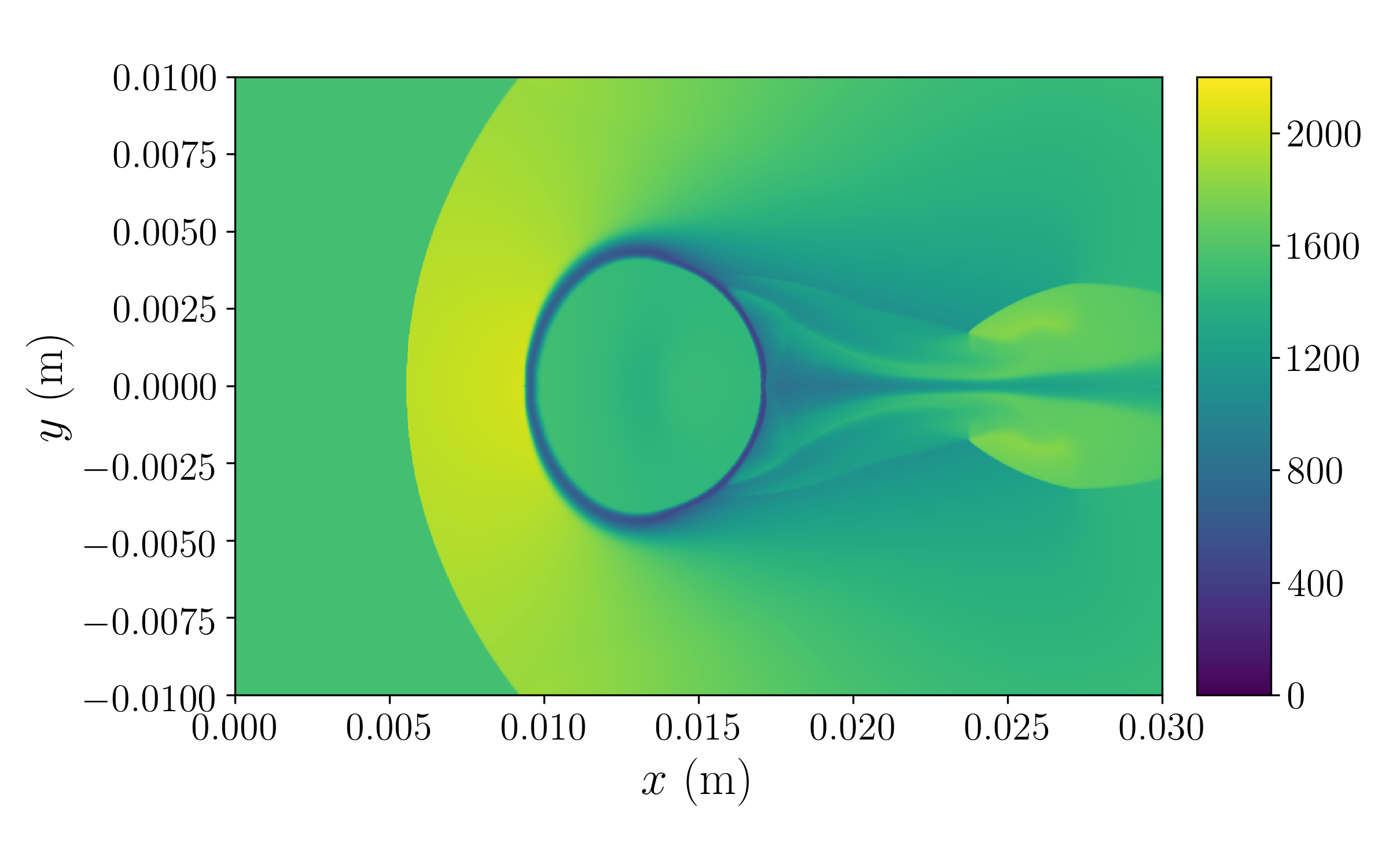}
\label{fig:compare_2D_Mach_10_shock_water_cylinder_sos_t3_HLLC}}
\subfigure[$t = 8\ \mu\mathrm{s}$, PP-WCNS-IS]{%
\includegraphics[width=0.45\textwidth]{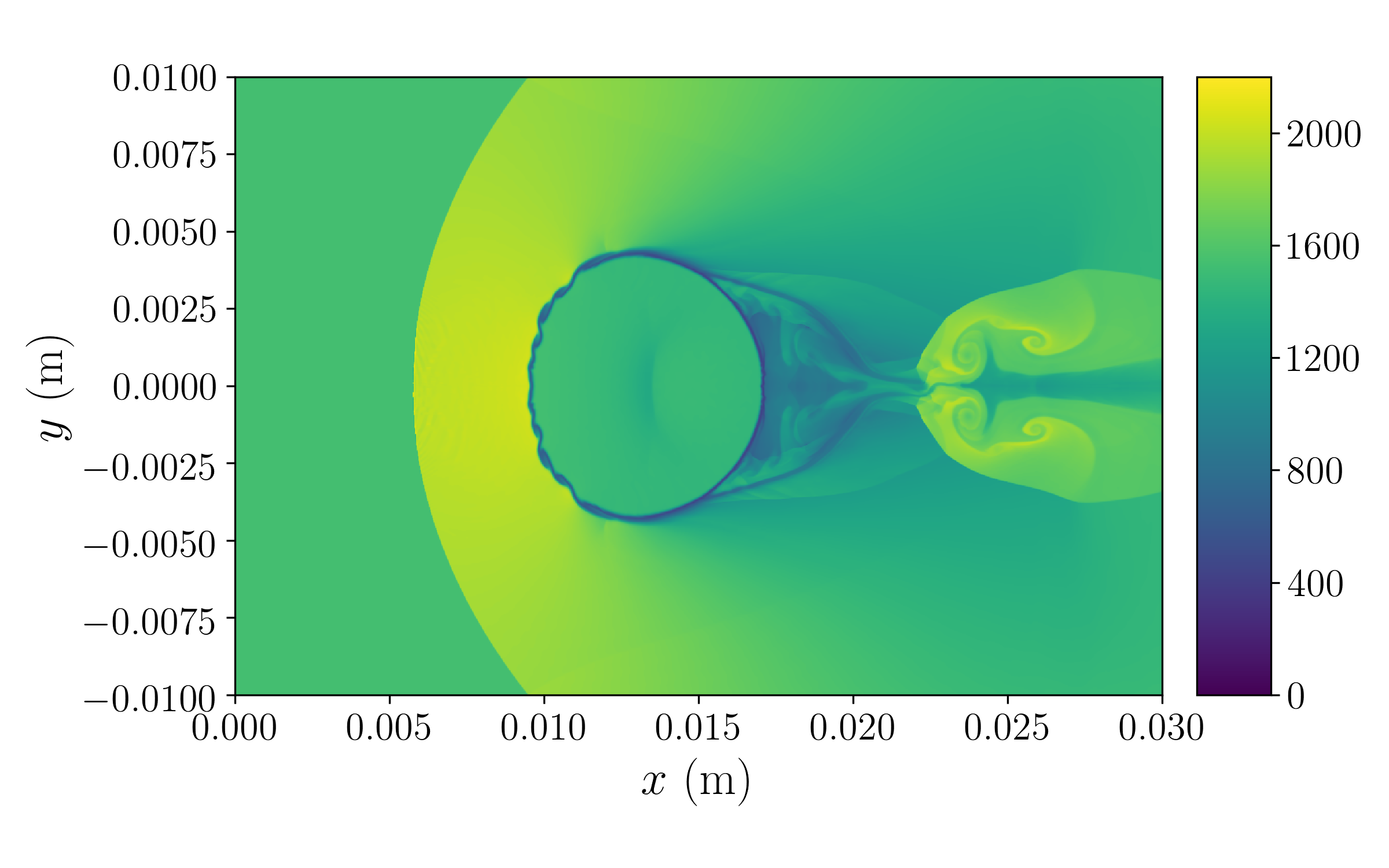}
\label{fig:compare_2D_Mach_10_shock_water_cylinder_sos_t3_WCNS5_IS_PP}}
\subfigure[$t = 16\ \mu\mathrm{s}$, HLLC]{%
\includegraphics[width=0.45\textwidth]{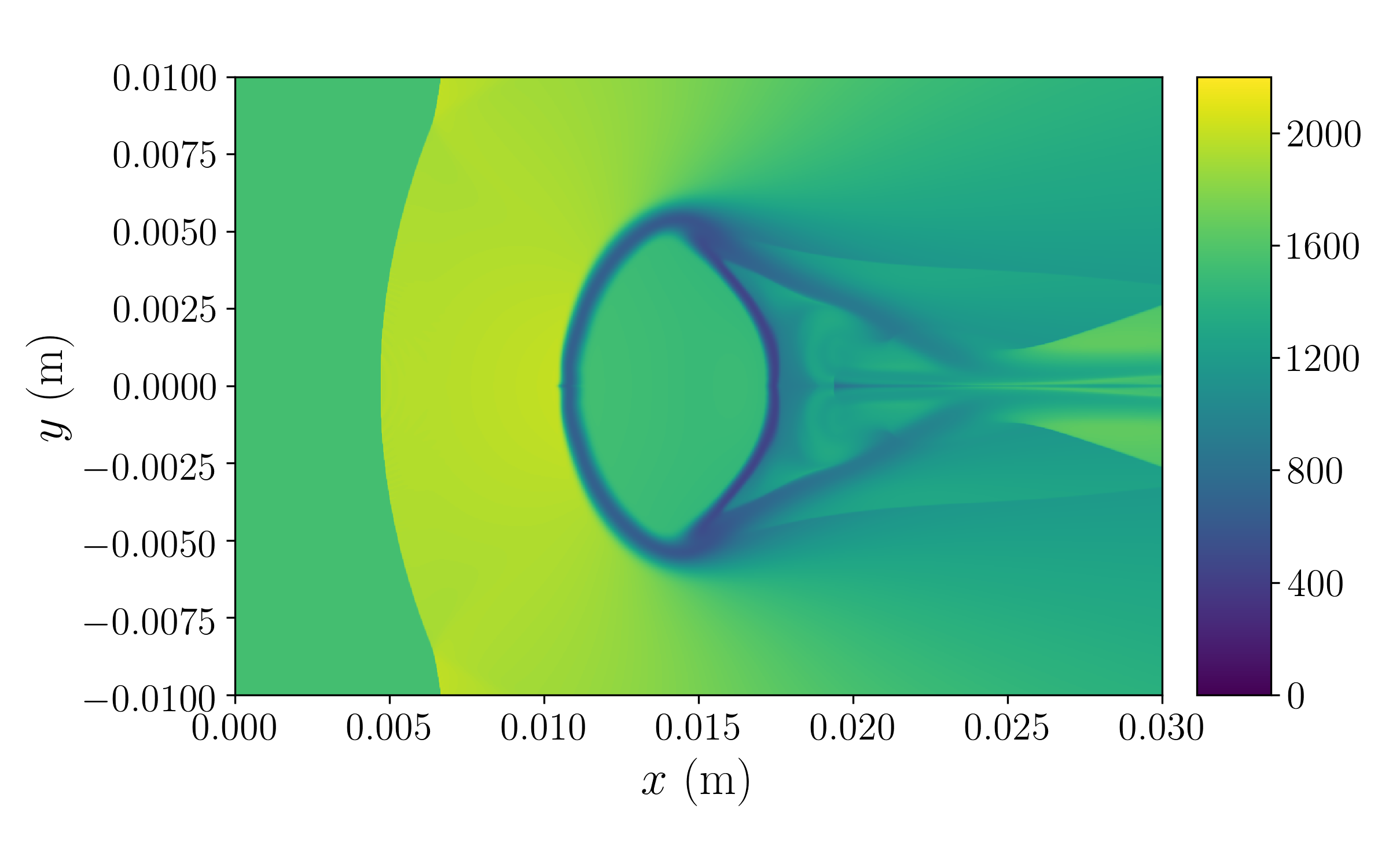}
\label{fig:compare_2D_Mach_10_shock_water_cylinder_sos_t4_HLLC}}
\subfigure[$t = 16\ \mu\mathrm{s}$, PP-WCNS-IS]{%
\includegraphics[width=0.45\textwidth]{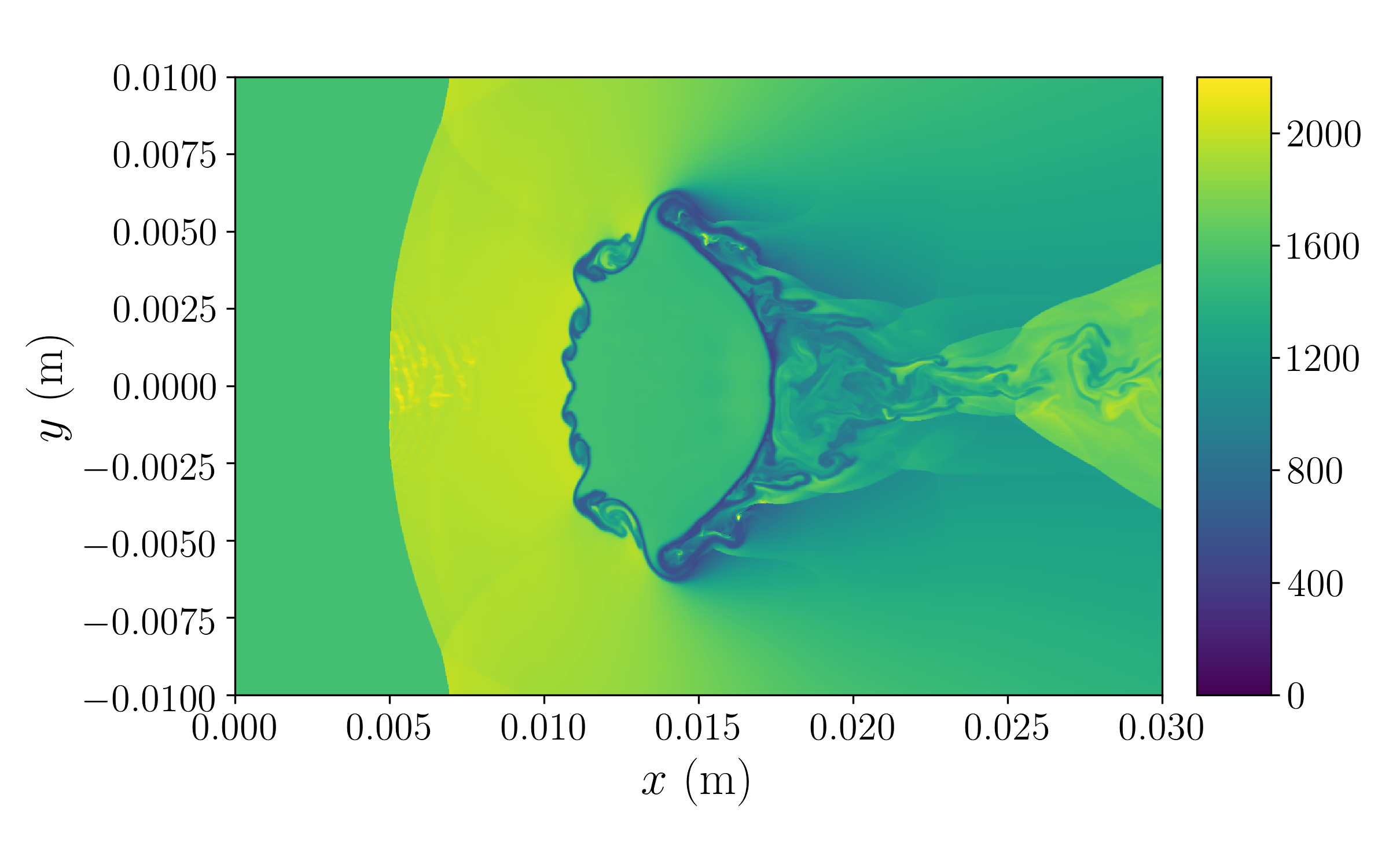}
\label{fig:compare_2D_Mach_10_shock_water_cylinder_sos_t4_WCNS5_IS_PP}}
\caption{Speed of sound of 2D Mach 10 shock water cylinder interaction problem.}
\label{fig:compare_2D_Mach_10_shock_water_cylinder_sos}
\end{figure}

\begin{figure}[!ht]
\centering
\subfigure[$t = 1\ \mu\mathrm{s}$, HLLC]{%
\includegraphics[width=0.45\textwidth]{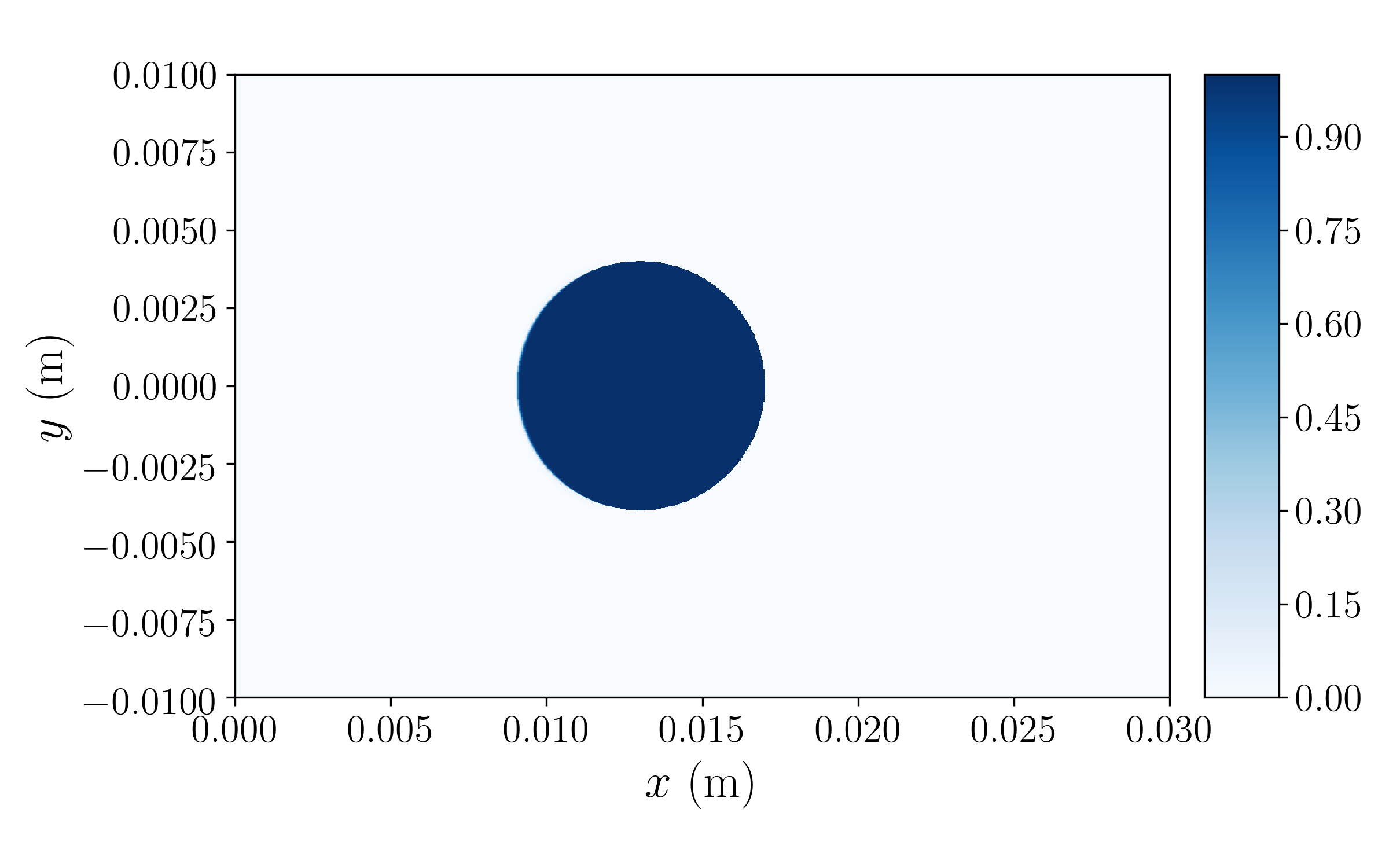}
\label{fig:compare_2D_Mach_10_shock_water_cylinder_alpha0_t1_HLLC}}
\subfigure[$t = 1\ \mu\mathrm{s}$, PP-WCNS-IS]{%
\includegraphics[width=0.45\textwidth]{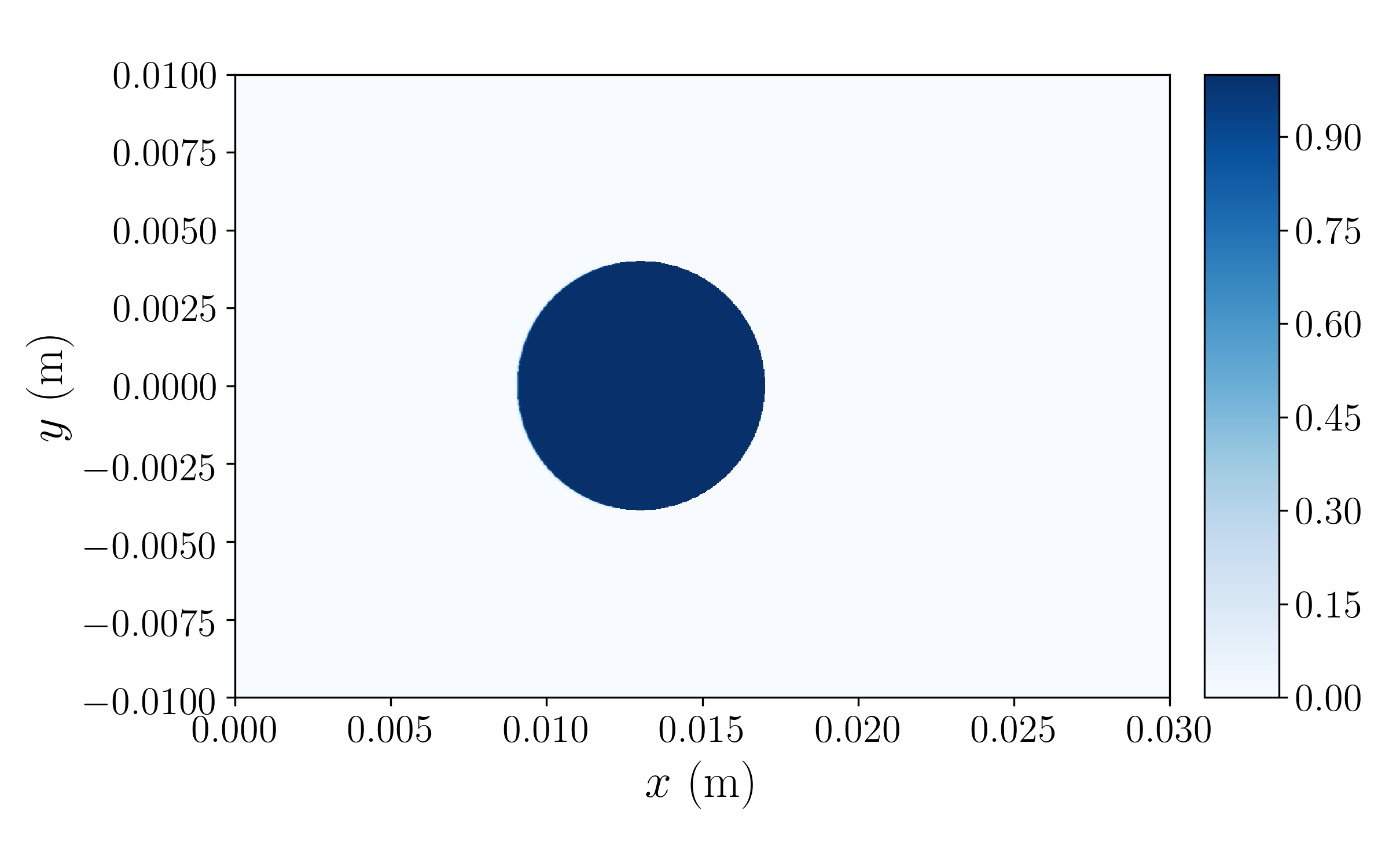}
\label{fig:compare_2D_Mach_10_shock_water_cylinder_alpha0_t1_WCNS5_IS_PP}}
\subfigure[$t = 4\ \mu\mathrm{s}$, HLLC]{%
\includegraphics[width=0.45\textwidth]{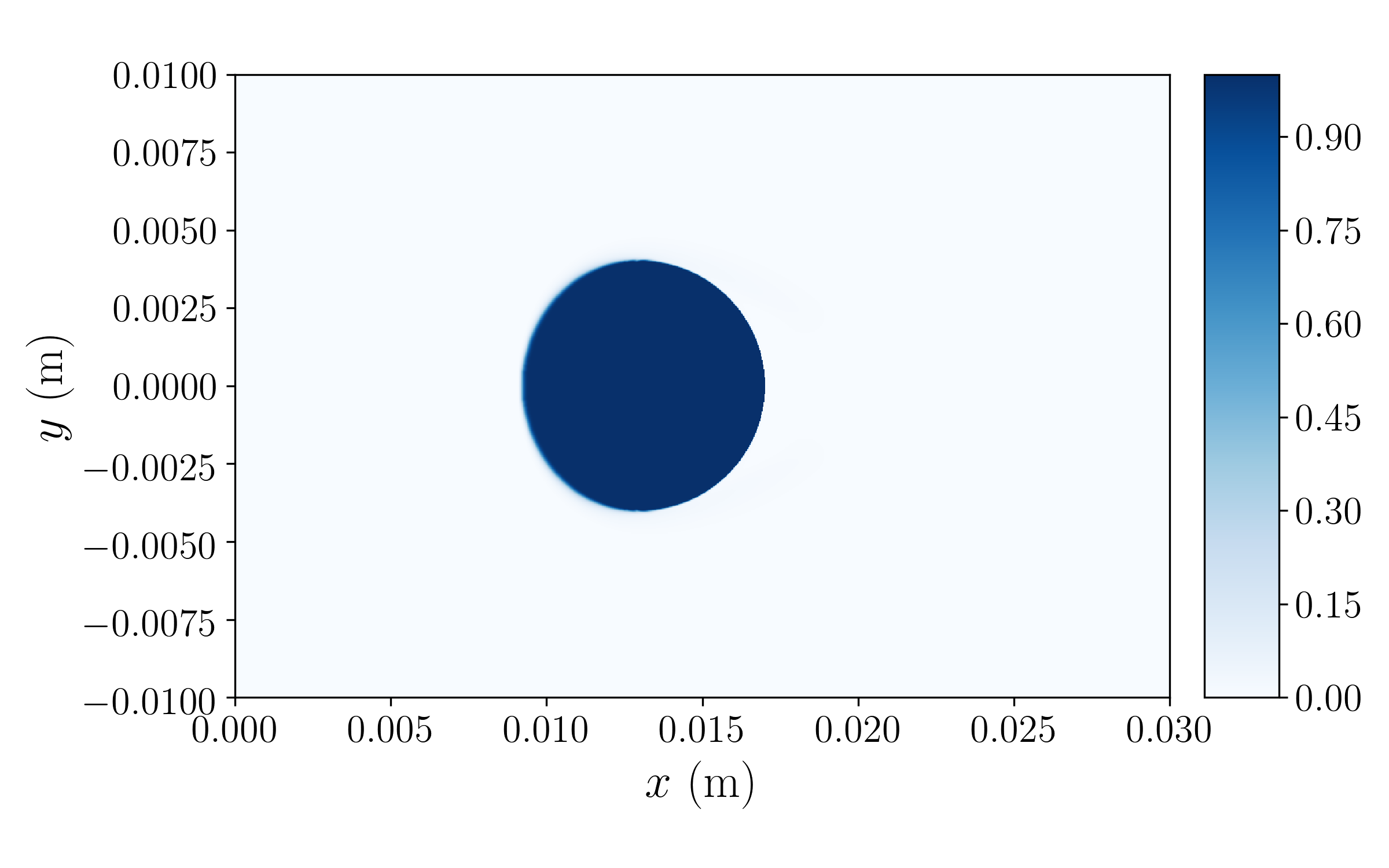}
\label{fig:compare_2D_Mach_10_shock_water_cylinder_alpha0_t2_HLLC}}
\subfigure[$t = 4\ \mu\mathrm{s}$, PP-WCNS-IS]{%
\includegraphics[width=0.45\textwidth]{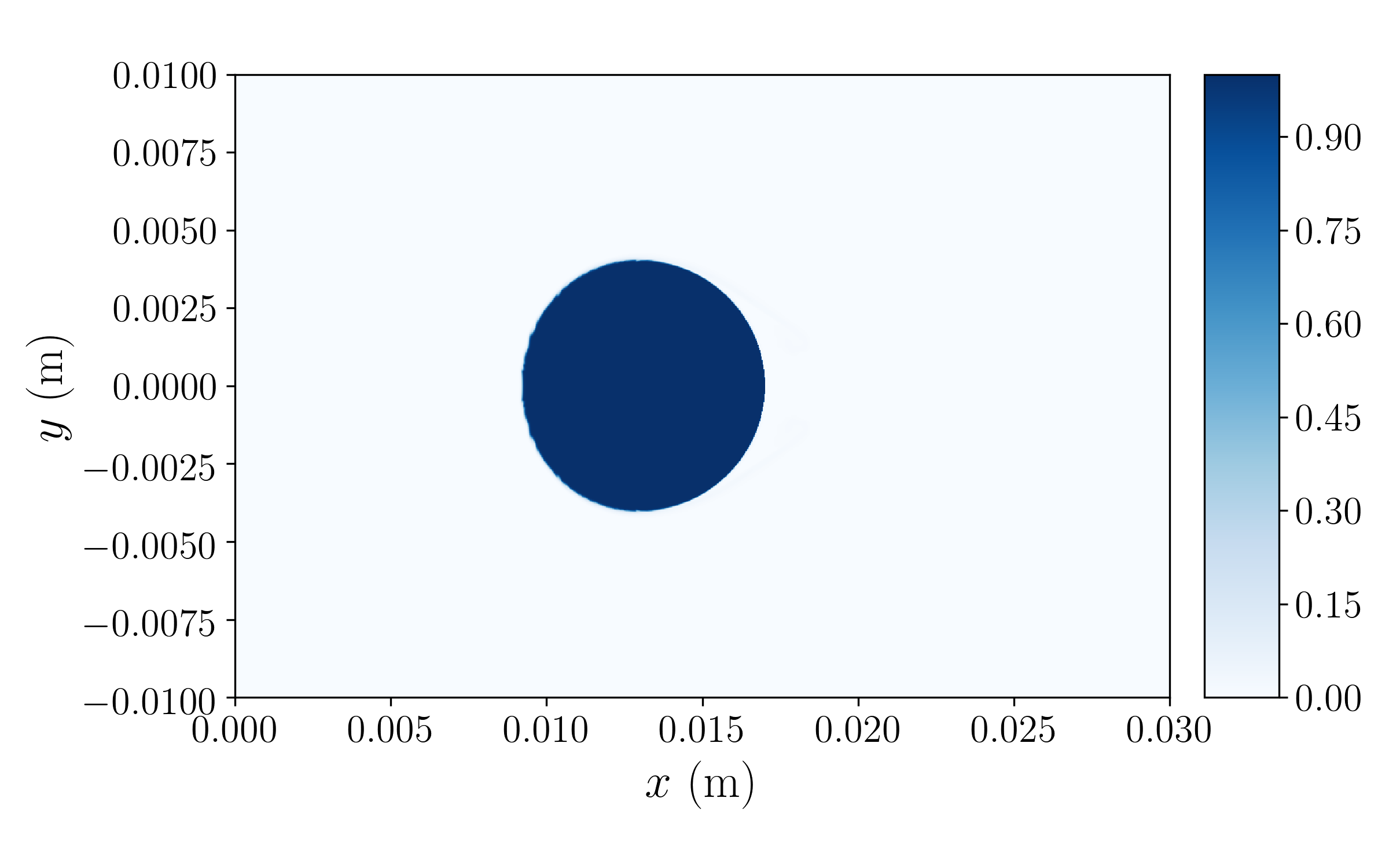}
\label{fig:compare_2D_Mach_10_shock_water_cylinder_alpha0_t2_WCNS5_IS_PP}}
\subfigure[$t = 8\ \mu\mathrm{s}$, HLLC]{%
\includegraphics[width=0.45\textwidth]{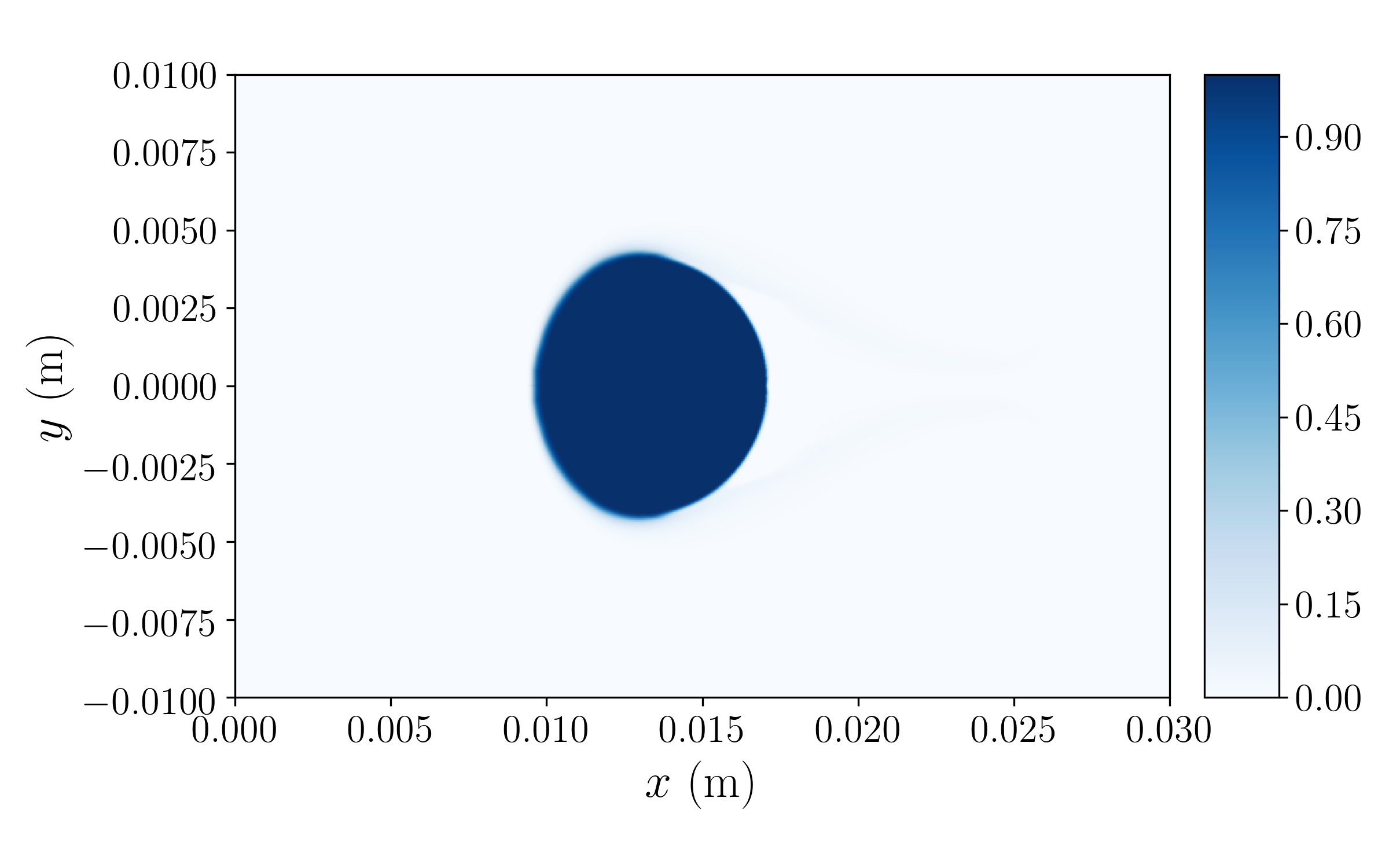}
\label{fig:compare_2D_Mach_10_shock_water_cylinder_alpha0_t3_HLLC}}
\subfigure[$t = 8\ \mu\mathrm{s}$, PP-WCNS-IS]{%
\includegraphics[width=0.45\textwidth]{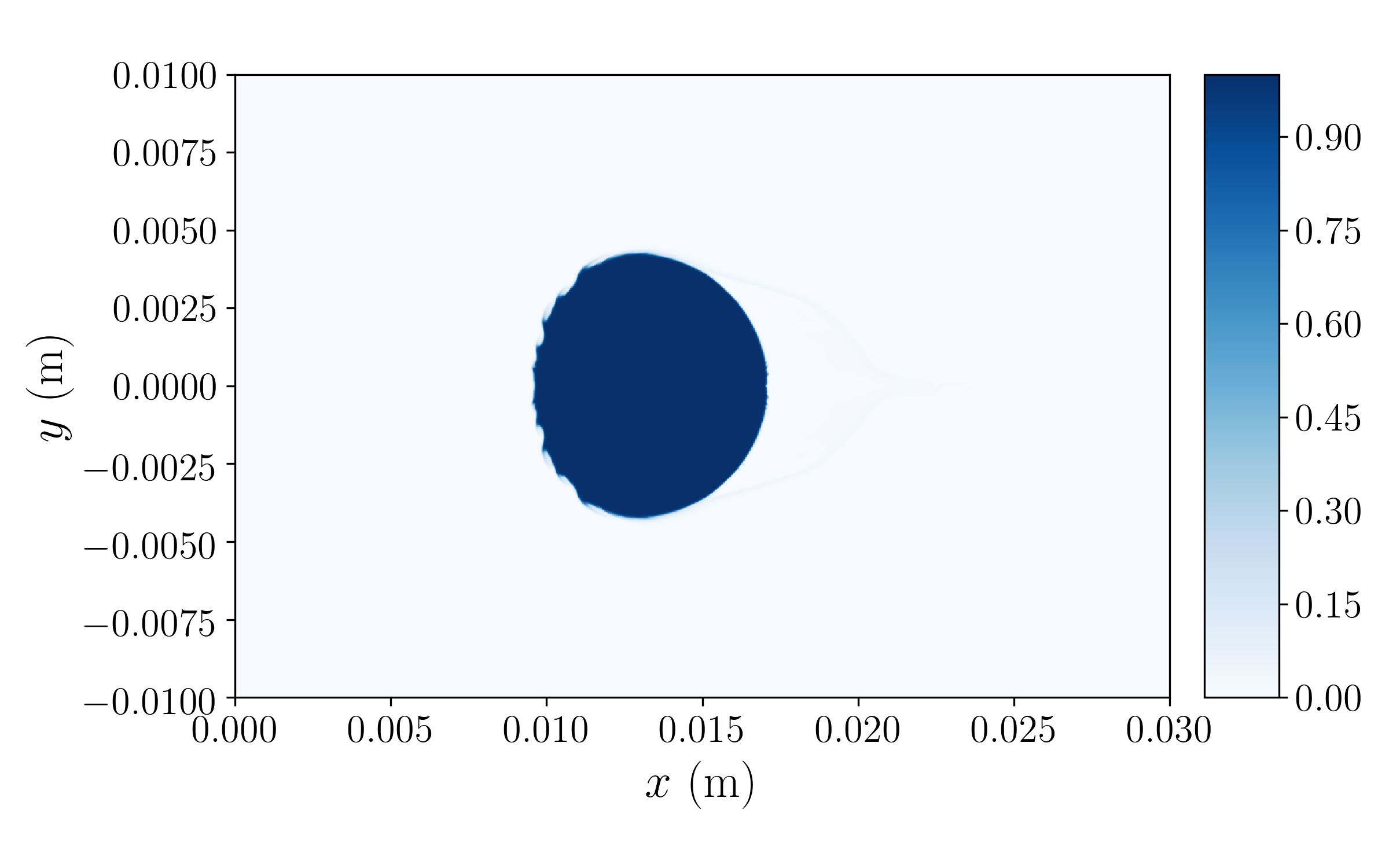}
\label{fig:compare_2D_Mach_10_shock_water_cylinder_alpha0_t3_WCNS5_IS_PP}}
\subfigure[$t = 16\ \mu\mathrm{s}$, HLLC]{%
\includegraphics[width=0.45\textwidth]{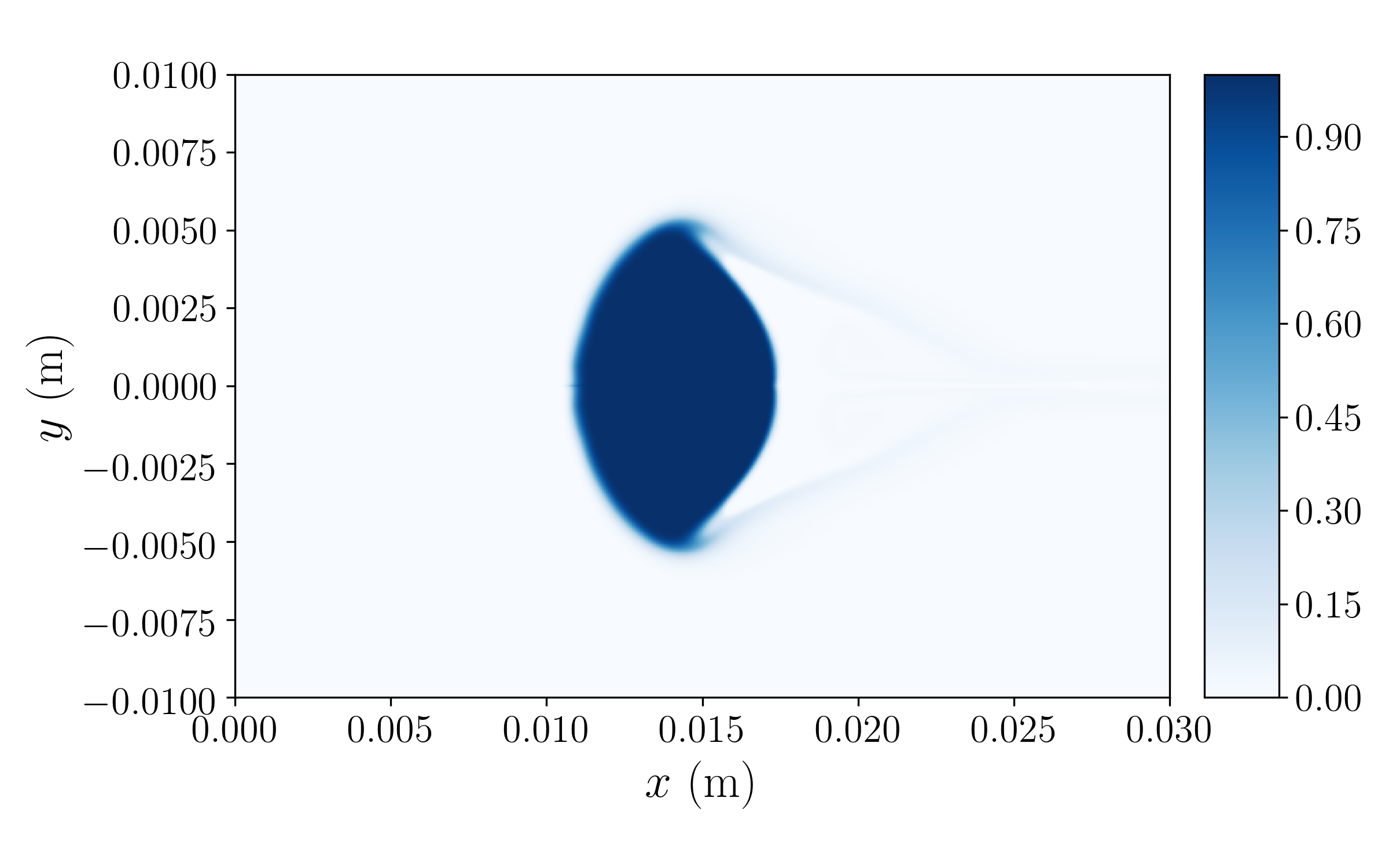}
\label{fig:compare_2D_Mach_10_shock_water_cylinder_alpha0_t4_HLLC}}
\subfigure[$t = 16\ \mu\mathrm{s}$, PP-WCNS-IS]{%
\includegraphics[width=0.45\textwidth]{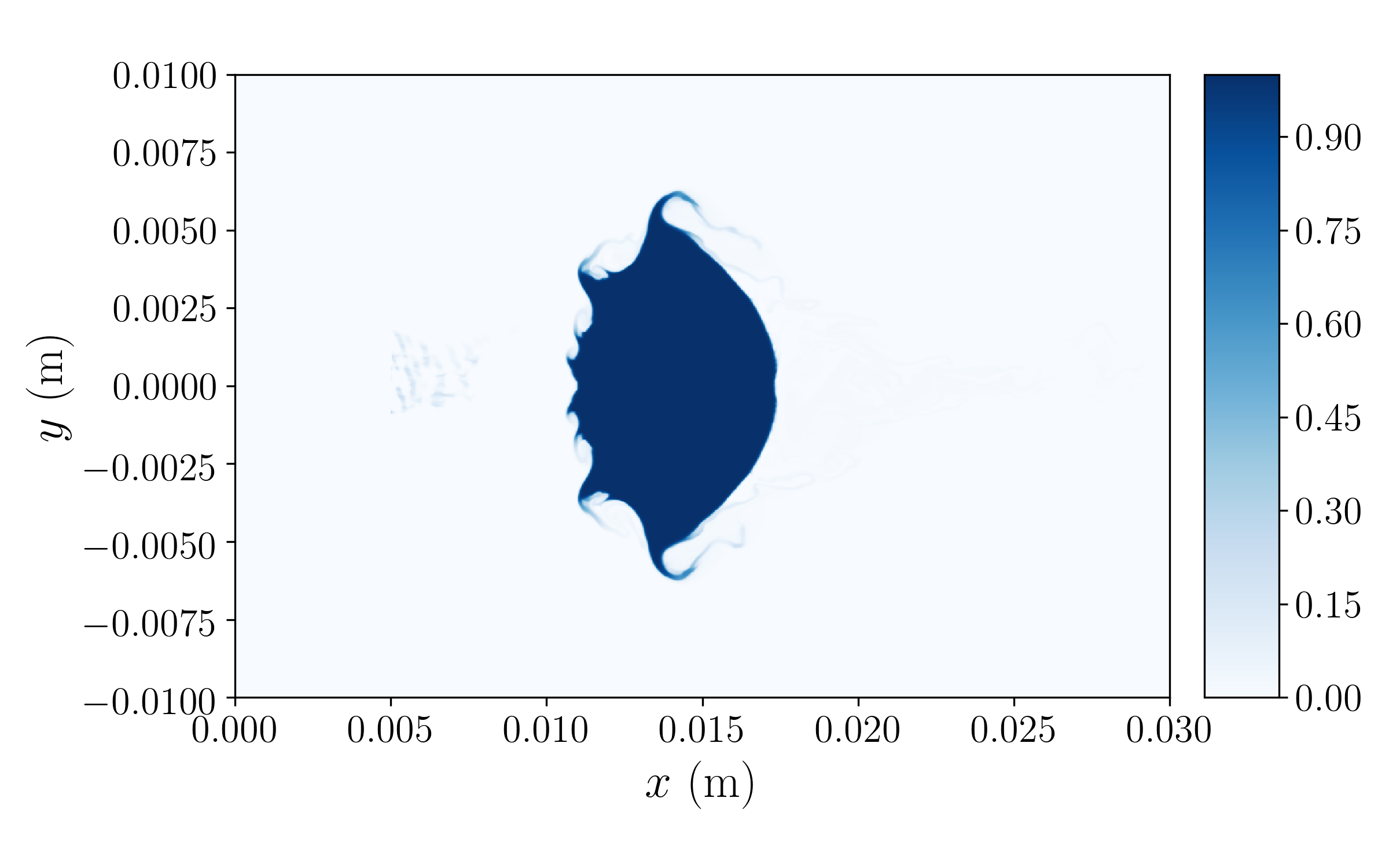}
\label{fig:compare_2D_Mach_10_shock_water_cylinder_alpha0_t4_WCNS5_IS_PP}}
\caption{Volume fraction of water of 2D Mach 10 shock water cylinder interaction problem.}
\label{fig:compare_2D_Mach_10_shock_water_cylinder_alpha0}
\end{figure}

This is a very extreme problem due to the high incident shock Mach number initially. The problem is simulated with the first order HLLC and the high-order PP-WCNS-IS. While the first order scheme has no numerical difficulty in this test case, numerical failure is experienced with WCNS-IS without the positivity- and boundedness-preserving limiters as the speed of sound becomes imaginary. This can happen at the strong incident shock, or at the low pressure regions created by the expansion waves inside the water column and at the counter-rotating vortices. The positivity- and boundedness-preserving limiters are necessary for the WCNS-IS scheme in this problem. Figure~\ref{fig:compare_2D_Mach_10_shock_water_cylinder_schl} shows the comparison of numerical schlieren defined as $\exp{\left( \left| \nabla \rho \right| / \left| \nabla \rho \right|_{\mathrm{max}} \right)}$ between the two schemes. From the figure, we can see that the PP-WCNS-IS has much thinner interface thickness over time compared with first order HLLC scheme and this is consistent with other test problems. Vortical features produced by the hydrodynamic instability due to baroclinic torque are observed at the interface for PP-WCNS-IS as time evolves. However, the first order scheme is too dissipative to produce the roll-up of the interfaces at the chosen grid resolution. The comparison of speed of sound between the two schemes can be seen in figure~\ref{fig:compare_2D_Mach_10_shock_water_cylinder_sos}. Finally, the volume fraction fields of both schemes are shown in figure~\ref{fig:compare_2D_Mach_10_shock_water_cylinder_alpha0}. It should be noted that the volume fraction field is also verified to be always bounded by the corresponding threshold chosen in the positivity- and boundedness-preserving limiters for PP-WCNS-IS.

\subsection{Two-dimensional Mach 100 water jet problem}

This test case is a multi-phase version of the popular Mach 2000 jet problem first proposed by~\citet{zhang2010positivity}. In this problem, a Mach 100 water jet enters a domain full of ambient air. The domain size is $\left[ 0, L \right] \times \left[ -0.25L, 0.25L \right]$, where $L = 1\ \mathrm{m}$ is chosen. The initial conditions of the ambient air are given by table~\ref{table:IC_2D_Mach_100_water_jet}. Constant extrapolation is used at top, bottom and right boundaries. The left boundary is described by Dirichlet boundary conditions given by table~\ref{table:BC_2D_Mach_100_water_jet}. The speed of the jet is $1.5\mathrm{e}{5}\ \mathrm{m\ s^{-1}}$, which is around Mach 100 with respect to the sound speed in the water jet. The computations are performed on a $1024 \times 512$ mesh.

\begin{table}[!ht]
  \begin{center}
    \begin{tabular}{@{}cccccc@{}}\toprule
    \addstackgap{\stackanchor{$\alpha_1 \rho_1$}{$(\mathrm{kg\ m^{-3}})$}} &
    \stackanchor{$\alpha_2 \rho_2$}{$(\mathrm{kg\ m^{-3}})$} &
    \stackanchor{$u$}{$(\mathrm{m\ s^{-1}})$} &
    \stackanchor{$v$}{$(\mathrm{m\ s^{-1}})$} &
    \stackanchor{$p$}{$(\mathrm{Pa})$} &
    $\alpha_1$ \\ \midrule
    $1.0\mathrm{e}{-8}$ & 1 & 0 & 0 & $1.0\mathrm{e}{5}$ & $1.0\mathrm{e}{-8}$ \\
    \bottomrule
    \end{tabular}
  \end{center}
  \caption{Initial conditions of 2D Mach 100 water jet problem.}
  \label{table:IC_2D_Mach_100_water_jet}
\end{table}

\begin{table}[!ht]
  \begin{center}
    \begin{tabular}{@{}c | cccccc@{}} \toprule
     &
    \addstackgap{\stackanchor{$\alpha_1 \rho_1$}{$(\mathrm{kg\ m^{-3}})$}} &
    \stackanchor{$\alpha_2 \rho_2$}{$(\mathrm{kg\ m^{-3}})$} &
    \stackanchor{$u$}{$(\mathrm{m\ s^{-1}})$} &
    \stackanchor{$v$}{$(\mathrm{m\ s^{-1}})$} &
    \stackanchor{$p$}{$(\mathrm{Pa})$} &
    $\alpha_1$ \\ \midrule
    \addstackgap{$y \in \left[ -0.05L, 0.05L \right]$}  & 1000 & $1.0\mathrm{e}{-8}$ & $1.5\mathrm{e}{5}$ & 0 & $1.0\mathrm{e}{5}$ & $1 - 1.0\mathrm{e}{-8}$ \\
    \addstackgap{otherwise} & $1.0\mathrm{e}{-8}$ & 1 & 0 & 0 & $1.0\mathrm{e}{5}$ & $1.0\mathrm{e}{-8}$ \\ \bottomrule
    \end{tabular}
  \end{center}
  \caption{Left boundary conditions of 2D Mach 100 water jet problem.}
  \label{table:BC_2D_Mach_100_water_jet}
\end{table}

\begin{figure}[!ht]
\centering
\subfigure[$t = 1\ \mu\mathrm{s}$, HLLC]{%
\includegraphics[width=0.45\textwidth]{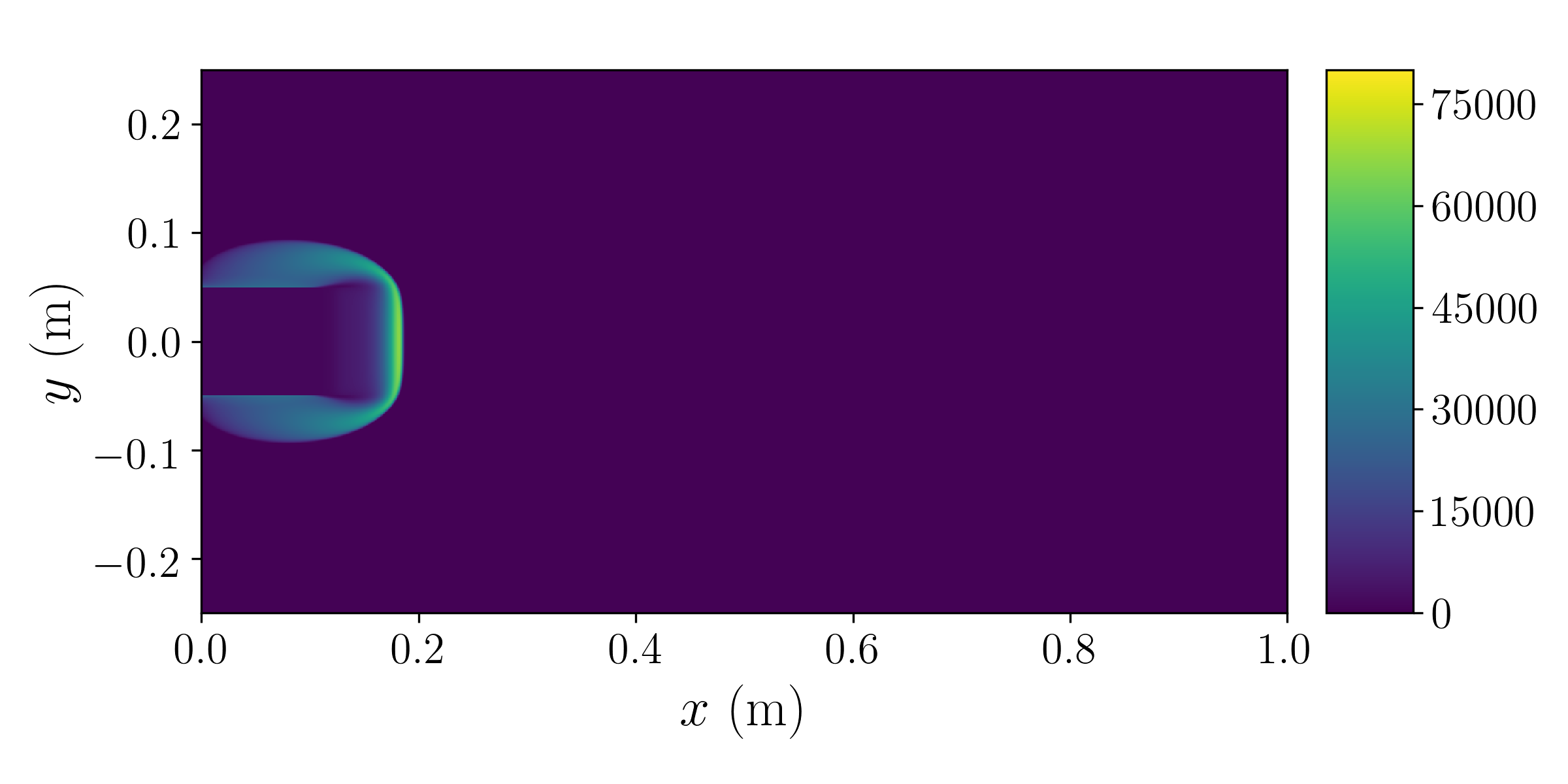}
\label{fig:compare_2D_Mach_100_water_jet_sos_t1_HLLC}}
\subfigure[$t = 1\ \mu\mathrm{s}$, PP-WCNS-IS]{%
\includegraphics[width=0.45\textwidth]{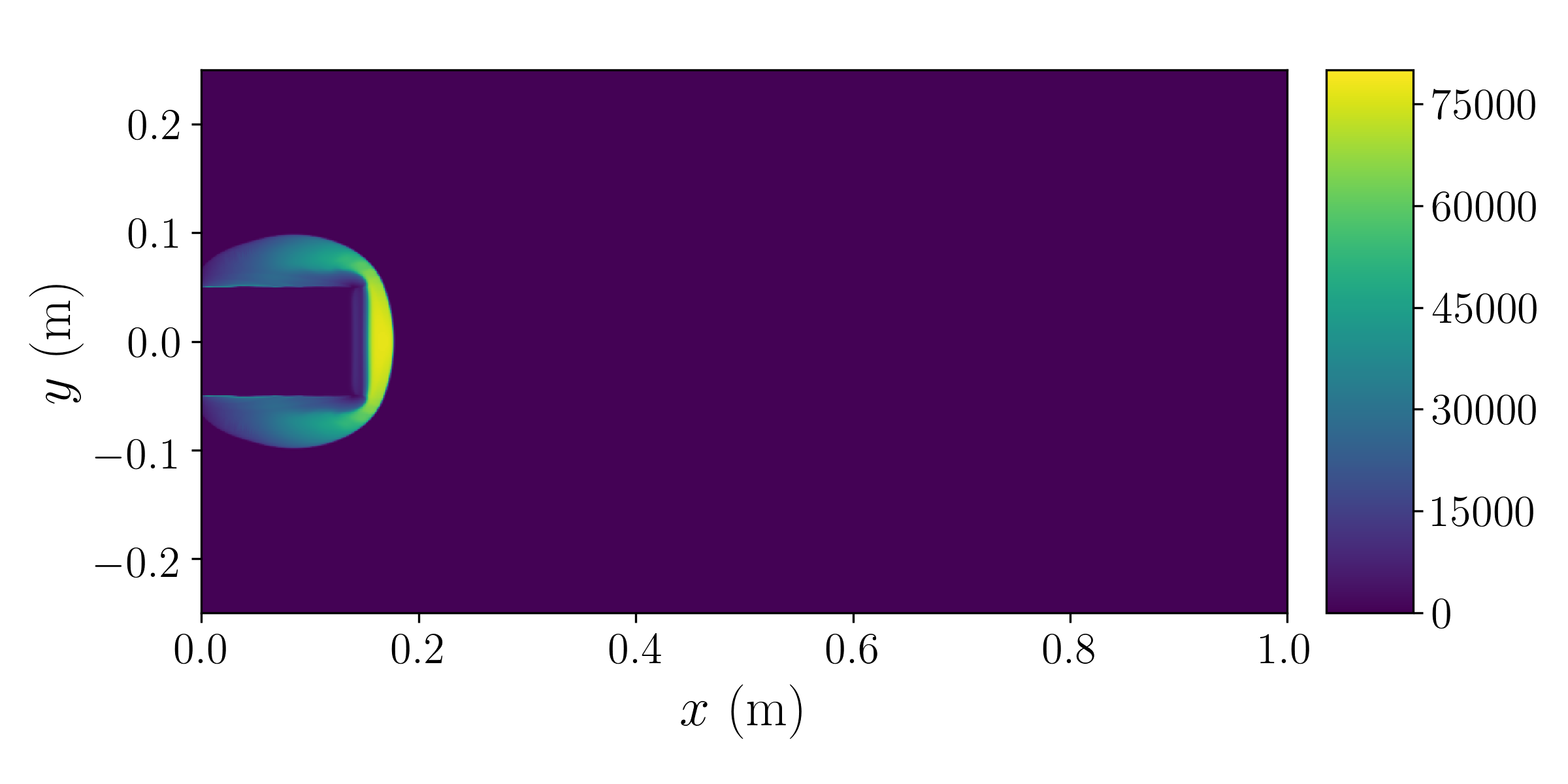}
\label{fig:compare_2D_Mach_100_water_jet_sos_t1_WCNS5_IS_PP}}
\subfigure[$t = 2\ \mu\mathrm{s}$, HLLC]{%
\includegraphics[width=0.45\textwidth]{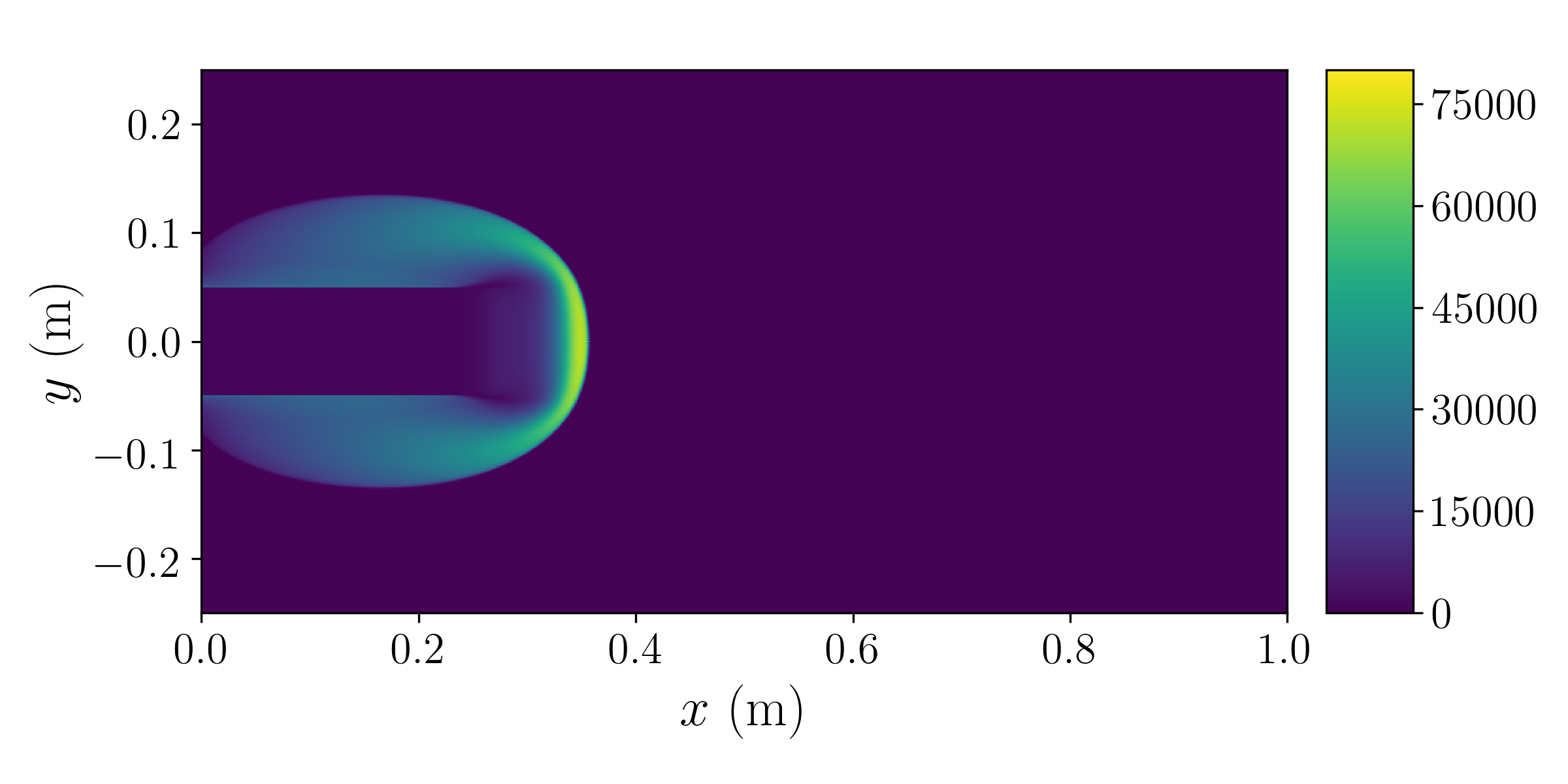}
\label{fig:compare_2D_Mach_100_water_jet_sos_t2_HLLC}}
\subfigure[$t = 2\ \mu\mathrm{s}$, PP-WCNS-IS]{%
\includegraphics[width=0.45\textwidth]{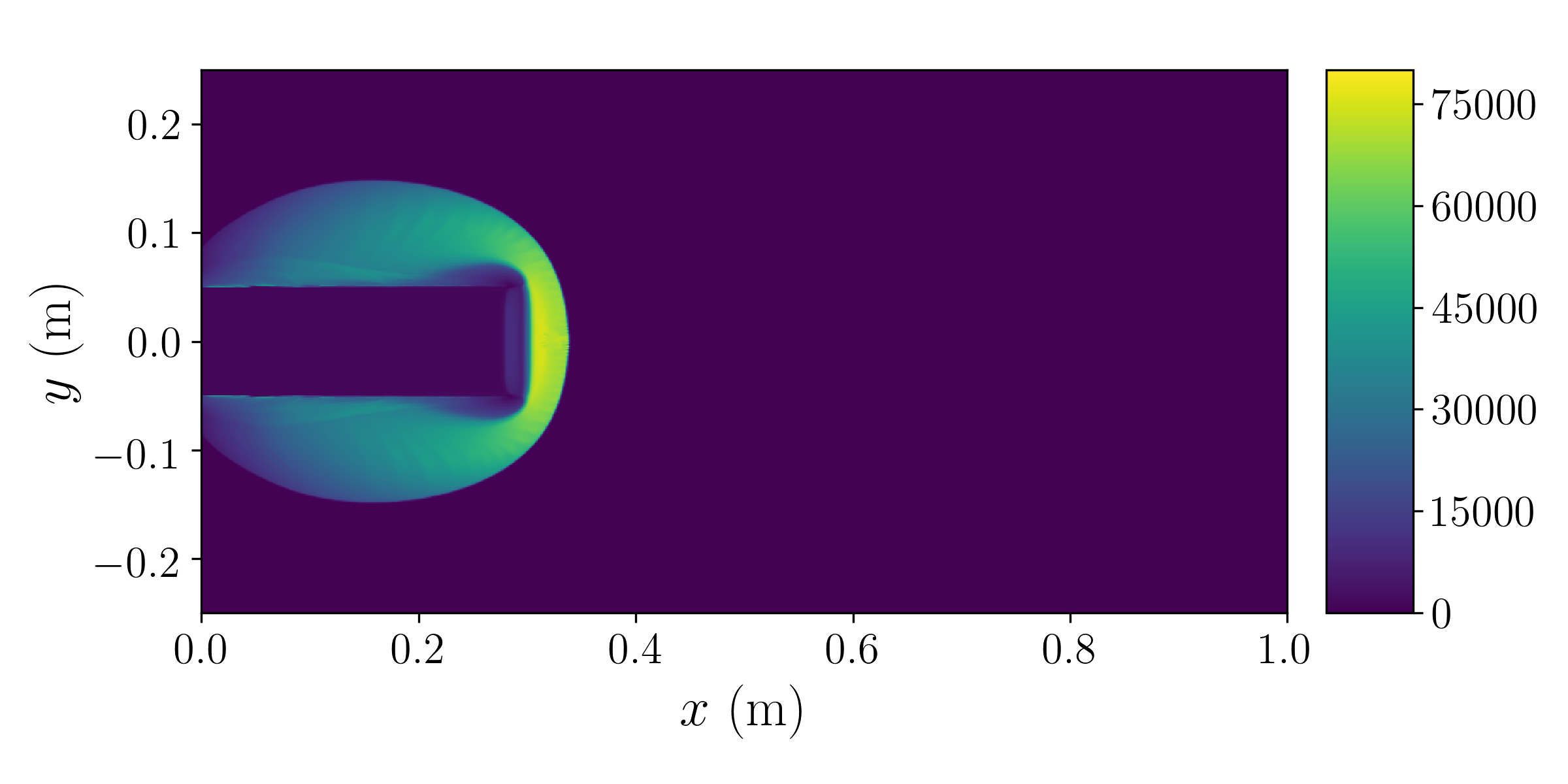}
\label{fig:compare_2D_Mach_100_water_jet_sos_t2_WCNS5_IS_PP}}
\subfigure[$t = 4\ \mu\mathrm{s}$, HLLC]{%
\includegraphics[width=0.45\textwidth]{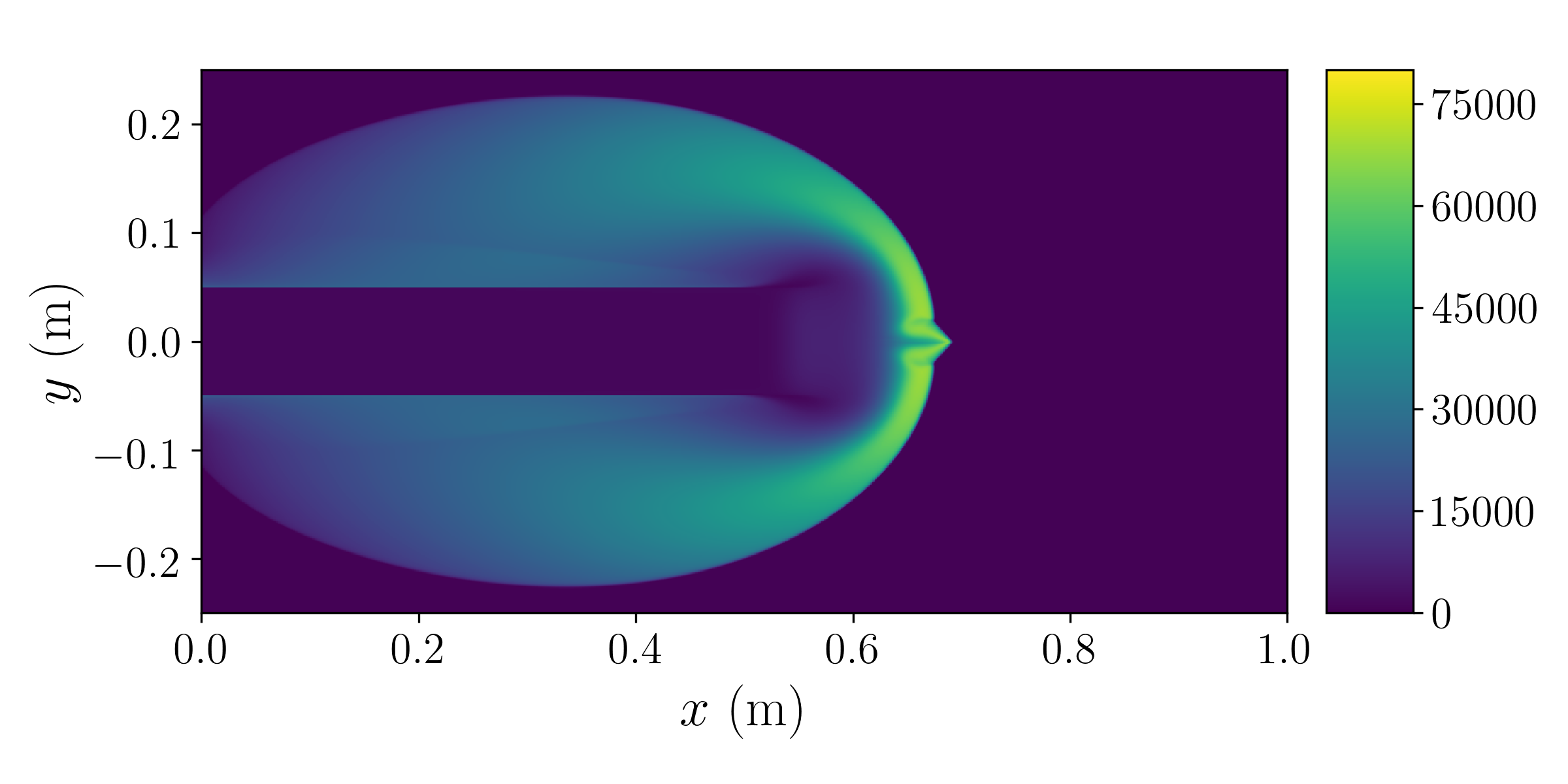}
\label{fig:compare_2D_Mach_100_water_jet_sos_t3_HLLC}}
\subfigure[$t = 4\ \mu\mathrm{s}$, PP-WCNS-IS]{%
\includegraphics[width=0.45\textwidth]{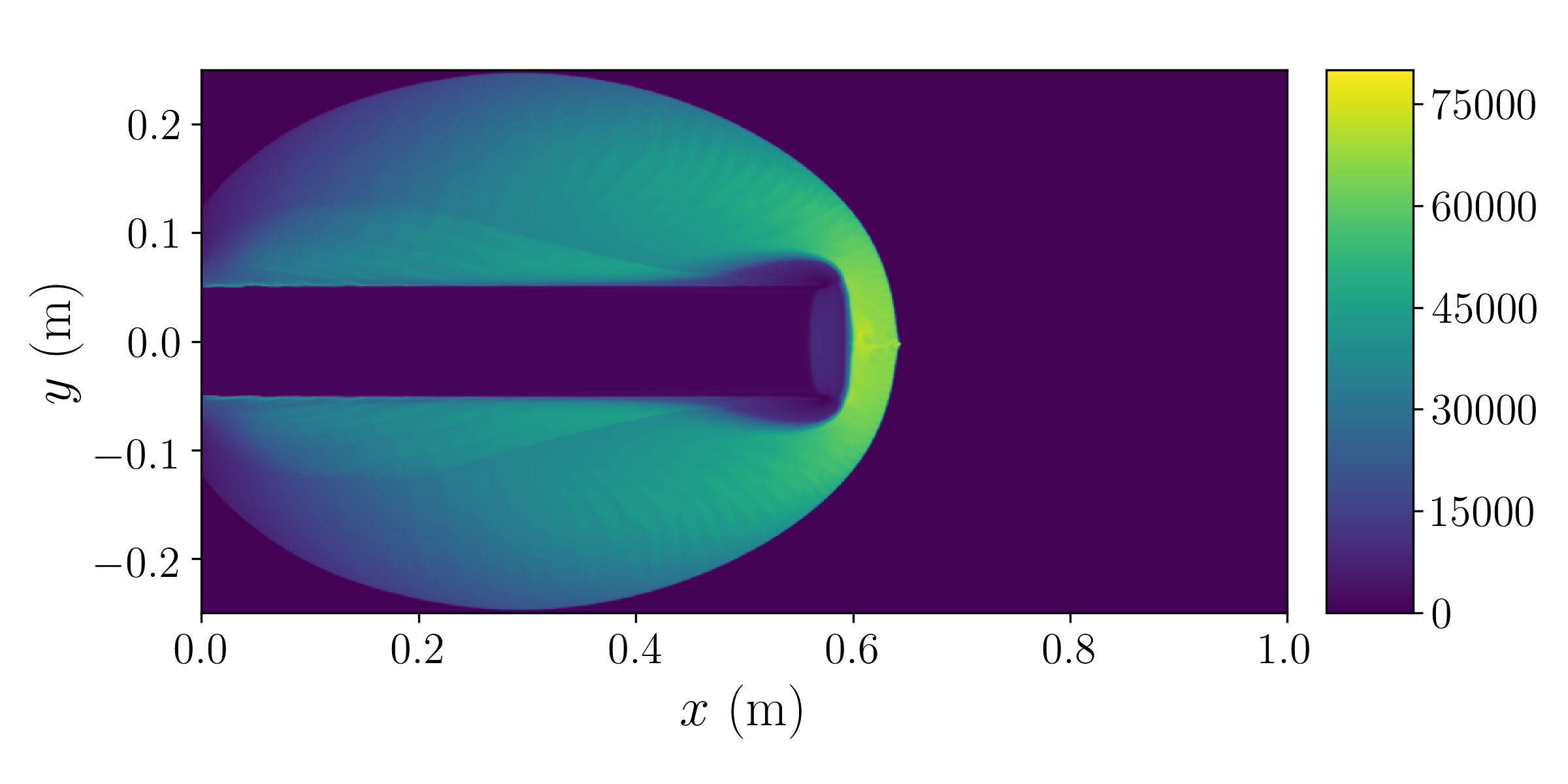}
\label{fig:compare_2D_Mach_100_water_jet_sos_t3_WCNS5_IS_PP}}
\subfigure[$t = 6\ \mu\mathrm{s}$, HLLC]{%
\includegraphics[width=0.45\textwidth]{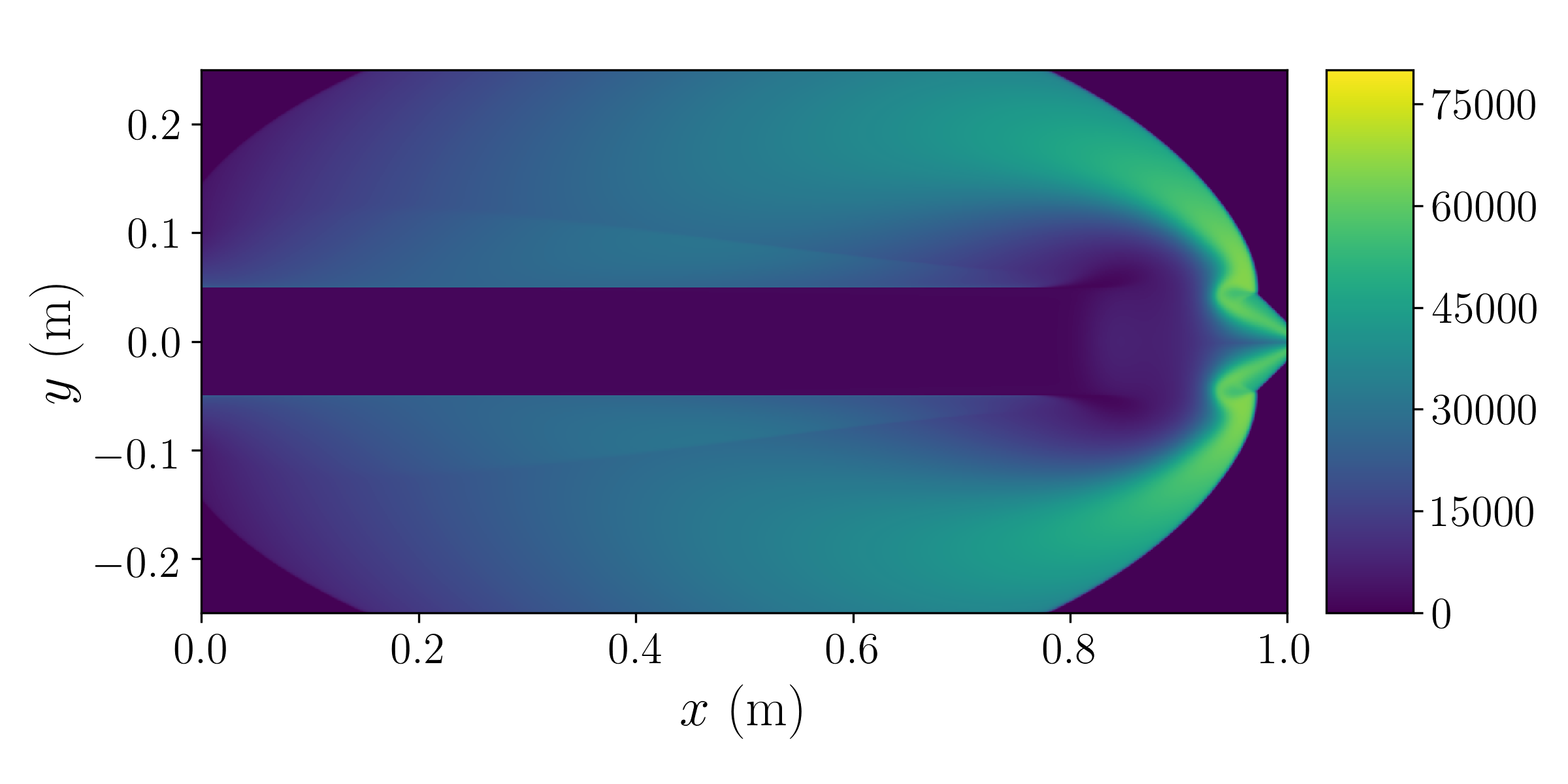}
\label{fig:compare_2D_Mach_100_water_jet_sos_t4_HLLC}}
\subfigure[$t = 6\ \mu\mathrm{s}$, PP-WCNS-IS]{%
\includegraphics[width=0.45\textwidth]{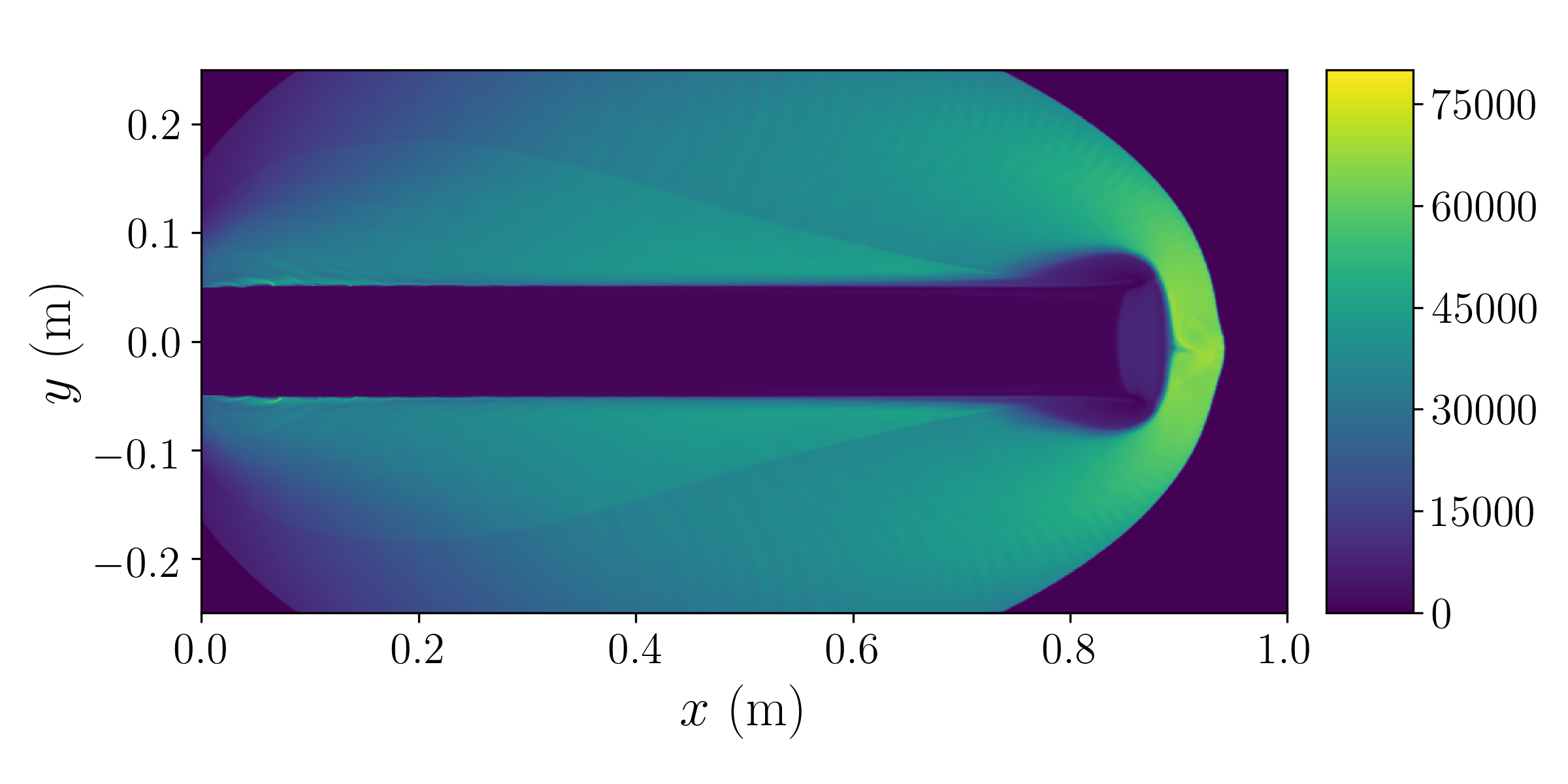}
\label{fig:compare_2D_Mach_100_water_jet_sos_t4_WCNS5_IS_PP}}
\caption{Speed of sound of 2D Mach 100 water jet problem.}
\label{fig:compare_2D_Mach_100_water_jet_sos}
\end{figure}

\begin{figure}[!ht]
\centering
\subfigure[$t = 1\ \mu\mathrm{s}$, HLLC]{%
\includegraphics[width=0.45\textwidth]{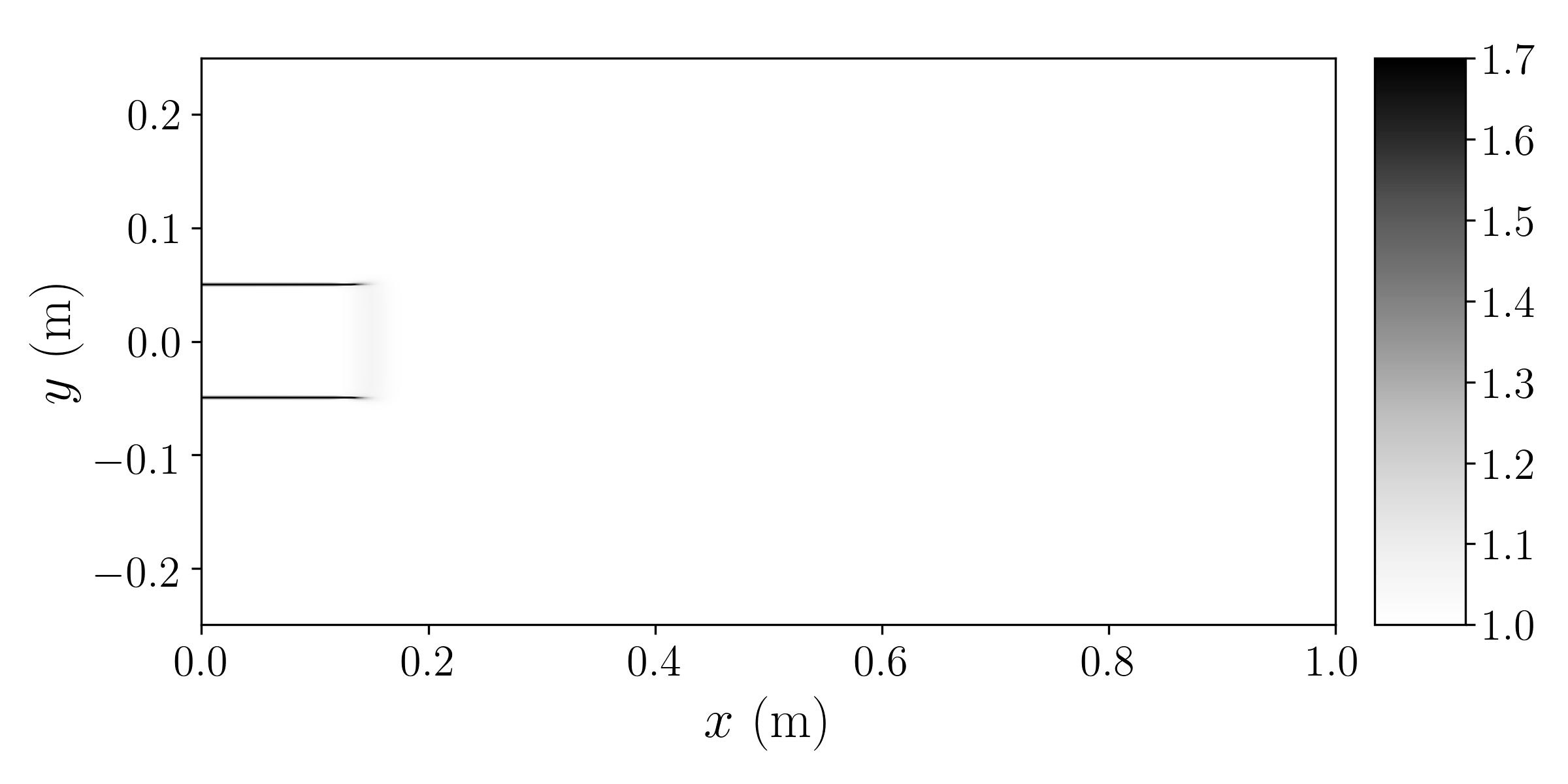}
\label{fig:compare_2D_Mach_100_water_jet_schl_t1_HLLC}}
\subfigure[$t = 1\ \mu\mathrm{s}$, PP-WCNS-IS]{%
\includegraphics[width=0.45\textwidth]{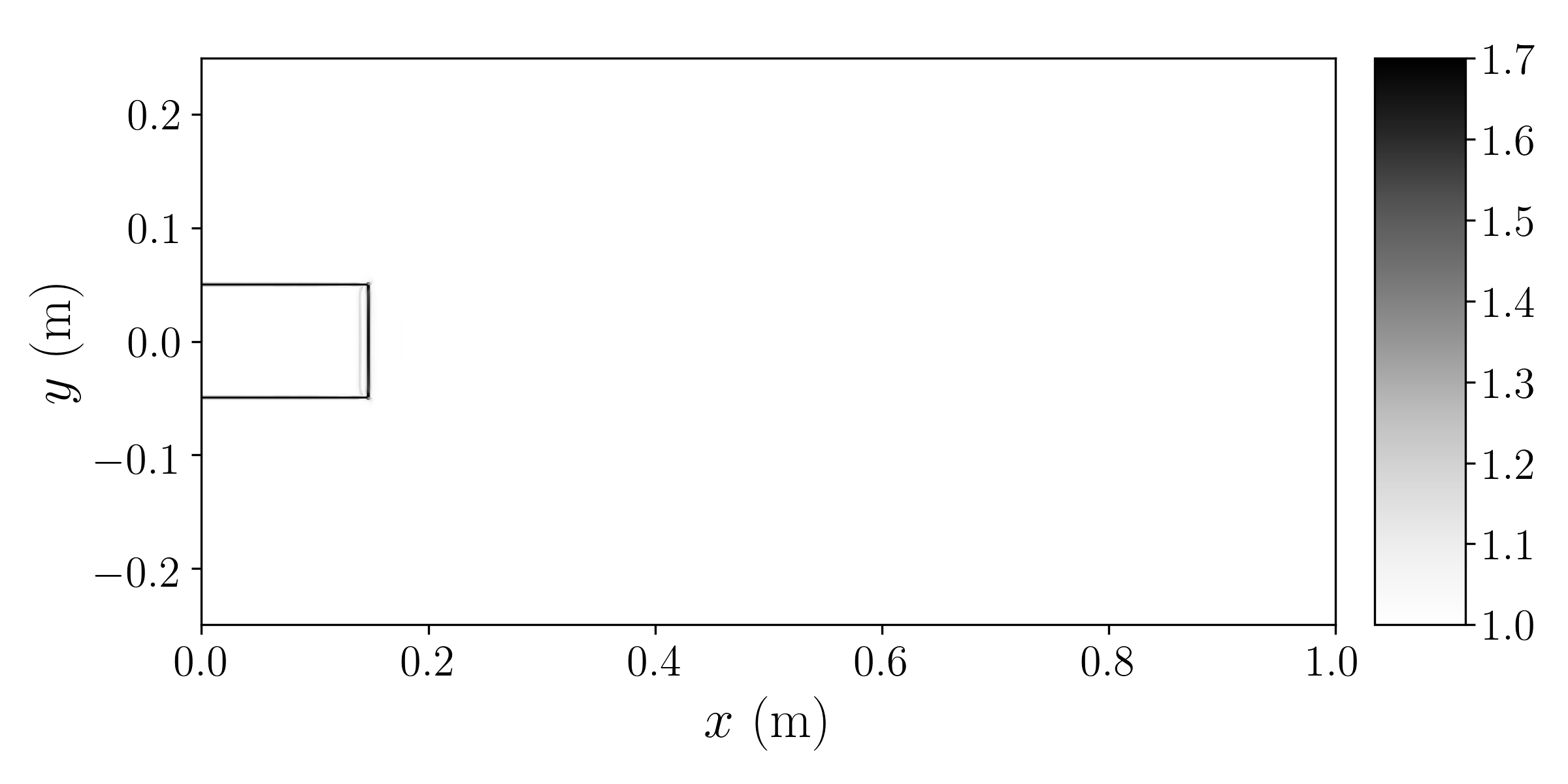}
\label{fig:compare_2D_Mach_100_water_jet_schl_t1_WCNS5_IS_PP}}
\subfigure[$t = 2\ \mu\mathrm{s}$, HLLC]{%
\includegraphics[width=0.45\textwidth]{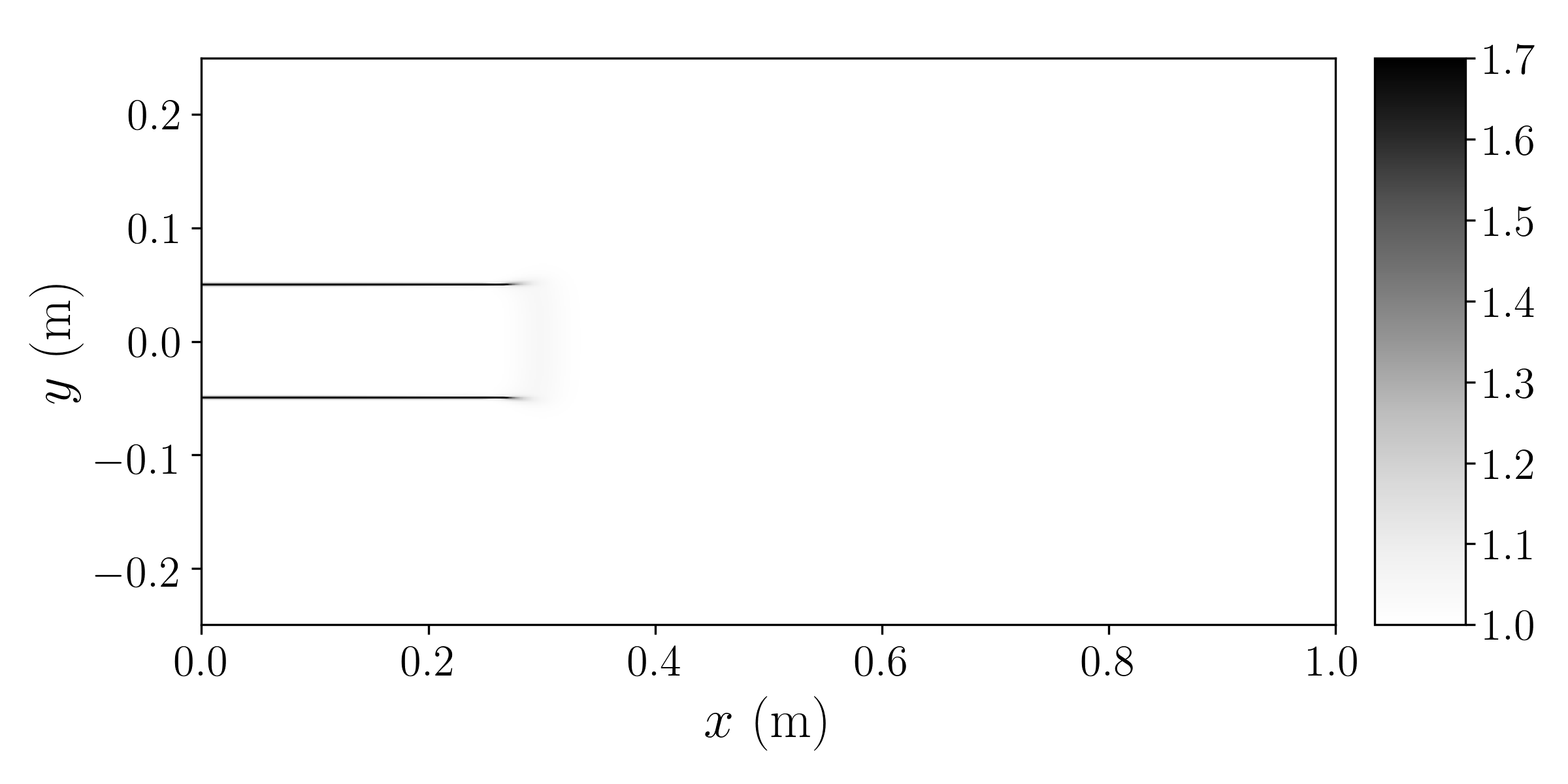}
\label{fig:compare_2D_Mach_100_water_jet_schl_t2_HLLC}}
\subfigure[$t = 2\ \mu\mathrm{s}$, PP-WCNS-IS]{%
\includegraphics[width=0.45\textwidth]{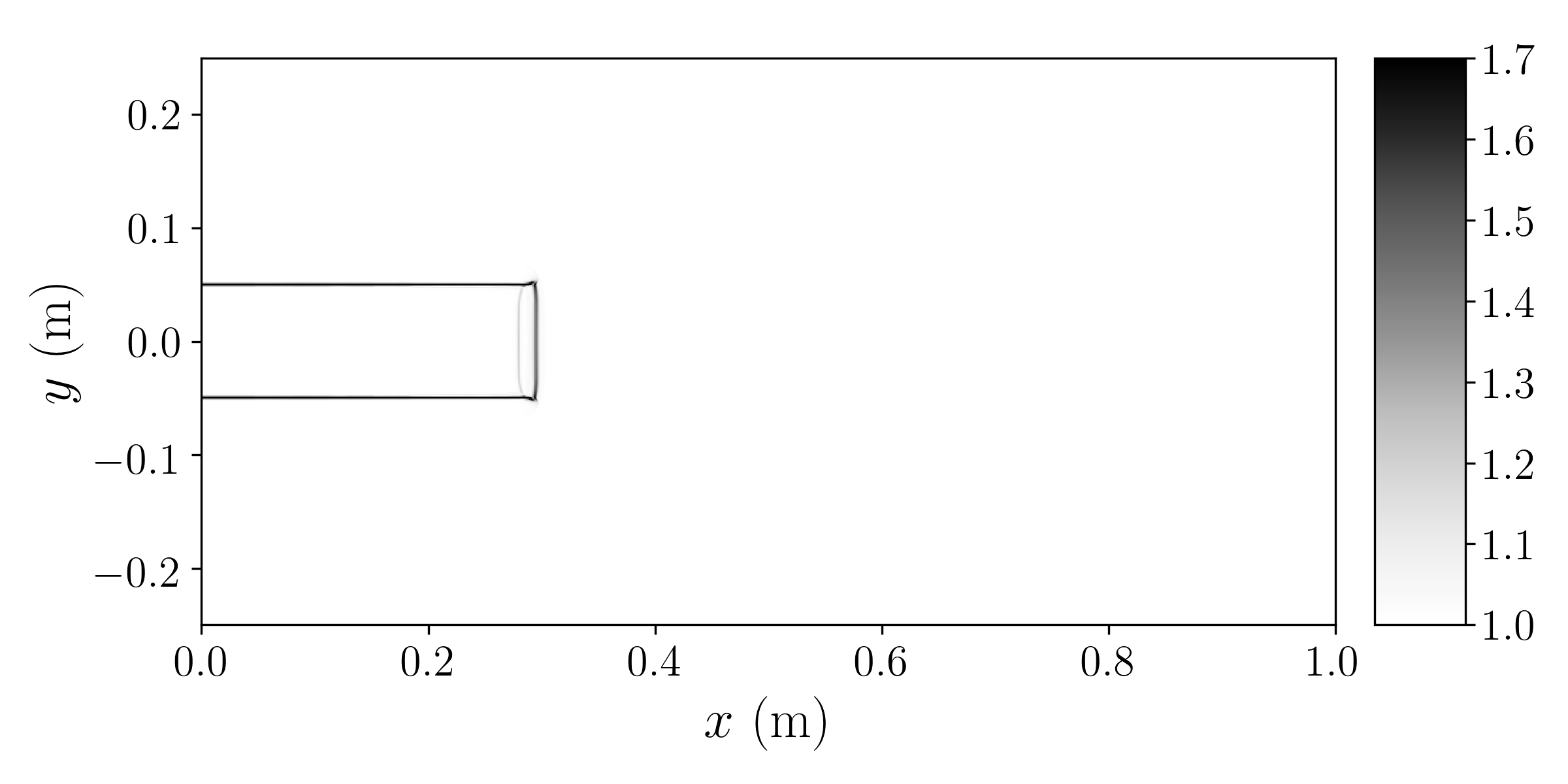}
\label{fig:compare_2D_Mach_100_water_jet_schl_t2_WCNS5_IS_PP}}
\subfigure[$t = 4\ \mu\mathrm{s}$, HLLC]{%
\includegraphics[width=0.45\textwidth]{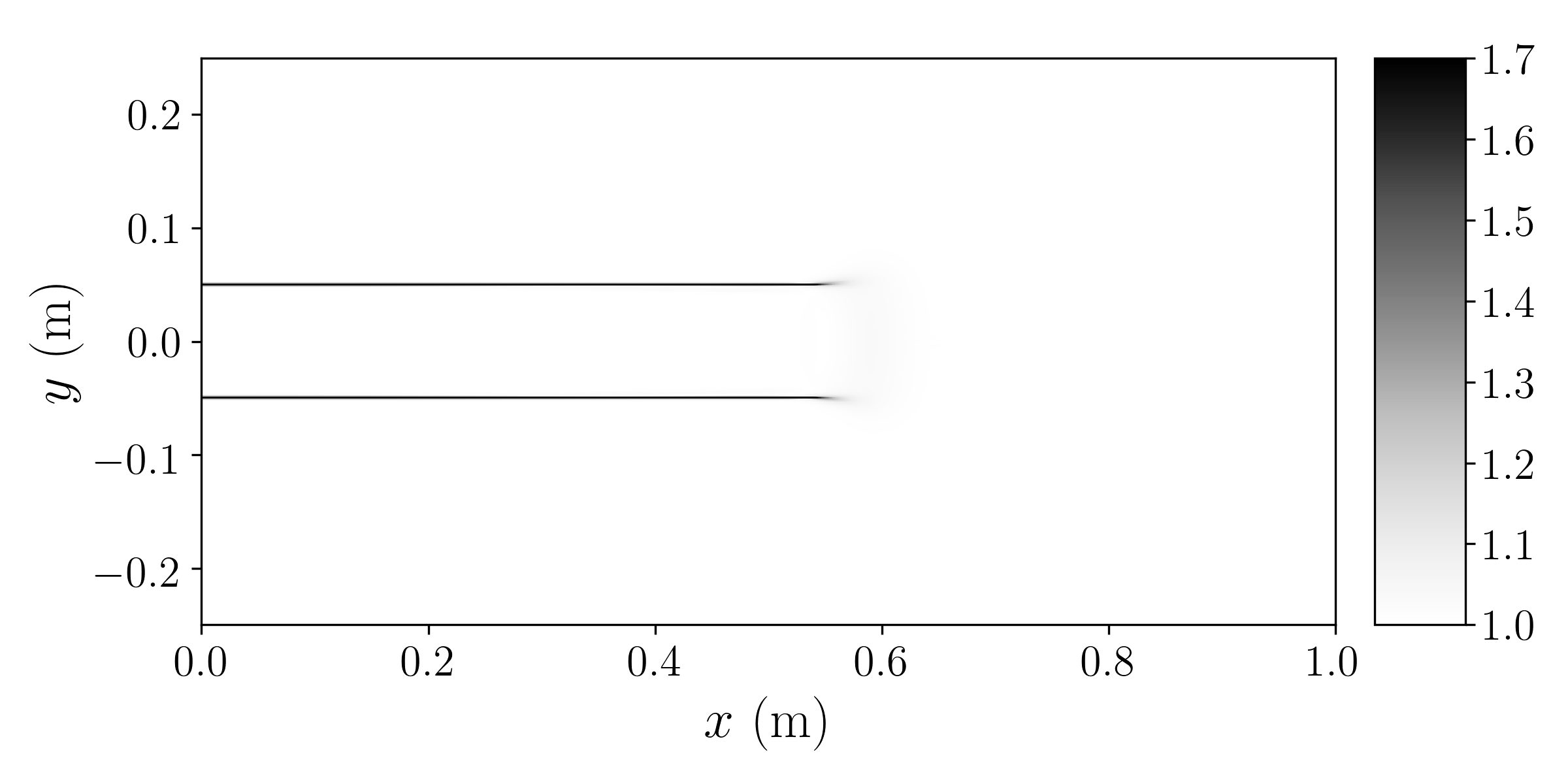}
\label{fig:compare_2D_Mach_100_water_jet_schl_t3_HLLC}}
\subfigure[$t = 4\ \mu\mathrm{s}$, PP-WCNS-IS]{%
\includegraphics[width=0.45\textwidth]{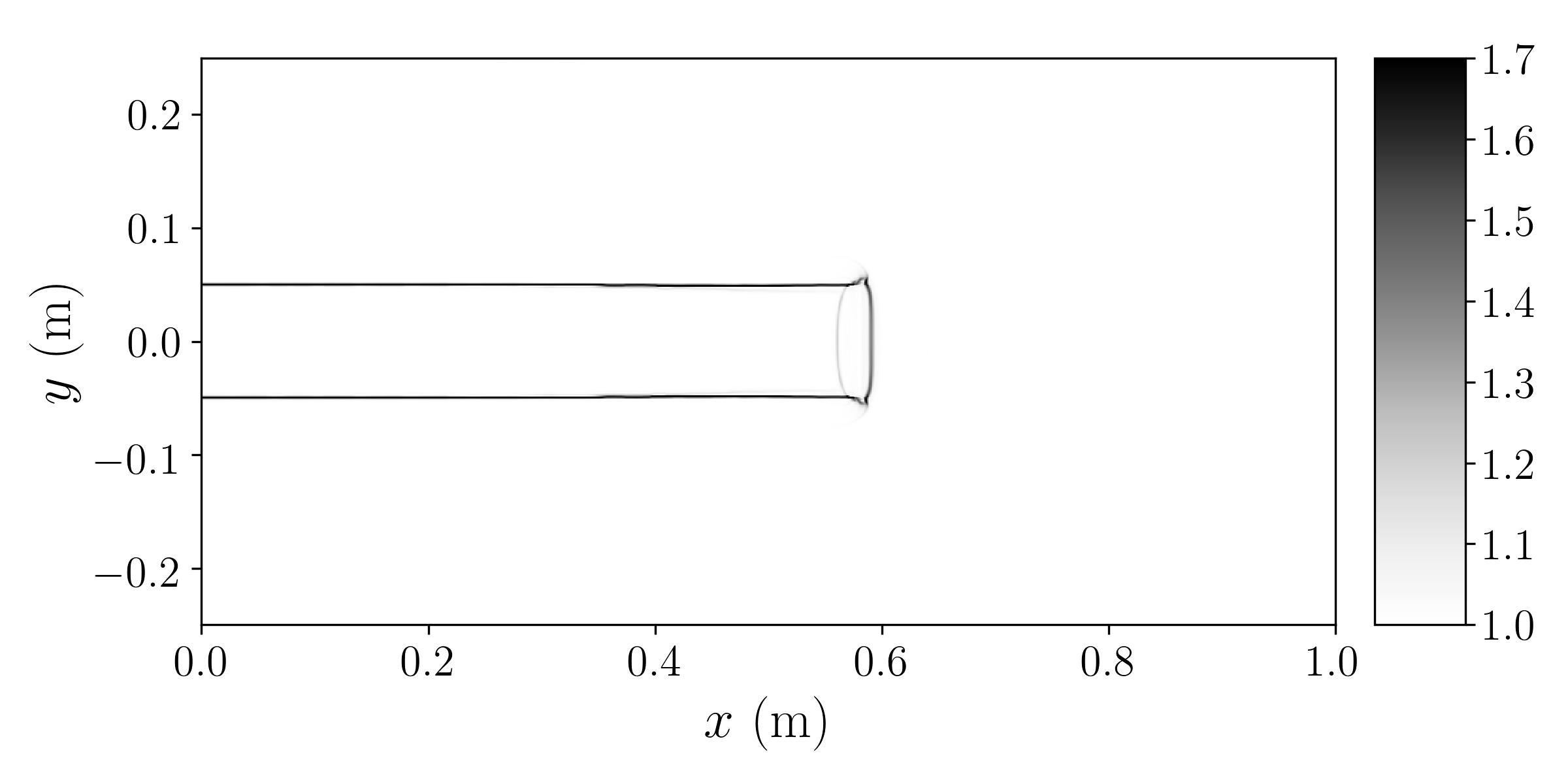}
\label{fig:compare_2D_Mach_100_water_jet_schl_t3_WCNS5_IS_PP}}
\subfigure[$t = 6\ \mu\mathrm{s}$, HLLC]{%
\includegraphics[width=0.45\textwidth]{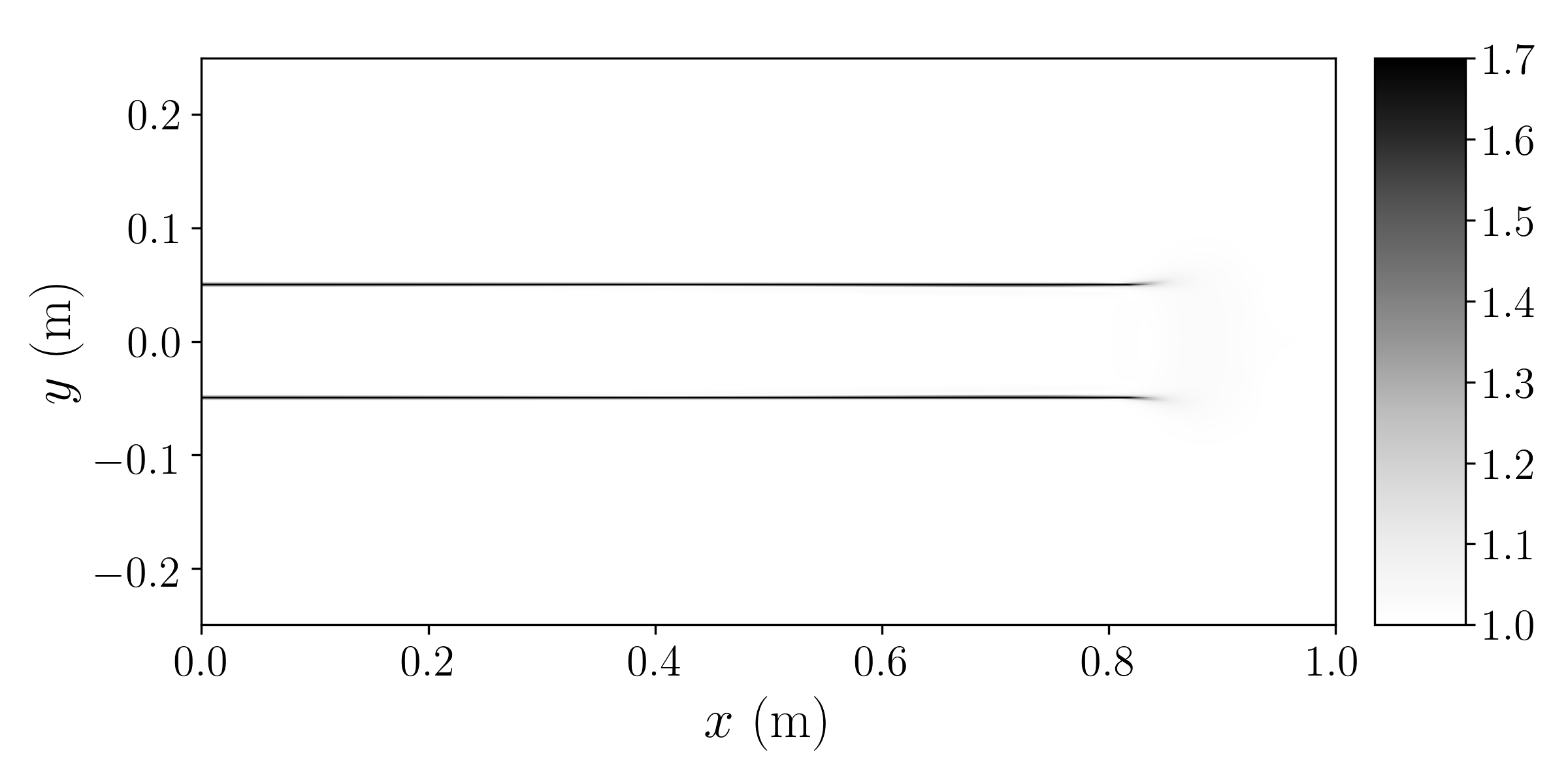}
\label{fig:compare_2D_Mach_100_water_jet_schl_t4}}
\subfigure[$t = 6\ \mu\mathrm{s}$, PP-WCNS-IS]{%
\includegraphics[width=0.45\textwidth]{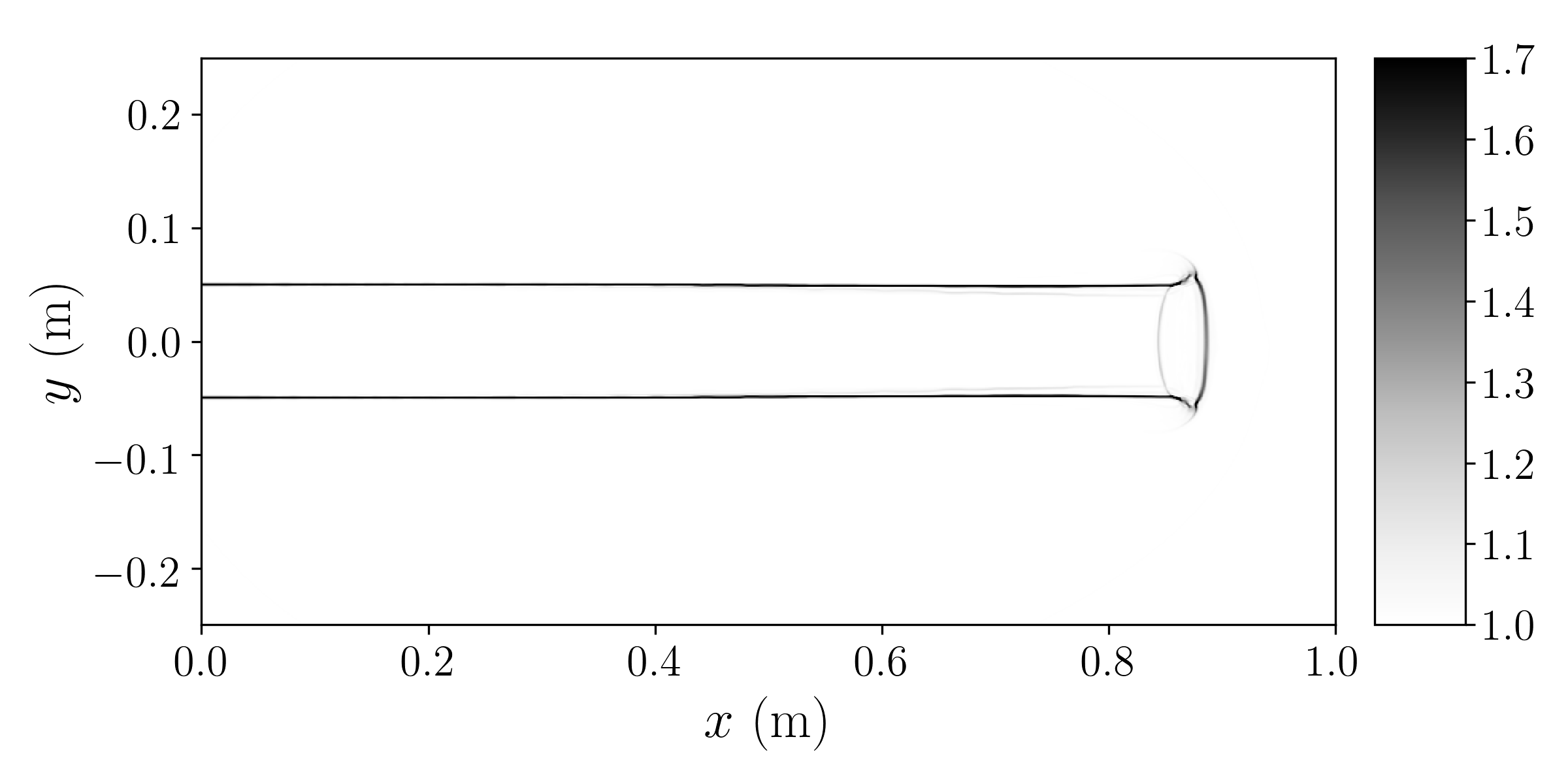}
\label{fig:compare_2D_Mach_100_water_jet_schl_t4_WCNS5_IS_PP}}
\caption{Numerical schlieren ($\exp{\left( \left| \nabla \rho \right| / \left| \nabla \rho \right|_{\mathrm{max}} \right)}$) of 2D Mach 100 water jet problem.}
\label{fig:compare_2D_Mach_100_water_jet_schl}
\end{figure}

\begin{figure}[!ht]
\centering
\subfigure[$t = 1\ \mu\mathrm{s}$, HLLC]{%
\includegraphics[width=0.45\textwidth]{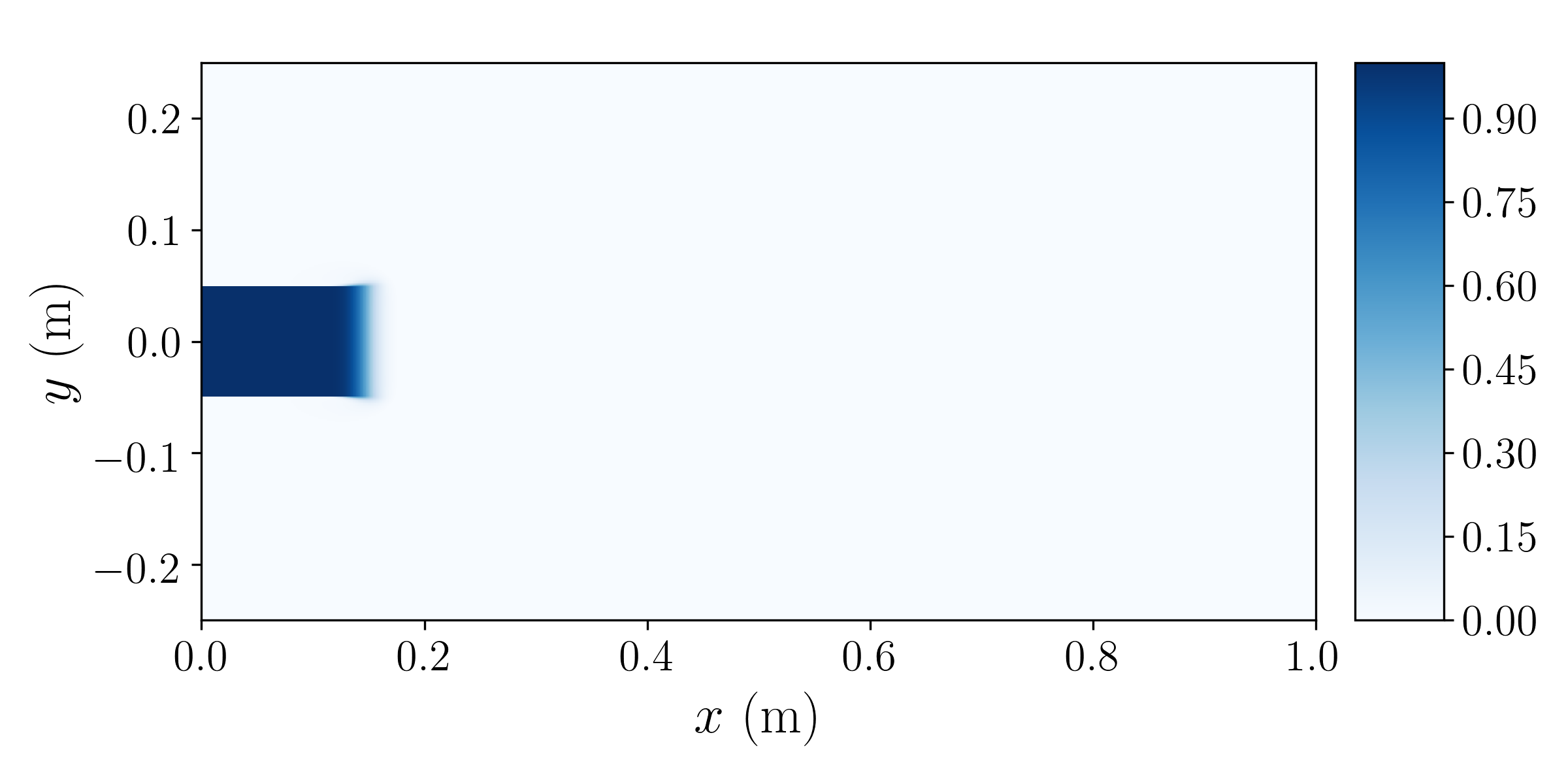}
\label{fig:compare_2D_Mach_100_water_jet_alpha0_t1_HLLC}}
\subfigure[$t = 1\ \mu\mathrm{s}$, PP-WCNS-IS]{%
\includegraphics[width=0.45\textwidth]{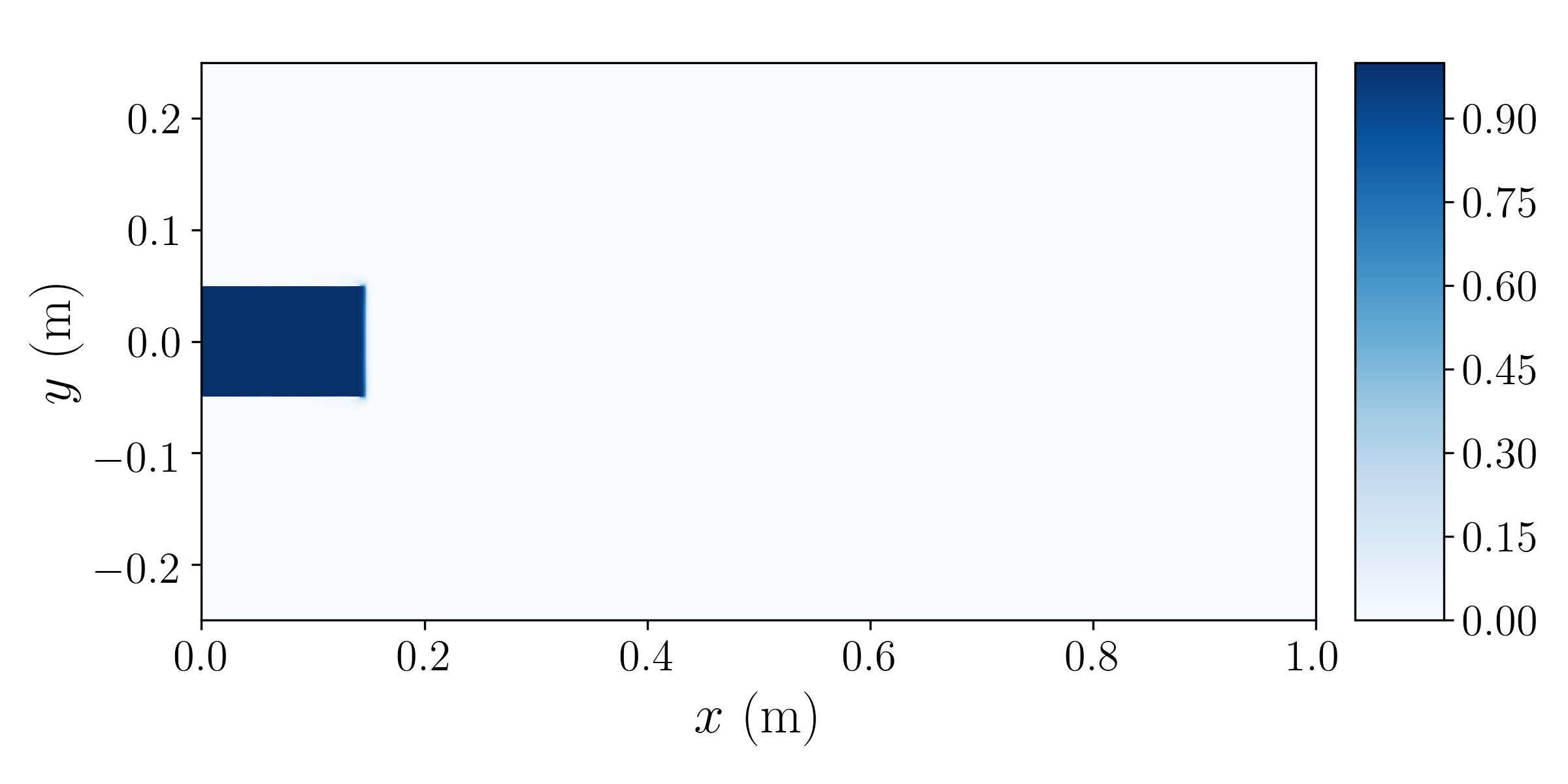}
\label{fig:compare_2D_Mach_100_water_jet_alpha0_t1_WCNS5_IS_PP}}
\subfigure[$t = 2\ \mu\mathrm{s}$, HLLC]{%
\includegraphics[width=0.45\textwidth]{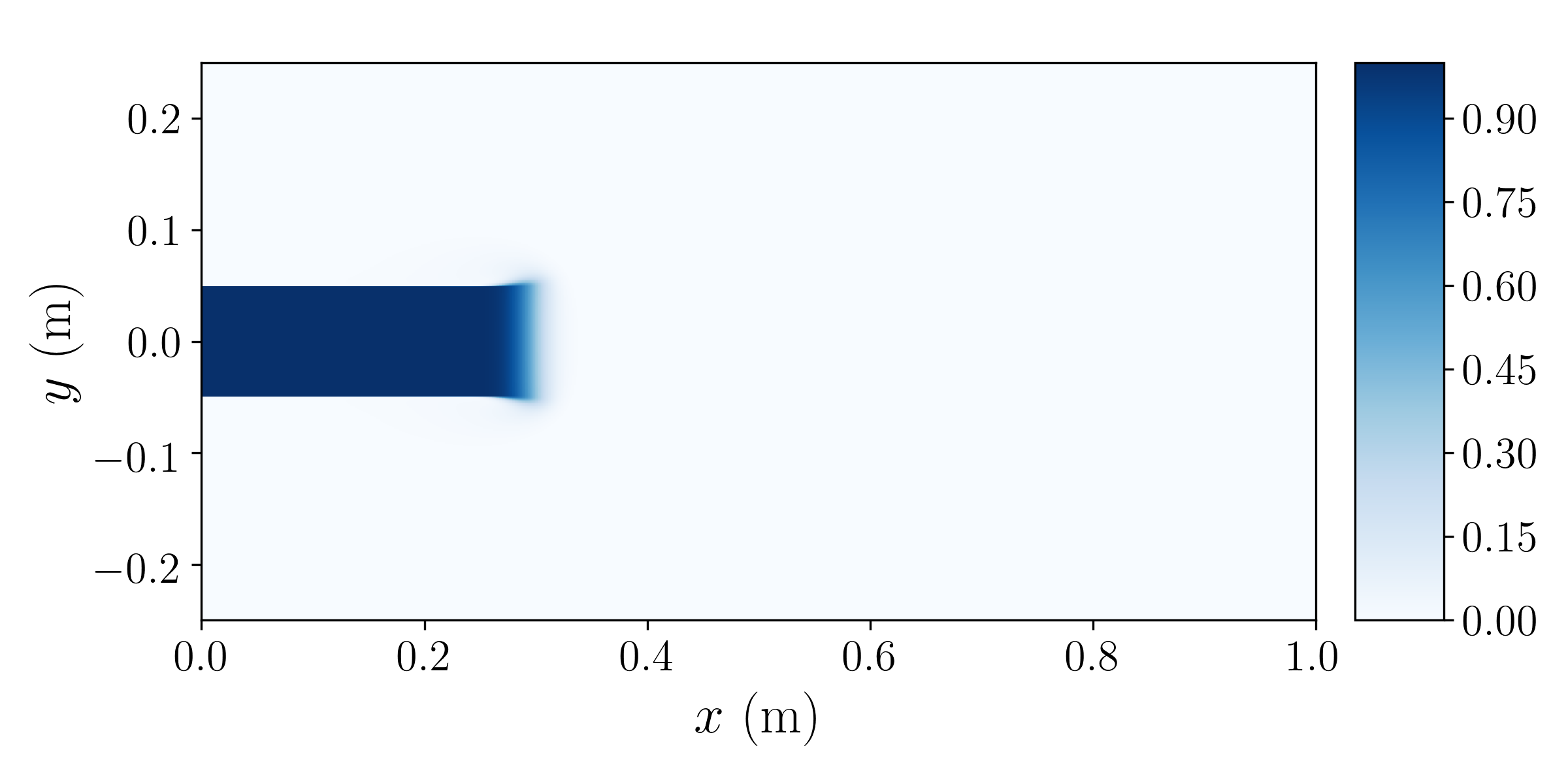}
\label{fig:compare_2D_Mach_100_water_jet_alpha0_t2_HLLC}}
\subfigure[$t = 2\ \mu\mathrm{s}$, PP-WCNS-IS]{%
\includegraphics[width=0.45\textwidth]{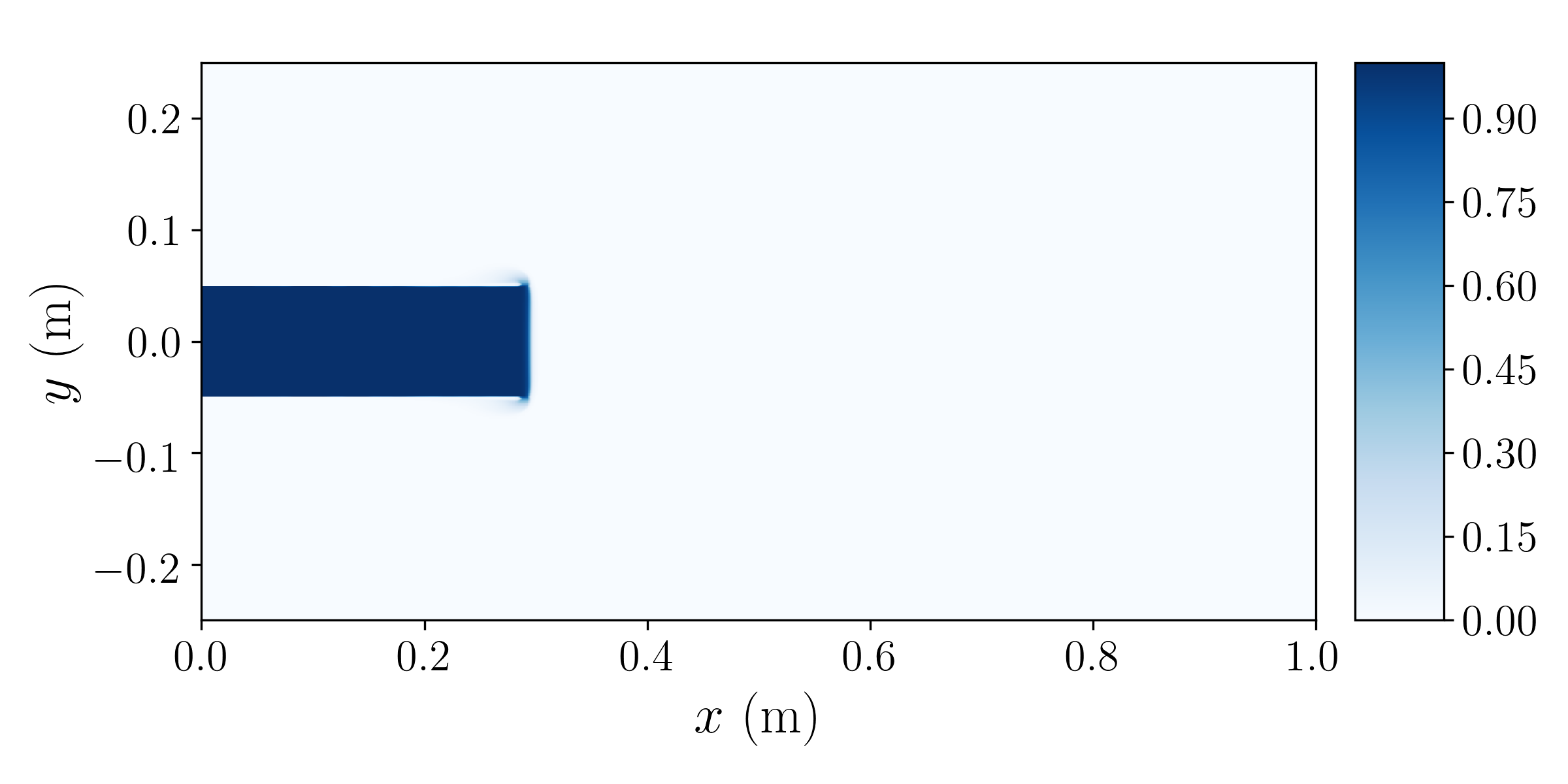}
\label{fig:compare_2D_Mach_100_water_jet_alpha0_t2_WCNS5_IS_PP}}
\subfigure[$t = 4\ \mu\mathrm{s}$, HLLC]{%
\includegraphics[width=0.45\textwidth]{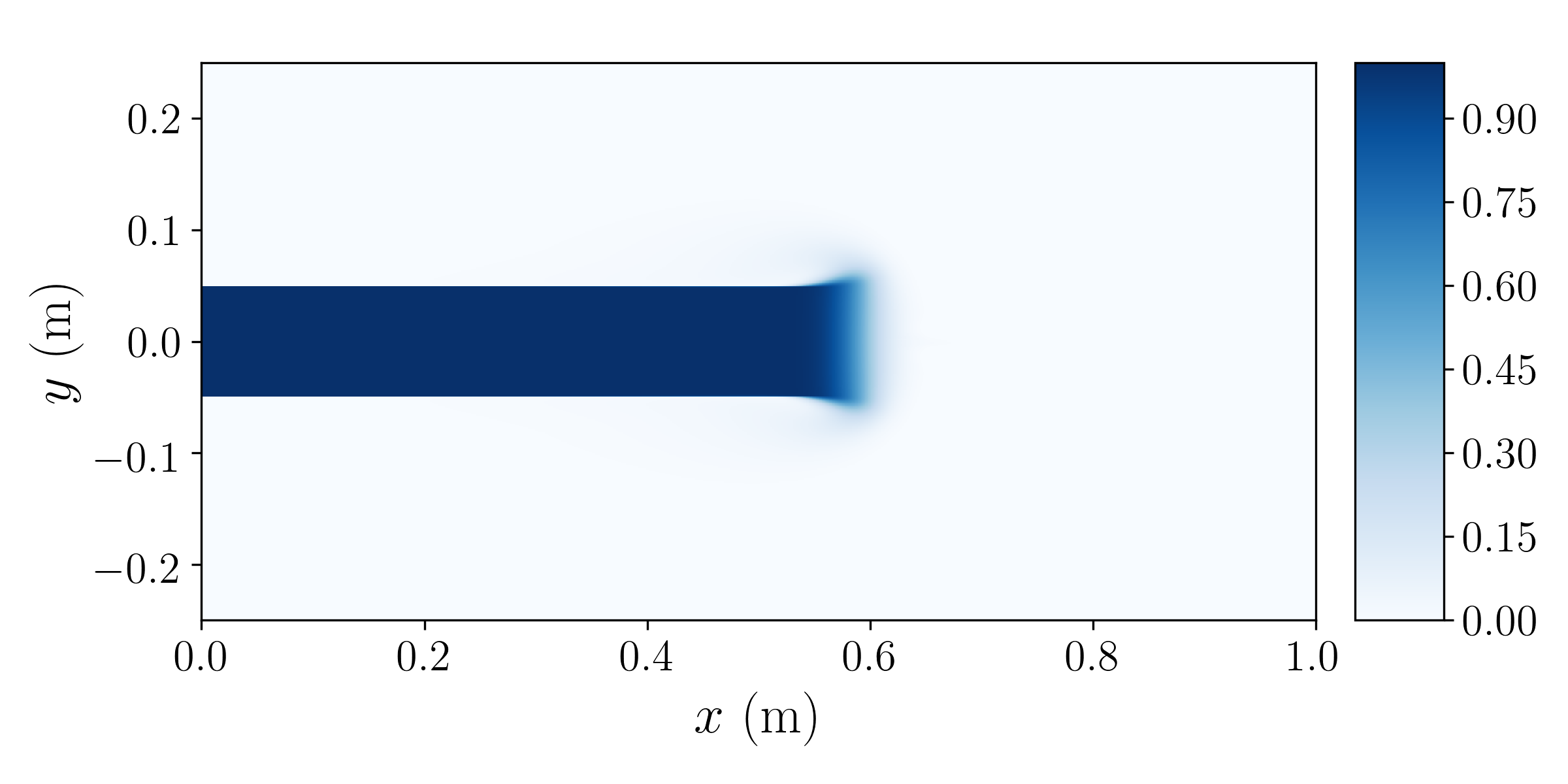}
\label{fig:compare_2D_Mach_100_water_jet_alpha0_t3_HLLC}}
\subfigure[$t = 4\ \mu\mathrm{s}$, PP-WCNS-IS]{%
\includegraphics[width=0.45\textwidth]{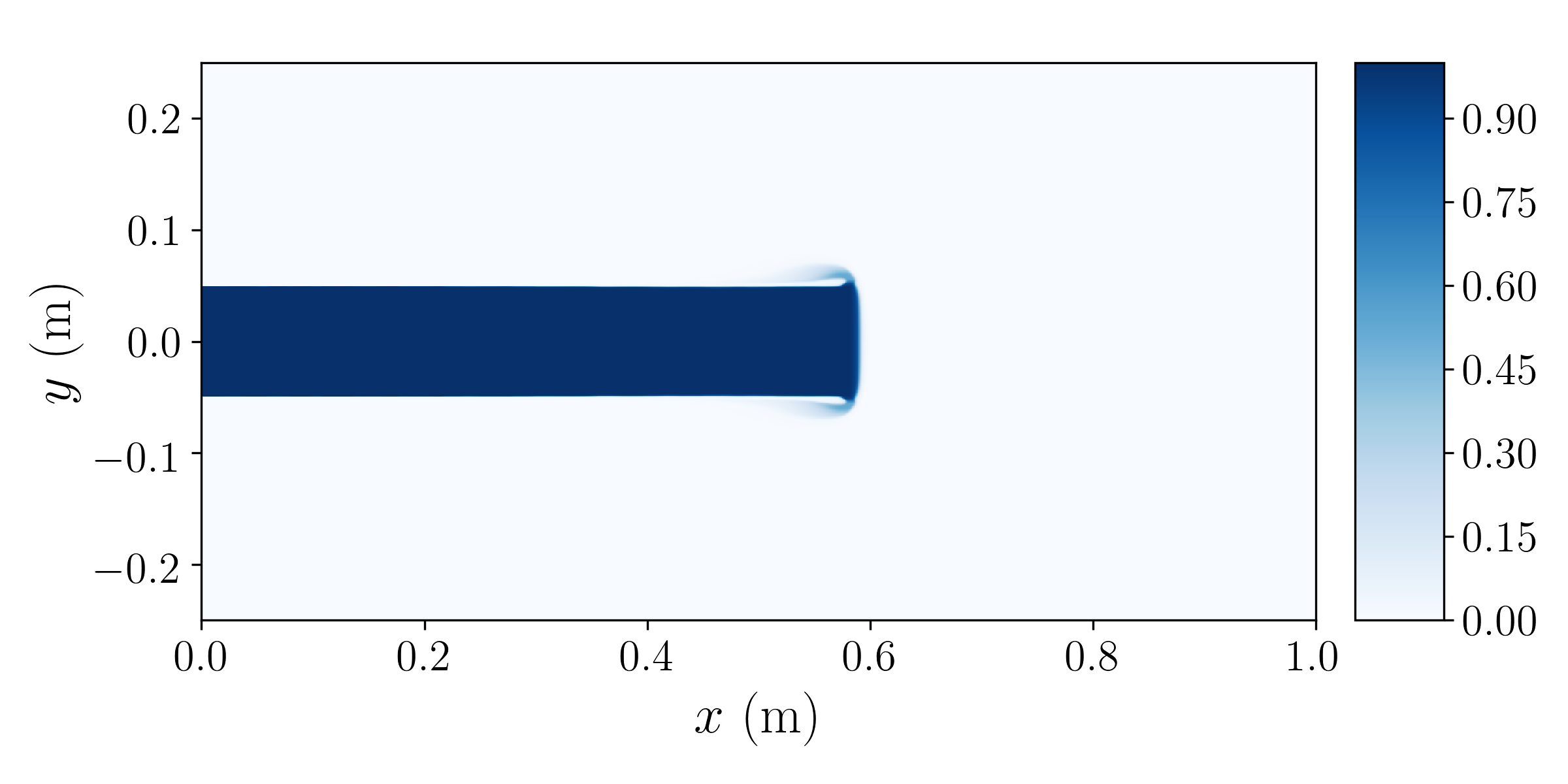}
\label{fig:compare_2D_Mach_100_water_jet_alpha0_t3_WCNS5_IS_PP}}
\subfigure[$t = 6\ \mu\mathrm{s}$, HLLC]{%
\includegraphics[width=0.45\textwidth]{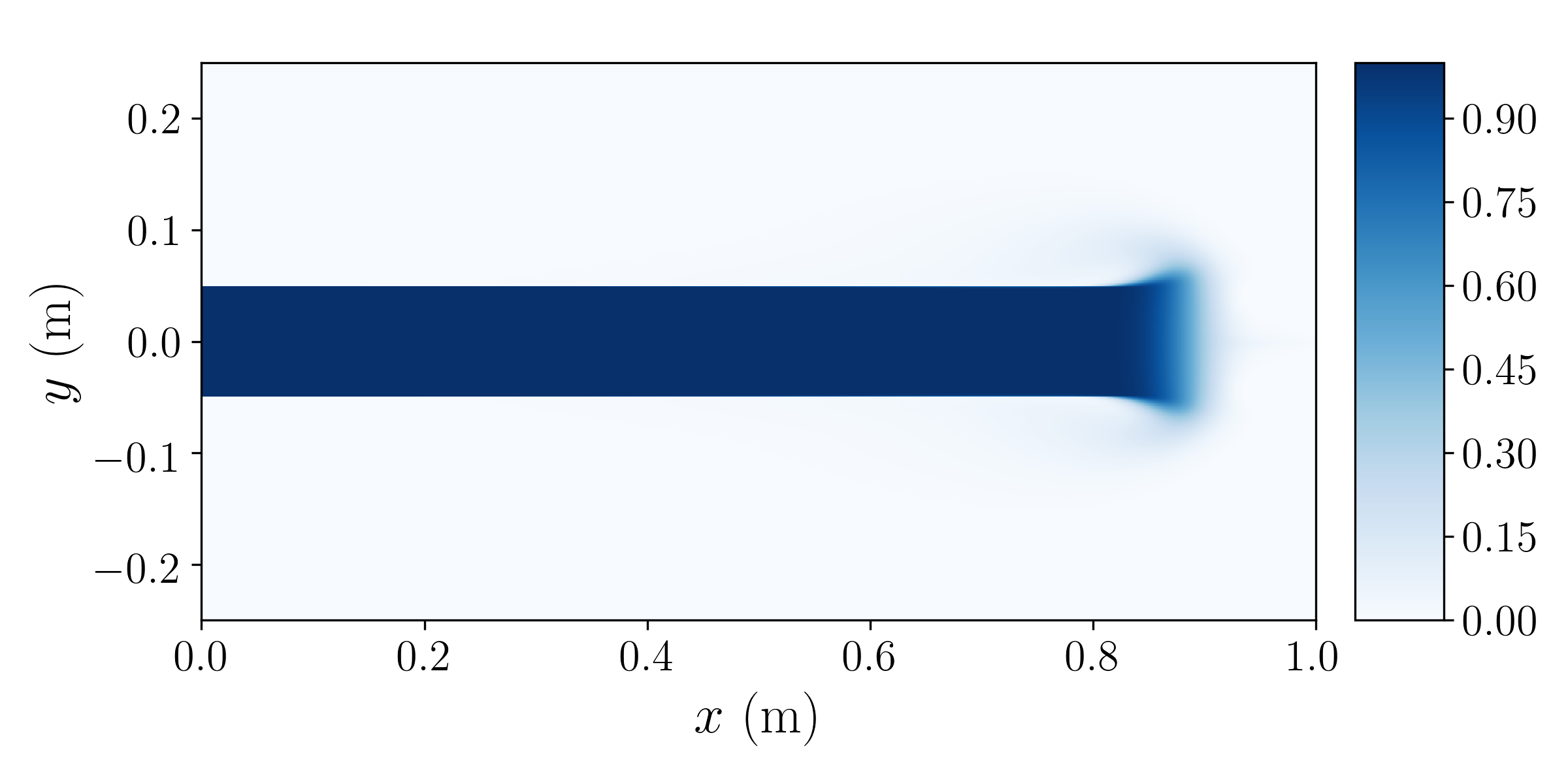}
\label{fig:compare_2D_Mach_100_water_jet_alpha0_t4_HLLC}}
\subfigure[$t = 6\ \mu\mathrm{s}$, PP-WCNS-IS]{%
\includegraphics[width=0.45\textwidth]{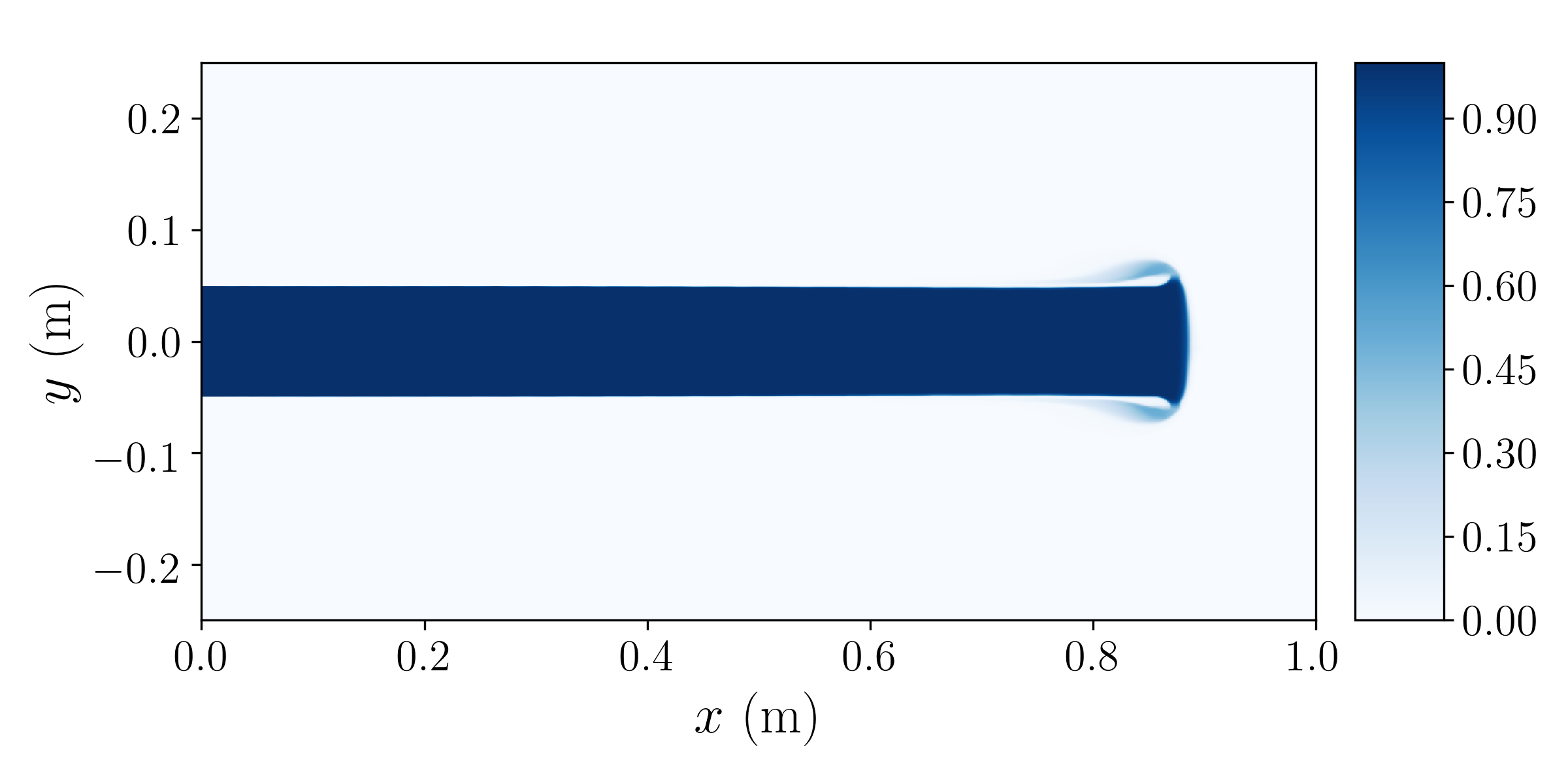}
\label{fig:compare_2D_Mach_100_water_jet_alpha0_t4_WCNS5_IS_PP}}
\caption{Volume fraction of water of 2D Mach 100 water jet problem.}
\label{fig:compare_2D_Mach_100_water_jet_alpha0}
\end{figure}

The comparison of speed of sound between the two schemes is shown in figure~\ref{fig:compare_2D_Mach_100_water_jet_sos}. Despite the large jump in sound speed across the bow shock ahead of the high speed water jet, none of the schemes fail due to the positivity-preserving properties of both schemes for sound speed. The main difference between the two schemes is at the water jet front where the interface at the water jet front produced by first order HLLC scheme is heavily smeared out while that of PP-WCNS-IS is reasonably captured with only a few grid points. There are also some small but obvious numerical artificts at the bow shock in the solutions computed with the first order scheme. In figure~\ref{fig:compare_2D_Mach_100_water_jet_schl}, the numerical schilren between the two schemes are compared. Since the interface water jet front is seriously diffused, the shape of the water-air interface at the water front cannot be visualized at all. Figure~\ref{fig:compare_2D_Mach_100_water_jet_alpha0} compares the volume fraction field of water at different times. The volume fraction field is verified to be bounded in PP-WCNS-IS. Similar to other fields, volume fraction interface at the water jet front is very diffused for the first order scheme compared to the high-order PP-WCNS-IS.

\section{Concluding remarks}

In this work, limiting procedures were proposed on a high-order finite difference scheme that can preserve the positivity of partial density of each phase and squared speed of sound and also the boundedness of the volume fractions in 1D and multi-dimensional gas-liquid two-phase problems under a mild assumption on the material properties of the gas and liquid. The procedures consist of two stages which limit the WENO interpolation and flux reconstruction respectively in the high-order WCNS-IS algorithm. Discrete conservation of solutions for the conservation equations in the five-equation model is still maintained even with the limiting. The overall positivity- and boundedness-preserving scheme, PP-WCNS-IS, was tested with different severe problems, under suitable CFL conditions. Comparison between the results of the first order HLLC scheme and PP-WCNS-IS showed the low dissipation and high resolution properties of the latter scheme while its robustness is also ensured. The positivity- and boundedness-preserving limiting procedures can also be potentially used with any conservative finite difference and finite volume schemes for gas-liquid two-phase flows. Future work includes generalization of the positivity- and boundedness-preserving limiters to more general equation of states and multi-phase flows with more than two species. There will also be future investigations on the use of the diffuse interface method for simulations of space vehicle launches with water-based sound suppression systems~\cite{vu2013multiphase,kiris2016computational}.

\section{Acknowledgments}

\noindent This work was partially supported by the NASA Exploration Ground Systems (EGS) program and the NASA Engineering and Safety Center (NESC). Computer time has been provided by the NASA Advanced Supercomputing (NAS) facility at NASA Ames Research Center. We also gratefully acknowledge Dr. Bruce T. Vu, Dr. Jeffrey A. Housman and Dr. Oliver M. Browne for valuable discussions.

\appendix

\section{Characteristic decomposition \label{appendix:char_decomp}}

The choice of variables for WENO reconstruction and interpolation is very critical to avoid spurious oscillations across discontinuities, especially across the material interfaces. It was shown in~\cite{johnsen2006implementation} that if conservative variables are chosen for WENO reconstruction, spurious oscillations will appear at material interfaces. Primitive variables were suggested~\cite{johnsen2006implementation,coralic2014finite,wong2017high} for reconstruction and interpolation in order to maintain pressure and velocity equilibria across interfaces. Furthermore, WENO reconstruction and interpolation of characteristic variables projected from primitive variables can avoid the interaction of discontinuities in different characteristic fields. To illustrate how the primitive variables are converted into characteristic variables, we follow previous works~\cite{coralic2014finite,wong2017high} by first rewriting the 2D governing equations in the quasi-linear primitive form:
\begin{equation} \label{eq:quasi-conservative_eqn}
    \frac{\partial{\Vv}}{\partial{t}} + \Av(\Vv)\frac{\partial{\Vv}}{\partial{x}} + \Bv(\Vv) \frac{\partial{\Vv}}{\partial{y}} = 0,
\end{equation}
where $\Vv$ is the vector of primitive variables. $\Vv$ and matrix $\Av$ are given by:
\begin{equation}
\begin{split}
    \Vv &= \begin{pmatrix} 
                    \alpha_1 \rho_1 \\
                    \alpha_2 \rho_2 \\
                    u \\
                    v \\
                    p \\
                    \alpha_1 \\
                    \end{pmatrix}, \quad
    \Av = \begin{pmatrix} 
                    u & 0 & \alpha_1 \rho_1 & 0 & 0              & 0  \\
                    0 & u & \alpha_2 \rho_2 & 0 & 0              & 0  \\
                    0 & 0 & u               & 0 & \frac{1}{\rho} & 0  \\
                    0 & 0 & 0               & u & 0              & 0  \\
                    0 & 0 & \rho c^2        & 0 & u              & 0  \\
                    0 & 0 & 0               & 0 & 0              & u  \\
                    \end{pmatrix}, \\
    \Bv &= \begin{pmatrix} 
                    v & 0 & 0 & \alpha_1 \rho_1 & 0              & 0  \\
                    0 & v & 0 & \alpha_2 \rho_2 & 0              & 0  \\
                    0 & 0 & v & 0               & 0              & 0  \\
                    0 & 0 & 0 & v               & \frac{1}{\rho} & 0  \\
                    0 & 0 & 0 & \rho c^2        & v              & 0  \\
                    0 & 0 & 0 & 0               & 0              & v  \\
                    \end{pmatrix} .
\end{split}
\end{equation}
The eigenvectors of the matrices $\Av$ and $\Bv$ have to be determined first in order to transform primitive variables to characteristic variables. The eigenvalue decompositions of the matrices are given by:
\begin{equation}
    \Av = \Rv_{\Av} \mathbf{\Lambda}_\Av \Rv_{\Av}^{-1}, \quad
    \Bv = \Rv_{\Bv} \mathbf{\Lambda}_\Bv \Rv_{\Bv}^{-1},
\end{equation}
where $\Rv_{\Av}$, $\Rv_{\Av}^{-1}$ and $\mathbf{\Lambda}_\Av$ are given by:
\begin{equation}
\begin{split}
    \Rv_{\Av} &= \begin{pmatrix}
        -\frac{\alpha_1 \rho_1}{2c} & 1 & 0 & 0 & 0 & \frac{\alpha_1 \rho_1}{2c}  \\
        -\frac{\alpha_2 \rho_2}{2c} & 0 & 1 & 0 & 0 & \frac{\alpha_2 \rho_2}{2c}  \\
        \frac{1}{2}                 & 0 & 0 & 0 & 0 & \frac{1}{2}  \\
        0                           & 0 & 0 & 1 & 0 & 0 \\
        -\frac{\rho c}{2}           & 0 & 0 & 0 & 0 & \frac{\rho c}{2}  \\
        0                           & 0 & 0 & 0 & 1 & 0  \\
      \end{pmatrix} , \quad
    \Rv_{\Av}^{-1} = \begin{pmatrix}
        0 & 0 & 1 & 0 & -\frac{1}{\rho c}                 & 0  \\
        1 & 0 & 0 & 0 & -\frac{\alpha_1 \rho_1}{\rho c^2} & 0  \\
        0 & 1 & 0 & 0 & -\frac{\alpha_2 \rho_2}{\rho c^2} & 0  \\
        0 & 0 & 0 & 1 & 0                                 & 0 \\
        0 & 0 & 0 & 0 & 0                                 & 1  \\
        0 & 0 & 1 & 0 & \frac{1}{\rho c}                  & 0  \\
      \end{pmatrix} , \\
    \mathbf{\Lambda}_{\Av} &= \begin{pmatrix}
        u - c & 0 & 0 & 0 & 0 & 0 \\
        0     & u & 0 & 0 & 0 & 0 \\
        0     & 0 & u & 0 & 0 & 0 \\
        0     & 0 & 0 & u & 0 & 0 \\
        0     & 0 & 0 & 0 & u & 0 \\
        0     & 0 & 0 & 0 & 0 & u + c \\
      \end{pmatrix} .
\end{split}
\end{equation}
$\Rv_{\Bv}$, $\Rv_{\Bv}^{-1}$ and $\mathbf{\Lambda}_\Bv$ have similar corresponding forms.

\section{Convergence analysis of the incremental-stencil WCNS\label{appendix:convergence}} 

The convergence analysis discussed in this section is the extension of the convergence analysis of~\citet{yan2016new} to HCS with WENO-IS interpolation. Assume that we have a 1D scalar hyperbolic conservation law of dependent variable $u$:
\begin{equation}
    \partial_t u + \partial_x F = 0 ,
\end{equation}
where $F = F(u)$ is the flux. Under the assumption that $\partial F(u) / \partial u > 0$ with perfect upwinding scheme, $\tilde{F}_{i+\half} = F\left( \tilde{u}_{i+\half,L} \right)$, where $\tilde{u}_{i+\half,L}$ is the left-biased WENO interpolated value of $u$. The subscript ``$L$" is dropped in this section for convenience.

After Taylor series expansion of the interpolation equations~\eqref{eq:interpolation_stencil_0}-\eqref{eq:interpolation_stencil_3} for the sub-stencils, we obtain:
\begin{equation}
  \tilde{u}^{k}_{i+\half} 
  = \left\{ \begin{array}{ll} 
      u_{i+\half} + A_i^{k} \dx^2 + \mathcal{O} \left( \dx^3 \right),
        &  \text{if }  k=0,1, \\ 
      u_{i+\half} + A_i^{k} \dx^3 + \mathcal{O} \left( \dx^4 \right),
        &  \text{if }  k=2,3,
    \end{array}\right. \label{eq:linear_interpolation_error_sub}
\end{equation}
where $A_i^{0}=u_i^{\prime\prime}/8$, $A_i^{1}=-3u_i^{\prime\prime}/8$, $A_i^{2}=-u_i^{\prime\prime\prime}/16$, and $A_i^{3}=-5u_i^{\prime\prime\prime}/16$. If we replace the nonlinear weights with the corresponding linear weights in equation~\eqref{eq:WENO_nonlinear_interpolation},
\begin{equation}
    \sum_{k=0}^{3} d_k \tilde{u}_{i+\half}^{k} = u_{i+\half} + B_i \dx^5 + \mathcal{O} \left( \dx^6 \right),  \label{eq:linear_interpolation_error_full}
\end{equation}
where $B_i=-3u_i^{(5)}/256$.

From equation~\eqref{eq:WENO_nonlinear_interpolation},
\begin{equation}
\begin{split}
    \tilde{u}_{i+\half}
        &= \sum_{k=0}^{3} \omega_k \tilde{u}_{i+\half}^{(k)} \\
        &= \sum_{k=0}^{3} d_k \tilde{u}_{i+\half}^{(k)} +
          \sum_{k=0}^{3} \left( \omega_k - d_k \right)
            \tilde{u}_{i+\half}^{(k)} .
\end{split}
\end{equation}
Using equations~\eqref{eq:linear_interpolation_error_sub} and \eqref{eq:linear_interpolation_error_full}, the equation above becomes:
\begin{equation}
\begin{split}
    \tilde{u}_{i+\half}
        &= u_{i+\half} + B_i \dx^5 + \mathcal{O} \left( \dx^6 \right) \\
        &\quad + \sum_{k=0}^{1} \left\{ \left( \omega_k - d_k \right)
          \left[ u_{i+\half} + A_i^{k} \dx^2 +
          \mathcal{O} \left( \dx^3 \right) \right] \right\} \\
        &\quad + \sum_{k=2}^{3} \left\{ \left( \omega_k - d_k \right)
          \left[ u_{i+\half} + A_i^{k} \dx^3 +
          \mathcal{O} \left( \dx^4 \right) \right] \right\} \\
        &= u_{i+\half} + B_i \dx^5 + \mathcal{O} \left( \dx^6 \right) +
          u_{i+\half} \sum_{k=0}^{3} \left( \omega_k - d_k \right) \\
        &\quad + \dx^2 \sum_{k=0}^{1} \left[ A_i^{k}
          \left( \omega_k - d_k \right) \right] +
          \dx^3 \sum_{k=2}^{3} \left[ A_i^{k}
          \left( \omega_k - d_k \right) \right] \\
        &\quad + \sum_{k=0}^{1} \left[ \left( \omega_k - d_k \right)
          \mathcal{O} \left( \dx^3 \right) \right] +
          \sum_{k=2}^{3} \left[ \left( \omega_k - d_k \right)
          \mathcal{O} \left( \dx^4 \right) \right] .
\end{split}
\end{equation}
The derivation is similar for $\tilde{u}_{i-\half}$. Therefore,
\begin{equation}
\begin{split}
    \tilde{u}_{i \pm \half} - u_{i \pm \half}
      &= B_i \dx^5 + u_{i \pm \half} \sum_{k=0}^{3} \left( \omega_k^{\pm} - d_k \right) +
      \dx^2 \sum_{k=0}^{1} \left[ A_i^{k} \left( \omega_k^{\pm} - d_k \right) \right] \\
      &\quad + \dx^3 \sum_{k=2}^{3}
      \left[ A_i^{k} \left( \omega_k^{\pm} - d_k \right) \right] +
      \sum_{k=0}^{1} \left[ \left( \omega_k^{\pm} - d_k \right)
      \mathcal{O} \left( \dx^3 \right) \right] \\
      &\quad + \sum_{k=2}^{3} \left[ \left( \omega_k^{\pm} - d_k \right)
      \mathcal{O} \left( \dx^4 \right) \right] +
      \mathcal{O} \left( \dx^6 \right) ,
\end{split}
\end{equation}
at midpoints $i+1/2$ and $i-1/2$. The superscript $\pm$ is added to $\omega_k$ to distinguish the values at the two different midpoints. Note that no $\pm$ is added to $d_k$, $A_i^k$ and $B_i$ since they have the same values at the two midpoints.
Using the fact that $\sum_{k=0}^{3} \omega_k^{\pm} = \sum_{k=0}^{3} d_k = 1$, we finally get:
\begin{equation}
\begin{split}
    \tilde{u}_{i \pm \half} - u_{i \pm \half}
    &= B_i \dx^5 + \dx^2 \sum_{k=0}^{1} \left[ A_i^{k}
    \left( \omega_k^{\pm} - d_k \right) \right] +
    \dx^3 \sum_{k=2}^{3} \left[ A_i^{k} \left( \omega_k^{\pm} - d_k \right) \right] \\
    &\quad + \sum_{k=0}^{1} \left[ \left( \omega_k^{\pm} - d_k \right)
    \mathcal{O} \left( \dx^3 \right) \right] +
    \sum_{k=2}^{3} \left[ \left( \omega_k^{\pm} - d_k \right)
    \mathcal{O} \left( \dx^4 \right) \right] \\
    &\quad + \mathcal{O} \left( \dx^6 \right) . \label{eq:u_tilde_expand}
\end{split}
\end{equation}

We will show that the WCNS-IS is fifth order accurate if $\omega_k^{\pm} - d_k = \mathcal{O} \left( \dx^4 \right)$. If we assume that $\omega_k^{\pm} - d_k = \mathcal{O} \left( \dx^4 \right)$, by Taylor series expansion,
\begin{equation}
\begin{split}
    \tilde{F}_{i \pm \half} &= F_{i \pm \half} + \frac{\partial F}{\partial u} \bigg|_{i \pm \half} \left( \tilde{u}_{i \pm \half} - u_{i \pm \half} \right) + \frac{1}{2} \frac{\partial^2 F}{\partial u^2} \bigg|_{i \pm \half} \left( \tilde{u}_{i \pm \half} - u_{i \pm \half} \right)^2 + \cdots \\
        &= F_{i \pm \half} + \frac{\partial F}{\partial u} \bigg|_{i \pm \half} \left( \tilde{u}_{i \pm \half} - u_{i \pm \half} \right) + \mathcal{O} \left( \dx^{10} \right) . \label{eq:F_tilde_expand}
\end{split}
\end{equation}
Substituting equations~\eqref{eq:u_tilde_expand} and \eqref{eq:F_tilde_expand} into the 1D version of equation~\eqref{eq:HCS6} (noticing $F$ is the scalar version of $\Gxv$ here) and using Taylor-sereis expanded equation~\eqref{eq:HCS6_Taylor_expand} (assuming $\psi=256/175$),
\begin{equation}
\begin{split}
    \widehat{ \frac{\partial F}{\partial x} } \bigg|_i
      &=\frac{\partial F}{\partial x} \bigg|_i + \mathcal{O} \left( \dx^8 \right) \\
      &\quad + \psi \left\{ \frac{\partial F}{\partial u} \bigg|_{i + \half} 
      \sum_{k=0}^{1} \left[ \left( \omega_k^{+} - d_k \right)
      \mathcal{O} \left( \dx^2 \right) \right] \right. \\
      &\quad \quad \quad \left. -
      \frac{\partial F}{\partial u} \bigg|_{i - \half} \sum_{k=0}^{1}
      \left[ \left( \omega_k^{-} - d_k \right) \mathcal{O}
      \left( \dx^2 \right) \right] \right\} \\
      &\quad + \psi \left\{ \frac{\partial F}{\partial u} \bigg|_{i + \half}
      \sum_{k=2}^{3} \left[ \left( \omega_k^{+} - d_k \right)
      \mathcal{O} \left( \dx^3 \right) \right] \right. \\
      &\quad \quad \quad \left. -
      \frac{\partial F}{\partial u} \bigg|_{i - \half} \sum_{k=2}^{3}
      \left[ \left( \omega_k^{-} - d_k \right)
      \mathcal{O} \left( \dx^3 \right) \right] \right\} \\
      &\quad + \psi \dx^4 B_i \left( \frac{\partial F}{\partial u} \bigg|_{i + \half} -
      \frac{\partial F}{\partial u} \bigg|_{i - \half}\right) \\
      &\quad + \psi \dx \left\{ \frac{\partial F}{\partial u} \bigg|_{i + \half}
      \sum_{k=0}^{1} \left[ A_i^{k} \left( \omega_k^{+} - d_k \right) \right] -
      \frac{\partial F}{\partial u} \bigg|_{i - \half}
      \sum_{k=0}^{1} \left[ A_i^{k} \left( \omega_k^{-} - d_k \right) \right] \right\} \\
      &\quad + \psi \dx^2 \left\{ \frac{\partial F}{\partial u} \bigg|_{i + \half}
      \sum_{k=2}^{3} \left[ A_i^{k} \left( \omega_k^{+} - d_k \right) \right] -
      \frac{\partial F}{\partial u} \bigg|_{i - \half}
      \sum_{k=2}^{3} \left[ A_i^{k} \left( \omega_k^{-} - d_k \right) \right] \right\} \\
      &\quad + \mathcal{O} \left( \dx^5 \right) . \label{eq:HCS6_Taylor_expand_w_WENO}
\end{split}
\end{equation}
Also, by Taylor series expansion,
\begin{align}
    \frac{\partial F}{\partial u} \bigg|_{i + \half} &=
    \frac{\partial F}{\partial u} \bigg|_{i} +
    \frac{\partial^2 F}{\partial u^2} \bigg|_{i} \left( u_{i + \half} - u_i \right) +
    \frac{1}{2} \frac{\partial^3 F}{\partial u^3} \bigg|_{i}
    \left( u_{i + \half} - u_i \right)^2 + \cdots , \\
    u_{i + \half} &= u_i + \frac{\dx}{2} \frac{\partial u}{\partial x} \bigg|_{i} +
    \frac{\dx^2}{8} \frac{\partial^2 u}{\partial x^2} \bigg|_{i} +
    \mathcal{O} \left( \dx^3 \right) .
\end{align}
Therefore,
\begin{equation}
\begin{split}
    \frac{\partial F}{\partial u} \bigg|_{i + \half} &=
    \frac{\partial F}{\partial u} \bigg|_{i} +
    \frac{\partial^2 F}{\partial u^2} \bigg|_{i}
    \left[ \frac{\dx}{2} \frac{\partial u}{\partial x} \bigg|_{i} +
    \frac{\dx^2}{8} \frac{\partial^2 u}{\partial x^2} \bigg|_{i} +
    \mathcal{O} \left( \dx^3 \right) \right] \\
    &\quad + \frac{1}{2} \frac{\partial^3 F}{\partial u^3} \bigg|_{i}
    \left[ \frac{\dx}{2} \frac{\partial u}{\partial x} \bigg|_i +
    \mathcal{O} \left( \dx^2 \right) \right]^2 + \mathcal{O} \left( \dx^3 \right) \\
    &= \frac{\partial F}{\partial u} \bigg|_{i} +
    \frac{\dx}{2} \frac{\partial^2 F}{\partial u^2} \bigg|_{i}
    \frac{\partial u}{\partial x} \bigg|_{i} +
    \frac{\dx^2}{8} \frac{\partial^2 F}{\partial u^2} \bigg|_{i}
    \frac{\partial^2 u}{\partial x^2} \bigg|_{i} + \frac{\dx^2}{8}
    \frac{\partial^3 F}{\partial u^3} \bigg|_{i}
    \left( \frac{\partial u}{\partial x} \bigg|_{i} \right)^2 \\
    &\quad + \mathcal{O} \left( \dx^3 \right) .
\end{split}
\end{equation}
Similarly, by Taylor series expansion,
\begin{equation}
\begin{split}
    \frac{\partial F}{\partial u} \bigg|_{i - \half} &=
    \frac{\partial F}{\partial u} \bigg|_{i} +
    \frac{\partial^2 F}{\partial u^2} \bigg|_{i}
    \left[ -\frac{\dx}{2} \frac{\partial u}{\partial x} \bigg|_{i} +
    \frac{\dx^2}{8} \frac{\partial^2 u}{\partial x^2} \bigg|_{i} +
    \mathcal{O} \left( \dx^3 \right) \right] \\
    &\quad + \frac{1}{2} \frac{\partial^3 F}{\partial u^3} \bigg|_{i}
    \left[ -\frac{\dx}{2} \frac{\partial u}{\partial x} \bigg|_i +
    \mathcal{O} \left( \dx^2 \right) \right]^2 +
    \mathcal{O} \left( \dx^3 \right) \\
    &= \frac{\partial F}{\partial u} \bigg|_{i} -
    \frac{\dx}{2} \frac{\partial^2 F}{\partial u^2} \bigg|_{i}
    \frac{\partial u}{\partial x} \bigg|_{i} +
    \frac{\dx^2}{8} \frac{\partial^2 F}{\partial u^2} \bigg|_{i}
    \frac{\partial^2 u}{\partial x^2} \bigg|_{i} +
    \frac{\dx^2}{8} \frac{\partial^3 F}{\partial u^3} \bigg|_{i}
    \left( \frac{\partial u}{\partial x} \bigg|_{i} \right)^2 \\
    &\quad + \mathcal{O} \left( \dx^3 \right) .
\end{split}
\end{equation}
As a result, equation~\eqref{eq:HCS6_Taylor_expand_w_WENO} is simplified to:
\begin{equation}
\begin{split}
    \widehat{ \frac{\partial F}{\partial x} } \bigg|_i &=
      \frac{\partial F}{\partial x} \bigg|_i \\
      &\quad + \psi \left\{ \frac{\partial F}{\partial u} \bigg|_{i + \half} 
      \sum_{k=0}^{1} \left[ \left( \omega_k^{+} - d_k \right)
      \mathcal{O} \left( \dx^2 \right) \right] \right. \\
      &\quad \quad \quad \left. - \frac{\partial F}{\partial u} \bigg|_{i - \half}
      \sum_{k=0}^{1} \left[ \left( \omega_k^{-} - d_k \right)
      \mathcal{O} \left( \dx^2 \right) \right] \right\} \\
      &\quad + \psi \left\{ \frac{\partial F}{\partial u} \bigg|_{i + \half}
      \sum_{k=2}^{3} \left[ \left( \omega_k^{+} - d_k \right)
      \mathcal{O} \left( \dx^3 \right) \right] \right. \\
      &\quad \quad \quad \left. - \frac{\partial F}{\partial u} \bigg|_{i - \half}
      \sum_{k=2}^{3} \left[ \left( \omega_k^{-} - d_k \right)
      \mathcal{O} \left( \dx^3 \right) \right] \right\} \\
      &\quad + \psi \dx^5 B_i \frac{\partial^2 F}{\partial u^2} \bigg|_{i}
      \frac{\partial u}{\partial x} \bigg|_{i} +
      \left[ \psi \dx \frac{\partial F}{\partial u} \bigg|_{i} +
      \mathcal{O} \left( \dx^3 \right) \right]
      \sum_{k=0}^{1} \left[ A_i^{k} \left( \omega_k^{+} - \omega_k^{-} \right) \right] \\
      &\quad + \left[ \frac{\psi \dx^2}{2} \frac{\partial^2 F}{\partial u^2} \bigg|_{i}
      \frac{\partial u}{\partial x} \bigg|_{i}
      + \mathcal{O} \left( \dx^4 \right) \right] \\
      &\quad \quad \left\{ \sum_{k=0}^{1}
      \left[ A_i^{k} \left( \omega_k^{+} - d_k \right) \right] +
      \sum_{k=0}^{1} \left[ A_i^{k} \left( \omega_k^{-} - d_k \right) \right] \right\} \\
      &\quad + \left[ \psi \dx^2 \frac{\partial F}{\partial u} \bigg|_{i} +
      \mathcal{O} \left( \dx^4 \right) \right]
      \sum_{k=2}^{3} \left[ A_i^{k} \left( \omega_k^{+} - \omega_k^{-} \right) \right] \\
      &\quad + \left[ \frac{\psi \dx^3}{2} \frac{\partial^2 F}{\partial u^2} \bigg|_{i} 
      \frac{\partial u}{\partial x} \bigg|_{i} +
      \mathcal{O} \left( \dx^5 \right) \right] \\
      &\quad \quad \left\{ \sum_{k=2}^{3}
      \left[ A_i^{k} \left( \omega_k^{+} - d_k \right) \right] +
      \sum_{k=2}^{3} \left[ A_i^{k} \left( \omega_k^{-} - d_k \right) \right] \right\} \\
      &\quad + \mathcal{O} \left( \dx^5 \right) .
\end{split} \label{eq:HCS6_Taylor_expand_w_WENO_simplified}
\end{equation}
It can be seen from equation~\eqref{eq:HCS6_Taylor_expand_w_WENO_simplified} that a sufficient condition for fifth order convergence is $\omega_k^{\pm} - d_k = \mathcal{O} \left( \dx^4 \right)$, which is already assumed earlier.

To prove that $\omega_k^{\pm} - d_k = \mathcal{O} \left( \dx^4 \right)$ is true, we can perform Taylor series expansion on the smoothness indicators given by equations~\eqref{eq:beta_0_IS}--\eqref{eq:beta_3_IS},
\begin{align}
\begin{split}
\beta_0 &= \left({u_{i}^{\prime}}\right)^{2} \dx^{2} +
  {u_{i}^{\prime}} {u_{i}^{\prime\prime}} \dx^{3} +
  \left(\frac{{u_{i}^{\prime}} {u_{i}^{\prime\prime\prime}}}{3} +
  \frac{\left({u_{i}^{\prime\prime}}\right)^{2}}{4}\right) \dx^{4}
  + \left(\frac{{u_{i}^{\prime}} {u_{i}^{(4)}}}{12} +
  \frac{{u_{i}^{\prime\prime}} {u_{i}^{\prime\prime\prime}}}{6}\right) \dx^{5} \\
  &\quad + \mathcal{O} \left( \dx^{6} \right) ,
\end{split}
\\
\begin{split}
\beta_1 &= \left({u_{i}^{\prime}}\right)^{2} \dx^{2} -
  {u_{i}^{\prime}} {u_{i}^{\prime\prime}} \dx^{3} +
  \left(\frac{{u_{i}^{\prime}} {u_{i}^{\prime\prime\prime}}}{3} +
  \frac{\left({u_{i}^{\prime\prime}}\right)^{2}}{4}\right) \dx^{4} \\
  &\quad + \left(- \frac{{u_{i}^{\prime}} {u_{i}^{(4)}}}{12} -
  \frac{{u_{i}^{\prime\prime}} {u_{i}^{\prime\prime\prime}}}{6}\right) \dx^{5}
  + \mathcal{O} \left( \dx^{6} \right) ,
\end{split}
\\
\beta_{01} &= \left({u_{i}^{\prime}}\right)^{2} \dx^{2} +
  \left(\frac{{u_{i}^{\prime}} {u_{i}^{\prime\prime\prime}}}{3} +
  \frac{13 \left({u_{i}^{\prime\prime}}\right)^{2}}{12}\right) \dx^{4} +
  \mathcal{O} \left( \dx^{6} \right) , \\
\begin{split}
\beta_2 &= \left({u_{i}^{\prime}}\right)^{2} \dx^{2} +
  \left(- \frac{2 {u_{i}^{\prime}} {u_{i}^{\prime\prime\prime}}}{3} +
  \frac{13 \left({u_{i}^{\prime\prime}}\right)^{2}}{12}\right) \dx^{4}
  + \left(- \frac{{u_{i}^{\prime}} {u_{i}^{(4)}}}{2} +
  \frac{13 {u_{i}^{\prime\prime}} {u_{i}^{\prime\prime\prime}}}{6}\right) \dx^{5} \\
  &\quad + \mathcal{O} \left( \dx^{6} \right) ,
\end{split}
\\
\begin{split}
\beta_3 &= \left({u_{i}^{\prime}}\right)^{2} \dx^{2} +
  \left(- \frac{2 {u_{i}^{\prime}} {u_{i}^{\prime\prime\prime}}}{3} +
  \frac{13 \left({u_{i}^{\prime\prime}}\right)^{2}}{12}\right) \dx^{4}
  + \left(\frac{{u_{i}^{\prime}} {u_{i}^{(4)}}}{2} -
  \frac{13 {u_{i}^{\prime\prime}} {u_{i}^{\prime\prime\prime}}}{6}\right) \dx^{5} \\
  &\quad + \mathcal{O} \left( \dx^{6} \right) .
\end{split}
\end{align}
The Taylor series expansion of reference smoothness indicator (equation~\eqref{eq:tau_5_IS}) gives:
\begin{equation}
    \tau_5 = \left({u_{i}^{\prime\prime\prime}}\right)^{2} \dx^{6} + \mathcal{O} \left( \dx^{8} \right) .
\end{equation}
Therefore,
\begin{equation}
    \frac{\tau_5}{\beta_k + \epsilon} = \mathcal{O} \left( \dx^{4} \right),
\end{equation}
provided not at critical points. As explained in \citet{borges2008improved} and \citet{wang2018incremental}, this is sufficient to have $\omega_k^{\pm} - d_k = \mathcal{O} \left( \dx^4 \right)$.


\bibliographystyle{abbrvnat}
\bibliography{bibtex_database.bib}

\end{document}